\documentclass[trackchanges]{aastex701}

\usepackage{bm}
\usepackage[page, titletoc]{appendix}
\usepackage{titlesec}
\usepackage{tabularx}
\usepackage{graphicx}
\usepackage{siunitx}
\usepackage{hyperref}
\usepackage{longtable}

\newcounter{finding}
\newcommand{\finding}[1]{%
  \refstepcounter{finding}%
  \vspace{10pt}\noindent\textbf{F\thefinding~-- #1}%
}

\titleformat{\section}[block]{\normalfont\large\bfseries}{\thesection.}{1em}{\centering}

\titlespacing{\section}{0pt}{*0}{*2}

\titleformat{\subsection}[hang]{\normalfont\large\bfseries}{\thesubsection}{1em}{}
\titleformat{\subsubsection}[hang]{\normalfont\normalsize\bfseries}{\thesubsubsection}{1em}{}

\titleformat{\paragraph}
  {\normalfont\bfseries\filcenter} 
  {\theparagraph}                  
  {1em}                            
  {}                               
\titlespacing{\paragraph}
{0pt}{3.25ex plus 1ex minus .2ex}{1.5ex plus .2ex}

\newcommand{\subsubsubsection}{\paragraph}

\defcitealias{Astro2020}{National Academies of Sciences, Engineering, and Medicine 2021}
\defcitealias{ROTAC2025FinalReport}{Roman Observations Time Allocation Committee \& Core Community Survey Definition Committees 2025}

\begin{document}

\title{4th TDAMM Workshop White Paper}

\author[orcid=0000-0001-9201-4706,sname='Kocevski']{Daniel Kocevski}
\affiliation{NASA Marshall Space Flight Center}
\email[hide]{daniel.kocevski@nasa.gov}  

\author[orcid=0000-0003-4768-7586,sname='Margutti']{Raffaella Margutti}
\affiliation{Department of Astronomy, University of California, Berkeley}
\email[hide]{rmargutti@berkeley.edu}  

\author[orcid=0000-0002-8262-2924,sname='Coughlin']{Michael Coughlin}
\affiliation{University of Minnesota}
\email[hide]{cough052@umn.edu}  

\author[orcid=0000-0003-2624-0056,sname='Fryer']{Christopher Fryer}
\affiliation{Los Alamos National Laboratory}
\email[hide]{fryer@lanl.gov}  

\author[orcid=0000-0002-8297-2473,sname='Alexander']{Kate Alexander}
\affiliation{Steward Observatory, University of Arizona}
\email[hide]{kdalexander@arizona.edu}

\author[orcid=0000-0002-6759-1701,sname='Andreoni']{Igor Lorenzo Andreoni}
\affiliation{Department of Physics and Astronomy, University of North Carolina at Chapel Hill}
\email[hide]{Igor.Andreoni@unc.edu}

\author[orcid=0000-0001-7346-5114, sname='Atteia']{Jean-Luc Atteia}
\affiliation{Univ. Toulouse, CNES, CNRS, IRAP, Toulouse, France}
\email[hide]{jean-luc.atteia@irap.omp.eu}

\author[orcid=0000-0001-8525-3442,sname='Aydi']{Elias Aydi}
\affiliation{Department of Physics and Astronomy, Texas Tech University}
\email[hide]{elias.aydi@gmail.com}

\author[orcid=0000-0003-4433-1365,sname='Baring']{Matthew Baring}
\affiliation{Department of Physics and Astronomy, Rice University}
\email[hide]{baring@rice.edu}

\author[orcid=0000-0002-7735-5796,sname='Bright']{Joe Bright}
\affiliation{Astrophysics, Department of Physics, University of Oxford}
\email[hide]{joe.bright@physics.ox.ac.uk}

\author[orcid=0000-0002-2942-3379,sname='Burns']{Eric Burns}
\affiliation{Department of Physics \& Astronomy, Louisiana State University}
\email[hide]{ericburns@lsu.edu}

\author[orcid=0000-0003-1673-970X,sname='Cenko']{Brad Cenko}
\affiliation{NASA Goddard Space Flight Center}
\email[hide]{brad.cenko@nasa.gov}

\author[orcid=0000-0002-3211-303X,sname='Curtis']{Sanjana Curtis}
\affiliation{Department of Physics, Oregon State University}
\email[hide]{curtsanj@oregonstate.edu}

\author[orcid=0000-0002-3137-4633,sname='DeColle']{Fabio DeColle}
\affiliation{Instituto de Ciencias Nucleares, Universidad Nacional Autónoma de México}
\email[hide]{fabio@nucleares.unam.mx}

\author[sname='Elliott']{Courey Elliott}
\affiliation{Department of Physics \& Astronomy, Louisiana State University}
\email[hide]{coureyelliott@gmail.com}

\author[orcid=0000-0002-0186-3313, sname='Fletcher']{Corinne Fletcher}
\affiliation{Universities Space Research Association}
\email[hide]{cfletcher@usra.edu}

\author[orcid=0000-0002-5605-2219,sname='Franckowiak']{Anna Franckowiak}
\affiliation{Astronomisches Institut, Ruhr-Universit\"at Bochum}
\email[hide]{franckowiak@astro.ruhr-uni-bochum.de}

\author[orcid=0000-0003-0191-2477,sname='Frohlich']{Carla Fr\"ohlich}
\affiliation{Department of Physics and Astronomy, North Carolina State University}
\email[hide]{cfrohli@ncsu.edu}

\author[orcid=0000-0002-0587-7042,sname='Goldstein']{Adam Goldstein}
\affiliation{Universities Space Research Association}
\email[hide]{agoldstein@usra.edu}

\author[orcid=0000-0002-1018-9383,sname='Habig']{Alec Habig}
\affiliation{Department of Physics and Astronomy, University of Minnesota Duluth}
\email[hide]{habig@neutrino.d.umn.edu}

\author[orcid=0000-0002-5698-8703, sname='Hammerstein']{Erica Hammerstein}
\affiliation{Department of Astronomy, University of California, Berkeley}
\email[hide]{ekhammer@berkeley.edu}

\author[orcid=0000-0002-9017-3567, sname='Ho']{Anna Y. Q. Ho}
\affiliation{Department of Astronomy, Cornell University}
\email[hide]{ayh24@cornell.edu}

\author[orcid=0000-0003-4253-656X,sname='Howell']{D. Andrew Howell}
\affiliation{Las Cumbres Observatory}
\affiliation{Department of Physics, University of California, Santa Barbara}
\email[hide]{dahowell@gmail.com}

\author[orcid=0000-0001-8551-2002, sname='Hu']{Chin-Ping Hu}
\affiliation{Department of Physics, National Changhua University of Education}
\email[hide]{cphu0821@gm.ncue.edu.tw}

\author[orcid=0000-0003-2868-489X, sname='Huang']{Xiaoshan Huang}
\affiliation{Theoretical Astrophysics, California Institute of Technology}
\email[hide]{xshuang@caltech.edu}

\author[orcid=0000-0002-0468-6025,sname='Hui']{Michelle Hui}
\affiliation{NASA Marshall Space Flight Center}
\email[hide]{c.m.hui@nasa.gov}

\author[orcid=0000-0002-5365-5444, sname='Humensky']{Brian Humensky}
\affiliation{NASA Marshall Space Flight Center}
\email[hide]{thomas.b.humensky@nasa.gov}

\author[orcid=0000-0002-6745-4790,sname='Kennea']{Jamie Kennea}
\affiliation{Schmidt Sciences}
\email[hide]{jkennea@schmidtsciences.org}

\author[orcid=0000-0001-7074-0539,sname='Kheirandish']{Ali Kheirandish}
\affiliation{Department of Physics \& Astronomy, University of Nevada, Las Vegas}
\email[hide]{ali.kheirandish@unlv.edu}

\author[orcid=0000-0003-0778-0321,sname='Lau']{Ryan Lau}
\affiliation{NSF NOIRLab}
\email[hide]{ryan.lau@noirlab.edu}

\author[orcid=0000-0002-2666-728X, sname='Miller']{M. Coleman Miller}
\affiliation{Department of Astronomy, University of Maryland}
\email[hide]{mcmiller@umd.edu}

\author[orcid=0000-0001-6350-8168,sname='Mockler']{Brenna Mockler}
\affiliation{Carnegie Observatories}
\email[hide]{brenna.mockler@gmail.com}

\author[orcid=0000-0002-6548-5622,sname='Negro']{Michela Negro}
\affiliation{Department of Physics \& Astronomy, Louisiana State University}
\email[hide]{michelanegro@lsu.edu}

\author[orcid=0000-0003-0440-7193,sname='Ness']{Jan-Uwe Ness}
\affiliation{European Space Astronomy Centre, European Space Agency}
\email[hide]{Jan.Uwe.Ness@esa.int}

\author[orcid=0000-0002-9700-0036,sname='O'Connor']{Brendan O'Connor}
\affiliation{McWilliams Center for Cosmology and Astrophysics, Department of Physics, Carnegie Mellon University}
\email[hide]{boconno2@andrew.cmu.edu}

\author[orcid=0000-0001-5624-2613,sname='Page']{Kim Page}
\affiliation{School of Physics \& Astronomy, University of Leicester}
\email[hide]{klp5@leicester.ac.uk}

\author[orcid=0000-0001-9227-8349,sname='Pritchard']{Tyler Pritchard}
\affiliation{NASA Goddard Space Flight Center}
\affiliation{University of Maryland, College Park}
\email[hide]{tyler.a.pritchard@nasa.gov}

\author[orcid=0000-0002-4744-9898,sname='Racusin']{Judith Racusin}
\affiliation{NASA Goddard Space Flight Center}
\email[hide]{judith.racusin@nasa.gov}

\author[orcid=0000-0001-6982-1008,sname='Radice']{David Radice}
\affiliation{Institute for Gravitation and the Cosmos, The Pennsylvania State University}
\affiliation{Department of Physics, The Pennsylvania State University}
\affiliation{Department of Astronomy and Astrophysics, The Pennsylvania State University}
\email[hide]{david.radice@psu.edu}

\author[orcid=0000-0002-8866-7891,sname='Roberts']{Christopher Roberts}
\affiliation{NASA Goddard Space Flight Center}
\email[hide]{christopher.j.roberts@nasa.gov}

\author[orcid=0000-0002-3500-631X, sname='Russell']{Dave Russell}
\affiliation{Center for Astrophysics and Space Science, New York University Abu Dhabi}
\email[hide]{dave.russell@nyu.edu}

\author[orcid=0000-0003-4102-380X, sname='Sand']{Dave Sand}
\affiliation{Steward Observatory, University of Arizona}
\email[hide]{dsand@arizona.edu}

\author[orcid=0000-0002-8249-8070, sname='Shawhan']{Peter Shawhan}
\affiliation{Department of Physics, University of Maryland}
\email[hide]{pshawhan@umd.edu}

\author[orcid=0000-0001-6279-0552, sname='Street']{Rachel Street}
\affiliation{Las Cumbres Observatory}
\email[hide]{rstreet@lco.global}

\author[orcid=0000-0001-9149-6707,sname='van der Horst']{Alexander van der Horst}
\affiliation{Department of Physics, The George Washington University}
\email[hide]{ajvanderhorst@email.gwu.edu}

\author[orcid=0000-0002-9867-6548, sname='Vandenbroucke']{Justin Vandenbroucke}
\affiliation{Department of Physics, University of Wisconsin--Madison}
\affiliation{Wisconsin IceCube Particle Astrophysics Center, University of Wisconsin--Madison}
\email[hide]{vandenbrouck@wisc.edu}

\author[orcid=0000-0002-9249-0515,sname='Wadiasingh']{Zorawar Wadiasingh}
\affiliation{NASA Goddard Space Flight Center}
\affiliation{University of Maryland at College Park}
\email[hide]{zorawar.wadiasingh@nasa.gov}

\author[orcid=0000-0001-6747-8509,sname='Yao']{Yuhan Yao}
\affiliation{Department of Astronomy, University of California, Berkeley}
\email[hide]{yuhanyao@berkeley.edu}

\author[orcid=0000-0002-7991-028X,sname='Younes']{George Younes}
\affiliation{NASA Goddard Space Flight Center}
\email[hide]{george.a.younes@nasa.gov}

\maketitle

\setcounter{footnote}{0}

\newpage
\tableofcontents

\newpage
\section{Executive Summary} 

\subsection{A New Era in TDAMM Discovery}

Time-Domain and Multi-Messenger Astrophysics (TDAMM) is entering a fundamentally new phase characterized by an unprecedented increase in the rate and diversity of astrophysical transient detections. The field is transitioning from a discovery-limited to a follow-up-limited era, driven by major investments across electromagnetic, gravitational-wave, and neutrino observatories. Upcoming facilities such as the Vera C. Rubin Observatory, the Nancy Grace Roman Space Telescope, and other wide-field survey instruments will produce a deluge of time-domain alerts, reaching millions of events per night. Simultaneously, upgrades to the gravitational-wave network (LVK O5 and beyond) and neutrino observatories (IceCube Gen2) will significantly increase the detection rates of non-electromagnetic messengers. New high-energy missions and expansions of the InterPlanetary Network (IPN) will further enhance discovery capabilities across the X-ray and gamma-ray regimes. This convergence of capabilities represents a transformative opportunity: for the first time, the community will routinely detect rare and high-impact events across multiple messengers. However, the scientific return from these discoveries will depend critically on the ability to rapidly identify, prioritize, and coordinate follow-up observations across a heterogeneous and globally distributed set of facilities.

\subsection{The Increasing Importance of TDAMM Infrastructure}

As a result of this rapid increase in the expected discovery rate, infrastructure is quickly evolving from a supporting role into a mission-critical component of TDAMM science. The modern TDAMM ecosystem relies on a layered architecture of alert distribution systems, data brokers, and coordination platforms that enable real-time response to transient events. Significant progress has been made in alert dissemination through systems such as GCN and SCiMMA, which provide low-latency, machine-readable alerts spanning gravitational waves, neutrinos, and electromagnetic signals. A growing ecosystem of alert brokers, Target and Observation Manager (TOM) systems, and marshaling platforms now supports real-time filtering, classification, and follow-up coordination. Early efforts toward cross-facility coordination, such as NASA's Astrophysics Cross-Observatory Science Support (ACROSS) and related initiatives, are beginning to aggregate observability, scheduling, and resource availability information. These efforts represent the first steps toward an integrated, interoperable system capable of orchestrating complex, multi-observatory campaigns. Despite this progress, the current infrastructure remains fragmented, with limited standardization and incomplete integration across facilities. As alert rates increase by orders of magnitude, infrastructure scalability, interoperability, and usability will become central determinants of scientific success.

\subsection{Critical Challenges}

The workshop identified three primary categories of challenges that could limit the scientific return of TDAMM investments: 1) infrastructure constraints in alert processing and coordination, 2) policy barriers that impede efficient data acquisition, and 3) emerging capability gaps driven by the aging of key missions.

\subsubsection{Infrastructure Limitations}

The imminent “alert tsunami” has the potential to overwhelm current systems. Rubin alone is expected to increase alert volumes by an order of magnitude, requiring brokers to operate at industrial scale with processing times on the order of tens of seconds. Without robust, automated classification and filtering, astronomers will be unable to effectively utilize the available data. Accessibility and usability of alert systems remain significant concerns, particularly for broad community participation. At the same time, effective coordination across facilities remains a major unresolved challenge. Observatory planning information, such as scheduling, visibility, and availability, is heterogeneous and lacks standardized, machine-readable formats. There is no unified framework for sharing this information, nor a centralized repository for follow-up observations and metadata. These limitations hinder the ability to execute coordinated, multiwavelength campaigns in real time.

Agencies should treat alert distribution and brokering systems as core scientific infrastructure, equivalent in importance to observatories themselves. This includes sustained investment in next-generation, high-throughput alert brokers, improvements in usability and accessibility, and support for advanced AI/ML-based classification systems. A long-term, strategic approach to funding this infrastructure development is essential to its success. Likewise, standardization of observatory metadata is also critical. Agencies should require machine-readable descriptions of facility status, visibility, and scheduling as core deliverables. Sustained support for cross-observatory coordination services and the development of a unified follow-up repository will enable more efficient and effective use of available resources.

\subsubsection{Policy Barriers}

Existing policies governing time allocation and data access are not always well aligned with the needs of TDAMM science. 
There are few comprehensive mechanism to request simultaneous observations across multiple facilities, and the speed and flexibility of Target of Opportunity/Director's Discretionary Time (ToO/DDT) processes can vary significantly across facilities. Rare, high-impact events are often disfavored in traditional proposal systems, and software development critical for enabling TDAMM science is generally considered underfunded and undervalued.

Policy reforms are needed to enable coordinated, multi-facility science. Agencies should expand joint ToO/DDT programs where possible, allow proposers to request cross-observatory data, and reduce barriers to rare-event science through mechanisms such as multi-cycle awards. Dedicated funding pathways for software development and joint theory/simulation programs are also essential. Importantly, agencies should formally recognize and support community-driven observing plans to facilitate pre-coordinated observations of the most scientifically interesting events. 

\subsubsection{Capability Gaps}

The community faces potential losses in key observational capabilities. The long-term future of rapid-response missions such as Swift is uncertain, and the field risks losing all-sky gamma-ray monitoring capabilities as missions like Fermi age. The IPN is at historically low capacity, and uncertainties in gravitational-wave upgrade schedules further complicate planning. Additionally, a severe bottleneck in spectroscopic follow-up threatens to limit the scientific interpretation of detected events.

Strategic investments are required to sustain and evolve these critical TDAMM capabilities. NASA should begin planning successors to key rapid-response and wide-field monitoring missions and support the expansion of the IPN, including placing additional nodes on future interplanetary and human spaceflight missions. Likewise, the NSF should ensure predictable upgrade paths for gravitational-wave and neutrino facilities. Across agencies, investments in rapid spectroscopic follow-up capacity are still urgently needed.

\subsection{Community Observing Plans}

A central outcome of the workshop is the further development of community observing plans, representing a paradigm shift from ad hoc responses to pre-planned, coordinated strategies for high-priority events. These plans are motivated by the recognition that certain rare, high-impact events require immediate, coordinated action that is difficult to realize within existing proposal frameworks. Community observing plans would establish pre-negotiated commitments from participating facilities, enabling rapid and scientifically optimized observations.

The white paper outlines the framework by which these plans could be implemented. Governance structures, such as a Trigger Advocate Committee, would oversee activation and execution of the plans. The plans would serve as living documents, regularly updated through workshops and community input. The framework emphasizes open science principles, including immediate public data release, assignment of DOIs, and explicit authorship and credit structures. Implementation could be coordinated through entities such as ACROSS, with potential funding models that support either science leadership teams with service responsibilities or dedicated infrastructure groups.

Eight high-priority source classes have been identified for potential implementation, each accompanied by detailed observing strategies in the white paper. For each class, the plans define the scientific motivation and triggering criteria, along with pre-defined decision trees to guide observational responses. They also outline the coordinated set of observations that would be needed to fully realize the scientific potential of these events.

\subsection{Conclusions}

Modern astrophysics is entering a new era of discovery that will place unprecedented strain on the existing TDAMM ecosystem. These challenges, however, are tractable, and targeted near-term investments can unlock disproportionate gains in the years ahead. By sustaining and scaling critical infrastructure, addressing emerging capability gaps, and modernizing policy frameworks to enable coordinated multi-facility science, funding agencies can ensure that the full value of recent and planned astrophysics investments is realized.

\newpage
\section{Consolidated Workshop Findings}

\subsection{Infrastructure Limitations}

\finding{Agencies should treat the alert generation, distribution, brokering, and classification ecosystem as mission-critical scientific infrastructure}: This includes sustained operational funding for scalable alert distribution systems (e.g., GCN, SCiMMA), next-generation brokers capable of Rubin-scale throughput, and robust, community-vetted event classification frameworks. These investments should prioritize usability, including low-barrier, user-facing interfaces that enable broad community participation without requiring specialized technical expertise. These systems should be supported as long-lived critical infrastructure, rather than short-term research projects supported through piecemeal funding.

\finding{Agencies and facility leadership should require standardized, machine-readable observatory metadata as a core operational deliverable}:\label{f:meta_data} New and existing facilities receiving federal support should be required to publish scheduling constraints, resource availability, instrument status, and operational limitations in API-accessible formats using common schemas. These data should be treated as essential infrastructure for time-domain science, enabling automated coordination, feasibility assessment, and cross-facility interoperability.

\finding{Agencies should sustain cross-observatory coordination platforms:}\label{f:coordination_platforms} Programs such as ACROSS and HEROIC should be supported as enduring infrastructure services, with capabilities that include unified visibility and feasibility calculations, integration with brokers and TOM systems, and support for automated observing workflows.

\finding{Agencies should support the development of a unified follow-up repository}:\label{f:follow-up_respository} To prevent researchers from manually reconstructing transient histories across fragmented archives, agencies should fund a centralized discovery layer that tracks the entire observational history of individual events.

\subsection{Policy Barriers}

\finding{Agencies should formalize multi-facility access through expanded joint observing opportunities:}\label{f:expand_joint_observing} Agencies should further expand and formalize joint observing opportunities by further establishing comprehensive, multi-facility ToO and DDT programs. These mechanisms should enable a single, integrated proposal that can span multiple space- and/or ground-based facilities to allow for a coherent suite of observations based on the data required to fully characterize and study a given source.

\finding{Review processes should recognize cross-facility dependencies as central to TDAMM science:}\label{f} Agencies and decision authorities should ensure that review guidance and evaluation criteria treat scientifically necessary external observations as indicators of mission relevance and ecosystem integration, rather than as an inherent mission weaknesses. TDAMM-focused missions should not be penalized for depending on complementary observations across wavelengths and messengers when such coordination is essential to realizing the full scientific value of the mission.

\finding{Facilities should lower barriers for rare event science:}\label{f:rare_event_barriers} Facilities should consider implementing or expanding multi-cycle triggering windows so that low-probability but high-reward events (e.g., nearby core-collapse supernovae) are not disadvantaged by standard annual proposal cycles because of their low probability of occurrence.

\finding{Agencies should formally recognize and enable pre-coordinated community observing plans:}\label{f:community_observing_plans} Funding agencies should embrace policy frameworks that allow observatories to pre-commit baseline observing resources for community defined classes of high-priority events, reducing reliance on ad hoc, facility-by-facility responses. This includes investigating means to fund analysis performed by teams or individuals in service of the community plans. Data collected as part of the community observing plan program should be immediately released to the community, with no proprietary period.

\finding{Agencies should establish sustained funding paths for TDAMM software development:}\label{f:software_development} Dedicated funding lines should be made available to support the development, maintenance, and operation of critical software, data services, and coordination platforms, recognizing these efforts as core scientific contributions.

\subsection{Capability Gaps}

\finding{Agencies should prioritize strategic TDAMM replacements:}\label{f:tdamm_replacement_missions} NASA should develop plans for a strategic high-energy monitor and rapid-response X-ray mission to replace the aging Swift and Fermi observatories. This includes prioritizing the development of next-generation, high-sensitivity X-ray observatories capable of providing deep imaging, spectroscopy, and precise localization of transient sources to replace Chandra and XMM-Newton.

\finding{NASA should sustain and expand the IPN:}\label{f:IPN}  NASA should routinely look for opportunities to add gamma-ray detection capabilities to interplanetary and heliophysics missions, specifically prioritizing nodes outside the ecliptic plane (e.g., on a Uranus Orbiter) to break geometric degeneracies in triangulation.

\finding{NSF should ensure predictable GW upgrades:}\label{f:GW_schedule} Stable support for the LIGO A+ upgrade is critical to ensure that peak detector sensitivity aligns with the operational lifetimes of complementary NSF, DOE, and NASA facilities and missions.

\finding{Agencies should support the development of additional spectroscopic capabilities:}\label{f:spectroscopy} Significant investment is required in moderate-aperture follow-up telescopes and dedicated space-based missions optimized for rapid-response optical and near-infrared (NIR) spectroscopy to address the spectroscopy bottleneck in future event classification. 

\finding{Agencies should encourage future missions to address cross-observatory coordination during formulation} Agencies should encourage, if not require, future TDAMM missions and federally supported observatories to address cross-observatory coordination in the design process. This should include early engagement with coordination frameworks such as ACROSS and related initiatives, definition of machine-readable interfaces for observability and operational status, and documentation of how the mission will support coordinated alerts, and follow-up requests.

\subsection{Community Observing Plans}



\finding{Agencies and participating facilities should establish a Trigger Advocate Committee to support the execution of the community observing plans}: This committee would help validate community observing plan activations and guide early-time response execution.

\finding{The community observing plans must be a living document}: The plans should be updated regularly through a body of subject matter experts and community and facility stakeholders.

\finding{Data collected as part of a community-defined observing plan should be made public immediately}: This requirement would preclude proprietary data-rights to any person or team.

\finding{Agencies and participating facilities should establish clear, well-defined credit mechanisms that recognize contributions across the full lifecycle of community observing plans}: This includes the development of observing strategies, execution of coordinated observations, analysis of resulting data, and citation of community data products.


\setcounter{finding}{0}

\newpage
\section{Introduction} \label{sec:introduction}

TDAMM astrophysics promises to expand our understanding of the Universe by enabling the study of astrophysical phenomena as dynamic, interconnected events observed across multiple messengers and wavelengths. By combining electromagnetic observations with gravitational waves, neutrinos, and cosmic rays, TDAMM opens fundamentally new windows on the most energetic and extreme processes in nature, from compact object mergers and relativistic jets to stellar explosions and accretion physics. The Astro2020 Decadal Survey recognized this transformative potential, prioritizing TDAMM as a central theme for the coming decade \citepalias{Astro2020}. The power of this approach lies not only in the breadth of physical insight it provides, but also in its inherently interdisciplinary nature, with implications spanning cosmology, fundamental physics, nuclear astrophysics, and particle physics

In response to this Decadal priority, NASA and the National Science Foundation (NSF) have convened a series of TDAMM workshops to define scientific priorities and identify challenges that impede the full scientific return of this emerging field.  The \href{https://science.nasa.gov/astrophysics/programs/physics-of-the-cosmos/time-domain-and-multi-messenger-astrophysics-workshop-2022/}{first TDAMM workshop}, held in Annapolis in 2022, brought together a broad cross-section of the community to identify key scientific questions and required capabilities \citep{TDAMMWorkshop2023}.  The workshop also highlighted the transient source classes most likely to yield transformative multi-messenger discoveries and underscored the importance of rapid, coordinated response across observatories.

The second workshop, \href{https://noirlab.edu/science/events/websites/MMA2023}{\textit{Windows on the Universe}}, held in Tucson in 2023, shifted focus toward the enabling infrastructure, alert systems, data access, and coordination tools, required to support this science at scale, directly informing new agency programs and community cyberinfrastructure investments \citep{Ahumada2024Windows}. The third workshop, \href{https://sites.google.com/view/3rd-tdamm-workshop/home}{\textit{Multidisciplinary Science in the Multimessenger Era}}, held in Baton Rouge in 2024, expanded the scope further, emphasizing the need for deep integration with nuclear physics, plasma physics, gravitational physics, and computation to fully interpret TDAMM observations and to adopt end-to-end, multi-diagnostic approaches drawn from other areas of physics \citep{TDAMM_BatonRouge_2024}.

The \href{https://science.nasa.gov/astrophysics/programs/physics-of-the-cosmos/community/fourth-tdamm-workshop/}{fourth TDAMM workshop} continued in this series, but marked a transition in emphasis. With the imminent start of the Vera C. Rubin Observatory’s Legacy Survey of Space and Time, the maturation of the global gravitational-wave and neutrino networks, and the arrival of new wide-field X-ray and gamma-ray missions, TDAMM is quickly becoming no longer limited by discovery capability. Instead, the dominant challenges are now operational: how to manage an unprecedented alert volume, how to coordinate responses across dozens of facilities with heterogeneous constraints, and how best to ensure that rare, high-impact events are recognized and acted upon in time. The fourth TDAMM workshop addressed these challenges by examining the technical and policy barriers that limit effective coordination and by proposing strategies for more rapid and efficient multi-facility responses across the evolving TDAMM ecosystem.

The fourth workshop was also tasked with defining the scientific drivers for pre-defined, community-driven observing plans for rare and scientifically valuable transients. As a central recommendation that emerged from the second workshop, and reinforced in the third, community observing plans were envisioned to enable coordinated, multi-facility observations without the delays and fragmentation inherent in traditional proposal-based systems. To further this concept, the workshop's Science Organizing Committee (SOC) identified eight source classes; gamma-ray bursts, tidal disruption events, X-ray binaries, supernovae, novae, magnetars, compact binary mergers, and high-energy neutrino sources, as areas where pre-coordinated strategies would provide the greatest scientific return. After soliciting input from the community, the SOC drafted community observing plans that defined triggering criteria, observational priorities, and coordination strategies that could be implemented to maximize the scientific return of rare, but scientifically valuable events.

This white paper summarizes the outcomes of the fourth TDAMM workshop and presents a path forward for realizing the full scientific potential of TDAMM astrophysics through the end of the decade. It summarizes the current TDAMM discovery ecosystem, outlines the key challenges identified by the community and presents strategies on how to address them.  The report also details the prioritized science cases and observing strategies developed during the workshop for the proposed community observing plans. The document is organized as follows: Sections \ref{sec:discovery} \& \ref{sec:intrastructure} describe the near-term discovery and infrastructure landscapes respectively; Section \ref{sec:challenges} presents the critical operational, technical, and policy challenges and associated findings; and Section \ref{sec:obs_plans_framework} outlines the framework for community coordination and observing strategies.  Appendix A presents an overview of the source classes that would most benefit from pre-planned observing plans and Appendix B presents the detailed community observing plans intended to serve as living resources for the community for the source classes identified by the SOC. Finally, Appendix C provides workshop information.


\newpage
\section{The TDAMM Discovery Landscape} \label{sec:discovery}

The field of TDAMM is entering a transformative era, defined by an unprecedented synergy between a new generation of powerful observatories and sophisticated data infrastructure. Ground-based facilities like the Vera C. Rubin Observatory are poised to generate a revolutionary deluge of optical data, issuing up to 10 million transient alerts per night. These will be complemented by triggers from non-electromagnetic observatories, including the LIGO-Virgo-KAGRA (LVK) gravitational-wave network and the IceCube Neutrino Observatory, each providing unique and complementary insights into the most energetic phenomena in the universe.

This rapidly evolving discovery landscape is further enabled by a diverse fleet of space-based observatories. Long-operating missions such as the Neil Gehrels Swift Observatory, the Fermi Gamma-ray Space Telescope, and the IPN continue to provide critical high-energy coverage and localization capabilities, while new wide-field X-ray missions such as Einstein Probe and SVOM are expanding the discovery space through sensitive, rapid-response observations and the Nancy Grace Roman Space Telescope promise to extend discovery into the near-infrared. Together, these facilities form a highly complementary, multiwavelength and multi-messenger ecosystem in which discovery and follow-up are increasingly interdependent.

This section provides an overview of the current and near-term TDAMM discovery landscape, with a focus on the capabilities that will define the next 3–4 years. The subsections that follow summarize key contributions from existing facilities, highlight newly operational missions, and outline upcoming observatories expected to come online in the near future. Together, these elements illustrate both the breadth of the emerging ecosystem and the critical role of coordinated, multi-messenger observations in maximizing the scientific return of this new era in TDAMM astronomy.

\subsection{The Optical and Infrared Data Revolution}

A central theme emerging across TDAMM facilities is the explosive growth in the volume of transient alerts and the speed at which they must be acted upon. Over the next several years, wide-field optical and infrared surveys will reach unprecedented depths, cadence, and sky coverage. These surveys are expected to drive the majority of transient triggers for the TDAMM ecosystem, producing both routine discoveries and rare events critical for multimessenger follow-up.

\subsubsection{Vera C. Rubin Observatory}

The Vera C. Rubin Observatory is poised to reshape the time-domain landscape through the Legacy Survey of Space and Time (LSST), a 10-year program designed to observe the entire southern sky roughly every few nights \citep{Ivezic2019_Rubin}. With its 6.67 m effective aperture and 3.2-Gpixel camera, LSSTCam, Rubin will survey the southern sky with a 9.6 deg² field of view and a regular revisit cadence, ultimately detecting millions of transients per night and producing the largest real-time alert stream ever generated in optical astronomy. These alerts, containing coordinates, magnitude, and other metadata of every source changing significantly in luminosity, will be processed and issued within 60 seconds of photon detection.

Rubin’s survey strategy is built around the LSST baseline cadence, which delivers multi-band imaging in u, g, r, i, z, and y filters with a typical visit per band every several nights.  A ``triplet" of observations within a given night is expected roughly $\sim$4\% of the time, which will enable the identification of rapidly evolving phenomena. The single-exposure depth of r$\sim$24.5 and coadded depth approaching r$\sim$27.5 will allow discovery of faint and distant transients, including high-redshift supernovae, rapidly fading kilonovae, and rare exotic phenomena.

The LSST Data Management pipeline will issue up to 10 million alerts per night, each containing difference-imaging photometry and machine-readable source features. This deluge of alerts is expected to fundamentally reshape optical TDAMM workflows. Instead of having individual teams or missions process this volume independently, the community will depend on large-scale, continuously operating alert brokers for real-time classification and event prioritization.

In the TDAMM context, Rubin will become a cornerstone facility for:

\begin{itemize}
    \setlength{\itemsep}{1pt}
    \setlength{\parskip}{1pt}
    \item Kilonova discovery following gravitational-wave triggers, including some events with large (\textgreater 100 deg$^2$) localization regions.
    \item Core collapse supernovae precursor activity, particularly shock cooling signatures within minutes to hours of explosion.
    \item Monitoring of optical counterparts to high-energy neutrino events.
    \item Discovery of rare, distant fast transients, including fast blue optical transients (FBOTs) and luminous fast-cooling events.
\end{itemize}

Rubin’s alert stream is tightly integrated with real-time community infrastructure. The AEON+ ecosystem, TOM toolkits, SkyPortal-based marshaling systems, and automated follow-up pipelines (e.g., via Gemini, LCO, SOAR, and other AEON sites) are being adapted specifically to support Rubin-scale operations. In addition, alert brokers must now incorporate probabilistic classification models that operate at LSST data rates, including machine-learning architectures capable of identifying precursor emission, rapidly evolving kilonova candidates, and anomalous or previously unseen classes of transients.

Rubin’s imminent arrival marks a transition from a discovery-limited epoch to a response-limited one. The challenge is not finding transients, but identifying the correct ones in time for follow-up, especially for multimessenger events where early optical information can constrain ejecta composition, energetics, and progenitor properties. As a result, Rubin is poised to become the primary driver of real-time optical discovery for the TDAMM community.

\subsubsection{The Nancy Grace Roman Space Telescope}

NASA’s Nancy Grace Roman Space Telescope will complement Rubin by delivering deep, stable, space-based imaging and NIR time-domain capabilities with a wide field of view unmatched among space observatories. Roman’s Wide Field Instrument (WFI) has a 0.28 deg$^2$ field of view, roughly 100 times that of Hubble, paired with diffraction-limited optics and exceptional photometric uniformity. This makes Roman uniquely suited for high-redshift transient discovery, precise cosmological measurements, and dust-penetrating time-domain studies that are inaccessible from the ground. 

While Roman will detect transients in all its survey modes, the planned High Latitude Time-Domain Survey (HLTDS) positions the mission as a major upcoming contributor to TDAMM discovery and classification of fast transients \citepalias{ROTAC2025FinalReport}. Operating in the NIR, Roman will detect transients obscured by dust, measure rest-frame optical emission at high redshift, and provide high-cadence, space-based light curves that will anchor cosmological studies. Roman’s stable PSF and calibration precision offer significant advantages for supernova cosmology, in particular, where minimizing systematics is critical for probing dark energy models. Roman will also extend the discovery space to early-universe explosions, capturing supernovae beyond the redshift reach of Rubin and enabling characterization of transients in dusty environments such as nuclear star-forming regions. These capabilities help complete the electromagnetic picture for multimessenger events that may occur behind significant extinction.

To enable prompt transient discoveries, the RAPID (Roman Alerts Promptly from Image Differencing) Project Infrastructure Team is tasked with delivering four key products \citep{Gandhi2026RomanRAPIDly}:
\begin{itemize}
    \setlength{\itemsep}{1pt}
    \setlength{\parskip}{1pt}
    \item Rapid image differencing of every new Roman image
    \item A prompt public alert stream of all transient and variable candidates
    \item Source match-files recording candidate light curves
    \item A forced photometry tool for investigating photometric history
\end{itemize}  

Together, Rubin and Roman will form a powerful optical–NIR discovery pair. Rubin will provided continuous wide-area monitoring and rapid transient identification, while Roman will supply precise NIR photometry, light curves, and host-galaxy characterization. This synergy is especially valuable for kilonovae, where NIR emission carries strong signatures of lanthanide-rich ejecta, and for Type Ia supernova cosmology, where NIR observations reduce intrinsic scatter and dust-related systematics.

\subsubsection{The Argus Array}

The Argus Array represents a complementary and highly distinctive element of the optical time-domain discovery landscape, occupying a parameter space between deep surveys such as Rubin and space-based facilities like Roman. Argus is a next-generation, ground-based, wide-field time-domain astronomy facility designed to deliver continuous, high-cadence monitoring over a substantial fraction of the northern sky, with capabilities optimized for fast, rare, and rapidly evolving transients that are central to TDAMM science \citep{Law2022_Argus}.

Argus consists of an array of 1,200 individual 28-cm telescopes that collectively form an 8-meter class instrument. The array will deliver simultaneous coverage of 8,000 square degrees per exposure, each telescope being equipped with a 102-megapixel CMOS detector with a 1.0 arcsecond pixel scale, summing to $\sim$122 Gpx in total per image\footnote{\url{https://argus.unc.edu/specifications}}.  The system is engineered for time-domain efficiency, achieving a $\sim$94\% duty cycle with sub-millisecond dead time between exposures and operating in 15-minute tracking intervals. A dual-band photometric system (anticipated to be modified B and R bands) will enable simultaneous two-color observations across the full field.

Argus’s observing strategy is explicitly optimized for multi-timescale variability and transient discovery. Cadences range from 1-second sampling during bright time to 1-minute sampling during gray and dark time, enabling continuous monitoring across timescales that are largely inaccessible to traditional survey designs. Over its planned five-year survey, Argus will accumulate roughly 20 million epochs at 1-second cadence and $\sim$300,000 epochs at 1-minute cadence for all northern sky regions. Real-time image subtraction and transient detection are integral to the pipeline, and all imaging, photometry, and transient alerts will be released publicly with no proprietary period. Construction is currently underway at a Northern Hemisphere site, with operations expected to begin in 2027.

Argus is particularly well matched to multi-messenger discovery and early characterization, combining continuous wide-area monitoring with high-cadence optical coverage that is critical for both gravitational-wave and gamma-ray transient science. By monitoring roughly half the sky each night, Argus can naturally tile the hundreds to thousands of square degrees typical of LVK gravitational-wave localizations as part of its baseline survey, providing effectively zero-latency optical coverage when a merger occurs within its instantaneous 8,000 square degree field of view and, otherwise, identifying counterparts within hours during nightly scans. The resulting dense light curves, often comprising hundreds of data points, are uniquely suited to capturing and modeling the rapidly evolving early blue emission of kilonovae on 12–24 hour timescales, with 1-minute sensitivity to $\sim$20.5 mag enabling detections out to $\sim$200 Mpc and stacked observations extending beyond $\sim$700 Mpc for events similar to GW170817. At the same time, the continuous, wide-field coverage enables a powerful and largely serendipitous gamma-ray burst capability: a substantial fraction of GRBs detected by Swift, Fermi, and future high-energy monitors will already be under optical surveillance at or near the moment of explosion, allowing Argus to detect prompt or very early afterglow emission and to track afterglow evolution from minutes to days post-burst. These observations provide critical constraints on relativistic jet breakout, early forward-shock physics, circumburst environments, and the possible emergence of associated supernova or kilonova emission.

In the broader optical–infrared ecosystem, Argus complements Rubin and Roman by prioritizing cadence and sky coverage over depth. Their combination is especially powerful for TDAMM sources whose scientific value depends critically on capturing the earliest phases of emission.

\subsubsection{High-Cadence, Wide-Field, and Specialized Survey Facilities}

Rubin, Roman, and Argus will define much of the optical and near-infrared discovery landscape in the coming years, but they will be complemented by a broader ecosystem of smaller, more specialized, and often more agile surveys. These facilities fill important gaps in cadence, wavelength coverage, sky accessibility, alert response, and operational flexibility. Their value lies not in matching the depth or scale of flagship surveys, but in covering parameter space that large facilities cannot always address efficiently, such as very bright transients, rapid follow-up of poorly localized alerts, near-infrared monitoring, northern or southern hemisphere coverage, and dense temporal sampling of selected fields.

High-cadence optical facilities provide one important class of complementary capability. The Large Array Survey Telescope (LAST) uses a modular array of small telescopes to provide wide-field, rapid-cadence visible-light monitoring, with a full-node field of view of roughly 355 deg² and sensitivity approaching 21st magnitude in stacked short exposures \citep{Ofek2023_LAST}. This architecture is well suited to fast transients, stellar flares, white dwarf transits, satellite glints, moving objects, and other rapidly evolving phenomena. By emphasizing cadence, modularity, and cost-effective deployment, LAST helps fill the gap between deep, slower surveys and continuous all-sky monitoring.

Several ongoing wide-field optical surveys will also remain important during the Rubin/Roman era. Pan-STARRS provides deep multi-band reference imaging and long temporal baselines over much of the sky, while ASAS-SN and ATLAS provide shallow, frequent monitoring of the bright transient sky \citep{Chambers2016_PanSTARRS,Shappee2014_ASASSN,Tonry2018_ATLAS}. These surveys are especially valuable for nearby, bright, or rapidly evolving events that may saturate deeper facilities or require immediate recognition before they fade. GOTO and BlackGEM add a more explicitly TDAMM-oriented capability, with robotic observing strategies designed to search large gravitational-wave localization regions and rapidly identify candidate optical counterparts \citep{Dyer2024_GOTO,Groot2019BlackGEM}. Together, these facilities help anchor the bright and fast end of the transient population, provide pre-discovery light curves, and support rapid triage of candidate counterparts.

Near-infrared facilities provide another critical complement to the optical discovery stream. WINTER is designed for wide-field near-infrared follow-up of kilonovae and other transients, using Y, J, and short-H band observations to probe emission that can be longer-lived and less viewing-angle dependent than optical kilonova light \citep{Frostig2021_WINTER,Frostig2025_WINTER}. PRIME, operating from the Southern Hemisphere, extends this capability into wide-field near-infrared monitoring of the Galactic bulge, including highly extincted regions inaccessible to optical surveys. PRIME will also support Roman microlensing science by providing simultaneous ground-based observations that enable parallax measurements and lens-mass constraints \citep{Kutyrev2023_PRIME,Sumi2025_PRIME}. More broadly, these infrared facilities help fill a major gap in the TDAMM ecosystem by enabling transient searches and follow-up in wavelength regimes where optical surveys are incomplete or biased.

\subsection{Ultraviolet Observations}

\subsubsection{Ultraviolet Transient Astronomy Satellite (ULTRASAT)}
ULTRASAT is a near-ultraviolet (230–290 nm) space observatory stationed in Geostationary Earth Orbit, designed specifically to revolutionize time-domain and multi-messenger astrophysics \citep{Shvartzvald2024ULTRASAT}. Its unprecedented 204-square-degree field of view allows it to avoid Earth occultations and stare continuously at the sky, dedicating 21 hours a day to a High-Cadence Survey taking 5-minute exposures. The remaining daily time is used to conduct deep all-sky and wide-area surveys. ULTRASAT features a rapid ToO mode that can interrupt normal operations to point at newly discovered transients in under 15 minutes. By combining this large area sky coverage, continuous monitoring, and real-time alert generation, the mission will serve as a premier discovery engine for early-stage supernovae, tidal disruption events, and kilonovae.

\subsubsection{The UltraViolet EXplorer (UVEX)}
UVEX is a NASA Medium-Class Explorer mission designed to provide crucial, currently missing observational capabilities essential for a broad range of astrophysics, with a particular emphasis on the time-domain ultraviolet sky \citep{Kulkarni2021UVEX}.  
UVEX carries a wide-field, two-band ultraviolet imager and a long,  and a multi-width slit spectrometer.
The imager features a $3.5^{\circ} \times 3.5^{\circ}$ field of view enabling simultaneous observations in both the far-UV  and near-UV bandpasses.
UVEX will perform the first cadenced, all-sky UV imaging survey of the sky, achieving a depth commensurate with modern wide-field optical and infrared facilities. UVEX promises to revolutionize our understanding of the dynamic universe via unprecedented rapid-turnaround ($\lesssim$3\,hr) UV spectroscopy.  Utilizing high-cadence time-domain surveys and a rapid Target of Opportunity (ToO) program, UVEX will provide prompt \emph{UV spectroscopic} follow-up of fast-evolving  transients. This rapid response is critical for probing the early-time environments, energetics, and emission processes of gravitational-wave-discovered compact object mergers, discovering hot and fast transients, and diagnosing the immediate aftermath of stellar explosions. On the spectroscopic side, UVEX will provide the first UV spectra of transients at $\delta t<2$\,d. On the photometric side, UVEX will provide a survey of transients in two UV channels (i.e., FUV and NUV). In both cases, UVEX will open a new window of transient discovery.

\subsection{Non-EM Messengers}

Time-domain and multimessenger astronomy is fundamentally anchored in the ability to detect signals beyond the electromagnetic spectrum. Gravitational-wave (GW) and neutrino observatories now operate as primary discovery engines in their own right, capable of triggering community-wide follow-up efforts and revealing physical conditions deep within compact-object mergers, stellar explosions, and other extreme phenomena. Over the next 3–5 years, both the International Gravitational-Wave Network (IGWN) and the IceCube Neutrino Observatory will expand their sensitivity, alert capabilities, and integration with the broader TDAMM ecosystem, while confronting operational and funding uncertainties that may impact long-term performance. 

\subsubsection{Gravitational Wave Detectors}

The global network of gravitational-wave detectors, comprising LIGO in the US, Virgo in Europe, and KAGRA in Japan, continues to be a cornerstone of multi-messenger astronomy. The separate LVK collaborations are planning to merge into a single, unified global consortium, IGWN, sometime in 2026. This transition aims to formalize joint operations, with an emphasis on improved calibration, shared alert standards, and consistent public communication channels. This is particularly important as the network prepares for increasingly sensitive observing runs and more complex multi-detector configurations. 

The fourth observing run (O4), which began in May 2023 and concluded in Nov 2025, provided the most sensitive GW detections to date. The O4 run saw higher duty cycles for the LIGO detectors, improvements in broadband strain sensitivity, and expanded sky coverage from the inclusion of Virgo and KAGRA (though at reduced sensitivity early in the run). While no confirmed binary neutron star (BNS) event with an electromagnetic counterpart was observed during O4, the run yielded 254 significant alerts and 5136 sub-threshold alerts. Numerous of these alerts, including sub-threshold candidates, prompted follow-up campaigns, validating community readiness and highlighting the need for continued automation in coordination systems. The O4 run has also yielded several important publications, including the first paper on the merger of a neutron star with an intermediate-mass compact object (a low-mass black hole or heavy neutron star) \citep{Abac2024CompactObjectNeutronStar}. O4 also saw several remarkably loud events that will allow for more detailed tests of General Relativity (GR) \citep{Abac2026GW250114Spectroscopy}.

The path to the next official run, O5, is the completion of the Advanced LIGO Plus (A+) upgrade. This is a comprehensive program designed to boost the LIGO detector sensitivities by up to a factor of $\sim$2 in range, corresponding to an even larger increase in sensitive volume\footnote{\url{https://emfollow.docs.ligo.org/userguide/}}. Assuming a 3 year duration of O5, new expected rates predict a median of 5 BNS public alerts over the first year, O5a, 10 over the course of the second year, O5b, and 15 over the course of the third, O5c. Key upgrades will include: improved vibration isolation, better control of stray light, new interferometric readout scheme (balanced homodyne), new mirror coatings to reduce thermal noise.

Due to funding and technical uncertainty, the possibility of an interim observing run before the full O5 sensitivity is achieved is being discussed. If it happens, such a ``bridge run" (or O4.5) would likely occur by the end of 2026 and might last approximately six months. Detectors would likely operate at sensitivity similar to O4, or only slightly better, as the major sensitivity gains come from the later phases of the A+ installation.

Multimessenger science, including low-latency alerts, will remain a central part of the program for the newly formed IGWN in the O4.5 and O5 era. This will include an emphasis on improving low-latency search pipelines for speed and robustness, reviewing and refining rapid response procedures, and improving classification probability estimates using multiple pipeline triggers.

\subsubsection{Neutrino Telescopes}

The IceCube Neutrino Observatory continues to be the world’s leading facility for detecting astrophysical neutrinos at TeV–PeV energies. IceCube functions as both a multimessenger alert producer, issuing real-time notifications of single high-energy neutrino candidates and TeV neutrino clusters in time and space, and an alert consumer, performing targeted searches in response to external triggers such as GRBs, gravitational-wave events, and bright X-ray transients. 

As an alert producer, IceCube currently generates around 30 alerts per year. A major improvement was implemented in September 2024 with a new online reconstruction procedure, resulting in significantly refined sky localization for track-like events from Muon neutrino interactions. The median 90\% localization area improved by a factor of $\sim$5, shrinking the area from 6.2 deg to 1.3 deg, with this information being relayed to the community with a latency of about 1 minute. As an alert consumer, IceCube operates a ``Fast Response Analysis" (FRA) pipeline to conduct rapid searches for coincident neutrino emission to transients reported by other observatories (LVK, Swift, Fermi, etc.). With a $>$99\% uptime and $4\pi$ sky coverage, IceCube is well-suited for this kind of follow-up to external events. During the most recent O4 run, the median latency for publishing these kinds of coincident alerts was reduced to 21 minutes, compared to 56 minutes in O3. The observatory is currently preparing for a rapid search for neutrinos from the upcoming outburst of the recurrent nova T Coronae Borealis (T CrB), which is expected to be brighter than the 2021 RS Ophiuchi outburst.

Recent years have shown that while associations between high-energy TeV-PeV neutrinos and electromagnetic sources remain rare, they are scientifically transformative when they occur. IceCube’s high-energy neutrino program has delivered several landmark results, including the identification of the first likely extragalactic neutrino source (TXS 0506+056) \citep{IceCube:2018dnn, IceCube:2018cha}, detection of neutrino emission from the Milky Way~\citep{IceCube:2023ame}, a 4.2$\sigma$ detection of neutrino emission from the Seyfert galaxy NGC 1068~\citep{IceCube:2022der}, constraints on hadronic acceleration in GRBs~\citep{IceCube:2023rhf}, and limits on neutrino emission from BNS mergers~\citep{IceCube:2023atb}. These results strengthen the role of neutrinos as a key complementary messenger, probing cosmic-ray acceleration and environments opaque to photons.%

In addition, IceCube is operating an MeV neutrino trigger that monitors for an increase in the noise rate caused by an MeV neutrino burst from a Galactic SNe~\citep{}. This information would be transmitted to SNEWs, where it will be time correlated with signals from other MeV neutrino detectors, such as Super-Kamiokande.

The first major enhancements to IceCube are currently being deployed and will add new calibration devices and updated optical modules. These upgrades will dramatically improve low-energy neutrino sensitivity, angular resolution, and photon-detection efficiency, enhancing the observatory's ability to detect neutrinos from nearby supernovae and from softer-spectrum transient sources. 

The next decade promises transformative discoveries in neutrino astronomy, driven by a new generation of observatories. The IceCube-Gen2 facility proposes to expand the instrumented volume by an order of magnitude \citep{Aartsen2021IceCubeGen2}. It will augment the existing facility with an enlarged in-ice optical array, a surface cosmic-ray detector, and a giant radio array for detecting the highest-energy neutrinos. Gen2 will be a wide-band observatory with sensitivity spanning six orders of magnitude in energy, from GeV to beyond EeV. This will allow it to probe the high-energy extension or cutoff of the astrophysical neutrino spectrum, test models of cosmogenic neutrino production, constrain the nature of ultra-high-energy cosmic ray accelerators and resolve the neutrino sky with significantly improved source detection capabilities. If realized, Gen2 will increase astrophysical neutrino detection rates significantly, while improving directional accuracy for high-energy events and providing deeper overlap with gamma-ray observatories and GW triggers. 

IceCube-Gen2 will be part of a growing global effort to build a network of large-scale neutrino telescopes. 
Other key facilities, such as the Cubic Kilometre Neutrino Telescope (KM3NeT) \citep{KM3NeT2025UltraHighEnergy}, are currently deployed and are already taking data with a partial detector configuration in the Mediterranean Sea and in Lake Baikal (Baikal-GVD) \citep{Safronov2021BaikalGVD}. The detection of the highest energy neutrino event ever detected, at 220\,PeV, with the KM3NeT ARCA detector reflects the science potential of future detectors. Additional detectors are planned in the Pacific Ocean (P-ONE)\citep{Agostini2020PONE} and the South Chinese Sea (TRIDENT)\citep{Ye2023TRIDENT}. Other detection techniques aim for the detection of Earth-skimming tau events or of radio signals from the Askaryan effect \citep{Aguilar2021RNOG, AlvarezMuniz2020GRAND, Thompson2023TAMBO, Brown2021Trinity} to expand the energy range to EeV energies.

\subsection{The High-Energy Electromagnetic Fleet}

A fleet of highly capable space-based observatories continues to provide the high-energy monitoring and rapid, multiwavelength follow-up crucial for detecting and characterizing newly discovered transients.

\subsubsection{Neil Gehrels Swift Observatory}

The Neil Gehrels Swift Observatory \citep{swift} has served as the rapid-response workhorse of the TDAMM ecosystem for more than two decades. Swift is a uniquely powerful multiwavelength platform, capable of observing in hard X-rays, soft X-rays, and the UV/optical spectrum with rapid-slewing capabilities. No other mission offers this combination of autonomous triggering, rapid repointing, and multi-band follow-up. Even after more than 21 years on orbit, Swift remains vital for discovery and characterization across TDAMM science. 

Swift has maintained an extremely open and heavily-used ToO program, receiving over 2,100 requests in 2025, many driven by optical surveys and new missions like Einstein Probe. This has made Swift’s ToO program a cornerstone of the multimessenger follow-up community. In recent years, Swift has introduced “Urgency 0” ToOs, requests executed as rapidly as possible by bypassing normal scheduling layers. Enabled by continuous commanding via the Tracking and Data Relay Satellite System (TDRSS) network, this new mode has allowed Swift to respond to external triggers with latencies of seconds.

The Urgency 0 system has achieved an on-target time of 82 seconds following a trigger from the CHIME FRB (Canadian Hydrogen Intensity Mapping Experiment Fast Radio Burst) detector. The mode has been used to perform automated, immediate follow-up of triggers from SVOM and Einstein Probe. The mission has also used the system to employ a novel strategy to slew the Burst Alert Telescope (BAT) into the probable location of an LVK ``Early Warning'' event before the predicted merger time.

Swift's orbit is decaying due to the current solar cycle, with a high likelihood of re-entry in the next 12 months. However, NASA has funded Katalyst Space Technologies to develop a reboost mission, which should launch in June 2026.  Successful implementation would give Swift a reprieve from re-entry, preserving one of NASA’s most productive time-domain assets into a period when the synergy with Rubin and Roman will be especially important. In order to prolong Swift's orbital lifetime and maximise the chances of reboost success, Swift suspended normal science observations on 10th February 2026, transitioning to purely drag-minimisation pointings.

\subsubsection{The Fermi Gamma-ray Space Telescope}

The Fermi Gamma-ray Space Telescope remains one of NASA’s premier discovery engines for high-energy time-domain and multi-messenger astrophysics. With its broad energy coverage and wide field of view, Fermi serves as both a survey instrument and a continuous monitoring of the dynamic gamma-ray sky. Its two instruments, the Large Area Telescope (LAT) and the Gamma-ray Burst Monitor (GBM), provide sensitivity from keV to GeV energies, allowing Fermi to detect prompt emission from explosive transients and track their high-energy evolution over timescales ranging from milliseconds to years. This combination of wide-field coverage, high duty cycle, and long operational baseline has made Fermi an indispensable component of the TDAMM ecosystem.

The LAT’s continuous all-sky survey mode and long-term monitoring capability in the MeV to GeV energy range have provided the identification and regular monitoring of thousands of transient and variable sources. The long-term monitoring of the gamma-ray sky by Fermi LAT played a crucial role in the first association of a high-energy neutrino detected by the IceCube neutrino observatory and the flaring blazar TXS 0506+056 \citep{2018Sci_IceCube-170922A}. LAT observations provided a vast catalog with which to compare neutrino arrival directions, and the long baseline monitoring of the source made it clear that TXS 0506+056 was undergoing its most luminous flaring episode in over a decade. In addition to transient detection, the precision timing of gamma-ray pulsars observed by the LAT effectively constitutes a space-based pulsar timing array, enabling studies of rotational stability, glitches, and long-term spin evolution that complement radio timing efforts \citep{2022Sci_pta}. In addition, the LAT’s decade-plus monitoring of bright blazars has begun to reveal candidate quasi-periodic oscillations in a subset of sources \citep{Penil_2020, Bhatta2022}. Such periodicity may signal the presence of supermassive black hole binaries, providing potential electromagnetic counterparts and precursor candidates for future low-frequency gravitational-wave detections with LISA.

The GBM complements LAT by acting as a near–all-sky monitor, observing roughly 60\% of the sky at any given time with an $\sim$85\% duty cycle. GBM detects hundreds of transients per year in the keV to MeV energy range, including gamma-ray bursts (GRBs), soft gamma-ray repeaters, magnetar giant flares, terrestrial gamma-ray flashes, solar flares, accreting binaries, and other high-energy phenomena. In the multi-messenger era, GBM continues to play a central role in following up gravitational-wave and neutrino alerts through both onboard triggers and more sensitive ground-based targeted searches, with the GBM’s detection of GRB 170817A alongside GW170817 representing a foundational milestone for TDAMM astronomy. The recent lowering of the onboard trigger threshold for short GRBs has enhanced the instrument's sensitivity to nearby neutron star mergers, while blind and externally seeded searches of downlinked data provide sub-threshold detections and upper limits within hours. These sub-threshold search capabilities have been utilized to provide additional high-energy context to recent detections by Einstein Probe, which has uncovered a population of soft X-ray transients that manifest in GBM as weak, untriggered signals, sometimes preceding GBM detections on the order of minutes. 

As time-domain discovery accelerates with facilities such as Rubin and Roman observatories, the number of externally discovered transients falling within the LAT and GBM fields of view will continue to grow. In this high-volume discovery environment, Fermi’s ability to retrospectively search continuous data streams for faint, sub-threshold emission will become increasingly essential for connecting electromagnetic discoveries to high-energy prompt emission and for maximizing the multi-messenger return of the broader TDAMM network.

\subsubsection{The InterPlanetary Network (IPN)}

For nearly five decades, the IPN has provided a unique high-energy transient localization capability by triangulating signals among gamma-ray detectors distributed across spacecraft throughout the solar system. The IPN has spanned over 50 missions and has been instrumental in advancing the discovery and study of GRBs and magnetars. The network relies on interplanetary baselines, utilizing missions far from Earth, predominantly those launched for planetary, heliophysics, and defense purposes, to achieve precise localizations and continuous and contiguous all-sky coverage.

The IPN will play an essential role in maximizing the science return of upcoming surveys by providing high-energy all-sky coverage and localizations capabilities. Combining IPN data with LIGO detections could significantly reduce the time needed for follow-up and counterpart identification. For example, in the case of GW170817, automated IPN could have produced a localization of 60 square degrees within an hour, which was twice the size of the final gravitational wave localization but achieved five times faster. Likewise, because the IPN provides total sky-coverage, a single IPN annulus could associate a GRB detection with a transient seen by the Rubin survey, enabling precise localization of GRBs without targeted follow-up, exceeding the rate of optical GRB positions found following Swift bursts. Finally, IPN localizations for the second brightest gamma-ray burst ever detected ultimately led to the first late-time infrared spectrum of a kilonova using the James Webb telescope. As surveys have begun across and beyond the electromagnetic spectrum, providing complete coverage of events found in those surveys is crucial to understanding cosmic explosions. 

A major modernization effort is currently underway that aims to fully automate the IPN workflows, standardize data formats, and reduce human-in-the-loop processing. The modernization effort will also result in the collation of high-energy transient alert streams from multiple missions, creating a single access point for real-time data from IPN. The IPN team is also taking the lead on developing a high-energy transient name server that will help the community ensure consistent event nomenclature across observatories, reduce ambiguity in transient alerts, and streamline rapid multi-mission follow-up. These efforts will also include the maintenance of a complete prompt GRB catalog (currently 15,000 events) and preservation of legacy data (e.g., Vela, Pioneer Venus Orbiter). The ultimate goal of these efforts is to treat the IPN as a single effective instrument, combining data directly for searches instead of processing information separately after the fact.

The period from 2015-2025 featured the fewest number of planetary detectors ever, with only Mars Odyssey serving as a distant spacecraft, resulting in response times measured in days. By the end of 2026, the network will expand to include four distant spacecraft (BepiColombo at Mercury, Psyche, Mars Odyssey, and the Martian moon explorer mission MMX-MEGANE), which will reduce the median response time for a first annulus to hours, dramatically increasing its utility for TDAMM science. Beyond 2026, the network is poised for a significant increase in detection capabilities. In the long term, the community is exploring the placement of IPN-capable detectors on several deep-space assets, including the Mars Telecommunications Orbiter, which would provide near-continuous communication and arc-minute annuli, along with the Uranus Orbiter and Probe flagship mission and a future solar polar orbiter. However, an IPN node on Mars Telecommunications Orbiter, the best opportunity for the foreseeable future, would require immediate prioritization by NASA. The inclusion of a node well out of the ecliptic plane is especially valuable: by breaking the near-coplanar geometry of the current network, it dramatically improves the triangulation baseline and reduces geometric degeneracies. In combination, the Uranus Orbiter and a high-inclination solar polar orbiter could provide the three-dimensional leverage needed to achieve sub-arcsecond localizations from gamma rays alone.

\subsubsection{Einstein Probe (EP)}

Launched in January 2024, EP is revolutionizing soft X-ray astronomy with its wide-field monitoring capabilities. Using novel lobster-eye optics, its Wide-field X-ray Telescope (WXT) covers an enormous 3,600 square degree field of view, or about 10\% of the sky.  The WXT operates in the 0.5–4 keV range and provides an angular resolution of $\sim$5 arc minutes (FWHM) and a sensitivity of $\sim$1 mCrab at 1 ks. This survey instrument is complemented by two higher-resolution Follow-up X-ray Telescopes (FXT). Each FXT utilizes a Wolter-1 optics and a pn-CCD detector to achieve a resolution of 24 arcseconds (HPD, on-axis) in the 0.3 to 10 keV range over a field of view of $\sim$1 degree. The spacecraft can autonomously slew to a new transient detected by the WXT and begin follow-up observations. Public alerts are issued via the General Coordinates Network (GCN) within minutes of an onboard trigger.

EP’s unique wide-field monitoring capabilities have opened a new window into the X-ray sky. Since launch, EP has detected over 6,000 known sources. It is responsible for a massive increase in observed transients: before EP, only 30-40 fast X-ray transients were known, but EP is detecting around 80 fast X-ray transients per year, totaling approximately 120 detected so far. Einstein Probe has also detected $\sim$40 GRBs, and has observed soft X-ray emission preceding the GRB which lasts significantly longer than the gamma-ray emission. Einstein Probe has also found several Tidal Disruption Event (TDE) candidates, including three that are jetted. Two candidates are associated with possible Intermediate Mass Black Holes (IMBH). One event, EP250702a, is a candidate IMBH-jetted TDE that may be disrupting a white dwarf. Another TDE (EP240222a) occurred 35 kpc away from the nucleus of a large galaxy, caused by an IMBH.

EP has associated fast X-ray transients with supernovae, such as EP240414a, which was linked to a Type Ic-BL supernova at z=0.4. This event, which had no associated GRB, is potentially a new type of fast X-ray transient caused by a weak relativistic jet interacting with a shell surrounding the progenitor star. The mission also provides extensive observations for the X-ray binary community, detecting outbursts from known and new objects, including neutron stars and black hole candidates. EP can detect nova outbursts in Be + White Dwarf binary systems in X-rays, even when they are outshone by the bright optical Be star.

\subsubsection{Space Variable Objects Monitor (SVOM)}

Launched in June 2024, SVOM is designed as an integrated system comprising a satellite with four onboard instruments and three dedicated ground-based telescopes. Its four onboard instruments include ECLAIRs and the Gamma-Ray Monitor (GRM), covering energy ranges of 4–150 keV and  15–5000 keV, the Micro-channel X-ray Telescope (MXT), a lobster eye optic covering 0.2–10 keV, and the Visible Telescope (VT), covering optical wavelengths. 

The primary scientific goal of SVOM is not focused on maximizing the number of detections, but rather on providing a complete characterization of GRBs and their early afterglows in great detail. SVOM achieves this through its anti-sun pointing orbit, which ensures that any transient it detects is on the night side of the Earth, enabling immediate follow-up observations from its ground segment. This system relies on fast communication to quickly transmit position data to the ground. Alerts are disseminated via GCN Notices, which are sent out in real time, typically within 15 seconds, via a network of approximately 50 VHF receiving stations worldwide.  ToO requests are uplinked to the spacecraft via the Chinese BeiDou system, typically on the order of a few minutes. 

As of October 1, 2025, SVOM had detected 184 GRBs onboard (159 with GRM, 65 with ECLAIRs) and dozens of Galactic X-ray bursts detected offline. Approximately 40\% of ECLAIRs-detected GRBs have a measured redshift. These detections include GRB 250314A, which was detected at z$\sim$7.3. Subsequent observations by the James Webb Space Telescope (JWST) showed that this event is compatible with an exploding star (a supernova) in the epoch of reionization. 

The SVOM team has been actively working with the Swift and Einstein Probe missions to ensure that fast X-ray follow-up of SVOM detected GRBs and in turn SVOM contributes optical follow-up for GRBs detected by Swift or Einstein Probe.

\subsubsection{The StarBurst Multimessenger Pioneer}

The StarBurst Multimessenger Pioneer is a highly sensitive, wide-field gamma-ray monitor optimized to detect the prompt emission from compact binary mergers \citep{Kocevski2024StarburstMission}. The instrument comprises 12 CsI(Tl) scintillation detectors that provide sensitivity over a 30–2000 keV energy range and an instantaneous field of view covering nearly the entire unocculted sky ($\sim$8 sr). With more than five times the effective area of Fermi-GBM, StarBurst is expected to detect approximately 200 short GRBs and roughly 800 long GRBs per year \citep{Woolf2023StarBurstSensorHead}. Its large effective area and wide sky coverage are expected to enable several additional joint gravitational wave–short GRB detections annually during the O5 observing run. Rapid communication of onboard detections and localizations will be achieved via TDRSS, enabling low-latency alert dissemination to the ground on timescales comparable to Fermi and Swift.

As one of NASA’s first Pioneers-class missions, StarBurst is part of a program specifically designed to operate with a higher risk tolerance than NASA's traditional science missions. This model is intended to leverage commercial services and commercial off-the-shelf (COTS) components wherever possible to reduce overall costs and accelerate mission development timelines. The mission is currently targeting a launch in early 2028 to coincide with the start of the Advanced LIGO O5 observing run, ensuring maximal overlap with the next major increase in gravitational-wave sensitivity. StarBurst will launch as a rideshare payload into low-Earth orbit (LEO), where it will complement and extend the gamma-ray burst (GRB) detection capabilities of Swift-BAT and Fermi-GBM.


Beyond its core GW–short GRB science objectives, StarBurst will serve as a key high-energy monitor within the broader TDAMM ecosystem. Its wide-field monitoring capabilities will provide critical coverage of long GRBs, magnetar giant flares, solar flares, terrestrial gamma-ray flashes, and potential gamma-ray counterparts to high-energy neutrinos. Operating in concert with gravitational-wave detectors, IceCube, Rubin, Roman, IPN, and future missions such as COSI, StarBurst will ensure continued gamma-ray coverage of TDAMM transients through the end of the decade.

\subsubsection{The Compton Spectrometer and Imager (COSI)}

Scheduled for launch in late 2027, COSI will be the first space-based MeV gamma-ray imaging spectrometer in more than two decades, restoring a critically missing capability in high-energy astrophysics. Operating in the poorly explored MeV band, COSI is optimized for high-resolution spectroscopy and imaging of nuclear gamma-ray emission, including radioactive decay lines from nucleosynthesis, electron–positron annihilation at 511 keV, nuclear de-excitation lines, MeV-band gamma-ray bursts, and magnetar flares, as well as extended diffuse gamma-ray structures \citep{Tomsick2023}. These measurements address fundamental questions about the origin of the elements, the physics of antimatter in the Galaxy, and the nuclear processes operating in extreme astrophysical environments.

COSI’s spectroscopic sensitivity will open a new observational window on nova and supernova nucleosynthesis, enabling direct measurements of freshly synthesized isotopes, while also probing both early- and late-time nuclear emission from compact-object mergers. In the TDAMM context, COSI will provide diagnostics of merger ejecta composition and energy deposition that are inaccessible at other wavelengths, complementing gravitational-wave observations and broadband electromagnetic follow-up. Although COSI is not designed as a primary burst monitor, it will deliver additional spectral information for transients initially identified by wide-field instruments such as Swift, Fermi, StarBurst, and IPN, substantially enhancing physical interpretation of these events.

COSI will also study nuclear transients events defined by time-variable gamma-ray line emission from radioactive decay and nuclear processes. By capturing the temporal and spectral evolution of these signals, COSI will enable direct, model-discriminating tests of nucleosynthesis pathways and energy-release mechanisms.

\subsection{Radio}

Radio observations will play an increasingly important role in the discovery landscape in the coming years, opening access to transient phenomena that are difficult to detect at other wavelengths. Unlike traditional radio facilities that primarily provide targeted follow-up of externally discovered events, the next generation of radio surveys will act as discovery engines in their own right. They will identify a broad range of radio transients, from fast radio bursts and pulsars to explosive and accretion-driven events. As optical and infrared surveys expand the transient discovery rate, radio surveys will provide a complementary view of the dynamic sky by sampling different emission mechanisms, timescales, and source environments.

\subsubsection{Square Kilometre Array (SKA)}

The Square Kilometre Array Observatory will be a transformative facility for radio time-domain discovery. SKA-Low, located in Western Australia, will operate from 50--350 MHz using aperture-array stations, while SKA-Mid, located in South Africa, will operate from 0.35--15.4 GHz using dish antennas. Together, these facilities will provide a major expansion in sensitivity, frequency coverage, survey speed, and angular resolution relative to current radio survey instruments \citep{Braun_2019}. SKA-Low will consist of 512 stations containing 131,072 antennas, with baselines extending to 74 km, while SKA-Mid will ultimately include 197 dishes, including the existing MeerKAT array, with baselines extending to 150 km. These configurations combine a dense central collecting area, which is especially valuable for pulsar and transient searches, with long baselines that provide the angular resolution needed to localize compact radio sources and connect them to astrophysical environments.

SKA will open a parameter space that is currently inaccessible. SKA-Low will be sensitive to coherent low-frequency emission from pulsars, magnetized stellar systems, plasma-interaction events, and potentially delayed low-frequency emission from explosive transients \citep{Fender_2015}. SKA-Mid will provide access to the broad GHz-frequency regime central to fast radio burst searches, pulsar discovery, compact synchrotron transients, radio supernovae, tidal disruption events, and accretion-driven variability \citep{Hashimoto_2020}. The combination of high-time-resolution search modes and repeated imaging surveys will allow the SKA to discover both fast radio bursts and slower radio variables, making radio astronomy a leading discovery channel in the coming years.

\subsubsection{MeerKAT and MeerKAT+}

MeerKAT provides the immediate path from the current southern radio survey landscape to the SKA discovery era. The array already consists of 64 highly sensitive dishes in South Africa and is one of the most capable radio interferometers currently operating in the Southern Hemisphere. Its location, observing modes, and technical heritage make it a direct precursor to SKA-Mid, and the existing MeerKAT dishes will be incorporated into the full SKA-Mid array. MeerKAT is already developing the observing strategies, calibration approaches, survey data products, and transient-search techniques that will scale into SKA operations. The upcoming MeerKAT+ expansion will strengthen this bridge by adding SKA-style dishes and extending array performance before full SKA-Mid operations. Additional collecting area will improve sensitivity to faint radio variables and transients, longer baselines will improve angular resolution, and broader receiver coverage will expand the frequency range available for survey and discovery programs. This makes MeerKAT+ an intermediate discovery platform that will improve southern radio transient surveys while preparing the community for the sensitivity, scale, and data-processing demands of SKA-Mid.

\subsubsection{Deep Synoptic Array (DSA)}

The Deep Synoptic Array will provide a complementary northern-hemisphere radio discovery capability. The array is being developed as a wide-field radio ``camera'' operating over roughly 0.7--2 GHz, with a design optimized for rapid, repeated imaging of large areas of the sky. Rather than functioning primarily as a pointed follow-up facility, the DSA is intended to operate as a dedicated survey telescope that repeatedly maps the viewable sky and produces science-ready image products in real time. This makes it a natural radio counterpart to next-generation optical and infrared synoptic surveys.

The DSA's discovery potential comes from combining wide-field imaging, high sensitivity, and fast-transient localization in a single system. Its repeated imaging surveys will be sensitive to slowly evolving radio transients and variables, including synchrotron emission from explosive events, accreting compact objects, active galactic nuclei, and other compact radio sources. At the same time, its high-time-resolution observing modes will enable large-scale searches for fast radio bursts and pulsars. This combination is particularly important for time-domain searches because it connects the fast-transient and slow-transient radio discovery spaces, which are often explored by different instruments and observing strategies. As a result, the DSA will provide both an all-sky radio reference data set and a discovery stream of radio-selected transients that can trigger multiwavelength characterization.

\subsubsection{CHIME and CHORD}

CHIME has already demonstrated the discovery power of wide-field, digitally formed radio surveys, particularly for fast radio bursts. Its stationary architecture and very large instantaneous field of view have enabled it to detect large numbers of FRBs and to establish the population-level properties of repeating and apparently non-repeating sources. The next stage of this discovery program is the CHIME/FRB Outriggers project, which adds widely separated stations in British Columbia, California, and West Virginia. These baselines will greatly improve localization precision, allowing many more bursts to be associated with host galaxies, local environments, and potential multiwavelength counterparts.

The Canadian Hydrogen Observatory and Radio-transient Detector will extend this approach into a more sensitive next-generation discovery instrument. CHORD is planned to operate over roughly 0.3--1.5 GHz, combining a large central array with remote outrigger stations. This architecture is designed to support hydrogen intensity mapping, pulsar science, and radio-transient discovery within a single survey facility. For TDAMM, CHORD will be especially important because it will improve the discovery rate and localization quality of fast radio transients while also providing sensitivity to variable and coherent emission across a broad low-GHz band.

\subsubsection{LOFAR2.0}

Low-frequency radio facilities will play a distinct role in the TDAMM discovery ecosystem because they probe emission mechanisms and propagation effects that are not accessible at optical, infrared, X-ray, or higher-frequency radio wavelengths. LOFAR is already the leading low-frequency radio array, operating across roughly 10--240 MHz with stations distributed across Europe. LOFAR2.0 is a major upgrade intended to extend the facility's scientific reach through improved hardware, digital processing, observing flexibility, and operational reliability. A key capability of the upgrade is the ability to use the low-band and high-band antennas simultaneously, increasing the effective field of view and sensitivity of the stations.

For time-domain discovery, low-frequency radio emission can reveal coherent bursts, magnetospheric emission, plasma interaction, delayed emission from explosive events, and propagation signatures such as dispersion, scattering, and Faraday rotation. These effects are especially relevant for pulsars, fast radio bursts, stellar radio bursts, space-weather-related phenomena, and compact objects embedded in dense or magnetized environments. Improved automation and data-handling capacity will make LOFAR2.0 better suited for systematic transient searches and rapid classification of variable low-frequency sources. In the broader TDAMM ecosystem, LOFAR2.0 will provide a northern low-frequency discovery capability that complements SKA-Low in the Southern Hemisphere.

\subsubsection{Murchison Widefield Array (MWA)}

The Murchison Widefield Array will continue to serve as an important southern low-frequency discovery facility and a precursor to SKA-Low. Operating from roughly 70--300 MHz in Western Australia, the MWA combines a very wide field of view with electronic pointing and flexible observing modes. These characteristics make it well suited to monitoring large areas of the sky for low-frequency variability, coherent bursts, solar and heliospheric events, and propagation effects through the ionosphere and interplanetary medium. Its location at the Murchison Radio-astronomy Observatory also places it in the same radio-quiet environment that will host SKA-Low.

Although the MWA predates the next generation of radio facilities, it remains highly relevant to the upcoming discovery ecosystem because it provides operational experience, survey data, calibration methods, and transient-search techniques directly applicable to SKA-Low. The low-frequency sky is technically challenging: ionospheric variability, bright diffuse foregrounds, source confusion, and radio-frequency interference all complicate transient discovery. Continued use of the MWA will help develop the observing strategies and analysis pipelines needed to identify real astrophysical variability in this regime. As SKA-Low comes online, MWA discoveries and methods will provide an important foundation for scaling low-frequency transient astronomy to much higher sensitivity and survey speed.



\subsubsection{ASKAP}

ASKAP shares a site with the MWA in Western Australia and was also built as a precursor to the Square Kilometre Array. The array consists of 36 12-m dishes at CSIRO's Murchison Radio-astronomy Observatory, with baselines extending to roughly 6 km. Its defining capability is the use of phased-array feeds on each antenna, which form multiple simultaneous beams and provide an instantaneous field of view of about 30 square degrees. This gives ASKAP an unusually high survey speed for an interferometric radio telescope, allowing it to map large areas of the southern sky while retaining the angular resolution needed to identify and localize compact radio sources.

For TDAMM, ASKAP is especially important because it bridges wide-field radio survey discovery and rapid transient localization. Its survey programs include continuum, spectral-line, polarization, and time-domain searches, and its rapid wide-field imaging capability has already demonstrated the value of radio surveys as discovery engines rather than only follow-up facilities. The Rapid ASKAP Continuum Survey has produced a large-area reference map of the radio sky, while time-domain programs such as VAST provide a framework for identifying variable and transient radio sources on timescales from seconds to years. These capabilities are well matched to synchrotron-emitting explosive transients, accreting compact objects, active galactic nuclei, stellar radio bursts, and other variable radio sources that may be missed or poorly characterized at optical, infrared, or high-energy wavelengths.

ASKAP has also played a major role in fast radio burst discovery and localization through the Commensal Real-time ASKAP Fast Transient survey. CRAFT uses ASKAP's wide field of view and real-time processing to search for short-duration radio bursts and trigger high-time-resolution data capture for localization and characterization. This has enabled ASKAP to localize FRBs to arcsecond precision and associate them with host galaxies, turning FRBs from a population-discovery problem into a multiwavelength and multi-messenger follow-up problem. In the broader TDAMM ecosystem, ASKAP provides a southern wide-field radio discovery capability that complements CHIME/CHORD at low-GHz frequencies, DSA in the northern sky, and SKA-Mid as the next-generation successor facility.

\newpage
\section{TDAMM Infrastructure and Tools} \label{sec:intrastructure}
This section shifts from the rapidly expanding discovery landscape to the infrastructure that enables TDAMM astrophysics. As the discovery rate of new transients accelerates, such infrastructure will become increasingly mission critical to maximizing the scientific return of new and existing facilities. Previous TDAMM workshops called for coordinated investments in archives, real-time alert systems, common data models, and analysis platforms that span agencies and messengers, as well as sustained support for the software and people who build and maintain them. NASA and NSF programs responded with a suite of initiatives: the modernization of the GCN into a scalable event distribution backbone and the development of additional community alert systems including SCiMMA; revitalization of long-standing networks like SNEWS; community brokers and marshalling platforms designed for Rubin-scale alert streams; and NASA’s ACROSS pilot, which aims to provide a coordination layer for NASA's space-based TDAMM assets and beyond.

Together, these efforts form an emerging TDAMM ``stack'' spanning alerts, classification, follow-up coordination, and analysis. This section provides an overview of this emerging ecosystem, outlining how these tools will help maximize the scientific return of new and existing facilities and providing the context for a discussion of the infrastructure challenges still faced by the community. 

\subsection{Alert and Messaging Systems}\label{subsec:alerts}

\subsubsection{GCN}\label{subsubsec:GCN}

The GCN\footnote{\url{https://gcn.nasa.gov}} remains the backbone of NASA’s real-time high-energy and multimessenger alert distribution. In recent years, GCN has undergone a major modernization effort, transitioning from the legacy socket, VOEvent, and email distribution system to a fully JSON-over-Kafka architecture designed for the LSST and IGWN eras. Kafka’s topic-based subscription model allows users to receive only the streams relevant to their science, enabling high-throughput, machine-readable, and fault-tolerant alert delivery. GCN Notices now provide normalized, structured metadata for transient events from NASA, NSF, and international partners, while Circulars, the community’s long-used channel for communicating scientific updates, have also been rebuilt with rich markdown support, editable submissions, self-service subscriptions, and powerful search tools. Together, they form a unified and increasingly modern interface for both automated systems and human-driven coordination.

\subsubsection{SCIMMA Messaging}\label{subsubsec:SCIMMA}

The NSF-funded Scalable Cyberinfrastructure for Multimessenger Astrophysics (SCIMMA)\footnote{\url{https://scimma.org}} initiative complements GCN with a cloud-native, community-oriented messaging layer. Its Hopskotch service is a flexible pub/sub system capable of distributing alerts, catalogs, light curves, spectra, and even FITS files at scale. Built on top of Hopskotch, HERMES introduces structured message types, a graphical message browser, and a well-defined API, allowing researchers to publish and ingest photometry or other data products in standardized formats. Importantly, HERMES can simultaneously transmit alerts to both GCN and the Transient Name Server (TNS), creating a seamless, multi-agency communication pipeline. Together, GCN and SCIMMA form the foundational alert architecture for next-generation TDAMM operations, enabling low-latency, cross-mission messaging at the scale required for Rubin and IGWN.

\subsection{Brokers and Marshalls} \label{subsec:Brokers}
The dramatic increase in alert volume from upcoming wide field surveys necessitates a sophisticated data infrastructure to bridge the gap between observatories and scientists. Alert brokers and marshals provide this essential function. Although closely linked, brokers and marshals play distinct and complementary roles. Brokers serve as the initial clearinghouse, ingesting raw alert streams from surveys to filter, cross-match with archival catalogs, and add value through machine learning classifications. While marshals act as software stacks for visualization, candidate vetting, human-in-the-loop refinement, modeling and analysis of targets and data, and the orchestration of follow-up observations and telescope triggering.

\subsubsection{Broker Systems}

\begin{table}[h]
    \label{Table:brokers}
    \centering
    \caption{Brokers that have been selected to receive the full Rubin-LSST stream of alerts.}
    \begin{tabular}{|>{\raggedright\arraybackslash}p{3cm}|>{\raggedright\arraybackslash}p{5cm}|>{\raggedright\arraybackslash}p{8cm}|}

        \hline
        \textbf{Broker} & \textbf{Lead Institution/Region} & \textbf{Primary Science Focus / Strength} \\
        \hline
        ALeRCE & Chile (University of Chile/MAS) & Rapid classification of transients (SNe, variable stars) using machine learning. \\
        \hline
        AMPEL & Germany (DESY/Humboldt) & Modular analysis platform for real-time multi-messenger astronomy. \\
        \hline
        ANTARES & USA (NOIRLab) & High-speed filtering and cross-matching with massive archival catalogs. \\
        \hline
        Fink & France (CNRS/IJCLab) & Community-driven broker built on Apache Spark; strong focus on multi-messenger and solar system. \\
        \hline
        Lasair & UK (Edinburgh/Oxford/Belfast) & User-friendly platform for creating custom filters and managing transient ``watches.'' \\
        \hline
        Pitt-Google & USA (University of Pittsburgh) & Uses Google Cloud infrastructure (BigQuery/PubSub) for massive scalability and easy user access. \\
        \hline
        BOOM / Babamul & USA (Caltech) & Designed as a lightweight, high-performance filtering service. \\
        \hline
    \end{tabular}
\end{table}

Seven brokers\footnote{\url{https://rubinobservatory.org/for-scientists/data-products/alerts-and-brokers}} were selected to receive the complete LSST (Legacy Survey of Space and Time) alert stream directly from Rubin. Each has a different scientific focus and provides various filtering and classification tools. Table \ref{Table:brokers} summarizes these brokers and their specific areas. Two additional brokers have been officially recognized as ``downstream" partners. They do not ingest the full alert stream but instead receive a pre-filtered subset from one of the full-stream brokers above to focus on niche science cases. These include:
\begin{itemize}
    \item \textbf{SNAPS}\footnote{\url{https://snaps.nau.edu}} (Solar System Notification Alert Processing System): Specifically designed to identify and classify moving objects (asteroids, comets) within our solar system.
    \item \textbf{POI}\footnote{\url{https://poibroker.uantof.cl}} (Point of Interest): A downstream broker associated with ANTARES, focused specifically on variable star research and feature calculation.
\end{itemize}

The Burst \& Outburst Observations Monitor (BOOM)\footnote{\url{https://www.ztf.caltech.edu/ztf-boom-babamul.html}} is an example of a U.S. based alert broker designed to address the massive data demands of the upcoming Vera C. Rubin Observatory era. BOOM is a high-performance software stack developed to succeed the legacy Kowalski system used by ZTF. While Kowalski reached its performance limits at approximately one million alerts per night, BOOM is specifically designed to handle the 10 to 20 times increase in volume expected from the Rubin’s alert streams. To achieve this, BOOM is built using Rust instead of Python, which significantly reduces CPU and memory usage while enabling higher levels of parallelization and lower latency. Its architecture is modular and distributed, utilizing specialized ``workers'' for alert ingestion, database management, machine learning, and filtering that can scale independently and horizontally based on computational load.

A major innovation within BOOM is the ``no-code'' filter builder UI, which allows astronomers to construct complex custom filters using a drop-down menu interface rather than writing direct database queries. This interface addresses a significant technical barrier found in legacy systems where users were required to have extensive, unrealistic levels of database-specific knowledge to filter data. BOOM also specializes in multi-survey processing, enabling it to simultaneously ingest and cross-match alert streams from ZTF, Rubin, and other instruments like DECam and LS4. This capability allows for continuous data enrichment, where objects from different surveys are matched by position to combine their light curves in real-time, providing better constraints on the rate of evolution for new transients. Classification is further enhanced by the AppleCiDEr multimodal machine learning framework, which averages results from various neural networks analyzing photometry, image cutouts, metadata, and spectra. These automated discovery tools, including the BTSbot, are fully integrated into the pipeline to trigger rapid follow-up observations through marshals like SkyPortal.

While BOOM refers to the underlying software stack, Babamul is its public-facing implementation as a publicly accessible Rubin alert broker. Babamul serves the scientific community by providing a set of static, phenomenological filters derived from over seven years of ZTF operational experience. Filter results are delivered with low latency via disjoint Kafka topics, and a dedicated documentation page provides auto-generated code to help researchers subscribe to specific data streams. The system is designed to be the primary entry point for users of all technical levels, offering a feature-rich front-end that facilitates deeper data analysis and seamless integration with existing marshals.

\subsubsection{Marshals}\label{subsubsec:marshalTOM}
TOM systems, also known as ``Marshals'', function as mission operations centers for science teams. Once a broker identifies a candidate of interest, the data is passed to a marshal which provides the cyberinfrastructure for science teams to manage their specific response. They integrate alerts, catalogs, follow-up planning, data ingestion, visualization, and team-based workflows into a single environment. The platforms are also capable of performing science-specific analysis and modeling.

There are two primary public software packages currently used by the community to build these systems:

\begin{itemize}
    \item \textbf{TOM Toolkit}\footnote{\url{https://lco.global/tomtoolkit/}}: Highly customizable; provides data visualization tools and built-in modules to interface with astronomical data services and telescope facilities; offers APIs to facilitate interfaces with user-preferred software; features a public Slack workspace for direct developer support \citep{street18}.
    \item \textbf{SkyPortal}\footnote{\url{https://skyportal.io}}: Designed to interface with brokers and telescopes; provides data visualization tools; used to build the Fritz system \citep{skyportal}. 
\end{itemize}

The diverse range of implementations of these packages illustrate the flexibility of these systems across different scientific domains and operational scales. Examples include the Supernova Exchange 2 (SNEx 2)\footnote{\url{http://supernova.exchange}}, which manages the Global Supernova Project by focusing on human-vetted workflows with interactive plots for photometry and spectroscopy \citep{Pellegrino2024, Howell2017AAS}. In contrast, the Black Hole TOM\footnote{\url{https://bh-tom2.astrouw.edu.pl}} coordinates a large ``Pro-Am'' collaboration of professional and citizen scientists, utilizing a background service to reduce data from 138 telescopes worldwide in real time \citep{Mikolajczyk2025}. Other specialized examples include GOATS\footnote{\url{https://goats.readthedocs.io/en/latest/index.html}} (Gemini Observing and Analysis of Targets System), which integrates NOIRLab tools and ``Dragon'' reduction packages, and the Follow-up Observations of Moving Objects (FOMO)\footnote{\url{https://github.com/lsst-sssc/fomo}}, which is tailored for the non-sidereal challenges of tracking comets discovered by the Rubin Observatory. Some systems, such as the Microlensing Observing Platform, even implement entirely automated workflows that perform real-time modeling to update target priorities without human input.

\subsection{Coordination \& Planning Initiatives}

\subsubsection{ACROSS}

NASA’s Astrophysics Cross-Observatory Science Support (ACROSS)\footnote{\url{https://science.data.nasa.gov/data-sites/across}} initiative is transforming isolated mission information and activities into fleet-wide capabilities that complement and enrich the broader TDAMM ecosystem. An outcome of the PhysCOS TDAMM strategic study \citep{2024FrASS..1101785H}, ACROSS provides a growing suite of open-source tools and data services and a dedicated staff who streamline multi-facility interactions, enabling mission teams and observers to rapidly respond to fleeting events beyond the grasp of any single facility. 
ACROSS capabilities span four complementary areas:

\begin{itemize}
    \item Science Feasibility Tools: An open-source API, web application, and Python client that aggregates mission scheduling, visibility, and pointing constraints from missions such as Swift, HST, JWST, IXPE, NuSTAR, and Fermi, with additional missions in progress. This allows researchers to query multi-mission observability from a single standard interface, enabling both manual assessment and automated planning workflows critical for rapid TDAMM follow-up. 
    
    \item Standardized Target-of-Opportunity (ToO) Resources: An open-source ToO API, web application and toolkit designed to provide missions – particularly cost-capped and extended phase missions with limited resources – with standardized, first-class capabilities for receiving, processing, and executing ToOs. This allows researchers to have enhanced visibility into multi-mission follow-up planning and leverages ACROSS science feasibility tools to further streamline the end-to-end ToO process for coordinated observations. 
    
    \item Coordination Services: Dedicated staff who serve as the interface between observers and mission operations teams for approved multi-facility observing programs, reducing coordination burden on individual missions while building institutional knowledge of cross-fleet operations that drives continuous improvements to follow-up response times and coordination workflow efficiency. 
    
    \item Community Engagement $\&$ Innovation Pathways: Support for the TDAMM Workshop series, a web portal providing centralized access to TDAMM tools, mission information and tutorials, and funding pathways that support researchers in extending ACROSS infrastructure and advancing TDAMM science investigations. 

\end{itemize}

ACROSS marks a fundamental shift in how NASA approaches time-domain and multi-messenger operations, building common interoperable science support ground infrastructure and transferable knowledge to ensure that NASA’s fleet operates as a progressively more cohesive and responsive observing system. 

\subsubsection{HEROIC}
SCIMMA’s Hopskotch Enablked Real-Time Observatory Information and Coordination (HEROIC)\footnote{\url{https://heroic.scimma.org}} tool addresses a long-standing challenge: telescope availability information is scattered across hundreds of webpages. HEROIC consolidates this into a single interface, allowing users to enter a target and immediately determine which telescopes, starting with NOIRLab and LCO, can observe it\citep{Howell2026}. This functionality is especially valuable in the minutes following a large GW localization, when actionable visibility information must be obtained rapidly.

\subsubsection{AEON+}

The efficacy of TOM systems is significantly enhanced by their integration with time-domain friendly observing infrastructure, such as the Astronomical Event Observatory Network (AEON)\footnote{\url{https://aeonplus.github.io}} Plus collaboration \citep{McCully2026}. AEON Plus brings together international facilities, including Gemini, Keck, and the Swift mission, to provide queue-scheduled or Target of Opportunity (ToO) operations that can be accessed through programmatic interfaces. By allowing TOM systems to request observations directly through code, these networks facilitate automated and scalable follow-up of rapidly evolving transients. This synergy between management software and global telescope networks enables a seamless transition from a survey alert to a fully characterized astrophysical discovery.

\subsection{Specialized Systems for High-Impact Events}

\subsubsection{TROVE}
Automated association and vetting of electromagnetic counterparts with poorly localized multi-messenger events (e.g., gravitational waves, neutrinos, gamma-ray transients, etc.) has proven challenging given the large localization regions and high volume of potential counterparts (e.g., \citet{Kilpatrick2021, Rastinejad_ApJ_2022}). The Tool for Rapid Object Vetting and Examination (TROVE)\footnote{\url{https://github.com/astro-trove/trove/}} fills this existing gap in publicly available software tools by ingesting multimessenger alerts and cross-matching with electromagnetic transients and public catalogs to compute a ranked probability score based on:

\begin{itemize}
    \setlength{\itemsep}{1pt}
    \setlength{\parskip}{1pt}
    \item 2D localization agreement
    \item Automated association of each electromagnetic transient with a host galaxy, providing a distance estimate and allowing for a calculation of the compatibility with the GW posterior
    \item Asteroid/variable star exclusion
    \item Photometric consistency with the expected electromagnetic counterparts 
\end{itemize}

The existing infrastructure is focused on the follow-up of gravitational waves. Future versions will incorporate neutrino, gamma-ray, and X-ray candidate scoring, and expanded machine-learning classifiers. TROVE’s web interface and API, now nearing a v1.0 release, are being designed for integration with TOM systems, ACROSS, and AEON workflows.

\subsubsection{SNEWS}
The SuperNova Early Warning System (SNEWS)\footnote{\url{https://snews2.org/}} is a global network that monitors neutrino detectors for coincident signals from a Milky Way (or nearby) core-collapse supernova \citep{Kara2024J}. Because neutrinos escape the stellar core hours before electromagnetic emission, SNEWS can provide the earliest possible alert for one of the most scientifically valuable events in astrophysics. SNEWS 2.0, now in development, modernizes the decades-old framework by rewriting the system in Python using SCIMMA’s Hopskotch messaging. Its goals include:

\begin{itemize}
    \setlength{\itemsep}{1pt}
    \setlength{\parskip}{1pt}
    \item Promptness ($<$1 hour latency)
    \item Positivity (false alarms $<$ 1 per century)
    \item Pointing (using triangulation and HEALPix maps analogous to GW localization)
\end{itemize}

Upon activation, SNEWS 2.0 will become one of the highest-priority alert channels in all of TDAMM, triggering rapid response across the global network of space- and ground-based observatories.

\subsubsection{NED-GWF}

Our understanding of GW events is greatly enhanced by identifying and studying their electromagnetic counterparts and host galaxies. The NASA/IPAC Extragalactic Database (NED)\footnote{\url{https://ned.ipac.caltech.edu/}} contains a rich and constantly growing compendium of galaxies, and provides essential data and tools to facilitate astrophysics research and exploration. NED's Gravitational Wave Follow-up (NED-GWF)
service aids in the searches for electromagnetic counterparts to GW events by providing host galaxy candidates inside an event's localization volume with several prioritization metrics, including star formation rate, stellar mass, specific star formation rate, etc. 

Within minutes after the IGWN issues an alert using the GCN, NED-GWF reacts by cross-matching in 3D the event's localization map with the galaxies in the corresponding volume and the following results are provided: 1) a list of all galaxies in the event's 90\% probability volume with prioritization metrics; 2) a  web page containing event information, a download feature for the full galaxy list, and visualizations (e.g., an all-sky image with the event's probability contours and locations of galaxies); and 3) the top 20 galaxies sorted by the joint 3D and WISE W1 luminosity probabilities (a proxy for stellar mass), available on the web page for all events and on GCN circulars for high-significance events. The galaxy lists are also available via API
on the same minute timescales for incorporation into other TDAMM workflows.  

NED-GWF uses a subset of NED containing over 2 million galaxies in the local Universe (D$\leq$1000 Mpc) called the Local Volume Sample \citep[NED-LVS;][]{cook23}
, where we have performed additional vetting, characterized the sample, derived physical properties, and quantified the completeness. The completeness relative to near-IR luminosities (which traces a galaxy's stellar mass) is roughly 100\% at D$<$30 Mpc and remains moderate (70\%) out to 300 Mpc. For brighter galaxies ($>L^{*}$), NED-LVS is $\sim$100\% complete out to 400 Mpc. For more details about NED-LVS see \citep{cook23}. These value-added aspects make NED-LVS a robust and significantly complete galaxy catalog appropriate for follow-up efforts to TDAMM events and other astrophysics research. The April 2026 version of NED-LVS includes for the first time fiducial galaxy angular diameters, which can be critical in time-domain astronomy to help identify the hosts of both poorly- and well-localized transient events. 

\newpage
\section{Critical Challenges and Findings} \label{sec:challenges}

The workshop identified a set of interlocking challenges faced by the TDAMM community that broadly fall into three categories: (1) infrastructure limitations associated with the unprecedented surge in alert rates and systemic barriers in data standardization and coordination that impede efficient cross-facility collaboration, (2) policy barriers that make comprehensive data collection difficult and often disfavor proposal selections that focus on rare, scientifically unique events, (3) growing operational vulnerabilities as key high-energy and multimessenger assets age, face uncertain upgrade paths, or lack long-term continuity plans. 

Together, these issues highlight the urgent need for sustained investment in scalable data infrastructure, modernized observatory operations, integrated policy mechanisms and community observing plans for high-priority sources, and next-generation space- and ground-based capabilities to ensure the TDAMM ecosystem can fully capitalize on the discoveries ahead.
\subsection{Infrastructure Limitations}
Facilities such as the Vera C. Rubin Observatory will generate alert streams orders of magnitude larger than those produced by current surveys, fundamentally changing how transient discovery and follow-up must be conducted. Successfully exploiting this discovery space will require alert generation, filtering, classification, and dissemination to operate at unprecedented scale. At the same time, a persistent, systemic challenge across the TDAMM ecosystem is the absence of standardized, centralized, and interoperable information flows. As discovery rates accelerate and the number of participating facilities grows, the lack of unified data structures and coordination mechanisms increasingly undermines the community’s ability to respond efficiently to transient events. Several issues, long recognized by the community, remain largely unresolved. These include:


\subsubsection{Alert Volume and Broker Capacity}

The transition to next-generation wide-field observatories has created a fundamental big-data challenge for TDAMM astrophysics. The Vera C. Rubin Observatory is expected to produce up to 10 million alerts per night, representing a 10–20× increase over current surveys such as the Zwicky Transient Facility (ZTF). Each alert must be generated, transmitted, and processed within roughly 60 seconds, forcing the community to sift through an enormous real-time data stream to isolate rapidly evolving events that are among the most impactful for multi-messenger science. This combination of massive volume and strict latency requirements is rapidly becoming one of the defining bottlenecks of the TDAMM era.

Filtering this deluge is the sole responsibility of alert brokers, which must operate at industrial scale. Existing systems, such as Kowalski, reach hard performance limits at roughly one million alerts per night for a single survey, far below Rubin’s expected load. This has driven the development of new architectures that emphasize high throughput, parallelism, and efficient memory use, such as the Rust-based BOOM broker. While falling short of mandating a particular language, the pattern the community endorsed at the workshop is a compiled high-performance core, Python-facing science interfaces (as Python remains the community's lingua franca for analysis), and filter-building interfaces that open the alert stream to users without database expertise. 
Achieving true Rubin-scale performance through such a system will require sustained engineering effort, rigorous optimization, and long-term operational support.

\subsubsection{Event Classification}

Coordinating follow-up across multiple surveys adds another layer of complexity. Cross-matching detections and combining light curves from heterogeneous facilities is challenging because data products differ widely in format, cadence, noise properties, and even basic schema (e.g., PSF photometry vs. forced aperture photometry). This heterogeneity complicates downstream analysis pipelines and increases the risk of misclassification. At the same time, missions like the Roman Space Telescope expect to be inundated with false positives, especially during high-cadence transient programs. Robust, scalable machine-learning classifiers, such as the AppleCiDeR deep-learning framework and other vetted models, will be essential for distinguishing genuine astrophysical transients from bogus detections or low-value candidates, enabling efficient use of scarce spectroscopic and follow-up resources. Workshop discussions emphasized that these production-level models will require sustained support for development, validation, transparency, performance characterization, and long-term maintenance.

\subsubsection{Fragmented Observatory Planning Information}

Coordinated, time-critical follow-up across the TDAMM ecosystem is fundamentally constrained by the lack of standardized, centralized, and machine-readable information describing observatory schedules, resource availability, and target observability. Scheduling data are currently published in highly heterogeneous formats—ranging from free-form text and static tables to bespoke web APIs and proprietary systems—forcing both astronomers and coordination tools to rely on ad hoc parsing, manual interpretation, and mission-specific expertise. At the same time, information about telescope and instrument status, including which facilities are operational, which instruments are mounted, and what observing modes are available, is fragmented across dozens of independent web pages with no common schema or authoritative access point. During rapidly evolving transient events, this fragmentation introduces critical delays as observers must manually assemble an incomplete and often outdated picture of which resources can respond.

These challenges are further compounded by the absence of a unified framework for calculating target visibility and practical observability across multiple facilities. Existing visibility tools vary widely in implementation and accessibility, relying on inconsistent assumptions, outdated software environments, or proprietary interfaces that limit interoperability and automation. Moreover, key concepts such as observability, which captures practical scheduling constraints (e.g., Swift’s ability to respond within seconds versus HST’s typical 48-hour turnaround), remain difficult to quantify in a standardized way. The lack of a unified visibility service complicates coordinated planning and reduces the efficiency of follow-up efforts. As a result, even when visibility is nominally achievable, the feasibility of executing a timely observation often remains unclear. Together, these deficiencies significantly increase coordination overhead, reduce the effectiveness of automated planning systems, and limit the community’s ability to respond efficiently to high-priority TDAMM events.

\subsubsection{Observatory Automation Limitations}

Many facilities still lack modern automation capabilities, such as queue scheduling, rapid Target-of-Opportunity modes, and programmatic interfaces for submitting observation requests, that are essential for seamless integration with TOM systems. In parallel, the broader ecosystem remains underdeveloped in automated, real-time infrastructure for event triage, prioritization, and cross-messenger classification, relying heavily on manual workflows that do not scale to the alert volumes expected from Rubin, LVK, IceCube, and future missions. This shortfall has created a growing mismatch between rapidly expanding discovery rates and the community’s ability to respond, leaving the TDAMM enterprise fundamentally response-limited unless sustained investment is made in automation and observatory-side modernization.

\subsubsection{Data Management Challenges}

The TDAMM ecosystem continues to be limited by fundamental shortcomings in how data are managed, integrated, and disseminated across facilities, wavelengths, and messengers. Core technical challenges, including the absence of interoperable data lakes, standardized metadata conventions, shared data models, and cross-archive compatibility, remain largely unresolved as new observatories such as Rubin, Roman, and others come online. To compound the issue, there is no comprehensive, centralized repository or discovery layer for follow-up observations, forcing researchers to manually reconstruct the observational history of individual transients by searching across numerous GCN circulars, mission archives, and the published literature. This fragmentation introduces inefficiencies, increases the likelihood of missing critical datasets and can also lead to redundant or duplicative observations as teams unknowingly repeat work already carried out elsewhere. Collectively, these effects can significantly hinder reproducibility, cross-comparison, and synthesis in multimessenger science. Without coordinated investment in standardized data products, unified query and access interfaces, and long-term data stewardship, these data management and dissemination gaps will increasingly constrain the scientific return of TDAMM science.

\subsubsection{Findings: Infrastructure Limitations}

\finding{Agencies should treat the alert generation, distribution, brokering, and classification tools as mission-critical scientific infrastructure}: This includes sustained operational funding for scalable alert distribution systems (e.g., GCN, SCiMMA), next-generation brokers capable of Rubin-scale throughput, and robust, community-vetted event classification frameworks. These investments should prioritize usability, including low-barrier, user-facing interfaces that enable broad community participation without requiring specialized technical expertise.  These systems should be supported as long-lived critical infrastructure, rather than short-term research projects supported through piecemeal funding.

Funding agencies should also ensure that at least one U.S.-led broker is demonstrably capable of Rubin-scale throughput. While no single language was currently seen as mandated by the community, workshop discussions endorsed the pattern of a compiled high-performance core, Python-facing science interfaces (as Python remains the community's lingua franca for analysis), and filter-building interfaces that open the alert stream to users without database expertise. Third, machine-learning classification frameworks are already essential for managing and triaging current and forthcoming alert rates; the need is not any single model but funding for the validation, transparency, performance characterization, and long-term maintenance of ML models operating in production streams. Finally, we caution against consolidation for its own sake: the diversity of the Rubin-selected brokers provides healthy redundancy and complementary science focuses, and the community's interest is best served by prioritizing interoperability standards among brokers, Marshals, and archives over convergence on any single tool.

The workshop consensus was that agencies should prioritize functions and interfaces rather than designate single winning implementations. The highest-priority investment is sustained operational support for the alert distribution, including GCN and SCiMMA, which constitutes the closest thing the ecosystem has to a single point of failure, together with completion of the migration to unified, machine-readable schemas. Second is ensuring that at least one U.S.-led broker is demonstrably capable of Rubin-scale throughput: existing Python-based systems have been shown to reach hard limits near one million alerts per night, while software stacks using compiled, memory-efficient languages have been shown to scale comfortably. We do not, however, recommend mandating a language; the pattern the community endorses is a compiled high-performance core, Python-facing science interfaces (as Python remains the community's lingua franca for analysis), and filter-building interfaces that open the alert stream to users without database expertise. Third, machine-learning classification frameworks are already essential for managing and triaging current and forthcoming alert rates; the need is not any single model but funding for the validation, transparency, performance characterization, and long-term maintenance of ML models operating in production streams. Finally, we caution against consolidation for its own sake: the diversity of the Rubin-selected brokers provides healthy redundancy and complementary science focuses, and the community's interest is best served by prioritizing interoperability standards among brokers, Marshals, and archives over convergence on any single tool.

\finding{Agencies and facility leadership should require standardized, machine-readable observatory metadata as a core operational deliverable}:\label{f:meta_data} New and existing facilities receiving federal support should be required to publish scheduling constraints, resource availability, instrument status, and operational limitations in API-accessible formats using common schemas. These data should be treated as essential infrastructure for time-domain science, enabling automated coordination, feasibility assessment, and cross-facility interoperability.

\finding{Agencies should sustain cross-observatory coordination platforms:}\label{f:coordination_platforms} Programs such as ACROSS and HEROIC should be supported as enduring infrastructure services, with capabilities that include unified visibility and feasibility calculations, integration with brokers and TOM systems, and support for automated observing workflows.




\finding{Agencies should support the development of a unified follow-up repository}:\label{f:follow-up_respository} To prevent researchers from manually reconstructing transient histories across fragmented archives, agencies should fund a centralized discovery layer that tracks the entire observational history of individual events.


\subsection{Policy Barriers}
Despite major advances in observational capabilities, the organizational and policy frameworks that govern access to facilities remain fragmented and at times inefficient. These systemic barriers significantly limit the community’s ability to respond rapidly and effectively to transient and multimessenger events. Several long-standing deficiencies in proposal structures, coordination mechanisms, and data access policies continue to impede high-impact science.

\subsubsection{Structural Barriers to Coordinated Observations} The current proposal and access framework for TDAMM science remains poorly suited to coordinated, multi-facility observing campaigns. Researchers pursuing multimessenger or pan-chromatic studies are often required to submit multiple independent proposals, frequently across different agencies or telescope allocation committees, to secure the full set of observations needed for a single science objective, with no mechanism to ensure coordinated approval. As a result, carefully designed programs are vulnerable to partial support or failure. When transient events occur, this fragmentation forces the community to rely on ad hoc ToO requests or rapidly assembled DDT proposals submitted in the midst of an event, introducing delays, redundant effort, and inconsistent responses that are incompatible with the timescales of rapidly evolving phenomena. The limited availability and uneven implementation of unified proposal opportunities and pre-approved, cross-facility coordination frameworks can constrain the community’s ability to mount timely, cohesive, and scientifically optimized responses to high-priority TDAMM events.

\subsubsection{The Perceived Mission Risk of Cross-Facility Dependencies} A recurring challenge for TDAMM-focused mission proposals is how review processes evaluate dependence on the broader multimessenger ecosystem. Review panels have at times treated reliance on complementary observations, such as optical, radio, or X-ray follow-up, as potential mission risks rather than intrinsic features of modern astrophysics. This creates a structural disincentive for proposals that are intentionally designed to operate within a multi-facility framework, and disproportionately impacts missions that aim to enable, rather than fully encapsulate, end-to-end science. In practice, no single observatory can independently realize the full scientific potential of TDAMM astronomy; progress on many key science cases depends on coordinated observations across wavelengths and messengers. Funding agencies and decision authorities should therefore ensure that review guidance and evaluation criteria explicitly recognize external scientific synergies as indicators of scientific relevance and ecosystem integration, rather than an inherent mission weakness. Providing guidance to review panels in this way and adjusting evaluation criteria accordingly would align policy with the collaborative reality of TDAMM astronomy.

\subsubsection{Challenges to Rare Event Observations} Science cases centered on rare but extraordinarily valuable events, such as ultra-long GRBs, nearby GRB–supernovae, tidal-disruption events with prompt emission, or exceptionally close core-collapse supernovae, are often disadvantaged in the existing proposal system. TAC evaluation frameworks often undervalue proposals aimed at low-probability but high-reward science, creating a structural bias against some of the most scientifically compelling transients. Events occurring less than once per year may be dismissed as unlikely to trigger, despite their potential to yield groundbreaking insights if captured.  When multiple facilities are required, the lack of widely available cross-agency or cross-observatory joint proposals further compounds the difficulty.

\subsubsection{Data Acquisition Latencies} For extremely fast-evolving transients, the existing proposal and scheduling systems frequently cannot respond quickly enough. For AT2018cow, DDT requests at facilities such as ALMA can take 1–2 weeks, far too long for these types of transient objects. For nearby core-collapse SNe like SN2023ixf, initial spectra were delayed by more than a day due to slow dissemination of the amateur discovery, resulting in missed early-time near-UV observations that are critical for constraining progenitor properties and shock physics. Beyond limiting human-driven response, such delays also prevent timely triggering of robotic and automated follow-up facilities across the electromagnetic spectrum, which are specifically designed to respond within minutes to well-reported events. Together, these cases illustrate how delays in reporting, approval, or scheduling can permanently and irreversibly limit the scientific return from rare, once-in-a-decade transient events.

\subsubsection{Under-resourced Software Development} Software creation and maintenance are consistently identified as being under-resourced by funding agencies and under-valued by the community. Sustained support is critical for highly essential tools like GCN, LSST-scale brokers, and science platforms. There is a critical need to provide clear, rewarding career paths and funding channels to retain software engineers and researchers who specialize in software, as many institutions do not recognize or reward contributions to infrastructure in tenure evaluation.

\subsubsection{Findings: Policy Barriers}

\finding{Agencies should formalize multi-facility access through expanded joint observing opportunities:}\label{f:expand_joint_observing} Agencies should further expand and formalize joint observing opportunities by establishing comprehensive, multi-facility ToO and DDT programs. These mechanisms should enable a single, integrated proposal that spans multiple space- and/or ground-based facilities to allow for a coherent suite of observations based on the data required to fully characterize and study a given source.

\finding{Review processes should recognize cross-facility dependencies as central to TDAMM science:}\label{f} Agencies and decision authorities should ensure that review guidance and evaluation criteria treat scientifically necessary external synergies as indicators of mission relevance and ecosystem integration, rather than as inherent mission weaknesses. TDAMM-focused missions should not be penalized for depending on complementary observations across wavelengths and messengers when such coordination is essential to realizing the full scientific value of the mission.

\finding{Facilities should lower barriers for rare event science:}\label{f:rare_event_barriers} Facilities should consider implementing and expanding multi-cycle triggering windows so that low-probability but high-reward events (e.g., nearby core-collapse supernovae) are not disadvantaged by standard annual proposal cycles because of their low probability of occurrence.

\finding{Agencies should formally recognize and enable pre-coordinated community observing plans:}\label{f:community_observing_plans} Funding agencies should embrace policy frameworks that allow observatories to pre-commit baseline observing resources for community defined classes of high-priority events, reducing reliance on ad hoc, facility-by-facility responses. This includes investigating means to fund analysis performed by teams or individuals in service of the community plans. Data collected as part of the community observing plan program should be immediately released to the community, with no proprietary period.

\finding{Agencies should establish sustained funding paths for TDAMM software development:}\label{f:software_development} Dedicated funding lines should be made available to support the development, maintenance, and operation of critical software, data services, and coordination platforms, recognizing these efforts as core scientific contributions.

\subsection{Capability Gaps \& Future Needs}
The coming years represent a pivotal moment for TDAMM, with major new facilities coming online even as several foundational assets approach end-of-life. This section identifies key operational risks and capability gaps that could limit the scientific return of upcoming multimessenger campaigns if not addressed in a coordinated and timely manner.

\subsubsection{Swift's Impending Re-entry} The Neil Gehrels Swift Observatory faces a period of acute operational risk due to its rapidly decaying low-Earth orbit. Current projections indicate a very high probability of atmospheric re-entry within the next 12 months if no intervention occurs. NASA has initiated and funded a reboost mission to raise Swift’s orbit and extend its scientific lifetime, reflecting the mission’s extraordinary value to the TDAMM community and to high-energy astrophysics more broadly. However, until the reboost is successfully executed, Swift’s near-term fate remains a significant concern. Its loss would remove the only facility capable of routinely providing arcsecond localizations of gamma-ray bursts within minutes, eliminating a cornerstone capability for rapid multimessenger follow-up. The situation underscores the fragility of the current TDAMM infrastructure and highlights the urgent need for sustained investment in both life-extension efforts and next-generation missions that can preserve and expand Swift’s unique role in the discovery ecosystem.

\subsubsection{Loss of High-Energy Discovery Capabilities} Swift and Fermi remain indispensable to multimessenger discovery, Swift for its rapid-response localization capabilities and Fermi for its exceptional all-sky coverage and broad energy range across GBM and LAT. However, both missions are well beyond their original design lifetimes, and their long-term operational futures are not guaranteed. New missions like StarBurst will significantly strengthen transient-detection capabilities in the near term and add a critical node to the InterPlanetary Network. Yet StarBurst is not designed to be a full replacement for either Swift or Fermi: it does not span the broad spectral reach of Fermi's GBM and LAT instruments, nor can it achieve Swift's precise gamma-ray and X-ray localizations. Moreover, as a Pioneers-class mission, StarBurst is formally funded for only a single year of science operations beyond commissioning, leaving its long-term contribution uncertain.

Sustaining the TDAMM ecosystem will ultimately require a next-generation, wide-field, all-sky high-energy monitor, ideally deployed outside low-Earth orbit to avoid Earth occultation, reduce background, and maximize sky coverage. This issue has been highlighted in multiple strategic and community TDAMM reports since 2020, including the NASA GW-EM Task Force Report, the Decadal survey, and the 1st, 2nd, and 3rd TDAMM Workshop white papers. Identified in every report is the need for a new US-led high-energy transient monitor. Such a mission is necessary not only to preserve but to expand upon the discovery capabilities that Swift and Fermi have provided for nearly two decades.

\subsubsection{Loss of X-ray Follow-up Capabilities} 
A parallel and equally critical gap exists in the availability of deep, high-sensitivity X-ray follow-up. While wide-field monitors have been essential for transient discovery, precise localizations and detailed late-time spectroscopic and temporal follow up have relied upon X-ray instruments and observatories such as Swift-XRT, Chandra, and XMM-Newton. All three of these missions are now operating far beyond their nominal lifetimes, and no clearly defined next-generation mission is poised to replace their capabilities in the near term, particularly in light of setbacks to proposed X-ray and gamma-ray missions such as the Advanced X-ray Imaging Satellite (AXIS) and the Transient High-Energy Sky and Early Universe Surveyor (THESEUS) missions. As a result, the TDAMM ecosystem faces a growing imbalance, where discovery rates are expected to increase dramatically, but the capacity for high-quality X-ray follow-up may stagnate or decline. Addressing this gap will require both maximizing the scientific return of existing facilities through increased ToO allocations and more flexible scheduling as well as renewed investment in future high-sensitivity X-ray observatories to ensure continuity of this neccessary capability.

\subsubsection{IPN Maintenance Risk} The IPN remains one of the most powerful tools for providing all-sky, high-precision localizations of relativistic transients, but its continued effectiveness depends on a set of heterogeneous spacecraft, several of which are well beyond their planned operational lifetimes. In particular, Konus-Wind, a cornerstone of the IPN for nearly three decades, was launched in 1994 and now represents a single-point vulnerability for gamma-ray timing triangulation at large baselines. Its longevity has been remarkable, but its eventual loss would materially degrade the community’s ability to localize short-duration high-energy transients to arcminute-scale annuli.

To ensure continuity of this critical capability, NASA and its international partners will need to sustain, refurbish, or strategically replace aging IPN nodes with modern all-sky high-energy monitors, preferably positioned well outside low-Earth orbit to preserve long baselines and mitigate Earth occultation. Without such forward planning, the IPN’s ability to support multimessenger discovery, early-warning science, and precise event localization will face increasing risk at exactly the time when TDAMM science demands the highest possible performance.

\subsubsection{GW Upgrade Uncertainty} Planning for the next phase of gravitational-wave discovery is complicated by uncertainty surrounding the schedule and scope of the A+ upgrade. A+ is expected to deliver a transformative increase in detector sensitivity—boosting the binary neutron star detection rate, shrinking localization regions, and dramatically enhancing the depth at which multimessenger counterparts can be found. However, the presidential budget request includes reductions that may delay or descope key components of the upgrade. Because the LVK network must synchronize major hardware installations with fixed multi-year observing cycles, even modest funding interruptions can ripple forward into substantial delays.

This uncertainty places pressure on preparations for the O5 observing run, which is envisioned as the first truly high-sensitivity multimessenger campaign of the current decade. If A+ installation is delayed or staged over a longer period, the network may enter O5 with reduced sensitivity, diminishing the number of detections and limiting the horizon of the GW detectors. This would be occurring at a time when complementary facilities, including Rubin, upgraded neutrino detectors, and new high-energy missions, are coming online. This is especially acute for the StarBurst mission, which was proposed and selected under the assumption of operational alignment with O5. StarBurst is designed to provide enhanced high-energy gamma-ray coverage precisely when the upgraded gravitational-wave network reaches peak sensitivity. A substantial shift in the O5 timeline could reduce or even eliminate the overlap between StarBurst’s prime science operations and the enhanced GW observing window, significantly limiting the mission’s ability to deliver its highest-impact multimessenger science.

Ensuring stable, predictable support for the A+ upgrade is therefore essential not only for gravitational-wave science, but for the entire TDAMM ecosystem that depends on reliable low-latency alerts, high detection rates, and precise sky maps to unlock the full scientific potential of multimessenger astronomy.
Spectroscopy Bottleneck: The availability of follow-up spectroscopy has already emerged as a critical bottleneck for time-domain and multimessenger science, and this challenge is projected to worsen substantially in the coming decade. The number of transient alerts from wide-field surveys such as the Vera C. Rubin Observatory will grow by orders of magnitude, while gravitational-wave detectors and high-energy monitors will continue to produce increasing volumes of events requiring rapid classification. Yet, the community’s current spectroscopic capacity, particularly in the optical and NIR, is insufficient to keep pace with this accelerating discovery rate.

Without a commensurate expansion of spectroscopic resources, many promising events will go unclassified, limiting the scientific return from national investments in facilities like Rubin, LIGO, IceCube, and next-generation space missions. Addressing this gap will require significant investment in new community-accessible, rapid-response optical/NIR spectroscopic capabilities, including dedicated follow-up telescopes, queue-scheduled observing modes, and instrument upgrades optimized for transient science.

\subsubsection{Spectroscopy Bottleneck} 

The availability of follow-up spectroscopy has already emerged as a critical bottleneck for time-domain and multimessenger science, and this challenge is projected to worsen substantially in the coming decade. The number of transient alerts from wide-field surveys such as the Vera C. Rubin Observatory will grow by orders of magnitude, while gravitational-wave detectors and high-energy monitors will continue to produce increasing volumes of events requiring rapid classification. Yet the community’s current spectroscopic capacity, particularly in the optical and NIR, is insufficient to keep pace with this accelerating discovery rate. 

Addressing this bottleneck requires both increased spectroscopic capacity and a fundamental shift toward rapid-response capabilities. The upcoming UVEX mission and the recently announced Lazuli Space Observatory represent important first steps toward alleviating this bottleneck through rapid-response, space-based spectroscopic capabilities. UVEX will provide UV spectroscopy and imaging from $\sim$200–300 nm, with a slitless spectroscopic capability at low resolution (R $\sim$ 100), which will help address this gap at shorter wavelengths. Lazuli will feature a 3-meter aperture and an integral field spectrograph spanning 400–1700 nm at low resolution (R $\sim$ 100–500), providing high-quality spectrophotometry across a broad wavelength range. Crucially, the ability for Lazuli to respond to targets of opportunity within hours allows it to capture many transient phenomena during their most diagnostically powerful phases.

 Ultimately the spectral resolution required for transient follow-up depends strongly on the underlying science case. Low-resolution spectroscopy ((R$\sim$100)–500) may be sufficient to characterize continuum shapes, featureless transients, and the broad spectral features of many non-interacting supernovae. However, (R$\sim$1000) is likely a more appropriate minimum for broadly useful UV spectroscopy, including distinguishing blended features in early-time tidal disruption events and other rapidly evolving transients. Resolutions of (R$\sim$1000–3000) would provide substantially greater diagnostic capability while retaining sufficient sensitivity for many extragalactic sources. Specialized cases, such as strongly interacting supernovae with dense forests of narrow far-UV lines, may require (R$>$5000), although the associated loss of sensitivity presents a significant challenge, especially for extragalactic sources. 


While UVEX and Lazuli alone cannot fully resolve the spectroscopy bottleneck, they illustrate the type of capability required to meaningfully alleviate it. Without a commensurate expansion of spectroscopic resources in these bands, many promising events will go unclassified, limiting the scientific return from national investments in facilities like Rubin, LIGO Scientific Collaboration, IceCube Neutrino Observatory, and next-generation space missions.


\subsubsection{Designing Future Missions for Fleet-Level TDAMM Operations}

Cross-observatory coordination infrastructure will be essential not only for maximizing the scientific return of existing facilities, but also for shaping the design of future TDAMM missions. As TDAMM science becomes increasingly dependent on rapid discovery, classification, localization, follow-up, and interpretation across multiple facilities, future missions should not be designed solely as isolated observatories. Instead, they should be conceived as components of a broader, coordinated observing ecosystem, with interfaces and operational concepts designed from the outset to enable fleet-level science. This approach could enable new distributed mission architectures that collectively provide the capability of a larger TDAMM facility. 

For example, a future high-energy transient system might distribute the functions of a Swift-like mission across multiple spacecraft, with one or more satellites providing wide-field discovery and others providing rapid localization or narrow-field follow-up. In such an architecture, the scientific capability would emerge from the coordinated behavior of the system as a whole, rather than from any single platform. This same principle could apply more broadly to constellations, hosted payloads, small missions, ground-based networks, and cross-agency facility partnerships designed to operate as an integrated response system.

Coordination frameworks such as ACROSS, HEROIC, AEON+, and related community infrastructure efforts can provide the operational layer needed to make these architectures practical. By aggregating information on observability, schedule constraints, instrument availability, and target priorities, these systems can reduce the logistical burden that would otherwise make distributed TDAMM operations difficult to execute. They can also provide common interfaces for communicating trigger criteria, follow-up priorities, and decision-tree outcomes across missions and facilities. For this reason, agencies should encourage future TDAMM missions and observatories to interface with cross-observatory coordination systems early in formulation, rather than treating such integration as an after-launch enhancement. These interfaces should be considered during mission design reviews and, where appropriate, captured in formal interface-control documentation, similar to the way missions currently define interfaces with archives and alert systems.

Early integration of future missions with coordination platforms would also have benefits beyond transient response. It would improve resource management by allowing missions to identify when their observations are uniquely valuable, when another facility can provide equivalent or superior coverage, and when coordinated observations are needed to realize the full scientific return of a target. It would also reduce duplication of effort and enable mission teams to design observing modes that are compatible with community-scale decision making. 

\subsubsection{Findings: Capability Gaps \& Future Needs}

\finding{Agencies should prioritize strategic TDAMM replacements:}\label{f:tdamm_replacement_missions} NASA should develop plans for a strategic high-energy monitor and rapid-response X-ray mission to replace the aging Swift and Fermi observatories. This includes prioritizing the development of next-generation, high-sensitivity X-ray observatories capable of providing deep imaging, spectroscopy, and precise localization of transient sources to replace Chandra and XMM-Newton.

\finding{NASA should sustain and expand the IPN:}\label{f:IPN}  NASA should routinely look for opportunities to  add gamma-ray detection capabilities to interplanetary and heliophysics missions, specifically prioritizing nodes outside the ecliptic plane (e.g., on a Uranus Orbiter) to break geometric degeneracies in triangulation.

\finding{NSF should ensure predictable GW upgrades:}\label{f:GW_schedule} Stable support for the LIGO A+ upgrade is critical to ensure that peak detector sensitivity aligns with the operational lifetimes of complementary NSF, DOE, and NASA facilities and missions.

\finding{Agencies should support the development of additional spectroscopic capabilities:}\label{f:spectroscopy} Significant investment is required in moderate-aperture follow-up telescopes and dedicated space-based missions optimized for rapid-response optical and NIR spectroscopy to address the spectroscopy bottleneck in future event classification. 

\finding{Agencies should encourage future missions to address cross-observatory coordination during formulation} Agencies should encourage, if not require, future TDAMM missions and federally supported observatories to address cross-observatory coordination in the design process. This should include early engagement with coordination frameworks such as ACROSS and related initiatives, definition of machine-readable interfaces for observability and operational status, and documentation of how the mission will support coordinated alerts, and follow-up requests.

\newpage
\section{Community Observing Plan Framework\label{sec:obs_plans_framework}}

The accelerating pace of discovery in TDAMM astronomy outlined in the previous sections demands a shift from ad hoc, event-by-event coordination to structured, community-wide preparedness for the most scientifically interesting events. The need for such predefined community observing programs is driven by the fact that certain events, particularly those with trigger rates significantly less than once per year, and which require pan-chromatic or multimessenger observations to fully characterize, are not well served by the current proposal process. These rare and transformative events often require rapid and coordinated response from multiple large, flagship facilities, which are often difficult to secure and execute through traditional proposal mechanisms.

The 2nd TDAMM Workshop underscored that solving this problem requires not only improved technical infrastructure to assist in coordinating such observations, but also agreed-upon, community-driven observing plans that can be activated immediately when exceptional events occur. The workshop findings suggested that the TDAMM community should collaborate with observatory staff to predefine such observing programs for the most exceptional sources, such as the next nearby gravitational wave counterpart or a Galactic supernova. 

These plans, which would essentially act as prewritten Director's Discretionary Time proposals, aim to ensure that coordinated observations across multi-mission, multiwavelength, and multi-messenger instruments can be streamlined and executed rapidly, reducing the latency between the event and observation, which is critical for capturing the most scientifically valuable, rapidly evolving phases of many types of transients. These plans are intended to bridge the space between the rapidly evolving discovery landscape and existing policy and operational constraints, ensuring that the community is well positioned to extract maximum scientific value when time-critical opportunities arise.

\subsection{Motivation}

As detailed in Section~\ref{sec:discovery}, the future TDAMM ecosystem will operate in an environment where  high-cadence, high-volume alert streams will be produced at rates far exceeding the community’s ability to respond, classify, and follow up. This ``alert tsunami'' is compounded by a myriad of systemic barriers that collectively hinder coordinated, time-critical observations. Meanwhile, the events of greatest scientific value tend to be disproportionately rare and fast-evolving. Without pre-negotiated strategies, critical opportunities, such as nearby core-collapse supernovae, exceptionally close gamma-ray bursts, bright kilonovae, or multi-messenger events producing gravitational waves, high-energy neutrinos, and gamma rays, risk being only partially observed or missed entirely due to duplicated effort, response delays, or mismatched priorities across facilities.

Community observing plans offer a practical and consensus-driven path to mitigating these challenges. Building on recommendations from the 2nd TDAMM Workshop for coordinated follow-up strategies and transparent prioritization frameworks, these plans define a base level of observational commitments that institutions and facilities agree to in advance. They outline triggering conditions, minimum data products, and source-specific observing strategies. By establishing these components ahead of time, and by aligning them with evolving TOM systems, broker capabilities, and observatory infrastructure, community observing plans transform what is currently an ad hoc, improvisational process into a scientifically optimized response. Documenting community consensus that a source class is highly important also encourages observatories to take the risk inherently associated with disruptive follow-up observations. This approach ensures that when high-value events occur, the TDAMM community can act quickly and coherently.

\subsection{Scope}
The community observing plans envisioned here are explicitly designed for rare, high-importance transient events that would benefit from rapid coordinated observations. These include sources such as nearby core-collapse supernovae, bright or unusual gamma-ray bursts, galactic magnetar giant flares, multi-messenger events involving gravitational waves or high-energy neutrinos, tidal disruption events with prompt emission, and other explosive phenomena where scientific return depends heavily on rapid, well-coordinated observations across the electromagnetic spectrum and potentially multiple messengers. These are precisely the events that strain current policy, scheduling, and coordination frameworks, and for which pre-negotiated community action can have the greatest impact.

The plans are intentionally scoped to define a baseline level of observations that maximize early scientific return and ensure uniform minimum coverage. This includes rapid localization, early multi-band imaging, prompt spectroscopy, or low-latency X-ray and gamma-ray follow-up depending on the source class. The baseline layer is not meant to replace or preclude deeper or longer-term programs. Instead, it is designed to enable them. Efforts such as longer-baseline observations, high-resolution spectroscopy, detailed monitoring campaigns, or facility-intensive follow-up should continue to be supported through supplementary proposals, standard TAC processes, and Director’s Discretionary Time. By separating early, time-critical actions from extended programs, community observing plans provide clarity for observatories while preserving the freedom of individual research teams to pursue data and funding for specialized science goals.

\finding{Community-defined observing plans should be limited in scope}:\label{f:scope} Community observing plans should be limited in scope to a baseline level of observations that maximize early scientific return, ensure uniform minimum coverage, and enable science extraction from later observations.
\subsection{Framework}
A successful community observing plan must function as an operational framework that observatories and research teams can activate when a high-priority transient event occurs. Such a framework requires four foundational components: (1) clear triggering criteria, (2) standardized communication pathways that operate across agencies and facilities, (3) predefined observational commitments, and (4) decision trees. Together, these elements make rapid coordination practical while ensuring that pre-planned responses remain appropriately scoped, rather than displacing the broader ToO opportunities available to the community.

\subsubsection{Triggering criteria}
Triggering criteria define when a community observing plan is activated. They must be concrete, unambiguous, and tied to measurable quantities such as:
\begin{itemize}
    \setlength{\itemsep}{1pt}
    \setlength{\parskip}{1pt}
    \item Localization area or distance estimate from a gravitational-wave skymap
    \item Gamma-ray fluence for high-energy transients
    \item  The presence of a temporally coincident multimessenger alert
    \item Brightness or color behavior in the earliest optical detections
    \item  Proximity indicators (e.g., ``high likelihood of being within 20 Mpc'')
\end{itemize}
These criteria allow facilities to quickly determine whether a particular alert qualifies as a community-priority event. Crucially, the criteria must remain flexible and revisable, reflecting evolving detector sensitivities, broker capabilities, and the TDAMM community’s changing scientific priorities.

\subsubsection{Communication Pathways}
Event discovery will continue to originate from a diverse set of systems, e.g., GCN notices/circulars, SCiMMA HERMES/Hopskotch, broker streams (BOOM, ANTARES), Astronomers Telegrams (ATels), etc.  Given the heterogeneity of observatory operations, a community observing plan must specify how information flows before, during, and after activation. To facilitate this, the framework must leverage:

\begin{itemize}
    \setlength{\itemsep}{1pt}
    \setlength{\parskip}{1pt}
    \item Which alert stream will likely disseminate the discovery trigger
    \item Which TOM systems and marshals are appropriate to enable cross-facility coordination, automated scheduling, and centralized metadata tracking
    \item Which, if any, queue-scheduling interfaces can be utilized to allow automated submission and execution of follow-up observations
    \item Common APIs and standardized schemas to reduce ambiguity and latency in communication between facilities
\end{itemize}

Identifying these communication pathways ensures that the relevant decision authorities receive timely updates on event status, priority shifts, and early classification results.

\subsubsection{Baseline Observational Commitments}
Once activated, each plan should define a minimal, pre-negotiated sequence of observations appropriate to the event class. These commitments are intended to optimize early-time science return and to ensure that essential data are collected even in cases where no PI-led proposal has yet been mobilized. Examples include:
\begin{itemize}
    \setlength{\itemsep}{1pt}
    \setlength{\parskip}{1pt}
    \item Low-latency X-ray and UV imaging and spectroscopy from space-based assets
    \item Immediate optical/NIR photometry for early temperature and expansion measurements
    \item Prompt low- and medium-resolution spectroscopy to establish classification or redshift
    \item Fast-response radio observations to probe shocks and circumstellar interactions
    \item Rapid multi-band imaging for color evolution and extinction estimation.
\end{itemize}

These baseline actions are intentionally modest in scope to ensure they are feasible across observatories, even during high-activity periods. They serve as the scientific ``floor'' upon which more extensive programs, such as deep spectroscopy, polarimetry, or long-duration monitoring, can be layered via supplemental proposals or ToO triggers.

\subsubsection{Decision Trees and Conditional Observing Strategies}
Each community observing plan must also include explicit decision trees that guide how observing strategies evolve as early data are acquired. Rare, high-impact transients frequently exhibit rapid and unexpected behavior, and the scientific value of follow-up depends critically on making informed choices within hours to days of discovery. Predefined decision logic reduces ambiguity, minimizes delays, and helps ensure that scarce resources are allocated where they provide the greatest scientific return.

These decision trees should encode conditional branches based on readily measurable properties from early observations, such as:
\begin{itemize}
    \setlength{\itemsep}{1pt}
    \setlength{\parskip}{1pt}
    \item  Relative brightness across filters (e.g., UV-bright versus NIR-bright behavior)
    \item  Early color evolution or temperature estimates
    \item  Rapid fading versus sustained emission
    \item  The presence or absence of high-energy or radio counterparts
    \item  Indications of heavy extinction or unusual spectral features
\end{itemize}

For example, a kilonova candidate that is initially UV-bright may motivate intensified ultraviolet and optical spectroscopy to probe early ejecta composition and shock-heated components, while a counterpart that is faint in the UV but bright in the near-infrared may instead trigger deeper NIR imaging and spectroscopy to constrain lanthanide-rich ejecta and r-process yields. Similarly, an unusually luminous optical transient might justify reallocating resources toward rapid spectroscopy and polarimetry.

Crucially, these decision trees should be defined in advance of any specific event and made explicit within the observing plan. They should specify not only which observations are recommended under different conditions, but also which observations may be deprioritized or halted as the event evolves. 

\subsection{Governance}
Effective governance is essential to ensuring that community observing plans remain scientifically justified and operationally feasible. Here we highlight some of the discussions at the workshop on how the observing plans could be implemented and updated over time.

\subsubsection{Decision Authority and Trigger Advocacy}
For community observing plans to function reliably in practice, the authority to activate a plan and to navigate its associated decision trees must be clearly established in advance. While automated systems and pre-established criteria can help identify candidate events, the rarity and scientific importance of these transients often necessitate informed human judgment, particularly when observations may disruptively impact multiple facilities.

Discussions at the workshop emphasized the need for something like a Trigger Advocate Committee: a small, standing group composed of representatives drawn from both NASA and the broader TDAMM community. This committee would be responsible for (1) confirming whether a newly discovered source meets the triggering criteria specified in a community observing plan, and (2) overseeing the execution of early-time decision logic as new data become available. 

To balance responsiveness with breadth of expertise, the Trigger Advocate Committee should be limited in size but diverse in composition, including expertise across relevant messengers, wavelengths, and observing modalities. Membership should be rotational, with fixed terms, to ensure continuity while maintaining community trust and broad representation. Importantly, decision authority should not require full committee consensus in time-critical situations; instead, a quorum or designated on-call advocates could be empowered to act within predefined bounds.

The process by which potential triggers are surfaced to the committee is equally critical. The community should be able to flag candidate events through multiple, well-defined pathways, including:
\begin{itemize}
    \setlength{\itemsep}{1pt}
    \setlength{\parskip}{1pt}
    \item Submission of structured alerts or annotations via broker systems or TOM platforms
    \item Targeted GCN Circulars or Notices explicitly referencing community observing plan criteria
    \item Messages distributed through machine-readable messaging systems (e.g., SCiMMA-supported channels)
    \item Direct submission through a centralized web portal or API endpoint designed for trigger advocacy, e.g., a super-DDT submission request
\end{itemize}

Given its emerging role as a cross-mission coordination hub, NASA’s ACROSS Initiative is well positioned to help facilitate the Trigger Advocate Committee and develop the centralized endpoint for trigger submission. ACROSS is already developing tools that provide fleet-level situational awareness, including schedule ingestion, observability calculations, and resource availability across NASA missions and partner facilities. Integrating trigger advocacy into this framework would allow decisions to be informed not only by scientific merit, but also by real-time operational feasibility and resource constraints.


\finding{Agencies and participating facilities should establish a Trigger Advocate Committee to support the execution of the community observing plans}: This committee would help validate community observing plan activations and guide early-time response execution.

\subsubsection{Observing Plan Revisions}

Building on the recommendations from the 2nd TDAMM workshop white paper and extended discussions at the 4th workshop, each observing plan must be a living document. Governance of observing plan updates can be facilitated through a standing community body, potentially coordinated jointly by NASA and NSF/NOIRLab, with representation spanning the various stakeholders in the TDAMM community. This community body would be tasked with implementing revisions to classification thresholds, minimum commitments, and operational workflows. This would ensure long-term adaptability as new missions and facilities come online and as legacy assets evolve or retire.

Revisions should be conducted on a predictable cadence, such as annually, with the flexibility to implement targeted updates following major TDAMM discoveries or infrastructure changes. Community workshops, such as future iterations of the TDAMM series, provide a natural venue for structured feedback, assessment of plan performance, and consensus-building around proposed modifications. Between workshops, interim updates could be circulated for public comment, ensuring transparency and broad community engagement.

\finding{The community observing plans must be a living document}: The plans should be updated regularly through a body of subject matter experts and community and facility stakeholders.

\subsubsection{Immediate Public Release of Data}

A central conclusion emerging from both the 2nd TDAMM workshop and discussions at the 4th workshop is that all data acquired as part of a community-defined observing plan must be made publicly available immediately, with no proprietary period. This principle applies not only to interim data products, quick-look analyses, and contextual information used to guide real-time decision making during an active observing campaign, but also to the final, science-ready data products.

Community observing plans are fundamentally distinct from traditional PI-led observing programs. They are triggered only for exceptionally rare, high-impact events, often with occurrence rates of far less than one per year, and are explicitly designed to mobilize shared community resources in order to maximize scientific return. In this context, proprietary data can be actively counterproductive. Any delay in data access inhibits rapid cross-facility coordination, limits the community’s ability to execute pre-defined decision trees, and undermines the core rationale for establishing community observing plans in the first place. The prompt release of data also reduces unnecessary duplication of effort and helps ensure that limited observing resources are deployed as efficiently as possible.

Finally, workshop participants stressed that the community observing plans are intended to benefit the full community, including early-career researchers, scientists at smaller institutions, and international partners who may not otherwise have access to proprietary data streams. 

\finding{Data collected as part of a community-defined observing plan should be made public immediately}: This requirement would preclude proprietary data-rights to any person or team.

\subsubsection{Publication, Authorship, and Data Credit}
The success of community observing plans will depend on the willingness of individuals, teams, and facilities to contribute effort in their development and implementation. Discussions at the 4th TDAMM workshop emphasized that such participation must be supported by clear, well-defined credit mechanisms that recognize contributions across the full lifecycle of a community observing plan, from plan development and real-time decision support to data acquisition, analysis, and final scientific interpretation. Importantly, the requirement for immediate public data release must be coupled with robust mechanisms for attribution and credit.

Community observing plans inherently distribute scientific labor across many roles. These include defining trigger criteria and decision trees, monitoring alert streams, performing rapid interim analyses that inform follow-up decisions, executing observations, generating and validating science-ready data products, and conducting the downstream analyses that lead to community-released results. To ensure sustainability of this model, all contributors to these activities must receive appropriate professional credit, regardless of whether their contributions result in traditional first-author publications.

A central finding that emerged from the 3rd workshop is that community observing plans themselves should be treated as citable scholarly products. Plans should be assigned persistent identifiers (e.g., DOIs) and referenced in subsequent publications in a manner analogous to major surveys, mission data releases, or community software packages. This establishes clear provenance for the observing strategy and acknowledges the collective intellectual investment that underpins coordinated response efforts.

Similarly, all data products generated as part of a community observing plan, including interim data products used to inform decision-tree branching, should be citable. DOIs should be issued for raw data deposits, reduced data products, and curated collections, following established practices used by archives such as ExoFOP, MAST and repositories such as Zenodo, which provide DOIs and attribution tracking. This enables proper attribution of both facilities and individuals, ensures long-term traceability, and allows contributions to be formally recognized in CVs, grant applications, and tenure packages.

To support fair authorship practices, the community should also adopt acknowledgment guidelines tailored to the collaborative nature of TDAMM observing plans. These guidelines should explicitly recognize:
\begin{itemize}
    \setlength{\itemsep}{1pt}
    \setlength{\parskip}{1pt}
    \item Facilities and instrument teams that provided data essential to implementing the observing plan
    \item  Individuals or teams who performed rapid analyses that informed triggering or prioritization decisions
    \item  Contributors responsible for producing, validating, or distributing community data products
    \item  Analysts who led or significantly contributed to the scientific interpretation of the final dataset
\end{itemize}

Finally, the TDAMM community, in collaboration with professional societies such as the AAS and APS, should work with journals to define clear expectations for citing community observing plans and associated data products. As with survey data or mission archives, journals should require appropriate citation of plan DOIs and data DOIs when community-generated products are used.


\finding{Agencies and participating facilities should establish well-defined credit mechanisms that recognize contributions across the full lifecycle of community observing plans}: This includes the development of observing strategies, execution of coordinated observations, analysis of resulting data, and citation of community data products.

\subsection{Funding Mechanisms}

Observing plans cannot be executed solely by volunteer effort. The workshop emphasized that the following activities may require funded personnel time:
\begin{itemize}
    \setlength{\itemsep}{1pt}
    \setlength{\parskip}{1pt}
    \item Monitoring alert streams for potential triggers
    \item Extracting event features, triaging false positives, and surfacing candidates that meet plan criteria
    \item  Executing ToOs across multiple observatories
    \item  Reducing and releasing public data products on short timescales
    \item Maintaining communication with observatory staff and the broader community
\end{itemize}
ACROSS staff may assist in coordination, but the scientific labor must remain community-led, supported by targeted funding mechanisms. To address this need, two complementary funding models emerged as particularly well aligned with the operational demands of community observing plans.

\subsubsection{ACROSS Science Leaders}
One proposed mechanism is the establishment of multi-year awards for graduate students or postdoctoral researchers based at non-NASA institutions, modeled loosely on existing programs such as FINESST or the Hubble Fellowship but with an explicit service component tied to TDAMM operations. These awards, which may correspond to less than one full-time equivalent effort, would support early-career scientists who combine independent research with defined responsibilities such as monitoring alert streams, supporting trigger evaluation, maintaining communication channels with the community, assisting with rapid data reduction, and helping organize future workshops or plan revisions. This model directly addresses workforce sustainability while embedding operational expertise within the community and providing clear professional recognition for service contributions.

\subsubsection{Community Infrastructure Teams}
A second, potentially complementary, mechanism is the support of multi-institution Community Infrastructure Teams through multi-year grants. These teams would be responsible for developing, maintaining, and operating the data pipelines, rapid-processing frameworks, and dissemination systems required to deliver science-ready data products on timescales of hours to days, depending on the observing plan. In addition to software development, these teams would commit to operational readiness during triggers, support evolving community standards, and ensure continuity beyond the term of individual fellowships. Workshop discussions emphasized that this model provides stability, distributes effort across institutions of varying size, and aligns naturally with existing investments in brokers, and TOM systems.

\subsection{Implementation}
\subsubsection{ACROSS as the Coordinating Entity for Community Observing Plans}
The ACROSS Initiative is uniquely positioned to serve as the implementer of community-defined observing plans. ACROSS was explicitly established to enable a whole-of-fleet approach to time-domain and multimessenger astrophysics by providing shared organizational, technical, and workforce infrastructure that spans NASA missions, U.S. ground-based facilities, and international partners. This mandate directly aligns with the operational requirements identified for successful execution of community observing plans.

A core requirement of these plans is real-time situational awareness: understanding which facilities are available, what their observing constraints are, and whether a given source is observable across multiple platforms. ACROSS is already developing science feasibility tools that ingest scheduling information, instrument state, and mission constraints to provide cross-observatory visibility and observability assessments for both space- and ground-based assets. These capabilities allow observing commitments defined in community plans to be translated into actionable, facility-specific execution pathways, reducing latency and minimizing duplication of effort.

ACROSS is also well suited to implement the triggering and monitoring infrastructure required for community observing plans. By integrating alert streams from established channels (e.g., GCN, brokers, and future machine-readable messaging systems) with its internal tools, ACROSS can develop automated monitoring and reporting systems that identify when predefined trigger criteria are met. In parallel, ACROSS-managed interfaces can enable community members to report candidate events they believe satisfy observing plan thresholds, ensuring that both automated and human-in-the-loop discovery pathways are supported. Equally important, through existing relationships with mission teams, observatory operators, and program offices, ACROSS can convene stakeholders to establish pre-negotiated observing commitments associated with approved community observing plans. 

Finally, ACROSS provides a natural administrative home for the funding mechanisms needed to sustain community observing plans. As discussed elsewhere in this white paper, execution of these plans requires funded effort for alert monitoring, coordination, data reduction, and rapid public release. ACROSS is well positioned to develop and administer targeted funding solicitations—both to support personnel directly engaged in executing observing plans and to maintain the underlying cyberinfrastructure, while ensuring that scientific leadership and decision-making remain community driven. Over time, this same framework can be extended beyond individual observing plans to support broader TDAMM infrastructure investments.


\subsubsection{Prioritized Science Cases}

The SOC identified eight classes of astrophysical sources that stand to benefit most from pre-established community observing plans. These source classes -- gamma-ray bursts, tidal disruption events, X-ray binaries, supernovae, novae, magnetars, compact binary mergers, and high-energy neutrino emitters -- are characterized by rapid evolution, extreme physical conditions, and the need for coordinated, low-latency observations across multiple wavelengths and messengers. For each class, the SOC further identified specific subclasses with exceptional scientific leverage, where the timely combination of space- and ground-based observations can yield transformational insight but is unlikely to be reliably achieved through traditional, facility-by-facility observing proposals.

Table \ref{Table:CommObsPlans} presents a summary of these prioritized subclasses. An overview of the current state of the field for each source class is presented in Appendix \ref{subsec:science_overviews} and the detailed observing plan for the high-priority subclasses is presented in Appendix \ref{subsec:obs_plans}.

\begin{table}[h]
    \centering
    \caption{Summary of High-Priority Subclasses for Community Observing Plans}\label{Table:CommObsPlans}
    \begin{tabular}{|>{\raggedright\arraybackslash}p{3cm}|>{\raggedright\arraybackslash}p{5cm}|>{\raggedright\arraybackslash}p{8cm}|}
        \hline
        \textbf{Source Class} & \textbf{High-Priority Subclasses} & \textbf{Scientific Rationale \& Key Characteristics} \\
        \hline
        Long Gamma-Ray Bursts (LGRBs) & Very Bright LGRBs, Ultra-Long GRBs, High-Redshift LGRBs, Merger-Origin LGRBs, VHE-Detected LGRBs, Low-Luminosity LGRBs / XRFs / FXTs & These subclasses probe extreme energetics, exotic progenitors like helium-core–black-hole mergers, the early universe, and jet composition through very-high-energy emission. \\
        \hline
        Tidal Disruption Events (TDEs) & Nearby TDEs, Jetted TDEs, Repeated TDEs, Off-nuclear TDEs & Prioritized to enable high-resolution spectral analysis of gas conditions, constrain jet launching mechanisms, and probe black hole/galaxy co-evolution with “wandering'' SMBHs. \\
        \hline
        X-ray Binaries (XRBs) & Predictable Outbursts, Exceptional Brightness, Nearby Systems & Focused on capturing repeatable accretion state cycles, utilizing photon-rich extreme regimes for polarimetry and fast timing, and anchoring population studies with proximity-based ground truth. \\
        \hline
        Supernovae (SNe) & Nearby Fast Blue Optical Transients, Very Nearby Core-Collapse SNe & Prioritized to study rare, long-lived central engines (e.g., AT2018cow) and to perform “autopsies'' of pre-explosion mass loss in local neighborhood events like SN2023ixf. \\
        \hline
        Novae & Nearby Fast Novae, Recurrent Novae & Focused on fast-evolving Galactic events where shock-powered emission maximizes signal-to-noise, and semi-predictable eruptions that serve as prime targets for TeV and neutrino detection. \\
        \hline
        Magnetars & Magnetar Giant Flares, Significant State Changes (Glitches/FRB Associations) & Targeted to probe quantum electrodynamics in super-critical fields, distinct magnetospheric regimes, relativistic outflows, and resolve the magnetar–FRB connection, probing distant possible multi-messenger signals. \\
        \hline
        Compact Binary Mergers (CBMs) & BNS Mergers, Neutron Star–Black Hole (NSBH) Mergers & These are the “standard sirens'' of cosmology; they are prioritized to constrain the neutron star equation of state and the cosmic origin of heavy elements. \\
        \hline
        High-Energy Neutrino Emitters & Galactic Core-Collapse SNe (MeV Neutrinos), High–Energy Neutrino Transients (TeV–PeV Neutrinos) & A Galactic SN is the highest-priority event; high-energy transients are prioritized when spatially/temporally coincident with flaring electromagnetic counterparts. \\
        \hline
    \end{tabular}
\end{table}

\newpage

\section*{Acknowledgements}

We would like to acknowledge the workshop participants, the Science Organizing Committee, the Local Organizing Committee, and all those who supported the workshop and the writing of this white paper. The workshop would not have been possible without the logistical, organizational, and technical support provided by the Universities Space Research Association and The University of Alabama in Huntsville. We gratefully acknowledge support from NASA and the National Science Foundation, whose funding made the workshop possible. We also thank NASA's Physics of the Cosmos Program and the broader TDAMM community for their continued engagement in the workshop series and for their contributions to the development of community observing plans, infrastructure recommendations, and implementation pathways for time-domain and multimessenger astrophysics.

\newpage

\renewcommand{\appendixpagename}{\centering\normalfont\large\textbf{APPENDIX}}
\begin{appendices}

\titleformat{\section}[block]{\normalfont\large\bfseries\centering}{Appendix \Alph{section}:}{0.5em}{}

\titlespacing{\section}{0pt}{*0}{*3}

\section{Science Overviews\label{subsec:science_overviews}}

This appendix compiles a set of science overviews developed as part of the 4th TDAMM workshop. In order to address the key observational requirements that underpin the community observing plans, the SOC identified a set of eight source classes that stand to benefit most from pre-coordinated, rapid, and multi-facility observations. The SOC then invited experts in these fields to present the current state of each of these science areas, focusing on their outstanding questions, observational challenges, and anticipated needs in the evolving time-domain and multi-messenger landscape. The material presented at the workshop forms the foundation of the science overviews included in this Appendix. The SOC has synthesized these contributions into the following sections, which summarize the current understanding of each source class and highlight the critical observational capabilities required to enable significant progress in the field. 

\newpage
\subsection{Gamma-ray Bursts}


The field of GRB astronomy is undergoing a significant paradigm shift, moving beyond historical classifications to embrace a far more complex and diverse reality. The traditional division of GRBs into ``short'' and ``long'' duration events, based on a two-second threshold \citep{kouveliotou1993}, is now understood to be an oversimplification that is both instrument-dependent \citep{bromberg2013} and insufficient to capture the true variety of progenitor systems. Recent discoveries have conclusively demonstrated that both compact object mergers and massive star collapses (collapsars; \citealt{woosley1993}) can produce signals across the duration spectrum. The landmark discoveries of kilonovae associated with long-duration GRBs (such as GRB 211211A) have dismantled the simple model that linked long GRBs exclusively to supernovae \citep{hjorth2012} and short GRBs to kilonovae \citep{li1998,berger2013,tanvir2013,Barnes2013,tanaka2013,berger2014, metzger2017, metzger2020, troja2023}.

New instruments, particularly Einstein Probe \citep{yuan2025}, are revolutionizing the field by uncovering a vast population of previously difficult-to-detect X-ray Flashes (XRFs) and Fast X-ray Transients (FXTs). These low-luminosity events, while individually faint, dominate the volumetric rate of cosmic explosions by a factor of $\sim 100$ and are conclusively linked to the collapse of massive stars. In parallel, GRBs continue to serve as unparalleled probes of the early Universe, allowing for studies of the epoch of reionization at redshifts greater than $\sim$8 \citep{kawai2006, tanvir2009, salvaterra2009, cucchiara2011, tanvir2012, cordier2025}. The advent of the James Webb Space Telescope (JWST) now offers the potential to study the supernovae associated with these distant events, providing a window into the nature of the Universe's first stars. For example, the supernova associated with GRB 250314A at $z\approx 7.3$ \citep{levan2025b,cordier2025}. 

Key open questions now focus on mapping the full diversity of progenitors, developing more robust real-time classification schemes (e.g., \citealt{jespersen2020, garcia2023,   negro2025}), determining the nature of the central engine (black hole vs. magnetar), and finding the theoretically predicted but elusive populations of ``orphan'' afterglows and failed jets (e.g., \citealt{nakar2002, rhoads2003}).

As new facilities expand the discovery space and provide increasingly detailed observations across the electromagnetic spectrum, the focus is shifting toward resolving several fundamental questions that now define the next phase of the field. These include determining the full range of progenitor systems capable of launching relativistic jets, developing more physically motivated real-time classification schemes, understanding the nature of the central engine that powers GRBs, and using high-redshift events to probe the earliest generations of stars. At the same time, wide-field surveys and sensitive X-ray monitors are expected to reveal long-predicted but largely undetected populations of transients—such as orphan afterglows, low-Lorentz-factor outflows, and failed jets—while clarifying the relationship between newly discovered classes of X-ray transients and the classical GRB population. The following section summarizes these key open questions that are shaping current research in GRB astronomy.

Despite significant progress in understanding GRB progenitors, jets, and emission mechanisms, critical knowledge gaps remain regarding jet launching, composition, structure, and the origin of diverse GRB classes. The proposed community observing plan advocates for a focused, community-driven effort targeting specific, high-impact LGRB subclasses that are most likely to yield transformative insights.

\subsubsection{The Breakdown of Traditional GRB Classification}

Historically, GRBs were classified into ``short'' and ``long'' based on a $\sim 2$ s duration threshold, a classification first identified in BATSE data \citep{kouveliotou1993}. This simple duration-based classification was foundational for decades. However, this duration‑based classification was understood as an observational convention that depends on detector bandpass and sensitivity, rather than to a single physical origin for each class \citep[e.g.,][]{bromberg2013}. Satellites with different energy bandpasses (e.g., BATSE, Swift, Fermi) yield different effective cutoffs between the two categories. Subsequent studies have shown that duration alone is insufficient to uniquely determine progenitor systems.

Even historically, the single duration divide has proven to be an inadequate model with early Swift era discoveries producing an initial tension 
\citep{gehrels2006, dellavalle2006, norris2006, galyam2006, hjorth2012}. Observations in recent years have solidified these earlier conclusions and proven that the underlying progenitors are far more varied and can produce signals that cross the two-second boundary 
\citep{Rastinejad2022, troja2022, levan2023, yang2024}, making the simple duration-based classification scheme incomplete. The true picture of GRB progenitors is turning out to be vastly more complex, with short GRBs appearing associated with collapsars 
\citep{Ahumada2021, rossi2022} and a wide range of theoretical channels capable of producing long GRBs, including those involving White Dwarfs (NS-WD, WD-BH, AIC), Intermediate-Mass Black Holes (IMBH-WD), and Helium Core-Black Hole (HeC-BH) mergers 
\citep{yang2022, troja2022, Rastinejad2022, gompertz2022, levan2024, chrimes2025, cheong2025, oconnor2025, levan2025, neights2026}. 
Some short gamma-ray bursts (sGRBs) may actually be misidentified extragalactic magnetar giant flares (MGFs), see, e.g.,  \citet{burns2021,trigg2024,2024Natur.629...58M,trigg25AA:GF,beniamini2025,2025MNRAS.537.1430S}.

Moving forward, the field requires improved classification schemes that can be implemented in real-time, potentially incorporating parameters like the multi-band hardness ratio, minimum variability timescale, and spectral lag alongside the prompt duration.

\subsubsection{Unraveling Progenitor Diversity and Associated Transients}

The direct association of GRBs with secondary, longer-wavelength transients like supernovae and kilonovae has been crucial for conclusively establishing their physical origins. The link between long-duration GRBs and the death of massive stars was firmly established with the association of GRB 980425 with supernova SN 1998bw 
\citep{galama1998, kulkarni1998, iwamoto1998}.  Long GRBs are consistently associated with a specific class of energetic, hydrogen-poor supernovae known as Type Ic-BL (e.g., \citealt{hjorth2012, cano2017}). Spectroscopic sequences of GRB-SNe, such as those for GRB 030329/SN 2003dh, show remarkable similarity to the SN 1998bw template \citep{Hjorth2003,Stanek2003,Matheson2003,Mazzali2003}. 

A key finding  is the lack of correlation between the energy of the gamma-ray jet and the peak magnitude or other properties of the associated supernova over seven orders of magnitude in energy (e.g., \citealt{Hjorth2013,cano2017, Srinivasaragavan2023}). This suggests the supernova explosion and the central engine powering the relativistic jet are largely disconnected phenomena. Adding to the complexity, deep searches for some long GRBs have yielded only upper limits, revealing a class of events without a detectable supernova, hinting at progenitor diversity even within the collapsar channel as early as 2006 (e.g., \citealt{dellavalle2006, fynbo2006}).

For the last few decades, evidence has mounted that short GRBs originate from a different channel: the merger of compact objects 
\citep{Blinnikov1984, Paczynski1986, Eichler1989, Narayan1992}, such as two neutron stars (NS-NS; \citealt{Ruffert1999, Rosswog2003, Rosswog2005}) or a neutron star and a black hole (NS-BH; \citealt{faber2006, shibata2011}). Late-time optical and near-infrared observations of short GRBs with the Hubble Space Telescope (HST) revealed late-time red excesses in their light curves (e.g., \citealt{berger2013, tanvir2013}) that could not be explained by the standard afterglow model \citep{meszaros1997, sari1998, wijers1999, granot2002}. These additional emission components were identified as kilonovae 
\citep{li1998, Barnes2013, kasen2017}: transients powered by the radioactive decay of heavy elements synthesized in the neutron-rich merger ejecta. About a dozen such kilonova candidates have been identified in association with short GRBs, solidifying their origin in compact object mergers (for recent reviews, see, e.g., \citealt{metzger2017, nakar2020, troja2023}). The discovery of GW170817 \citep{abbott2017}, GRB 170817A \citep{Goldstein2017, Savchenko2017}, and its associated kilonova AT2017gfo (e.g., \citealt{coulter2017, drout2017, arcavi2017, evans2017, Soares2017, pian2017, troja2017,Kasliwal2017,kasen2017,Chornock2017,Perego2017,Valenti2017,Utsumi2017,Andreoni2017}) provided incontrovertible proof of this association of short duration GRBs to neutron star mergers, solidifying the earlier lower significance kilonova associations claimed in the literature. 

The most significant recent disruption to the GRB classification model has been the discovery of kilonovae following two long-duration GRBs, GRB 211211A \citep{Rastinejad2022, troja2022, yang2022} \& GRB 230307A \citep{levan2024, yang2024}. These two nearby ($z \sim 0.076$ and $z \sim 0.065$, respectively) and extremely bright long-duration GRBs were conclusively shown to be followed by kilonovae, not supernovae, including spectroscopic classification with the James Webb Space Telescope for GRB 230307A \citep{levan2024, gillanders2025}. This discovery definitively demonstrated that the duration of the prompt gamma-ray signal is on its own not a reliable indicator of the progenitor type. Compact object mergers can evidently produce long-lasting (tens of seconds) central engine activity, as suggested earlier by, e.g., GRB 060614 \citep{dellavalle2006, galyam2006}.  It is worth noting, though, that these two bursts are among the brightest GRBs ever detected by Swift and Fermi, respectively. Their discovery may be subject to a selection effect, and ultimately such merger-driven long GRB-kilonova associations may be exceedingly rare.

\subsubsection{GRBs as Probes of the Early Universe}

The immense luminosity of GRBs makes them detectable across cosmological distances, providing a unique tool to study the early Universe and the Epoch of Reionization. GRBs can be seen at redshifts where the first galaxies and stars were forming. In the Swift era, 12 GRBs have been discovered at $z > 6$, less than one billion years after the Big Bang. GRB 090423A was detected at a redshift of $z \approx 8.2$ \citep{tanvir2009, salvaterra2009}, making it one of the first objects ever observed from that era. Furthermore, the afterglows of high-redshift GRBs act as background light sources, allowing astronomers to measure fundamental properties of the early cosmos, including the neutral hydrogen fraction, the UV escape fraction from early galaxies, and the metallicity of the intergalactic and interstellar medium (e.g., \citealt{cucchiara2011, campana2011, totani2014, melandri2015, Salvaterra2015, tanvir2019, levan2025b, levan2025a}). They also offer a potential, though as-yet unrealized, pathway to identifying the Universe's first generation of stars (Population III stars).

Recently highlighting the great promise of high-z GRBs was the SVOM detection of GRB 250314A \citep{cordier2025}. Discovered at a redshift of $z \approx 7.3$ \citep{cordier2025}, this event's observed duration of under 10 seconds corresponds to an intrinsically short duration in its rest frame, again highlighting the unreliability of observed duration for classification; though this is likely a tip-of-the-icerberg selection effect (e.g., \citealt{kocevski13, moss2022, moss2026}). Follow-up imaging with JWST revealed a spectral energy distribution consistent with a Type Ic-BL supernova \citep{levan2025b}, like SN 1998bw. If confirmed with further imaging, this would be the highest-redshift supernova ever discovered, demonstrating the power of GRBs to pinpoint stellar explosions that would otherwise be missed by deep-field surveys.

\subsubsection{The Emerging Landscape of X-ray Flashes and Fast X-ray Transients}

A growing frontier in GRB astronomy involves soft-spectrum events such as X-ray flashes (XRFs; \citealt{heise2001, sakamoto2004, sakamoto2005}) and fast X-ray transients (FXTs; \citealt{jonker2013, alp2020, quirola2023}). Defined observationally as events with a peak energy below $\sim 20$ keV, XRFs were historically linked to GRBs \citep{heise2001}. While many are simply high-redshift GRBs whose peak energy has been shifted into the X-ray band \citep{sakamoto2004, pelangeon2008, wei2025}, a population of intrinsically soft XRFs exists \citep{Srinivasaragavan2025b, jiang2025, liu2025}.
Events like GRB 060218, a low-luminosity GRB with a low peak energy, were followed by a Type Ic-BL supernova, firmly connecting them to the collapsar channel \citep{pian2006, soderberg2006, campana2006}.

These low-luminosity GRBs and XRFs are estimated to dominate the volumetric rate of relativistic explosions by a factor of $\sim 100$ over classical, high-luminosity GRBs \citep{soderberg2006, virgili2009}. 
Their faintness means they can only be detected in the local Universe, explaining their low discovery rate until the launch of a sensitive wide-field survey mission like the Einstein Probe \citep{yuan2025}.

Einstein Probe, with its soft X-ray capabilities, is now discovering these events in greater numbers \citep{yuan2025}. The redshift distribution of EP-discovered FXTs solidly matches that of long GRBs, reinforcing the conclusion that the majority are produced by collapsars \citep{oconoor2025a}. Several EP transients (e.g., EP240414a, EP240801a, EP241021a, EP250108a, EP250827b; \citealt{vandalen2025, vanhoof2026, eyles2026, Rastinejad2025, Srinivasaragavan2025a, Srinivasaragavan2025b, srivastav2025, quirola2026}) have been associated with Type Ic-BL supernovae. These events show a distinct ``early blue excess'' in their optical light curves \citep{vandalen2025, srivastav2025, Srinivasaragavan2025b}, a feature not previously seen in  Swift GRBs, which may point to emission from a jet cocoon or interaction with an extended stellar envelope
\citep{sun2025,vandalen2025, srivastav2025, Hamidani2025, Gianfagna2025,  eyles2026, Rastinejad2025, Srinivasaragavan2025a, Srinivasaragavan2025b, zheng2025,li2026}.

\subsubsection{The Puzzle of Ultra-Long Gamma-Ray Bursts}

Ultra-long GRBs (ULGRBs), with prompt emission lasting thousands to tens of thousands of seconds, present some of the most challenging puzzles in the field \citep[e.g.,][]{tikhomirova2005, gendre2013, levan2014, ioka2016}. ULGRBs are defined by their extreme duration ($>$ a few thousand seconds; \citealt{levan2014}), which implies a central engine that can remain active for an exceptionally long time. Their detection is hampered by the observing strategies of low-Earth orbit satellites and the limited background stability, often requiring inter-satellite networks (like the IPN) to characterize.

While at least one ULGRB has been associated with a supernova \citep{greiner2015}, confirming a collapsar origin, the extreme duration suggests a peculiar progenitor. Theories include the collapse of a blue supergiant star \citep{meszaros2001, stratta2013, nakauchi2013, ioka2016}, a magnetar central engine, the merger of a helium core with a black hole (HeC-BH merger;  \citealt{fryer1998, woosley2012, neights2026}), or the tidal disruption event (TDE) of a star by a black hole (either a ``micro-TDE'' or an IMBH-WD TDE or IMBH-MS TDE; \citealt{perets2016, levan2025, oganesyan2025, oconnor2025, li2026, eyles2026, beniamini2025, an2025, granot2026}).

Detected at a redshift of $z \approx 1.036$ \citep{gompertz2026}, GRB 250702B has been nicknamed ``The Longest of All Time'' GRB \citep{levan2025, neights2026}. It triggered multiple gamma-ray instruments over a period of 25,000 seconds \citep{neights2026}. Its properties are more consistent with a ULGRB than a relativistic TDE \citep{levan2025, oconnor2025}. However, its separating factor from other previous ULGRBs is its initial soft X-ray detection by the Einstein Probe \citep{li2026} nearly a day before the gamma-ray triggers \citep{neights2026}. Crucially, X-ray observations revealed short-timescale ($\sim$ 1.5 kilosecond) flaring activity as late as two days after the initial trigger, providing direct evidence of a prolonged, active central engine lasting for at least three days \citep{oconnor2025}. Proposed progenitors include blue-supergiant collapsars, magnetars, helium-core–black-hole (HeC-BH) mergers, and various tidal-disruption (micro-TDE) scenarios \citep{levan2025, oganesyan2025, oconnor2025, carney2025, li2026, eyles2026, beniamini2025b, an2025, neights2026, granot2026}. Distinguishing among these models remains an active area of theoretical and observational work. 

\subsubsection{Open Questions}

Despite decades of progress, GRB astronomy is defined today by a set of fundamental open questions. These include:

\begin{itemize}
    \item 
{\bf Progenitor Diversity}: What is the complete inventory of stellar and compact object systems that can produce long-duration gamma-ray signals and launch relativistic jets?
    \item 
{\bf Improved Classification}: How can the community move beyond the simplistic long vs. short paradigm? Can new prompt emission metrics like minimum variability timescale be implemented in real-time to better classify events and guide follow-up observations?
    \item 
{\bf Central Engine Nature}: For any given GRB, what determines whether the central engine is a black hole or a rapidly spinning magnetar, and what is the population fraction of each?
    \item 
{\bf The First Stars}: Can JWST observations of supernovae associated with z > 5 GRBs reveal signatures that differentiate them from local events? Could this be a viable path to identifying Population III star progenitors?
    \item 
{\bf Missing Populations}: Where are the theoretically predicted populations of ``orphan" afterglows (from jets viewed off-axis), ``dirty fireballs" (low Lorentz factor outflows), and completely failed jets? The Einstein Probe and the Vera C. Rubin Observatory are expected to be key facilities in searching for these transients.
    \item 
{\bf Connecting Transient Classes}: What is the physical relationship between the new class of Fast X-ray Transients being discovered by the Einstein Probe and the ``classical" fast X-ray transients discovered serendipitously over the past decades by observatories like XMM-Newton and Chandra?

\end{itemize}

\newpage
\subsection{Tidal Disruption Events}
TDEs represent the dramatic destruction of a star by the immense gravitational forces of a supermassive black hole \citep[SMBH;][]{Hills75, Rees88}. These transient phenomena, which typically occur in the centers of galaxies, provide a unique laboratory for studying dormant black holes, the physics of super-Eddington accretion, and the effects of general relativity. Initially predicted to be bright soft X-ray sources \citep[e.g.,][]{Rees88}, observations have revealed that a majority of TDEs are discovered in the optical and ultraviolet spectra, characterized by a smooth rise in brightness over weeks and a power-law decline over months \citetext{e.g., \citealp{vanVelzen21}; \citealp{Hammerstein23}; see also \citealp{Gezari2021} and \citealp{Mockler2025} for reviews}.

The field is marked by significant observational diversity. X-ray emissions, when present, exhibit a wide range of behaviors, from appearing early or late relative to the optical peak to showing rapid, sub-hour variability \citep[e.g.,][]{Guolo23}. Furthermore, radio emissions are detected in approximately 40\% of TDEs, typically years after the initial event, implying that mass ejection is a common outcome \citep{Cendes2024}. Rare but powerful ``jetted TDEs'' emit across the entire electromagnetic spectrum and are potential sources of high-energy particles \citep[e.g.,][]{levan11, Cenko2012, Andreoni2022}.

TDEs serve as powerful probes. Their light curves and spectra can be used to estimate the mass of the central black hole, providing an independent channel for BH demographics, particularly in the $10^5$--$10^8 M_\odot$ range \citep{Evans1989, Strubbe2011, lodato09, guillochon13, Kesden2012, mockler_weighing_2019, ryu_measuring_2020, mummery24}. The event timeline is broadly split into two phases: an early, complex period dominated by the unique feeding mechanism of a highly eccentric debris stream, where processes like stream self-intersection generate precursor emissions \citep[e.g.,][]{Huang2024}; and a later phase where a more conventional accretion disk forms and dominates the emission.

TDE research is motivated by key open questions in astrophysics, including:

\begin{itemize}
    \setlength{\itemsep}{1pt}
    \setlength{\parskip}{1pt}
    \item How did the MBHs at the centers of galaxies form?
    \item Where are the IMBHs?
    \item How efficient is super-Eddington accretion?
    \item How do MBH disks form and launch jets?
    \item How does the large-scale evolution of a galaxy connect to the properties of the stars and MBH at its center?
\end{itemize}

These questions drive current research into constraining the black hole mass function with TDEs \citep[e.g.,][]{yao_tidal_2023}, determining the overall mass budget in a flare (how much mass is accreted versus ejected) \citep{lu_self-intersection_2020}, modeling the complex transition from the initial debris stream to a stable accretion disk \citep{miller_flows_2015, bonnerot_first_2021, steinberg_stream-disk_2024, Huang2024, andalman_resolving_2025}, determining the efficiency of emission \citep{dai_unified_2018, mockler_energy_2021, huang_bright_2023}, understanding the properties and conditions of the rare TDEs that launch relativistic jets \citep{kara_relativistic_2016, pasham_birth_2023, yao_-axis_2024}, and exploring the connection between host galaxy properties and the rates and attributes of stars disrupted in their centers \citep{kochanek_abundance_2016, mockler_evidence_2022, miller_evidence_2023, hinkle_most_2025}. As new observatories like the Rubin Observatory come online, the expected deluge of TDE discoveries promises to transform these events from astrophysical curiosities into powerful tools for population studies and precision physics.

\subsubsection{The Nature of Tidal Disruption Events}
A TDE occurs when a star passes too close to a supermassive black hole. The process unfolds through a distinct sequence of events:

\begin{enumerate}
    \item \textbf{Orbital Perturbation}: In the dense environment of a galactic nucleus, a star orbiting the central black hole can be scattered onto a highly eccentric, nearly parabolic orbit through multi-body interactions or two-body relaxation.
    \item \textbf{Tidal Disruption}: As the star approaches the pericenter of its new orbit, it experiences extreme tidal forces from the black hole. If the pericenter is within the tidal radius, the gravitational gradient across the star overcomes its self-gravity, stretching, compressing, and ultimately shredding it into a long, elongated stream of debris.
    \item \textbf{Debris Evolution}: The stellar debris retains a spread of orbital energies. In a full disruption, approximately half of the debris is on unbound trajectories and is ejected from the system, while the other half remains gravitationally bound to the black hole.
    \item \textbf{Emission and Accretion}: The bound debris falls back toward the black hole over a period of months to years. This ``fallback" material eventually forms a disk, transiently feeds the dormant black hole at a rate that can easily exceed its Eddington limit. Both the process of forming the disk (circularization) and the disk itself can produce large amounts of energy, powering a luminous flare of electromagnetic radiation. This period of accretion disk formation and evolution is when TDEs are typically observed.
    \item \textbf{Possible Jet Formation}:  In some events, accretion onto the black hole can potentially launch a relativistic jet along its rotational axis. This likely requires rapid spin, strong magnetic fields, and a super-Eddington accretion flow, enabling energy extraction via processes such as the Blandford–Znajek mechanism. When aligned with our line of sight, these “jetted TDEs” appear as bright, non-thermal X-ray and gamma-ray transients, providing a powerful probe of jet formation and extreme particle acceleration.
\end{enumerate}
The study of TDEs is a rapidly evolving field. The cumulative number of discoveries has grown significantly over the past three decades, and upcoming facilities like the Rubin Observatory are expected to discover thousands of such events per year \citep{bricman_prospects_2020}, enabling detailed population studies for the first time.

\subsubsection{Observational Characteristics}

TDEs are multi-band emitters, though their appearance across the electromagnetic spectrum is highly diverse and has challenged initial theoretical predictions.

\begin{itemize}

\item \textbf{Optical and UV Emission}:  While early models predicted TDEs would be primarily soft X-ray sources with temperatures around 100 eV, the vast majority of discoveries have been made in the optical and ultraviolet (UV) bands. These optical TDEs exhibit distinct characteristics:
\begin{itemize}
    \item \textbf{Light Curves}: A smooth rise to peak brightness over a few weeks, followed by a longer decline over several months that often follows a $\sim t^{-5/3}$ power law.
    \item \textbf{Spectra}: A broad, blue continuum consistent with a black body temperature in the optical and UV of $>10,000$ Kelvin. Superimposed on this continuum are prominent emission lines of hydrogen, helium, and occasionally metals (most commonly nitrogen), which are used to classify TDE subtypes.
    \item \textbf{Temperature Evolution}: While individual events show significant variation (including both cooling and heating phases), the overall optical/UV blackbody temperature of the TDE population does not evolve drastically, typically staying within an order of magnitude.
\end{itemize}

\item \textbf{X-ray Emission Diversity}: The X-ray behavior of TDEs is remarkably varied and often decoupled from the optical/UV emission. Many optically discovered TDEs lack a detectable X-ray counterpart \citep[e.g.,][]{vanVelzen21, Hammerstein23}. When X-rays are detected, their light curves can show diverse evolutions: some appear late relative to the optical peak, others exhibit flat evolution, and a few are detected very early, even preceding the optical rise \citep[e.g.,][]{Guolo23}. Some TDEs display rapid X-ray variability on sub-hour timescales, as captured by instruments like NICER. Others have been shown to have late-time X-ray variability associated with quasi-periodic eruptions \citep[e.g.,][]{Nicholl24}. A rare subclass of TDEs produces extremely bright X-ray emission associated with the launch of a relativistic jet \citep[e.g., AT\,2022cmc, Sw J1644;][]{levan11, Cenko2012, Andreoni2022}.

\item \textbf{Radio Detections and Outflows}: Late-time radio observations have become a crucial tool for understanding the full lifecycle of TDEs. Radio emission is detected in approximately 40\% of TDEs, suggesting that outflows are a common byproduct of the disruption and accretion process \citep[e.g.,][]{alexander_multi-wavelength_2025}. This radio emission is typically detected a few years after the optical peak, significantly extending the observable duration of the event. The host galaxies of radio-emitting TDEs appear similar to those of the general optical TDE population.

\end{itemize}

\subsubsection{The TDE Timeline: From Debris Stream to Accretion Disk}
The evolution of a TDE is marked by a transition from a unique, stream-fed geometry to a more traditional accretion disk. The primary phases include:

\begin{itemize}
    \item \textbf{Early Phase: Dissipation and Emission Precursors}
    
    The initial phase is defined by the highly eccentric, asymmetric debris stream. This period is difficult to observe, but theoretical work and simulations have identified several key dissipation processes that could power the early emission before a coherent disk forms:
        \begin{itemize}
            \item \textbf{Nozzle Shock}: As the debris stream passes through the pericenter, it is vertically compressed, which can generate a shock \citep{steinberg_stream-disk_2024, Huang2024, andalman_resolving_2025, hu_converged_2026}.
            \item \textbf{Stream Self-Intersection}: Due to relativistic apsidal precession, the returning debris stream can collide with itself \citep{jiang_prompt_2016}. The location of this intersection is highly sensitive to the black hole's mass, occurring farther out for lower-mass BHs ($\leq 10^{6.5} M_\odot$) and closer in for higher-mass BHs \citep{dai_soft_2015}. This process is considered a primary mechanism for initiating accretion and driving outflows \citep[e.g.,][]{piran_disk_2015, lu_self-intersection_2020, bonnerot_first_2021, huang_bright_2023}.
            \item \textbf{Stream-Disk Shock}: The incoming stream can plunge into and shock against the circularizing gas that is beginning to form a disk \citep{andalman_tidal_2022, steinberg_stream-disk_2024, Huang2024}.
        \end{itemize}
    These processes may be responsible for observed pre-peak phenomena \citep{huang_bright_2023}, such as ``optical bumps'' in the rising light curve or the early, bright X-ray flares seen in events like AT2022dsb.

    \item \textbf{Late Phase: Disk-Dominated Emission}
    
    In the later stages, most TDEs are believed to form a long-lived accretion disk \citep{van_velzen_late-time_2019}. The emission during this phase can be modeled to constrain properties of the system, including the black hole's mass and spin \citep{wen_continuum-fitting_2020, mummery24}. This phase can also host other transient phenomena. A prime example is AT2019qiz, a nearby TDE that began exhibiting Quasi-Periodic Eruptions (QPEs), repeating soft X-ray flares, years after its initial optical peak \citep{Nicholl24}. These QPEs provide an independent probe of the late-time disk, constraining its size and surface density long after the main TDE has faded \citep{franchini_quasi-periodic_2023, linial_qpes_2025}. Understanding the transition from the early, wind/outflow-dominated phase to the late, disk-dominated phase is a critical goal in TDE research.
\end{itemize}

\subsubsection{TDEs as Probes of Fundamental Astrophysics}


TDEs offer a rare opportunity to study the onset of accretion onto otherwise quiescent black holes, serving as laboratories for fundamental physics that span from stellar dynamics to relativistic accretion and feedback.  These fundamental processes include:

\begin{table}[h]
    \caption{Summary of TDE system properties.}
    \label{tab:TDEprops}
    \begin{tabular}{|>{\raggedright\arraybackslash}p{3cm}|>{\raggedright\arraybackslash}p{11.5cm}|}
    \hline
        \textbf{Property} & \textbf{Typical Value / Characteristic} \\
        \hline
        Timescale & 1--3 years (optical/X-ray), extended to several more years by radio. \\
        \hline
        BH Mass ($M_{\rm BH}$) & Preferentially $10^5$--$10^8 M_\odot$.\\
        \hline
        Accretion Rate & Initially Super-Eddington, with fallback rates that can keep smaller BHs above the Eddington limit for months. \\
        \hline
        Radiation Efficiency & Highly variable, showing a much larger spread than typical Active Galactic Nuclei (AGN). \\ 
        \hline
    \end{tabular}
\end{table}

\begin{itemize}
    \item \textbf{Constraining the Black Hole Mass Function and finding IMBHs}: TDEs are powerful tools for measuring black hole masses. They preferentially occur around lower mass $M_{\rm BH}$ below $\sim 10^8 M_\odot$. This is coincident with the portion of the BH mass function that is hardest to constrain through traditional direct measurement methods, but that is most sensitive to the formation scenarios of the black holes \citep[e.g.,][]{greene_intermediate-mass_2020}. Their rates are expected to peak around $\sim 10^6 M_\odot$ due to dynamical arguments based on the density profiles of stars at the centers of galaxies and the size of the radius for tidal disruption given the black hole’s gravity \citep{stone_rates_2016, pfister_enhancement_2020}. Above $10^8 M_\odot$, the rates drop off sharply, as a black hole larger than $\sim 10^8 M_\odot$ will swallow a solar-type star whole without disrupting it outside the event horizon \citep{van_velzen_mass_2018}. TDE candidates have also been observed in the IMBH mass range (at or below $\sim 10^5 M_\odot$), both off-nuclear \citep[e.g., AT2024tvd,][]{yao_massive_2025}and in the centers of dwarf galaxies \citep[e.g., AT2020neh,][]{angus_fast-rising_2022}.  While their rates are expected to be at least a factor of 10 lower than those of TDEs around $\sim 10^6 M_\odot$ black holes \citep{chang_rates_2025} and their optical luminosities are expected to be lower given the black hole’s Eddington limits, it is still expected that $>100$ will be observed with LSST \citep[][]{bricman_prospects_2020, graham_zwicky_2019, yao_tidal_2023}. Several methods can be used to estimate $M_{\rm BH}$ from a TDE, providing an independent channel to study black hole demographics:

    \begin{itemize}
        \item \textbf{Host Galaxy Scaling Relations}: Using established correlations between galaxy properties and central black hole mass.
        \item \textbf{Light Curve Modeling}: Combining the observed light curve with theoretical fallback models.
        \item \textbf{UV Plateau Modeling}: Using the UV luminosity during a plateau phase, combined with disk models.
    \end{itemize}

\end{itemize}

Other probes of fundamental physics include:

\begin{itemize}
    
    \item \textbf{Super-Eddington Accretion}: The rapid fallback of stellar debris (e.g., half a solar mass in half a year onto a $10^6 M_\odot$ black hole) drives mass flow rates to the black hole well above the Eddington limit \citep[]{Hills75, Rees88}. This super-Eddington state distinguishes TDEs from most AGNs, leading to different observational signatures, such as generally shallower X-ray spectra. A key open question is what fraction of this infalling mass actually accretes onto the black hole versus being expelled in an outflow.
    
    \item \textbf{Disk formation and Jet launching}: For TDEs that occur in quiescent galaxies, we can watch the MBH accretion disk form in real time. Simulations show that shocks can circularize material, leading to the formation of a MBH accretion disk on timescales of weeks to months \citep{jiang_prompt_2016, bonnerot_first_2021, steinberg_stream-disk_2024, huang_bright_2023}. At early times this disk will often be fed at super-Eddington rates, but should transition to sub-Eddington on timescales of months to years \citep[e.g.,][]{guillochon_hydrodynamical_2013}.
    
    \item \textbf{Black Hole - Galaxy Coevolution}: Black hole–galaxy coevolution is encoded in the reciprocal relationship between the growth of massive black holes (MBHs) and the dynamical and stellar properties of their host galaxies, and TDEs provide a uniquely sensitive probe of this connection. Because TDEs preferentially occur in galaxies hosting black holes in the $10^5$–$10^8,M_\odot$ range \citep[e.g.,][]{stone_rates_2016,hannah_counting_2024, chang_rates_2025}, precisely where traditional dynamical measurements are challenging, they offer an independent avenue to constrain the black hole mass function and test formation pathways, including the occupation fraction and growth of intermediate-mass black holes. Moreover, TDE rates depend on the central stellar density, orbital structure, and dynamical processes, directly linking event demographics to the evolutionary state of galactic nuclei \citep[e.g.,][]{stone_rates_2016, chen_tidal_2011, french_host_2020, hammerstein_tidal_2021, melchor_tidal_2024}. The accretion episodes triggered by TDEs, often reaching super-Eddington regimes \citep{mockler_energy_2021}, can also drive winds and relativistic jets \citep{alexander_multi-wavelength_2025, yao_-axis_2024}, injecting energy and momentum into the surrounding environment and potentially contributing to feedback processes that regulate star formation and nuclear gas supply. At the same time, correlations between TDE properties and host galaxy characteristics, such as the over-representation of post-starburst (“E+A”) galaxies \citep[e.g.,][]{french_tidal_2016}, suggest that recent star formation history and dynamical restructuring play a critical role in setting disruption rates. In this way, TDEs serve not only as signposts of otherwise quiescent black holes, but as direct tracers of the coupled evolution of black holes and their host galaxies across cosmic time, bridging small-scale accretion physics with large-scale galaxy evolution.

\end{itemize}

\subsubsection{Key Case Studies}
\begin{itemize}
    \item \textbf{Nearby TDE} (AT2019qiz): As one of the closest TDEs detected, it has provided exceptionally detailed data \citep{nicholl_outflow_2020, hung_discovery_2021, patra_spectropolarimetry_2022, short_delayed_2023, wu_torus_2025}. Early spectroscopy revealed blue-shifted emission lines, indicating a fast-expanding outflow \citep{nicholl_outflow_2020}. It was also the first TDE in which low-ionization absorption lines were detected \citep{hung_discovery_2021}. The subsequent discovery of QPEs from its remnant disk has made it a benchmark for studying late-time TDE evolution \citep{Nicholl24}.
    \item \textbf{Jetted TDEs} (e.g., Sw J1644, AT2022cmc): These rare events are characterized by powerful relativistic jets that emit across the electromagnetic spectrum, from radio to X-rays  \citep[][]{levan11, Cenko2012, Andreoni2022, pasham_birth_2023, yao_-axis_2024}. They are crucial for understanding how and when black holes can launch jets.
\end{itemize}

\subsubsection{Open Questions}
The study of TDEs is poised to answer fundamental questions in astrophysics. Key research areas include:
\begin{itemize}
    \item \textbf{The Mass Budget}: Combining optical, X-ray, and late-time radio data to determine the ultimate fate of the stellar debris: how much mass is accreted, how much remains in the disk, and how much is ejected in outflows?
    \item \textbf{Jet Launching Mechanism}: Why are jets so rare in TDEs? What disk conditions are required to launch them, especially given that simulations struggle to produce strongly magnetized disks from the debris?
    \item \textbf{Disk Formation and Eccentricity Damping}: How does the highly eccentric debris stream circularize to form an accretion disk? Understanding this process is key to connecting the early and late phases of the event.
    \item \textbf{Probing General Relativity}: Can the signatures of precursors, which are sensitive to apsidal and Lense-Thirring precession, be used as a direct diagnostic of GR to break the degeneracy between black hole mass and spin in models?
    \item \textbf{Understanding Precursor Emission}: How does the pre-peak color evolution of TDEs trace the formation of the reprocessing layer that is thought to produce the bulk of the optical light?
    \item \textbf{Investigating Peculiar Events}: What can be learned from unusual events like off-center TDEs (suggesting wandering black holes or inspiraling secondary systems), featureless TDEs, partial disruptions, and other repeating transients?
\end{itemize}

\subsubsection{Under-Explored Parameter Space}
Despite tremendous progress in recent years, significant gaps remain in our observational data, particularly during the earliest stages of the disruption. The primary limitation in exploring these spaces is early-time classification. Most TDEs are currently identified near or after their peak brightness, by which time the critical data regarding disk formation and initial shocks has already been lost. The following details the primary under-explored parameter spaces and their importance for understanding the physics associated with TDEs.

\begin{itemize}
    \item \textbf{Pre-Peak X-ray Observations and Variability}: One of the most critical gaps in TDE research is the lack of pre-peak X-ray data. The existence and formation of an accretion disk are best probed at soft X-ray wavelengths \citep[$\sim 0.1-10$ keV, e.g.][]{thomsen_dynamical_2022, Huang2024, giron_multigroup_2026}. Short cadence X-ray monitoring at early times would allow researchers to study stream shocks, disk formation timescales, and disk stability. High-quality early X-ray spectra could also provide independent constraints on black hole mass and spin \citep[e.g.,][]{miller_flows_2015, wen_continuum-fitting_2020}, which are currently difficult to determine from optical data alone. Observations of early-time X-ray hardness ratios, which may show harder emission pre-peak before softening, could reveal the transition from emission from shocks to a super-Eddington disk.

    \item \textbf{Pre-Peak X-ray Observations and Variability}: One of the most critical gaps in TDE research is the lack of pre-peak X-ray data. The existence and formation of an accretion disk are best probed at soft X-ray wavelengths ($\sim 0.1-10$ keV).
    
    \begin{itemize}
        \item \textbf{Disk Formation}: Short cadence X-ray monitoring at early times would allow researchers to study stream shocks, disk formation timescales, and disk stability.
        \item \textbf{Energetics and Spin}: High-quality early X-ray spectra could provide independent constraints on black hole mass and spin, which are currently difficult to determine from optical data alone.
        \item \textbf{Accretion Physics}: Observations of early-time X-ray hardness ratios, which may show harder emission pre-peak before softening, could reveal the transition from a reprocessed source to an optically thin disk.
    \end{itemize}
    
    \item \textbf{Early-Time Ultraviolet Photometry and Spectroscopy}: While TDEs are known to be ``very blue" events, we possess very few instances of pre-peak UV data. This has led to gaps in our knowledge of several key properties, including:
    \begin{itemize}
        \item \textbf{Temperature Evolution}: Pre-peak UV photometry is essential for constraining the temperature evolution of the event. Limited evidence suggests that temperatures may be significantly higher pre-peak than at the peak of the light curve, but more data is needed for population studies.

        \item \textbf{Spectral Gaps}: There has never been a confirmed UV spectrum of a TDE captured during the rise to peak. UV spectra are vital because TDEs often peak in the far UV, meaning these observations are necessary to understand the total bolometric luminosity and energetics.

        \item \textbf{Star Composition and Outflows}: UV line ratios (such as Nitrogen to Carbon) provide superior constraints on the composition of the disrupted star compared to optical lines. Furthermore, broad absorption lines (BALs) in the UV are the primary indicators of high-velocity outflows, helping to distinguish between different emission mechanisms
    \end{itemize}

    \item \textbf{Early and Late-Time Polarimetry}: Measurement of polarization is a nascent field within TDE research and represents a major underexplored frontier, allowing for the investigation of properties including:
    \begin{itemize}
        \item \textbf{Emission Mechanisms}: Constraining polarization at both early and late times would help identify the source of emission.

        \item \textbf{Geometry}: Theory suggests that early-time emission from stream collisions should be highly asymmetric (and thus more polarized), whereas a mature, ``puffy" super-Eddington disk would likely be more spherical and show less polarization
    \end{itemize}

    \item \textbf{Early-Time Radio Follow-up}: While late-time radio monitoring has become common, radio observations during the initial optical flare (the first 100 days) are still rare.
    \begin{itemize}
        \item \textbf{Outflow Launching}: Detections or even upper limits during this phase are crucial for determining when and how many outflows are launched. This data differentiates between outflows driven by early disk formation (shocks) versus those driven by later accretion state changes.
    \end{itemize}

    \item \textbf{Short Cadence Late-Time X-ray and QPEs}: The parameter space of late-time variability remains sparsely sampled.
    \begin{itemize}
        \item \textbf{QPEs}: Recent discoveries have linked QPEs, rapid, repeating X-ray flares, to TDE host galaxies.
        \item \textbf{Compact Objects}: Short cadence X-ray monitoring years after the initial event can reveal these QPEs, which may be signatures of extreme mass ratio inspirals (EMRIs) or specific disk instabilities, providing a window into the stellar populations and dynamics in the galactic nucleus.
    \end{itemize}

    \item \textbf{Sub-Day Optical Variability}: Most current surveys lack the cadence to probe sub-day optical variability.
    \begin{itemize}
        \item \textbf{Diffusion Timescales}: High-cadence observations are required to constrain the diffusion timescales of the emitting source and the internal variability of the disruption process.
    \end{itemize}

\end{itemize}

\newpage
\subsection{X-ray Binaries}

XRBs are binary star systems comprising a compact object—either a black hole or a neutron star—accreting matter from a companion star. They are broadly categorized into Low-Mass X-ray Binaries (LMXBs), where the companion is a low-mass star ($< 1\text{--}3\,M_\odot$; e.g., \citealt{bahramian2023}), and High-Mass X-ray Binaries (HMXBs), featuring a high-mass companion ($> 1\text{--}3\,M_\odot$; e.g., \citealt{fornasini2023}).

XRBs are among the most extreme objects in the universe. They are the brightest sources in the X-ray sky and were instrumental in providing the first dynamical evidence for the existence of black holes \citep[e.g.,][for Cygnus X-1]{paczynski1974,orosz2011}. Their extreme compactness and gravitational force make them unique laboratories for testing General Relativity. Furthermore, they exhibit some of the fastest known variability, launch the fastest material flows in our galaxy via relativistic jets \citep[e.g.,][]{fender2006,bright2020,vandeneijnden2021}, and can accelerate particles to Terascale (TeV) and even Petascale (PeV) energies, potentially contributing to the cosmic ray spectrum \citep{albert2007,cooper2020,lhaaso2025}. Additionally, they are small-scale analogues to active galactic nuclei and understanding their properties and evolution has important implications when considering black holes at all mass scales \citep{merloni2003,falcke2004}. 

LMXBs are typically transient, undergoing dramatic outbursts with similar properties to those predicted by the Disk Instability Model \citep{hameury2020}. During these events, black hole LMXBs transition through distinct accretion states, mapped on an X-ray ``hardness-intensity diagram", which encodes the properties of the inner accretion disk as an outburst progresses (although other properties, such as observed X-ray (quasi-)periodicity, are also key to determining the spectral state of a source). Intrinsically linked to the inner disk properties is the production of winds, jets, and ejections, the properties of which are strongly correlated with the accretion state. This disc-jet connection is observed to scale from stellar-mass black holes in XRBs to supermassive black holes in AGN, suggesting a universal mechanism \citep{merloni2003,falcke2004}. HMXBs, often persistent sources, are typically powered by accretion from the strong stellar wind of a massive companion onto a magnetized neutron star, creating X-ray pulsars. The study of XRBs provides critical insights into accretion and outflow physics, energy feedback into the interstellar medium, as well as the progenitors of gravitational wave sources.

\subsubsection{Classification of X-ray Binaries}

XRBs are binary systems where a compact object, either a neutron star (NS) or a black hole (BH), accretes matter from a companion star. The classification of an XRB is determined by the mass of this companion star.

\begin{itemize}
    \item \textbf{LMXBs}: LMXBs feature a low-mass companion star with a mass of less than approximately 1-3 M$_\odot$. Accretion primarily occurs via Roche lobe overflow, where the companion star expands to fill its gravitational potential well, spilling matter through the first Lagrange point (L1) and forming an accretion disk around the compact object \citep[e.g.,][]{paczynski1971}. In addition to this mass transfer mechanism, magnetic fields can play a significant role in governing the accretion flow, particularly in systems hosting magnetized neutron stars. In such cases, the inner accretion disk may be truncated by the compact object’s magnetosphere, channeling material along magnetic field lines onto the magnetic poles and producing pulsations. Magnetic stresses within the disk itself are also thought to drive angular momentum transport \citep[e.g., via the magnetorotational instability; ][]{balbus1991}, enabling accretion to proceed. They can produce outflows in the form of disc winds and relativistic jets and ejections. Systems with prominent jets are sometimes referred to as ``microquasars.". Most LMXBs are transient sources, undergoing dramatic outbursts that can last from days to years. A minority are persistent X-ray sources. Orbital periods typically range from a few hours to several days. Subsequently the radial extent of the accretion disk varies significantly between sources. Subclasses include BH-LMXB, NS-LMXB, Accreting Millisecond X-ray Pulsars (AMXPs), Transitional Millisecond Pulsars (tMSPs), Ultracompact XRBs (UCXBs), Very Faint X-ray Transients (VFXTs).

    \item \textbf{HMXBs}: HMXBs feature a high-mass companion star with a mass greater than approximately 1-3 M$_\odot$. Accretion is often driven by the strong stellar wind from the massive companion star. In some cases, Roche lobe overflow or accretion from a ``decretion disc" around the companion can occur. In addition to the massive star and compact object, HMXBs may feature an accretion disc and corona. Outflows include powerful stellar winds, disc winds, and jets. Due to the persistent stellar wind from the companion, most HMXBs are persistent X-ray sources. HMXBs have a wide range of orbital periods, and for systems with neutron stars, the spin period is often measurable. Subclasses include BH-HMXB, NS-HMXB, Be-XRBs, Supergiant XRBs (wind or disk-fed), Wolf-Rayet XRBs, Superfast X-ray Transients (SFXTs), Ultraluminous X-ray Sources (ULXs).

\end{itemize}

\subsubsection{The Extreme Nature of X-ray Binaries} 

XRBs are defined by their extreme physical properties, which make them crucial objects for astrophysical study.

\begin{itemize}

\item {\bf Brightest Objects in the X-ray Sky}: XRBs are the brightest point sources in the X-ray sky. Transient XRBs can increase in brightness by over six orders of magnitude, reaching peak fluxes of up to 42 Crab (e.g., A0620-00), significantly outshining the brightest persistent sources. In X-ray images of galaxies like M51, the majority of point sources are XRBs.

\item {\bf First Dynamical Evidence for Black Holes}: Observations of XRBs in quiescence provided the first definitive, dynamical proof of black holes. By measuring the extreme radial velocity curve of the companion star orbiting an unseen object, astronomers could use Kepler's laws to infer a mass too great for a neutron star, as was famously done for Cygnus X-1 and A0620-00.

\item {\bf Most Compact Objects Known}: XRBs host objects with the highest known compactness (Mass/Radius ratio) and gravitational force. This makes them ideal laboratories for testing General Relativity, allowing for measurements of spacetime warping and gravitational redshift through effects like the relativistic broadening of iron emission lines. A stellar-mass black hole exerts a stronger gravitational force at its event horizon than a supermassive black hole.

\item {\bf Fastest Varying Objects Known}: XRB emission varies on extremely short timescales. Analysis of power spectra reveals quasi-periodic oscillations (QPOs) on top of broadband noise. The orbital frequency of matter at the Innermost Stable Circular Orbit (ISCO) provides a theoretical limit, with frequencies of $\sim$220 Hz for a non-spinning 10 M$_\odot$. black hole and up to 1615 Hz for a maximally spinning one. Neutron star XRBs can exhibit even faster variability, with kHz QPOs observed in sources like Sco X-1.

\item {\bf Launch the Fastest Flows in Our Galaxy}: XRBs launch powerful, relativistic jets. Some discrete ejections exhibit apparent superluminal motion. These jets can deposit a significant fraction of the accretion energy into the Interstellar Medium (ISM). The XRB SS433 is a unique case, producing precessing, baryonic jets (containing hydrogen and helium) traveling at 0.26 times the speed of light (0.26c). XRBs offer a unique advantage over AGN in that jet evolution can be studied on human timescales (seconds to years) rather than millions of years.

\item {\bf Particle Acceleration to TeV and PeV Energies}: XRBs are powerful particle accelerators. The jets of SS433 are seen impacting the surrounding supernova remnant W50, creating hotspots that emit TeV gamma-rays as detected by H.E.S.S. More impressively, the HMXB V4641 Sgr has been observed by HAWC and LHAASO to produce gamma-rays up to 800 TeV, which requires the acceleration of particles to PeV energies. This discovery suggests that XRBs could be a source for the ``knee" in the cosmic ray spectrum.

\end{itemize}

\subsubsection{Physics of Low-Mass X-ray Binaries}

LMXBs are accreting systems in which a neutron star or black hole draws material from a low-mass companion star via Roche-lobe overflow, forming an accretion disc. Their observational phenomenology is governed by the interplay between accretion physics, radiative processes, and outflows, producing a rich set of time-variable behaviors across the electromagnetic spectrum.

\begin{itemize}
    \item \textbf{Transient Outbursts and the Disk Instability Model}: The transient behavior of most LMXBs is described by the Disk Instability Model (DIM). In the quiescent state, a cold accretion disc fills with matter transferred from the companion star. When the temperature at a certain radius in the disc becomes high enough to ionize hydrogen, a thermal-viscous instability is triggered. This instability causes heating fronts to propagate through the disc, leading to a rapid increase in the mass accretion rate onto the compact object. This manifests as a multiwavelength outburst, with the optical flux rising first as the outer disc heats up, followed several days later by the X-ray flux as the matter reaches the inner regions.
    
    \item \textbf{The State Cycle of Black Hole LMXBs}: During an outburst, black hole LMXBs evolve through a series of distinct accretion states, which can be tracked on a Hardness-Intensity Diagram (HID).
    
    \begin{itemize}
    
        \item \textbf{Hard State}: The outburst begins in the hard state. The X-ray spectrum is dominated by a power-law component, attributed to Compton up-scattering of soft photons in a hot corona. This state is characterized by high fractional RMS variability, QPOs, and a ``steady" compact jet detectable from radio to infrared wavelengths. Cold disc winds may also be present.
        
        \item \textbf{State Transition}: As the source brightens, it transitions towards the soft state. During this transition, the QPOs change abruptly, the jet spectrum evolves, and bright radio flares are often observed as discrete blobs of plasma are ejected.
        
        \item \textbf{Soft State}: In the soft state, the X-ray spectrum is dominated by thermal emission from the accretion disc. The variability is low, the core radio jet is quenched (though emission from earlier ejections may persist), and hot, highly ionized disc winds are observed in X-ray absorption lines.
        
        \item {\bf Return to Quiescence}: As the outburst fades, the system transitions back through the hard state before returning to quiescence.
    
    \end{itemize}
    
    \item \textbf{Disc-Jet Coupling and Unification}: The HID demonstrates a clear coupling between the accretion flow and jet output. In the hard state, there is a strong correlation between the radio (jet) and X-ray (accretion flow) luminosities. This relationship forms the basis of the Fundamental Plane of Black Hole Activity, a scaling relation that connects stellar-mass black holes in XRBs with supermassive black holes in AGN, suggesting a single, scale-invariant jet formation process. While neutron star XRBs also have jets, they are systematically fainter in radio than black hole systems for a given X-ray luminosity, a puzzle that remains an open question.
    
    \item \textbf{Neutron Star LMXB Phenomena}: NS LMXBs exhibit unique phenomena related to the neutron star's solid surface and magnetic field. These include:
    
    \begin{itemize}
        
        \item {\bf Type I X-ray Bursts}: These are thermonuclear flashes on the neutron star surface. The burst emission irradiates the accretion disc, causing a delayed optical/infrared echo that can be used to map the geometry of the system.
        
        \item {\bf Burst Oscillations}: Millisecond-period oscillations can be observed during Type I bursts, providing information about the neutron star's properties.
        
        \item {\bf Spin and Orbital Evolution}: Precise timing of accreting millisecond X-ray pulsars, like SAX J1808.4-3658, allows for the measurement of the neutron star's spin evolution over decades. Likewise, long-term monitoring can reveal anomalous changes in the binary's orbital period.
    
    \end{itemize}
\end{itemize}

\subsubsection{Physics of High-Mass X-ray Binaries}

HMXBs are complex systems dominated by the interaction between a magnetized neutron star and the environment of its massive companion.

\begin{itemize}
    
    \item {\bf Accretion Geometry}: Accretion from a stellar wind is complex, forming structures like an accretion wake and a photoionization wake. For highly magnetized neutron stars, matter is channeled by the magnetic field onto the magnetic poles, forming an accretion column.
    
    \item {\bf X-ray Pulsars}: The channeling of accretion onto the poles, combined with the neutron star's rotation, produces beamed X-ray emission, making the system an X-ray pulsar. The geometry of this beamed emission can be a ``pencil beam" or a ``fan beam."
    
    \item {\bf Classification}: HMXBs are classified using the Corbet diagram, which plots the neutron star spin period against the binary orbital period. This diagram separates distinct populations, such as Be-XRBs (which tend to have longer orbital periods) and wind-fed supergiant systems.

    \item {\bf Spectra and Outbursts}: HMXB X-ray spectra can be highly obscured by intrinsic absorption from the dense stellar wind. Be-XRBs can exhibit quasi-regular short outbursts due to interactions between the neutron star and the decretion disc of the Be-type companion star.
    
    \item {\bf Extreme Case: Cygnus X-3}: This unique HMXB contains a Wolf-Rayet companion with an exceptionally strong stellar wind. The interaction of a relativistic jet with this dense wind leads to particle collisions that produce high-energy gamma-rays and potentially neutrinos, with some IceCube events showing coincidence with gamma-ray flares from the source.

\end{itemize}

\subsubsection{The Role of X-ray Binaries in the Universe}

XRBs are critical tools for understanding a wide array of fundamental astrophysical processes, including:

\begin{itemize}

    \item {\bf Compact Object Science}: They provide the primary means of measuring the masses and spins of stellar-mass black holes and neutron stars.
    
    \item {\bf General Relativity in Strong Gravity}: Their extreme gravity allows for direct tests of General Relativity in the strong-field regime.
    
    \item {\bf Accretion and Outflow Physics}: They are the best laboratories for studying the physics of accretion discs, coronae, and the launching of jets and winds, with changes occurring on observable timescales.
    
    \item {\bf Energy Injection and Feedback}: The powerful jets and winds from XRBs inject significant energy into the ISM, providing a form of stellar-scale feedback analogous to AGN feedback.
    
    \item {\bf Particle Acceleration}: They are confirmed sites of particle acceleration to extreme (PeV) energies.
    
    \item {\bf Gravitational Wave Physics}: They are progenitors of merging compact objects detected by LIGO/Virgo and potential sources of continuous gravitational waves for future detectors.

\end{itemize}

\subsubsection{Under-Explored Parameter Space}

A significant portion of the XRB population and behavior remain under-explored or difficult to detect. This includes:

\begin{itemize}
    
    \item \textbf{The Hidden Population of Galactic Black Holes}: A major gap in our current understanding is the vast population of quiescent black holes. It is estimated that there are more than 10$^{5}$ black holes in our galaxy, but the vast majority remain invisible because they are not actively accreting. These objects are typically only discovered when they enter a period of transient outburst, during which their X-ray flux can increase by nearly six orders of magnitude. Historically, only 5–10 such outbursts are detected annually by all-sky monitors, leaving a significant portion of the galactic black hole population unmapped.
    
    \item \textbf{Transient Sub-classes and Faint Phenomena} Several sub-classes of XRBs inhabit parameter spaces that have only recently begun to be explored. These include:
    
    \begin{itemize}
    
        \item {\bf VFXTs}: These sources are characterized by outbursts that do not reach the high luminosities (close to the Eddington limit) typical of standard transients.
        
        \item {\bf Transitional Millisecond Pulsars (tMSPs)}: This rare class of objects demonstrates the ability to swing between being a radio pulsar and an accreting Low-Mass X-ray Binary (LMXB), revealing a complex evolutionary link.
        
        \item {\bf Ultracompact XRBs}: Often involving a compact object accreting from a white dwarf, these systems are significant as potential gravitational wave sources for future missions like LISA.

    \end{itemize}

    \item \textbf{The Frontier of Rapid Temporal Variability}: The parameter space of high-frequency timing is critical for understanding the innermost regions of accretion discs. Matter orbiting at the ISCO can reach frequencies of 220 Hz for non-spinning black holes and over 1600 Hz for maximally-spinning ones. Neutron star XRBs can exhibit kilo-Hertz Quasi-Periodic Oscillations (kHz QPOs) exceeding 1000 Hz. Capturing the very earliest stages of outbursts is another area of active development; for example, the XB-NEWS pipeline was designed to automatically identify precursors and measure accretion variability between outbursts.
    
    \item \textbf{High-Energy Particle Acceleration}: A surprising and newly explored frontier is the ability of XRBs to act as PeVatrons. While it was long known that XRBs launch relativistic jets, recent observations of sources like V4641 Sgr and SS 433 have detected gamma-ray emission up to 800 TeV. This requires particles to be accelerated to PeV energies, suggesting that XRBs may contribute to the ``knee" in the cosmic ray spectrum—a role previously thought to be reserved for more massive galactic structures.
    
    \item \textbf{Unresolved Physics of Jets and Winds}: Despite decades of study, several fundamental questions regarding outflows remain open:
    
    \begin{itemize}
    
        \item {\bf Jet Disparity}: It remains an open question why neutron star jets are consistently fainter in radio wavelengths than black hole jets at similar X-ray luminosities.
        
        \item {\bf Wind Launching}: The specific mechanisms that launch blueshifted absorption line winds—which are seen in both black hole and neutron star systems—continue to be debated.
        
        \item {\bf Accretion Geometry}: In HMXBs, the exact geometry of X-ray emission from the accretion column (such as ``pencil beam" vs. ``fan beam" models) remains uncertain
    
    \end{itemize}
    
\end{itemize}

\subsubsection{Future Needs and Capabilities}

Advancing the study of XRBs requires a multifaceted approach involving new observational hardware, technical instrumental upgrades, and enhanced community coordination to capture the rapid, multiwavelength evolution of these systems

\begin{itemize}

    \item \textbf{Missions and Observatories}: Several current and upcoming missions are essential for expanding the known population of XRBs and exploring new physical regimes. These include:
    
    \begin{itemize}
    
        \item {\bf Discovery of Quiescent Systems}: Identifying systems not currently in outburst is a high priority. The Roman Space Telescope is expected to discover quiescent systems through its galactic bulge survey, while the Square Kilometre Array (SKA) and ngVLA will enable proper motion studies to detect these faint sources.
        
        \item {\bf Extragalactic Surveys}: The LSST (Legacy Survey of Space and Time) will be critical for discovering long-period XRBs both within our galaxy and in extragalactic environments.
        
        \item {\bf GW Astrophysics}: Future missions like LISA are cited for their capability to detect gravitational waves from ultracompact binaries, such as white dwarfs accreting onto neutron stars or black holes.
        
        \item {\bf Continuous Monitoring}: The ARGUS mission is highlighted for its potential to provide all-sky optical monitoring, while maintaining wide-field X-ray monitors (like MAXI) and improving X-ray timing (like NICER) remains crucial.
    
    \end{itemize}
    
    \item \textbf{Technical and Instrumental Enhancements}: Advancing XRB science will require targeted instrumental improvements to address current observational limitations and enable higher-fidelity measurements. These include:

    \begin{itemize}

        \item \textbf{Bright Source Handling}: A major limitation in current X-ray detectors is ``pile-up" and telemetry issues caused by exceptionally bright sources. Future facilities like SKA and ngVLA will require sub-arraying capabilities, allowing astronomers to use only a portion of the array to prevent saturation when high sensitivity is unnecessary.
        
        \item {\bf High-Resolution Timing and Spectroscopy}: Scientists require a global network of sub-millisecond observations and better absolute clocks to understand the mechanisms of jet launching. Additionally, high-quality broad-band X-ray spectroscopy is needed to accurately distinguish between neutron star and black hole candidates.
        
        \item {\bf Polarimetry}: Building on the success of IXPE, future missions must continue providing X-ray polarization to constrain the geometry and orientation of accretion flows
    
    \end{itemize}

    \item \textbf{Operational Frameworks and Coordination}: Because the most impactful XRB science relies on strictly simultaneous multiwavelength data, there is a need for significant improvements in how the community coordinates observations, including:

    \begin{itemize}
    
        \item {\bf Automated Alert Systems}: There is a critical need for the automated dissemination of alerts in machine-readable formats (such as VOEvents) to enable rapid follow-up by robotic telescopes. Currently ATels are still relied upon quite heavily.
        
        \item {\bf Formal Coordination Channels}: Current ad-hoc coordination often suffers from high failure rates when observatory schedules shift. The creation of formal channels for multi-facility synchronization and potentially ``locking" schedules for major facilities (like JWST) to facilitate simultaneous campaigns would be of substantial benefit.
        
        \item {\bf Shared Calibration}: Implementing shared calibration strategies for monitoring facilities would increase time efficiency and improve the consistency of long-term data sets
    
    \end{itemize}
\end{itemize}

\newpage

\subsection{Novae}

The scientific understanding of novae has undergone a fundamental paradigm shift, moving from a classical model of simple thermonuclear eruptions to a more complex and dynamic picture where energetic shocks are a primary driver of the event's luminosity, even at optical wavelengths \citep{2017NatAs...1..697L, 2020NatAs...4..776A}. The catalyst for this transformation was the unexpected discovery of high-energy GeV gamma-ray emission from novae by the Fermi-LAT space telescope \citep{2010Sci...329..817A, 2016ApJ...826..142C, 2018A&A...609A.120F, 2021ApJ...910..134G}. This finding revealed that nova eruptions host powerful shocks capable of accelerating particles to relativistic speeds.

Subsequent analysis indicates that these shocks could possess a luminosity that rivals the total bolometric output of the nova itself \citep[$\sim$10$^{33}$ erg~s$^{-1}$;][]{2015MNRAS.450.2739M, 2020NatAs...4..776A}. The dense environment of the nova ejecta causes this immense shock energy to be radiated away, rather than dissipated into adiabatic expansion, contributing significantly to the nova's emission across the electromagnetic spectrum, including in visible light. This discovery of ``radiative shocks" as a key power source explains previously puzzling features in nova light curves and has been confirmed by observations of simultaneous flaring in both optical and gamma-ray bands \citep{2020NatAs...4..776A}.

\subsubsection{Novae as Astrophysical Laboratories}

Novae are now recognized as premier astrophysical laboratories in our Galactic neighborhood. Their relative proximity and frequency allow for detailed, multiwavelength studies of fundamental physical processes, including:

\begin{itemize}
    \item {\bf Shock Physics}: Investigating particle acceleration efficiency in conditions applicable to other shock-powered energetic transients like supernovae.
    \item {\bf Common Envelope Physics}:  Observing complex, multi-stage mass ejections that serve as real-time analogs for common envelope evolution.
    \item {\bf Nucleosynthesis}: Constraining models of isotope and lithium production in the universe.
    \item {\bf Multi-Messenger Astrophysics}: Presenting the potential for novae to be the next class of detectable neutrino sources. The anticipated eruption of the nearby recurrent nova T Coronae Borealis offers a rare and imminent opportunity to test these new theories and potentially achieve the first-ever neutrino detection from a nova, solidifying their status as multi-messenger phenomena.
\end{itemize}

\subsubsection{The Standard Model of Nova Eruptions}

Prior to the Fermi era, novae were (mostly) considered well-understood phenomena. The standard model described them as stellar eruptions occurring in binary star systems composed of a white dwarf and a companion star (either a main-sequence star or an evolved giant). Primary components of the model included:

\begin{itemize}
    \item {\bf Mechanism}:  The white dwarf's powerful gravity accretes hydrogen-rich material from its companion. This material accumulates on the white dwarf surface, where pressure and density increase until they trigger a thermonuclear runaway (TNR) at the base of the accreted layer.

\item {\bf Eruption}:  The TNR leads to a massive eruption that ejects the accreted envelope at high velocities, ranging from 500 to 5,000 km~s$^{-1}$.

\item {\bf Luminosity Source}:  The eruption's luminosity was believed to be powered entirely by thermal emission from remnant nuclear burning on the hot surface of the white dwarf. This process sustains a near-Eddington luminosity of approximately 10$^{38}$ erg~s$^{-1}$.

\item {\bf Multiwavelength Evolution}:  As the opaque ejecta expand and thin, the peak of the spectral energy distribution shifts to higher energies. An observer sees the nova peak first in optical wavelengths, followed by a peak in the ultraviolet, and finally in super-soft X-rays as the nuclear burning on the white dwarf surface becomes directly visible.

\item {\bf Recurrence}:  After the eruption, accretion resumes \citep[potentially while the TNR is still ongoing;][]{2018MNRAS.480..572A}, building material for a future eruption. While typical recurrence periods are on the order of at least thousands of years, a subset known as `recurrent novae' have outbursts on human-observable timescales of decades \citep{2010ApJS..187..275S, 2014ApJ...788..164P, 2016ApJ...833..149D}. Novae are common, with 10 to 15 discovered in our Galaxy each year \citep{2017ApJ...834..196S, 2022ApJ...937...64K}.

\end{itemize}

\subsubsection{A Paradigm Shift: The Discovery of GeV Shocks in Novae}

The launch of the Fermi Gamma-ray Space Telescope \citep{2004SPIE.5488..763V, 2022hxga.book...29T} and its Large Area Telescope \citep[LAT;][]{2009ApJ...697.1071A} fundamentally challenged the standard nova model. The instrument's detection of GeV gamma-ray emission from multiple novae was a shocking discovery, as there is no mechanism in the purely thermal model to produce such high-energy photons.
This discovery provided unambiguous evidence for the presence of powerful shocks within the nova ejecta. These shocks accelerate charged particles to relativistic speeds through diffusive shock acceleration, where particles gain energy by repeatedly crossing the shock front.
The resulting high-energy photons can be produced through two primary channels, which current gamma-ray spectra cannot definitively distinguish:

\begin{itemize}
    \item {\bf Leptonic}: High-energy electrons produce gamma-rays.
\item {\bf Hadronic}: High-energy protons (hadrons) interact to produce neutral pions, which then decay into gamma-rays.
\end{itemize}

The location and formation mechanism of these shocks are thought to depend on the nature of the binary system, with two possible scenarios:

\begin{itemize}
    \item {\bf External Shocks}: In systems with a red giant companion, the nova ejecta violently collide with the dense circumbinary medium previously created by the giant star's stellar wind \citep{2010Sci...329..817A}.

\item {\bf Internal Shocks}:  In systems with a main-sequence companion, which lack a dense external medium, shocks are believed to be internal to the ejecta. This occurs when the nova produces multiple phases of mass loss, such as a faster outflow catching up to and colliding with a slower, earlier ejection \citep{2014Natur.514..339C, 2015MNRAS.450.2739M}. 

\end{itemize}

\subsubsection{Radiative Shocks: A New Power Source for Nova}

While hints of shocks existed in pre-Fermi data, such as hard X-ray emission and irregular optical light curves with jitters and flares, the Fermi discovery highlighted how energetically dominant these shocks could be.

\subsubsubsection*{The Shock Energy Budget}

The observed gamma-ray luminosity (L$_{\gamma}$) of novae is typically 10$^{35}$ to 10$^{36}$ erg~s$^{-1}$, which is about 0.1-1\% of the total bolometric luminosity (L$_{\rm Bol}$) of the nova. However, the Diffusive Shock Acceleration mechanism is inefficient. Models suggest that only a few percent of the total shock energy is used to accelerate particles, and of that, only about 20\% is emitted in the Fermi energy band. This implies that the observed gamma-ray luminosity represents only $\sim$1\% of the total power of the shock (L$_{\rm sh}$).
This calculation leads to a stunning conclusion: L$_{\rm sh}$ must be approximately  10$^{38}$ erg~s$^{-1}$, a value that rivals the entire bolometric luminosity of the nova traditionally attributed to thermonuclear burning \citep{2020NatAs...4..776A}.
While this energy likely originates primarily from the kinetic energy of the interacting ejecta, the binary orbital motion may also contribute if it plays a role in expelling the nova envelope \citep{1990ApJ...356..250L, 2014Natur.514..339C, 2026NatAs..10..271A}. In this case, momentum transferred from the binary motion could provide an additional contribution to the shock energy budget.

\subsubsubsection*{From Shock Power to Visible Light}

In the extremely dense environment of the early nova ejecta, the cooling time for the shock-heated gas is very short. Consequently, the immense energy from the shocks is radiated away immediately rather than being dissipated through adiabatic expansion. These radiative shocks act as a powerful, additional source of emission that contributes directly across the spectrum, including to the visible light that defines the nova outburst \citep{2015MNRAS.450.2739M}. Within this framework, the shock-heated gas reaches temperatures of 10$^6$--10$^7$~K, implying that the intrinsic emission is predominantly in the X-ray regime. However, because these shocks occur early in the eruption (during the first few days to weeks), they propagate into a dense environment. As a result, much of the X-ray emission is expected to be absorbed by the surrounding ejecta and subsequently reprocessed to lower energies, enabling it to escape in the optical and ultraviolet bands and contribute significantly to the observed visible emission \citep{2017NatAs...1..697L, 2020NatAs...4..776A}.

\subsubsubsection*{Observational Confirmation}

This new paradigm, where shocks significantly power the nova's optical light, is supported by compelling observational evidence.

\begin{itemize}
    \item {\bf V5856 Sgr}: Discovered in 2016, the optical and gamma-ray light curves showed a distinct correlation, with both rising and dipping in unison, suggesting a shared physical origin \citep{2017NatAs...1..697L}.
    \item {\bf V906 Car}: Discovered in 2018, V906 Car provided definitive proof. Simultaneous, powerful flares were observed in both optical and gamma-ray bands \citep{2020NatAs...4..776A}. During these flares, the nova's total bolometric luminosity doubled, demonstrating that shocks can indeed power a significant fraction of the visible emission.
\end{itemize}

\subsubsection{Novae as Premier Astrophysical Laboratories}

The revelation that novae are complex, shock-powered events has repositioned them as ideal laboratories for studying a wide range of astrophysical processes. Their brightness, proximity, and multiwavelength nature make them uniquely accessible.

\subsubsubsection*{Shock Physics}

For shock physics, novae provide a nearby environment to study particle acceleration and radiative shock dynamics in detail. Understanding these processes has direct implications for interpreting more distant and energetic shock-powered transients, such as Type IIn Supernovae, stellar mergers, and FBOTs.

\subsubsubsection*{Common Envelope Physics}

Novae can also play an important role in our understanding of common envelope physics. The internal shock model requires multiple, complex outflows, challenging the idea of a single, impulsive ejection. Evidence suggests that the binary's orbital motion may help expel the envelope in stages, creating a ``common envelope" phase. Unlike common envelope evolution in planetary nebulae, which occurs over millennia, this process in novae evolves on observable timescales of weeks to months.

\begin{itemize}
    \item {\bf Spectroscopic Evidence}: Spectra of many novae show co-existing absorption features corresponding to slow outflows (hundreds of km~s$^{-1}$) and later, faster outflows \citep[thousands of km~s$^{-1}$;][]{2020ApJ...905...62A}.

    \item {\bf Direct Imaging}: High-resolution imaging with facilities like the CHARA Array has resolved nova ejecta in the first days and weeks of eruption:

    \item {\bf V1674 Her (a very fast nova)}: Imaged just 2-3 days after its eruption, the nova revealed a clear departure from spherical symmetry, showing two distinct expanding components \citep{2026NatAs..10..271A}.
    
    \item {\bf V1405 Cas (a slow nova)}: Initially imaged 55 days into its eruption, it showed a large ($\sim$1 AU) expanded envelope with little surrounding ejecta \citep{2026NatAs..10..271A}. After gamma-rays were detected weeks later (signaling shock interaction), subsequent imaging revealed a new, extended structure, providing direct evidence for delayed and complex mass ejection.

\end{itemize}

\subsubsubsection*{Fireball and Nucleosynthesis}

The initial thermonuclear flash is predicted to produce a brief (few hours) X-ray/UV burst (the “fireball phase”), which was observed definitively for the first time from nova YZ Ret by eROSITA \citep{2022Natur.605..248K}. Novae are known cosmic factories for several key isotopes, including $^{13}$C, $^{15}$N, $^{26}$Al, and $^{22}$Na \citep{1998ApJ...494..680J, 2016PASP..128e1001S}, and are hypothesized to be the main producers of Lithium in the galaxy \citep{2015ApJ...808L..14I}. 

Future missions like ULTRASAT (for UV flashes) and COSI (for detecting nuclear lines directly) will provide unprecedented constraints on nucleosynthesis models.

\subsubsubsection*{Multi-Messenger Astrophysics}

The hadronic model for gamma-ray production predicts that novae should also be sources of high-energy neutrinos.

\begin{itemize}
    \item {\bf Neutrino Potential}:  If gamma-rays originate from neutral pion decay, charged pions produced in the same interactions will decay to produce neutrinos, making novae potential multi-messenger sources.

\item {\bf Current Status}: Thus far, searches for a correlated neutrino signal from Fermi-detected novae using IceCube have yielded only upper limits \citep{2023arXiv230715372T, 2023JCAP...03..015G}.

\item {\bf High-Energy Frontier}:  The 2021 eruption of the recurrent nova RS Ophiuchi was detected as a bright source of very-high-energy (TeV) gamma-rays by the H.E.S.S. \citep{2022Sci...376...77H} and MAGIC \citep{2022NatAs...6..689A} telescopes, confirming that novae can accelerate particles to extreme energies.

\item {\bf The T Coronae Borealis Opportunity}:  The recurrent nova T Coronae Borealis (T CrB), located only $\sim$0.8~kpc away, is predicted to erupt in the near future. It will reach 2nd magnitude, making it easily visible to the naked eye, although will then fade rapidly. Its proximity and expected power make it the prime candidate for becoming the first nova ever detected in neutrinos \citep{2025arXiv250707096T}, which would herald a new era of multi-messenger nova science.

\end{itemize}

\subsubsection{Under-Explored Parameter Space}

Observations revealing the presence of energetic shocks capable of accelerating particles to relativistic speeds have opened an under-explored parameter space regarding how these shocks contribute to the total energy budget and multiwavelength appearance of novae. A critical area of current research involves determining the parameters that dictate shock luminosity, including:

\begin{itemize}
    \item {\bf Luminosity Scaling}: The shock luminosity is proportional to the density and the velocity of the ejecta (L$_{sh} \propto \rho v^2$).

\item {\bf Radiative Shocks}: While total bolometric luminosity (L$_{\rm Bol}$) is near Eddington levels ($\sim 10^{38}$ erg~s$^{-1}$), shock energy can rival this value if the shocks are radiative rather than adiabatic. This means shocks may provide an additional source of visible emission, altering our fundamental understanding of nova light curves.

\item {\bf Gamma-ray Correlation}: Observed gamma-ray luminosity (L$_\gamma$) is typically 0.1--1\% of L$_{\rm Bol}$. Understanding the efficiency of particle acceleration and whether the emission is hadronic or leptonic remains a key part of this under-explored space.

\item {\bf Connection between shocks and dust formation}: The shock heated gas behind the shock could cool down very rapidly, creating ideal environment for dust to condensate while being shielded from the harsh radiation originating on the surface of the white dwarf \citep{2017MNRAS.469.1314D}.

\end{itemize}

\subsubsection{Future Needs and Strategic Imperatives}

The discovery of powerful, radiative shocks has revolutionized the study of novae.  Novae are now understood to be true multiwavelength and potential multi-messenger laboratories in our Galactic backyard. To continue advancing this field, the following are needed:

\begin{itemize}
    \item {\bf Sustained Observations}:  More observations across the electromagnetic spectrum are critical. The continued operation of key facilities like the Fermi Gamma-ray Space Telescope is essential for detecting the high-energy signatures of shocks. Future gamma-ray facilities may allow us to detect shock emission from extragalactic novae as well.
\item {\bf Advanced Modeling}:  More sophisticated theoretical work and modeling are required to fully understand the interplay between thermonuclear burning, complex mass loss, and shock physics.
\item {\bf Interdisciplinary Collaboration}:  Enhanced collaboration between observers, theorists, and researchers studying other shock-powered transients (e.g., supernovae) will be crucial for synthesizing a complete picture of these dynamic events.

\end{itemize}

\newpage

\subsection{Supernovae}

The advent of the Vera C.\ Rubin Observatory's LSST
marks a pivotal moment for supernova science, promising an unprecedented increase in the
discovery rate of transient events, potentially by a factor of 10 or 100 for rare phenomena.
This will transform once-in-a-decade observations into common occurrences. However, the
LSST's Wide-Fast-Deep survey cadence of three days is fundamentally insufficient for
capturing the most critical scientific clues about supernova progenitors and explosion
physics, which often manifest on timescales of less than two days.
 
The most profound advancements will come from a ``Top of the Pyramid'' or ``Tip of the
Spear'' scientific philosophy. This approach prioritizes the intensive, high-cadence,
multiwavelength study of individual, nearby supernovae to acquire ``game-changing'' data
that drives theoretical understanding. Achieving this in the LSST era necessitates a hybrid
strategy that leverages LSST's discovery power but relies on a coordinated, automated
ecosystem of ground- and space-based follow-up facilities.
 
Pathfinder projects like the DLT40 survey and strategic initiatives such as the ``Shadowing
LSST'' DECam program, PASSTA, and the Global Supernova Project are pioneering the
necessary infrastructure. Success will depend on developing unified alert brokers to merge
disparate data streams, increasing automation for rapid response, overcoming data access
limitations like the LSST's ``80-hour rule,'' and fostering large-scale, worldwide
collaborations to secure world-class data sets.
 
\subsubsection{The LSST Era: An Unprecedented Opportunity and a Fundamental
Challenge}
 
The upcoming LSST era is poised to revolutionize time-domain astronomy. With its
3.2-gigapixel camera and 9.6 square-degree field of view, the survey will generate an
enormous volume of data, on the order of 10 million alerts per night. This will
dramatically increase the sample size of observed supernovae, opening a vast new
discovery space. The key opportunities made available through these new capabilities include:
  
\begin{itemize}
    \item \textbf{Massive Volume Increase:} Phenomena observed rarely today, such as the
    early light curve excess of SN~2023ixf, will become far more common. LSST's depth will
    allow such features to be detected out to 200 megaparsecs (Mpc), a significant expansion
    from the current ${\sim}7$~Mpc window for similar events.
 
    \item \textbf{Probing New Physics:} The sheer number of discoveries will enable robust
    statistical studies and the detection of rare precursor activities, such as small outbursts in
    the years preceding a core-collapse supernova explosion.
\end{itemize}
  
Despite its discovery power, the LSST's primary Wide-Fast-Deep (WFD) survey has a
three-day cadence. This poses a significant problem for supernova science, as many of the
most revealing physical signatures evolve on much shorter timescales. The consensus
within the community is that while LSST data on its own will be valuable, it is insufficient
for the detailed, early-time science required to understand supernova physics. This
necessitates a robust ecosystem of rapid follow-up resources.
 
\subsubsection{``Top of the Pyramid'' Science Philosophy}
 
To navigate the deluge of LSST data, advancements will come from a ``Top of the
Pyramid'' philosophy, visualized as a pyramid representing different approaches to
supernova studies.
 
\begin{itemize}
    \item \textbf{Base of the Pyramid:} A large number of supernovae for which limited,
    mediocre data is collected, primarily for statistical sample papers.
 
    \item \textbf{Middle of the Pyramid:} Nearby, bright, or exotic supernovae that receive
    good data, often resulting in single-object publications.
 
    \item \textbf{Pinnacle of the Pyramid:} A select few events where extraordinary,
    ``world-class'' data is obtained, revealing new clues about explosion mechanisms and
    progenitor stars. This ``game-changing'' data drives and tests theoretical models and
    inspires new work down the pyramid.
\end{itemize}
 
The core argument is that the intensive study of individual supernovae drives the field. The
goal in the LSST era is to expand the ``pinnacle'' of the pyramid, making the acquisition of
world-class data on unique events a more frequent occurrence.
 
\subsubsection{Key Scientific Frontiers in Early-Time Supernova Observation}
 
The first hours and days after a supernova explosion are critical for constraining physical
models. The pursuit of high-cadence, early-time data is focused on several key scientific
questions, which vary by supernova subclass.
 
\subsubsubsection*{Type Ia Supernovae: Progenitor Clues}


Early multi-band light curves and spectra of Type~Ia SNe offer direct tests of progenitor
scenarios.
 
\begin{itemize}
    \item \textbf{Early Light Curve Excesses:} Theoretical models predict that interactions
    between the supernova ejecta and a non-degenerate companion star (in a
    single-degenerate scenario) can produce an excess or ``bump'' in the light curve. These
    features occur on timescales of less than two days, requiring discovery and follow-up on
    sub-day cadences.
 
    \item \textbf{Rapid Spectral Evolution:} Extremely early spectra can reveal extraordinary
    features. SN~2021aefx, for instance, showed a silicon velocity of
    30{,}000~km\,s$^{-1}$ and decelerated by ${\sim}9{,}000$~km\,s$^{-1}$ in a single
    day. Its spectrum was so unusual it was initially misclassified as a Type~Ic.
 
    \item \textbf{multiwavelength Constraints:} Combining optical data with space-based
    observations is crucial. SN~2021aefx was the first Type~Ia to be observed by JWST
    between 2.5--5 microns, providing new insights into iron-group elements, emission
    geometry, and ionization structure. Early \textit{Swift} UV data is also critical for testing
    shock-interaction models.
\end{itemize}
 
\subsubsubsection*{Core-Collapse Supernovae: Pre-Explosion Environment}


For core-collapse supernovae, early observations act as an ``autopsy'' of the material
surrounding the star just before it died.
 
\begin{itemize}
    \item \textbf{Flash Spectroscopy:} Very early spectra can show ``flash features''---recombination lines from circumstellar material (CSM) ionized by the initial shock
    breakout. The evolution of these features reveals the composition, density, and extent of
    the CSM, providing clues about the progenitor star's mass loss.
 
    \item \textbf{Unprecedented Light Curve Features:} SN~2023ixf, a nearby supernova in
    M101, revealed a never-before-seen early excess in its light curve, compiled from amateur
    astronomer data. The physical origin (shock breakout cooling, CSM interaction) remains
    unknown, but LSST will be able to detect such features out to 200~Mpc.
 
    \item \textbf{Intra-Night Evolution:} Recent observations of SN~2024ggi demonstrated
    that spectroscopic features can evolve significantly over just a few hours. High-resolution
    spectra taken eight hours apart revealed narrow lines evolving, potentially tracing the
    ${\sim}40$~km\,s$^{-1}$ wind of the red supergiant progenitor. This underscores the
    need for observational cadences much shorter than one day for special events.
 
    \item \textbf{Precursor Outbursts:} Some massive stars exhibit ``burps'' or outbursts in
    the years before their final explosion. LSST's deep, repeated imaging of the sky is
    perfectly suited to systematically detect this precursor activity.
\end{itemize}
 
\subsubsection{An Ecosystem for Rapid, Automated Follow-Up}
 
Achieving ``Top of the Pyramid'' science in the LSST era requires a coordinated ecosystem
of surveys and facilities built for speed and automation. 
 
\subsubsubsection*{Pathfinder Models: The DLT40 Survey}


The DLT40 survey serves as a pathfinder for such a system, demonstrating a successful hybrid approach. Key details include:
 
\begin{itemize}
    \item \textbf{Core Mechanics:} Uses 0.4\,m telescopes for discovery, incorporating
    real-time data from other surveys (ATLAS, ZTF).
 
    \item \textbf{Automation:} Employs machine learning for real-time alerts and automated
    follow-up. A high ML score on a new source can trigger additional imaging and, within
    15~minutes, automatically trigger spectroscopic observations with facilities like the Las
    Cumbres Observatory's FLOYDS spectrographs, with no human in the loop.
 
    \item \textbf{Rapid Response:} Has developed APIs for real-time triggering of major
    facilities, including the 6.5-meter MMT, enabling rapid follow-up on large telescopes.
\end{itemize}
 
\subsubsubsection*{Strategic Programs for the LSST Era}


Several large-scale projects are being launched to provide the necessary high-cadence
follow-up data. These programs emphasize collaboration and immediate public data release,
with PASSTA making all spectra public on Wiserep and TNS within ${\sim}24$ hours.
 
\begin{table}[h]
    \centering
    \caption{Summary of Strategic Programs for the LSST Era}
    \label{tab:lsst}
    \begin{tabular}{|>{\raggedright\arraybackslash}p{3cm}|>{\raggedright\arraybackslash}p{5cm}|>{\raggedright\arraybackslash}p{8cm}|}

\hline
\textbf{Program} & \textbf{Key Features} & \textbf{Goal}\\
\hline
 
\textbf{Shadowing LSST} &
Approved NOIRLab survey using DECam for 80 nights. &
Observe LSST fields on subsequent nights to create a $<$1-day cadence, specifically to
find early SN features like the 2023ixf excess out to 80~Mpc. \\[6pt]

\hline
\textbf{PASSTA} &
A SOAR large survey program with ${\sim}60$ nights. &
Focus on spectroscopic follow-up of young SNe within ${\sim}50$~Mpc discovered by
LSST. \\[6pt]

\hline
\textbf{Global Supernova Project} &
Long-running collaboration of $>$200 astronomers. &
Obtain high-quality light curves and spectra using the 25 robotic telescopes of the Las
Cumbres Observatory and ${\sim}30$ other facilities. \\[6pt]

\hline
\textbf{Gemini Large Programs} &
Coupled programs on Gemini. &
Provide optical, NIR, and high-resolution spectroscopy for nearby supernovae discovered
by these surveys. \\

\hline
\end{tabular}
\end{table}
 
 
These dedicated follow-up systems are expected to dramatically increase the observation rate of key
early-time phenomena, as summerized in Table \ref{sc_overview:sne:rates}.
 
\begin{table}[h]
    \caption{Summary of Strategic Programs for the LSST Era}
    \label{sc_overview:sne:rates}
    \begin{tabular}{|>{\raggedright\arraybackslash}p{10.5cm}|>{\raggedright\arraybackslash}p{4cm}|}

    \hline
    \textbf{Phenomenon} & \textbf{Projected Annual Rate }\\
    \hline
     
    \hline
    SN~2023ixf-like excesses (Normal Core Collapse) & ${\sim}75$ per year \\[4pt]
    
    \hline
    SN~IIb early excesses                            & ${\sim}15$ per year \\[4pt]
    
    \hline
    SN~Ia with early excesses                        & ${\sim}250$ per year \\[4pt]
    
    \hline
    Precursor Outbursts (SN~IIn)                     & ${\sim}15$ per year \\[4pt]
    
    \hline
    Precursor Outbursts (SN~Ibn)                     & ${\sim}1$ per year \\
     
    \hline
    
\end{tabular}
\end{table}

\subsubsection{Future Needs and Strategic Imperatives}
 
To fully realize the scientific potential of the LSST era, several critical needs must be
addressed by the community. These include:
 
\begin{itemize}
    \item \textbf{Unified Alert Streams:} The current ecosystem of transient alerts (from ZTF,
    ATLAS, LSST, etc.) operates in ``standalone silos.'' There is a deep need to merge the
    streams, especially for nearby galaxies ($<$100--200~Mpc), through the creation of a
    ``Nearby Galaxy Broker.''
 
    \item \textbf{Improved Data Access:} The LSST ``80-hour rule,'' which can delay access to
    images, is considered not ideal and potentially detrimental for young nearby supernova
    science that relies on immediate data for rapid follow-up.
 
    \item \textbf{Widespread Automation:} The community is urged to be brave and automate
    triggering to obtain higher cadence data and discover new phenomena. Manual
    intervention is too slow for the most time-sensitive science.
 
    \item \textbf{Enhanced Collaboration:} Success is contingent on coordination. Scientists are
    encouraged to join a collaboration and get a world-class data set to pool resources and
    expertise effectively. \textit{Swift}'s Priority Zero Target of Opportunity program is an
    excellent model for NASA to emulate.
\end{itemize}

\newpage
\subsection{Magnetars}

Magnetars serve as unique laboratories for exploring fundamental physics under extreme conditions \citep{Lamb1982SurfaceMagnetospheric, Katz1982PhysicalProcesses, Liang1984MagneticFlare, hading06RPPh, turolla15RPPh}. Decades of observation have established them as sources of various high-energy transient events, including a distinct population of GRBs known as MGFs accessible to current large FOV gamma-ray monitors from nearby galaxies \citep{roberts21Natur, burns2021, mereghetti24Natur}. MGFs are almost indistinguishable from the cosmological short GRB population.

A central theme of recent research is the definitive link between magnetars and Fast Radio Bursts (FRBs) \citep{lorimer07Sci}. The magnetar SGR 1935+2154 is the first and only source directly observed to produce both FRB-like radio emission and associated X-ray and soft gamma-ray activity, confirming that at least a fraction of repeating FRBs originate from these objects \citep{Bochenek20:1935, chime2020:1935, 2020ApJ...898L..29M, li21NatAs}. This occurred during the intense 2020 burst storm from the source \citep{younes20ApJ1935}; the FRB-associated X-ray burst displayed a considerably different spectral shape compared to the rest of the bright bursts detected during the same epoch \citep{younes2021NatAs, ridnaia21NatAs}. Additionally, high-cadence monitoring of SGR 1935+2154 has revealed a complex interplay between its rotational dynamics, X-ray outbursts, and radio emission \citep{Younes-2023-NatAs}. Notably, its 2022 outburst was associated with two distinct "spin-up" glitches bracketing the FRB event, suggesting that internal neutron star dynamics (glitches) can trigger the conditions necessary for both intense X-ray activity and powerful radio bursts \citep[][see also \citealt{Younes-2023-NatAs}]{hu24Natur:1935}.

Further discoveries include the potential identification of extragalactic magnetar giant flares \citep[e.g.,][]{roberts21Natur, burns2021, mereghetti24Natur, trigg25AA:GF, beniamini2025} and the emergence of a new population of long-period radio transients, some of which may represent an evolved class of magnetars \citep{hurley23Natur, rea24ApJ:lprt, wang25Natur, cooper24MNRAS, beniamini23MNRAS}. These advancements underscore the critical importance of prompt follow-up, high-cadence, and long-term monitoring campaigns for understanding these enigmatic and powerful cosmic objects.

\subsubsection{Discovery and Fundamental Properties}

Magnetars represent a distinct class of isolated neutron stars powered by the decay of ultra-strong magnetic fields ($\gtrsim10^{14}$ G), rather than rotational energy \citep{Paczynski1992AcA, duncan92ApJ, thompson95MNRAS, thompson96ApJ}. The study of magnetars began with the detection of unusual high-energy events, and historically they are intimately tied to the field of GRBs. A powerful GRB detected in 1979 exhibited distinct properties: it was spectrally softer than typical GRBs, exhibited a pulsating tail, and was followed by repeating bursts \citep{mazets79Natur}. This led to the discovery of a class of objects known as Soft Gamma Repeaters (SGRs). Throughout the 1980s, before the cosmological nature of GRBs and multiple classes were understood, magnetars were a leading candidate for the nature of GRBs. 

SGRs were identified as isolated neutron stars with a unique set of observational properties that distinguish them from conventional pulsars. They are characterized by long spin periods (typically 1–12 s) a and rapid spin-down rates, with a period derivative ($\dot{P}$) of approximately 10$^{-13}$ to 10$^{-11}$ s$^{-1}$, and persistent X-ray luminosities that exceed their rotational energy losses \citep{kouveliotou98Natur, kouveliotou99ApJ}. This energy budget, along with their burst phenomenology, points to magnetic field decay and crustal stress as the dominant power source.  Inferred from their spin properties, their surface magnetic fields are immense, exceeding 10$^{14}$ Gauss (G). These characteristics led to the ``magnetar" model, which posits that their energetic emissions are powered by the decay and dissipation of their extraordinarily high magnetic fields. Magnetars are generally isolated, lacking binary companions, and exhibit complex timing behavior including glitches and strong timing noise.




\subsubsection{Observational Characteristics}

Magnetars are among the most dynamically active neutron stars, producing a wide range of high-energy fast (seconds to milliseconds) transient phenomena spanning many orders of magnitude in energy. This activity is typically categorized into three broad categories:

\begin{itemize}
    \setlength{\itemsep}{1pt}
    \setlength{\parskip}{1pt}
    \item \textbf{Short Bursts}:  10$^{37}$ – 10$^{41}$ ergs
    \item \textbf{Intermediate Flares}:  10$^{41}$ – 10$^{43}$ ergs
    \item \textbf{Giant Flares}:  $>$10$^{44}$ ergs, up to 10$^{47}$ ergs
\end{itemize}


Giant flares are the most extreme manifestations of magnetar activity and exhibit a characteristic structure consisting of a brief, intense initial spike followed by a long-lasting, pulsating tail modulated at the neutron star spin period. These events are thought to originate from large-scale magnetic reconnection or catastrophic crustal failure \citep{hurley05Natur, palmer05Natur}. Observations have revealed quasi-periodic oscillations in flare tails, potentially linked to global seismic or magneto-elastic oscillations of the neutron star interior, providing a unique probe of dense matter physics \citep{israel05ApJ, strohmayer05ApJ}.

Magnetar outbursts, which can last weeks to months, are characterized by dramatic increases in X-ray luminosity followed by gradual decay \citep{cotizelati18MNRAS}. These events are generally interpreted as the result of magnetic energy deposition in the crust and magnetosphere, leading to heating, particle acceleration, and evolving surface emission. Observationally, outburst light curves are well described by multi-component decay models, and their energetics and evolution appear correlated with magnetic field strength and internal structure.

\subsubsubsection*{Giant Flares}

Giant flares are the most energetic events produced by magnetars and have several key features:

\begin{itemize}
\item \textbf{Pulsating Tail}:  They are typically followed by a long-lasting, quasi-thermal pulsating tail following an initial prompt spike.
\item \textbf{Nucleosynthesis}:  Detailed calculations suggest that the giant flare from SGR 1806+20 might be a site of r-process nucleosynthesis \citep{patel25ApJ}.
\item \textbf{QPOs}:  QPOs have been detected in both giant flares and short bursts, potentially linked to crustal or magneto-elastic oscillations.
\item \textbf{Ionospheric Disturbance}:  The 2004 giant flare from SGR 1806+20 was so powerful that it significantly disturbed Earth's ionosphere \citep{inan07GeoRL}.
\end{itemize}

\subsubsubsection*{Extragalactic Giant Flares}

Recent observations suggest that MGFs are a distinct population of GRBs and have been detected from other galaxies \citep{roberts21Natur, burns2021, mereghetti24Natur, trigg25AA:GF}.

\begin{itemize}
    \item \textbf{Identified Candidates}: GRB 200415A (in galaxy NGC 253) and GRB 231115A (in galaxy M82) are considered strong candidates for MGFs, both showing a hard initial spike followed by a softening tail.
    \item \textbf{Occurrence Rate}: Analysis of short GRBs indicates an occurrence rate for MGFs of approximately $5.5^{+4.5}_{-2.7} \times 10^{5}\ \mathrm{Gpc}^{-3}\ \mathrm{yr}^{-1}$ \citep{beniamini2025, trigg25arXiv251023367T}.
    \item \textbf{Fast X-ray Transients}: Magnetar flares are also strong candidates to explain some fast X-ray transients.
\end{itemize}

Magnetars are increasingly recognized as contributors to multiple transient populations. In particular, magnetar giant flares from nearby galaxies are now understood to form a subpopulation of short gamma-ray bursts, often indistinguishable from cosmological events without precise localization. Recent observations have identified strong candidates for extragalactic magnetar flares, including events associated with nearby galaxies such as NGC 253 \citep{roberts21Natur} and M82 \citep{mereghetti24Natur}. These detections suggest that magnetars contribute significantly to the observed short GRB population at low redshift, with implications for event rate estimates and population synthesis \citep{beniamini2025}. In addition, magnetar activity has been proposed as an explanation for some classes of fast X-ray transients, further broadening their role in the transient sky.

Prompt follow-up observations of magnetar outbursts have significantly clarified their behavior. Magnetars with lower quiescent luminosities tend to exhibit larger relative increases in brightness during outbursts, while their X-ray light curves are generally well described by either plateau-decay profiles or combinations of exponential components \citep{cotizelati18MNRAS}. A connection with magnetic field strength is also evident: high-field magnetars typically display slower, more prolonged decay phases. Although there is no strong correlation between magnetic field strength and the peak luminosity reached during an outburst, a more robust relationship exists between the field strength and the total energy released. Taken together, these results consistently indicate that the dissipation of magnetic energy is the primary driver of magnetar outburst activity.

\subsubsection{The Magnetar–Fast Radio Burst (FRB) Connection}

One of the most significant recent breakthroughs in the field has been the establishment of a direct connection between magnetars and FRBs. FRBs are millisecond-duration radio transients of extragalactic origin with energies up to $\sim10^{42}$ ergs. While their origins were long uncertain, observations of the Galactic magnetar SGR 1935+2154 provided the first definitive link between magnetars and FRB-like emission. A high energy burst was detected by Insight-HXMT and INTEGRAL IBIS simultaneously with an FRB from the source, with quasi-period features aligned with radio pulses \citep{chime2020:1935, Bochenek20:1935, 2020ApJ...898L..29M, li21NatAs, ridnaia21NatAs}. The X-ray and soft gamma-ray emission was spectrally spectrally different from other high energy  bursts in the storm, likely indicating a peculiar viewing geometry or quasi-polar burst \citep{younes2021NatAs}. This suggests peculiar conditions are required for the generation of FRBs in magnetars and could explain their observed rarity when compared to the far more common X-ray bursts. 



High-cadence high-energy monitoring of SGR 1935+2154, particularly with the NICER instrument, has provided an unprecedentedly detailed view of its complex behavior. NICER observations in 2020 detected an intense bursting period, or "burst storm," with over 217 bursts in 1120 seconds, occurring approximately half a day before the landmark FRB detection \citep{younes20ApJ1935}. In October 2020, the magnetar entered a radio pulsating phase, producing emissions much fainter than the FRB. A possible "spin-down" glitch (a sudden decrease in rotation speed) was detected around this time \citep{Younes-2023-NatAs, zhu23SciA}.


The magnetar reactivated in October 2022, leading to another FRB detection and remarkable rotational dynamics. Again, a burst storm was detected 2.5 hours before the FRB. The burst occurrence rate decreased significantly after the FRB event. High-cadence monitoring also revealed two strong spin-up glitches (sudden increases in rotation speed). The first occurred ~4.5 hours  before the FRB, and the second occurred ~4.5 hours after it. The total rotational energy released during this glitching period is estimated to be $6.5 \times 10^{41}$ ergs \citep{hu24Natur:1935}.


Glitches provide a probe into the interior structure of neutron stars. They are thought to occur when the superfluid component in the star's crust, which is detached from the rest of the star, suddenly couples and transfers its angular momentum. The temporal correlation between the burst rate and persistent emission enhancements in SGR 1935+2154 suggests that both are driven by the same energy deposition event beneath the star's surface.

The detailed timing of the 2022 event suggests a causal chain: a glitch triggers crustal and magnetospheric disturbances, leading to a burst storm and an intermediate flare. These events do not appear to trigger the FRB directly but instead alter the magnetospheric environment, creating conditions that are favorable for the subsequent emission of a powerful radio burst \citep{hu24Natur:1935}. This is supported by spectral analysis showing rapid softening after an intermediate flare that precedes the FRB \citep{hu25ApJ}.

The emerging picture suggests that magnetars are a dominant progenitor class for repeating FRBs, with their complex magnetic environments providing the conditions necessary for coherent radio emission. However, the diversity of FRB properties indicates that additional channels or evolutionary pathways may also contribute.

\subsubsection{Summary of Magnetar Observations by NICER and NuStar}

In addition to the case of SGR 1935+2154, the NICER mission has been instrumental in monitoring magnetar activity since 2017, enabling many of the recent time-domain breakthroughs.

\subsubsubsection*{Pulse peak migration in SGR 1830-0645}

Through near-daily NICER observations during the first 37 days of the outburst decay, the soft X-ray emission of SGR 1830-0645 exhibited a clear migration of pulse peaks \citep{younes22ApJ:1830,younes22ApJ:ppm}.

\begin{itemize}
    \item \textbf{Pulse Profile Evolution}: The initially complex, triple-peaked pulse profile gradually shifted and merged into a simplified, single-peaked shape. Such peak merging has never been observed before for a magnetar.
    \item \textbf{Possible Theoretical Interpretations}: One possible origin for this surface evolution is the tectonic-like, plastic motion of the neutron star's crust. The inferred speed of this crustal motion is estimated to be ~100 m/day, which strongly constrain the density and depth of the driving region. Alternatively, the hot spots could be powered by particle bombardment from a twisted external magnetic field. The observed pulse peak migration would then be a combination of field-line footpoint motion and evolving surface radiation beaming \citep{lander15MNRAS, gourgouliatos21MNRAS, younes22ApJ:ppm, kojima22ApJ, stefanou25MNRAS}.
    \item \textbf{Phase-Dependent Short Bursts}: NICER detected 84 short bursts, with their occurrence phases preferring the peak of the pulsed persistent emission. This is the first time such a strong phase correlation has been observed, suggesting that the burst emission region is located at a very low altitude and the triggering mechanism is connected to the surface active zone.
\end{itemize}

\subsubsubsection*{Link between magnetars and high-B-field pulsars: Swift J1818.0-1607}

This source was discovered following the detection of a typical magnetar-like hard X-ray burst by Swift-BAT. Follow-up radio and NICER observations confirmed its radio-loud nature, a periodicity of 1.36 s (the shortest among the magnetar population), and a characteristic age of $\sim$470 years \citep{esposito20ApJ, lower20ApJ, hu20ApJ:1818}.

Swift J1818.0-1607 is the first magnetar discovered that shows a peak X-ray luminosity that is lower than its spin-down luminosity. The radio pulsar nature of the source at the time of X-ray activity is also uncommon, marking the sixth magnetar to show radio pulsed emission among the approximately 30 known magnetars to date. These properties suggest that Swift J1818.0-1607 serves as a crucial link bridging regular magnetars and high B-field rotation-powered pulsars, with implications for neutron star and magnetar formation rates in the Galaxy \citep{beniamini19MNRAS, sautron25ApJ}.


\subsubsection{Future Directions}

Recent discoveries have expanded the known phenomenology of magnetars and suggest the existence of previously unrecognized populations. In particular, long-period radio transients with spin periods beyond the traditional pulsar “death line” may represent evolved magnetars or related systems. These objects could be evolved magnetars, white dwarf pulsars, or binary systems. Some of these sources have now been detected in X-rays, namely, ASKAP J1832$-$0911 has shown a transient, pulsating X-ray emission at the radio pulse period. This strongly indicates a connection to high-energy emission processes \citep{wang25Natur}, and that these sources are largely more energetic than previously believed, with high energy behavior crucial for understanding source constraints and physics.

At the same time, given the energetic events associated with magnetars, such as crustal fractures and magnetic reconnection, they are considered plausible sources of gravitational waves (GW). This is likely only detectable by LVK in a galactic magnetar giant flare. The LVK collaboration searched GEO 600 data for a GW signal associated with the 2022 SGR 1935+2154 FRB and glitch epochs (LIGO, Virgo, and KAGRA were not operational at the time). No significant GW emission was detected \citep{abac24ApJ}. An upper limit on the GW-to-radio energy ratio was derived as $< 10^{14} – 10^{16}$. Future 3rd generation GW observatories like Cosmic Explorer or Einstein Telescope might be more promising for nearby extragalactic events \citep{2021ApJ...918...80M}.

\subsubsection{Open Questions}

Despite significant progress, key questions remain, including:

\begin{itemize}
    \setlength{\itemsep}{1pt}
    \setlength{\parskip}{1pt}
    \item What physical conditions determine whether a magnetar produces FRBs? What fraction of extragalactic FRBs have magnetar progenitors?
    \item What is the energy dissipation mechanism in magnetars from the inner crust, to outer crust, to magnetosphere?
    \item How do magnetic field geometry and evolution govern burst energetics?
    \item What is the true contribution of magnetars to the short GRB population? How are magnetars connected to other multiwavelength and multi-messenger transient phenomena such as SLSNe, core-collapse SNe, and the long-period radio transient population? 
    \item How do magnetars form, and evolve over time? What is their relationship to the rest of the neutron star population? What is their birth rate and primarily formation channels?
\end{itemize}

\subsubsection{Future Needs}

Advancing the study of magnetars will require coordinated, multiwavelength observational strategies with rapid response and high cadence. Key priorities include:

\begin{itemize}
    \setlength{\itemsep}{1pt}
    \setlength{\parskip}{1pt}
    \item Simultaneous radio and high-energy monitoring to capture FRB-associated activity
    \item Sensitive wide-field X-ray and gamma-ray surveys to identify and localize bursts
    \item Long-term timing campaigns to probe rotational dynamics and internal structure
    \item Improved localization capabilities to distinguish magnetar flares from cosmological transients
\end{itemize}

New radio surveys are being developed to enhance the detection of radio transient phenomena with excellent localization, such as DSA, CHORD, BURSTT, CHIME/FRB, and SKA. 

\newpage
\subsection{Compact Binary Mergers (CBMs)}

CBMs of neutron stars and black holes serve as unparalleled laboratories for exploring fundamental physics, from the state of matter at supranuclear densities to the cosmic origin of heavy elements. Multi-messenger astronomy, combining GW and electromagnetic (EM) observations, is the essential tool for deciphering these events. The outcome of a merger—whether it promptly collapses to a black hole or forms a transient or stable neutron star remnant—is the primary determinant of the resulting EM counterpart, particularly the kilonova.

Mass ejection, the engine of the kilonova, is a complex process with multiple, interacting components, including dynamical ejecta from the merger itself and subsequent winds from a post-merger accretion disk. A significant challenge in the field is that these components are typically modeled by different research groups using distinct simulation technologies; a unified, self-consistent simulation that captures all relevant physics, or surveys relevant parameters with accurate uncertainty quantification, is currently lacking. This fragmentation leads to major uncertainties, especially in the predictive models, or ``fits," for dynamical ejecta from neutron star-neutron star mergers, which are described as highly unreliable. In contrast, models for post-merger disk mass are considered more robust.

The future of the field lies with next-generation GW observatories like Cosmic Explorer and the Einstein Telescope, which are projected to detect tens to thousands of merger events per day. This wealth of data promises to revolutionize the understanding of GRBs, enable precision measurements of remnant disk masses via GWs, and potentially provide the tightest constraints to date on the neutron star equation of state through simultaneous GW and GRB observations. Despite significant progress, critical questions remain regarding the nature of matter inside neutron stars, the specific progenitors of GRBs, and the full nucleosynthetic output of mergers, underscoring the concurrent need for both more observational data and more advanced theoretical models.

\subsubsection{Core Scientific Objectives in Merger Research}

The study of compact binary mergers is driven by a set of fundamental, yet unanswered, questions in astrophysics and physics. These inquiries form the foundation for current and future research programs. These key questions include:

\begin{itemize}
    \item \textbf{The Nature of Matter}: What is the composition of matter inside neutron stars? Are they made of neutrons, protons, electrons, and muons, or do exotic states like quarks exist?
\item \textbf{Origin of Elements}: What fraction of the universe's r-process elements are produced in neutron star mergers?
\item \textbf{GRB Progenitors}: Which types of binary mergers produce GRBs, and can they power both short-duration and long-duration GRBs?
\item \textbf{Cosmological Tools}: Can CBMs be utilized as standard sirens for cosmological measurements?
\item \textbf{Tests of Gravity}: Does general relativity break down in the extreme-gravity environment of a merger?

\end{itemize}

\subsubsection{Merger Outcomes and Remnants}

The result of a compact binary merger is not monolithic and depends critically on the type of objects involved and their physical properties. The primary goal of multi-messenger astronomy is to try to distinguish the outcome of mergers with the multi-messenger emission to constrain these unknown properties. The nature of the remnant is a critical diagnostic of the binary's initial parameters and the underlying equation of state of nuclear matter.

\subsubsubsection*{Neutron Star-Neutron Star (NS-NS) Mergers}

The outcome is primarily determined by the ratio of the total system mass to the maximum possible mass of a non-rotating neutron star, a value set by the equation of state.

\begin{itemize}
    \item \textbf{Prompt Collapse}: If the total mass is significantly larger than the maximum mass threshold, the remnant collapses immediately into a black hole. This scenario is expected to produce very little dynamical ejecta and a fainter EM counterpart because the ``bounce" mechanism is absent.

\item \textbf{Massive Neutron Star Remnant}: If the total mass is below the threshold, a massive neutron star remnant is formed. These remnants are classified by their lifespan:
\begin{itemize}
    \item \textbf{Short-Lived (10--20 ms)}: The remnant survives briefly before losing angular momentum to gravitational waves and collapsing to a black hole.
\item \textbf{Long-Lived}: Black hole formation is either significantly delayed (occurring due to other processes like accretion disk effects) or avoided entirely, potentially leaving a stable neutron star.

\end{itemize}

\end{itemize}

\subsubsubsection*{Black Hole-Neutron Star (BH-NS) Mergers}

The outcome depends on the properties of both objects: the neutron star's mass and radius, and the black hole's mass and spin.

\begin{itemize}
    \item \textbf{Plunge/Swallowing}: If the black hole is very massive, it will swallow the neutron star whole. This is unlikely to result in a bright counterpart because little to no matter escapes to power an EM signal.
\item \textbf{Tidal Disruption}: If the black hole is less massive and not spinning rapidly, the merger can proceed to form a system where part of the neutron star material is ejected into space (powering a kilonova), while another part remains bound in an accretion disk around the final black hole.

\end{itemize}

The original classification of remnants as ``super massive'' or ``hyper massive'' has been discouraged as not being indicative of the system's evolution. The ``short-lived'' vs. ``long-lived'' distinction is now preferred.

\subsubsection{Mass Ejection Mechanisms and Kilonovae}
The electromagnetic emission known as a kilonova is powered by the radioactive decay of heavy elements synthesized in material ejected during and after the merger. Understanding the kilonova requires modeling multiple, distinct ejecta components launched by different physical processes.

\begin{table}[h]
    \centering
    \caption{Summary of the Ejecta Components from a Compact Binary Merger}
    \label{tab:cbm_ejecta}
    \begin{tabular}{|>{\raggedright\arraybackslash}p{3cm}|>{\raggedright\arraybackslash}p{5cm}|>{\raggedright\arraybackslash}p{8cm}|}
    
        \hline
        \textbf{Ejecta Component} & \textbf{Launch Mechanism} & \textbf{Key Characteristics} \\
        \hline
        Dynamical Ejecta & 
        Tidal forces and shocks during the merger. & 
        Includes a fast tail launched by shock waves from the ``bounce" of the colliding stellar cores.\\
        \hline
        Shocked Ejecta & 
        Material squeezed out from the collision interface. &
        A very fast component whose detection would offer insight into the shock dynamics and equation of state.\\
        \hline
        Neutrino-Driven Winds &
        Neutrino heating from the hot remnant disk and central object. &
        A significant contributor to the post-merger ejecta.\\
        \hline
        Spiral-Wave Wind &
        Angular momentum transport by spiral density waves in the remnant disk. &
        Ejects material from the disk over longer timescales.\\
        \hline
        Magnetized Winds &
        Magnetic fields amplified in the remnant. &
        Can drive powerful outflows, but are difficult to model without ``overdoing" the ejecta mass.\\
        \hline
        Disk Recombination &
        Instabilities in the disk as accretion rates fall. &
        The disk can become inefficiently cooled, leading to instabilities that unbind the remaining material.\\
        \hline     

    \end{tabular}
\end{table}


The various ejecta components do not exist in isolation and their interaction is crucial. For instance, the dynamical ejecta, while often subdominant in mass compared to wind ejecta, is opaque and significantly reprocesses the radiation emitted from the inner regions.

A major problem in the field is the fragmentation of simulation efforts. The researchers and simulation technologies used to model the dynamical ejecta are typically different from those used for post-merger wind components. As a result, ``we don't currently have a simulation that we find, we have [all components]. So our knowledge is somewhat mix-wise.''

\subsubsection{Challenges in Numerical Modeling and Data Interpretation}

Accurately mapping binary properties to observable signals requires robust theoretical models, but there are significant pitfalls in the current state of the art. There is a strong caution against using generalized fitting functions (``fits'') found in the literature to predict the mass and velocity of dynamical ejecta from NS-NS mergers. Many fits are created by combining data points from a wide variety of simulations from the entire literature. This is described as a ``really, really bad idea'' because these simulations often use different physics (e.g., some include neutrino transport while others do not). This mixing of inconsistent data leads to systematic errors. For example, simulations that neglect neutrino effects tend to over predict both the velocity and mass of the ejecta, while simulations with neutrinos yield much lower velocities. Overall, a constant value for the ejected mass is often a better predictor than these complex, often unreliable fits.

A more reliable method is to use fits developed from targeted, self-consistent simulation campaigns, such as those run specifically to model GW170817. These can be trusted for systems with parameters close to those simulated but should not be extrapolated broadly. In contrast, models for the remnant disk mass are considered to ``work a little bit better.'' This is because the disk mass constitutes a large fraction (up to 20--40\%) of the total remnant mass, making it numerically easier to capture accurately compared to the dynamical ejecta, which can be less than 1\% of the total mass.

\subsubsection{Electromagnetic Emission from Binary Black Hole Mergers}

While the focus is often on mergers involving neutron stars, there are emerging theories for how binary black hole (BBH) mergers could produce EM counterparts. These ideas are less developed and require significant theoretical work. Proposed scenarios include:

\begin{itemize}
    \setlength{\itemsep}{1pt}
    \setlength{\parskip}{1pt}
   \item Mergers occurring within the gas-rich disks of AGN.
    \item The merger of binary supermassive black holes (binary AGNs).
    \item EM emission driven by superradiant instabilities or from charged black holes.
\end{itemize}

These models are in early stages and need better theoretical predictions to guide observational searches.

\subsubsection{The Future: Next-Generation Gravitational Wave Observatories}

The next era of GW astronomy, enabled by observatories planned for the 2030s and 2040s, will fundamentally change the landscape of compact merger science. Next-generation observatories include the Cosmic Explorer in the United States and the Einstein Telescope in Europe. These instruments will be sensitive enough to see almost the entire universe's worth of NS-NS mergers, with expected event rates of ``tens to hundreds or maybe thousands per day."

This dramatic increase in detection rate and sensitivity will enable novel scientific inquiries that are impossible today.

\begin{itemize}
    \item \textbf{Probing Post-Merger Physics}: Cosmic Explorer will be able to detect the high-frequency GW signal from the post-merger remnant itself. This includes oscillations from a massive neutron star remnant at 2-4 kHz and the final ringdown of a newly formed black hole.

\item \textbf{Constraining the Neutron Star Radius}: There is a tantalizing possibility of linking GW frequencies from the remnant directly to QPOs observed in the gamma-ray signal of some short GRBs. If this connection is validated by simultaneous GW and GRB observations, ``it would be the tightest constraint on the radius of neutron stars that we have to date."

\item \textbf{Measuring Remnant Disk Mass}: The GW ringdown frequency of a black hole formed in a merger is perturbed by the surrounding accretion disk. Analysis of the numeric data indicates that for a sufficiently strong signal (SNR of 5), ``it would be possible to measure the disk mass with a precision of 10\%.'' This provides a direct test of GRB engine theories.

\end{itemize}

Multi-messenger observations, theory, and simulations have collectively advanced our understanding of compact objects significantly. However, many foundational questions remain. The path forward requires a dual-track approach: acquiring more high-quality observational data while simultaneously developing more sophisticated and comprehensive theoretical models to interpret that data.

\newpage

\subsection{Neutrinos}

After more than a decade of observation, the field of high-energy neutrino astronomy is transitioning from the detection of a diffuse astrophysical flux to the identification of its origins~\citep[see][for recent reviews]{Halzen:2022pez, Arguelles:2024ncf}. A compelling body of evidence now points toward AGN, particularly those with cores obscured to gamma-rays, as a primary source of cosmic neutrinos. The Seyfert 2 galaxy NGC 1068 has been identified as the first steady point source of high-energy neutrinos~\citep{2022Sci...378..538I, IceCube:2026hzq, IceCube:2024dou}, with a significance that is now well-established. Its high neutrino flux, contrasted with a much lower observed gamma-ray flux, strongly suggests that neutrinos are produced in a dense, optically thick environment near the central supermassive black hole~\citep[see e.g.][]{Murase:2019vdl,Inoue:2019yfs,Eichmann:2022lxh}.

Searches for similar sources are yielding promising results. Catalog-based searches targeting bright Seyfert galaxies in both the Northern and Southern skies have revealed excesses from several other AGN, including NGC 4151, the second most significant steady source candidate~\citep{IceCube:2024dou,IceCube:2026hzq}. 
The population of Seyfert galaxies could thus be a significant contributor to the diffuse neutrino flux, especially at 1-100\,TeV energies~\citep[see e.g.][]{Kheirandish:2021wkm, 2026arXiv260318214T,2025ApJ...989..215F,Murase:2019vdl,2026arXiv260101533K,2026PhRvD.113b3019S, Carpio:2026xkf}. These findings are bolstered by comparisons between the diffuse gamma-ray and neutrino flux, indicating a neutrino source population of gamma-ray obscured sources like AGN cores~\citep[see e.g.][]{Murase:2019vdl}.

In parallel, a significant neutrino flux originating from the Galactic plane has been identified with a significance exceeding $5\sigma$~\citep{IceCube:2023ame,Thiesmeyer:2025qgo}. While the signal is concentrated towards the Galactic Center, its precise nature, whether from discrete sources such as supernova remnants or a truly diffuse origin from cosmic-ray interactions with interstellar medium, remains an open question. Other potential neutrino sources, such as Core Collapse Supernovae (CCSNe) and GRBs, are also under intense scrutiny. While prompt emission from GRBs has been strongly constrained~\citep{Abbasi:2022whi}, choked-jet~\citep{He:2018lwb,2016PhRvD..93h3003S} and interacting CCSNe~\citep{Murase:2010cu, Kheirandish:2022eox, Waxman:2024njn} represent a promising high-energy neutrino source class~\citep{IceCube:2023esf}. In addition, a Galactic supernovae would result in a high-statistics neutrino signals at MeV energies~\citep{2011A&A...535A.109A}.

The future of the field is bright, with the next-generation observatory, IceCube-Gen2, poised to increase the current detector volume by nearly an order of magnitude. This facility, part of a growing global network of neutrino telescopes, will provide unprecedented sensitivity from GeV to EeV energies, enabling definitive source identification and offering a new window into the most extreme accelerators in the universe~\citep{IceCube-Gen2:2020qha}.

\subsubsection{The High-Energy Neutrino Flux and Candidate Sources}

The observed high-energy neutrino flux is composed of several components that dominate at different energy levels. At the lowest energies, atmospheric neutrinos from cosmic-ray interactions in Earth's atmosphere are the dominant component. As energy increases, the astrophysical component, first identified over a decade ago, becomes the primary contributor~\citep{IceCube:2013low}. At the highest energies, searches are underway for a cosmogenic neutrino flux, predicted to be generated by the interaction of ultra-high-energy cosmic rays with the cosmic microwave background (CMB)~\citep{IceCubeCollaborationSS:2025jbi}.

The measured spectrum of cosmic neutrinos exhibits distinct features, notably a harder spectrum at energies below $\sim30$\,TeV compared to higher energies where the spectrum is well characterized by a power law~\citep{IceCube:2025tgp}. Comparisons with extragalactic gamma-ray background indicate that a majority of the neutrino sources should be gamma-ray absorbed~\citep{Murase:2015xka}.  For example, the ``disk-corona" model for AGN, where protons are accelerated in the magnetized corona above the accretion disk, provide environments that are naturally optically thick to GeV-TeV gamma-rays~\citep{Murase:2019vdl, Inoue:2019yfs,Eichmann:2022lxh}. MeV gamma-ray measurements will be necessary to obtain a complete picture of the multi-messenger emission from these sources~\citep{Murase:2019vdl}.

For an astrophysical object to be a significant source of TeV-PeV cosmic neutrinos, it must meet two primary conditions:
\begin{itemize}
    \item \textbf{Cosmic Ray Acceleration:} The source must be capable of accelerating cosmic rays (protons and nuclei) to energies exceeding 1\,PeV ($10^{15}$ eV).
    \item \textbf{Beam Dump Environment:} The source must possess a dense environment of gas or radiation that acts as a ``beam dump'', facilitating interactions with the accelerated cosmic rays to produce pions, which subsequently decay into neutrinos and gamma rays.
\end{itemize}


Based on these criteria, several classes of astrophysical objects are considered leading candidates for neutrino production:

\begin{itemize}
    \item \textbf{AGN}: Powered by supermassive black holes, their cores, accretion disks, and relativistic jets are promising sites for particle acceleration and interaction~\citep[see][for a review]{Murase:2022feu}.
    \item \textbf{Supernova Remnants (SNRs)}: The expanding shock fronts of exploded stars are known to accelerate cosmic rays within our galaxy~\citep{Simon:2025axa}.
    \item \textbf{GRBs:} These cataclysmic explosions, associated with collapsing massive stars or merging neutron stars, produce the most powerful relativistic outflows known~\citep{Kimura:2022zyg}.
    \item \textbf{X-ray Binaries:} Systems containing a compact object (neutron star or black hole) accreting matter from a companion star~\citep{Kantzas:2023oww, Carpio:2025arz, Kuze:2025wda}.
    \item \textbf{Core Collapse Supernovae (CCSNe):} The collapse of massive stars, particularly when the resulting shockwave interacts with dense circumstellar material~\citep{Murase:2010cu, Kheirandish:2022eox, Waxman:2024njn}.
\end{itemize}

\subsubsection{Active Galactic Nuclei: A Primary Neutrino Source Class}
A growing body of evidence implicates AGN, particularly a subclass with gamma-ray obscured cores, as a major source of the observed astrophysical neutrino flux. IceCube has identified the active galaxy NGC 1068 as the first significant steady source of high-energy neutrinos~\citep{2022Sci...378..538I}. NGC 1068 is a nearby Seyfert 2 galaxy, characterized by a heavily obscured nucleus. It is one of the best-studied AGN and played a key role in developing AGN unification schemes. Its core is Compton-thick, with a column density of $\sim10^{25}$\,cm$^{-2}$, and it is luminous in both X-ray and infrared, indicating high star formation. A crucial finding, though, is that the observed neutrino flux from NGC 1068 is significantly higher than the gamma-ray flux measured by telescopes like Fermi-LAT. Models based on the observed gamma-ray flux cannot account for the neutrino signal. This discrepancy strongly implies that the neutrinos are produced in a region that is optically thick to gamma-rays. The gamma-rays produced alongside the neutrinos are absorbed within the source, while the neutrinos escape unimpeded. This points to the neutrinos originating deep within the AGN core, likely in the vicinity of the accretion disk and corona ($< 100$ Schwarzschild radii)~\citep{Murase:2022dog, Carpio:2026xkf, Eichmann:2026kvj}.



Using intrinsic X-ray flux as a proxy for neutrino production, searches have been extended to catalogs of bright Seyfert galaxies, revealing further evidence for AGN as a source class~\citep{IceCube:2026hzq,IceCube:2024dou}. While accumulating signals point to gamma-ray obscured AGN, the exact nature of the emission (steady vs. episodic) is not yet determined. The energy ranges of contributing events also appear to vary between sources, with NGC 7469's excess dominated by $>100$\,TeV events~\citep{Sommani:2024sbp}, while NGC 1068's spectrum is shifted to lower energies.

\subsubsection{Galactic Neutrino Emission}

IceCube has unambiguously identified a neutrino flux originating from the plane of the Milky Way.An initial 10-year analysis identified the Galactic component at $4.5\sigma$~\citep{IceCube:2023ame}, with newer analyses increasing the significance to over $5\sigma$~\citep{Thiesmeyer:2025qgo}. While most of the signal originates from the direction of the Galactic Center, a key unresolved question is whether this flux is from a population of discrete sources or is truly diffuse, arising from the interaction of cosmic rays with interstellar gas and radiation. Current analyses cannot distinguish between different emission models. However, tests favor a diffuse-like origin or contributions from many faint sources over emission concentrated in known source catalogs (like SNRs, pulsars or Pulsar Wind Nebulae), as the significance of those catalog searches is consistent with expectations from the diffuse template search. X-ray binaries like Cygnus X-3 have also been identified as potential galactic candidates sources, and the strongest time-dependent excess for a binary was correlated with an X-ray outburst from V404 Cyg. However, the predicted flux levels from these objects are generally not high enough to be conclusively identified by current experiments~\citep{IceCube:2022jpz}.

\subsubsection{Transient and Episodic Sources}

The study of transient and episodic sources represents a critical frontier in neutrino astronomy, shifting the focus from steady-state emitters to the most violent, time-variable phenomena in the universe. While sources like NGC 1068 have been established as steady neutrino emitters, current research indicates that episodic emissions cannot be ruled out for many classes of AGN. Transient sources are characterized by brief, high-intensity bursts of energy, whereas episodic sources may exhibit multiple ``flares'' or periods of activity over longer durations.
\begin{itemize}
    \item \textbf{Core Collapse Supernovae (CCSNe)} are expected to produce neutrinos in two distinct phases:
    \begin{itemize}
        \item \textbf{Thermal MeV Neutrinos:} Emitted over $\sim10$ seconds during the core collapse, extracting the gravitational binding energy of the progenitor star~\citep[see][for a review]{Janka:2017vlw}.
        \item \textbf{High-Energy (GeV-PeV) Neutrinos:} Produced over days to months when the supernova shockwave interacts with dense CSM shed by the progenitor star before its explosion~\citep{Murase:2010cu, Kheirandish:2022eox,Waxman:2024njn}.
    \end{itemize}
    In addition, a supernova within the Milky Way would be a spectacular neutrino event, with detectors like IceCube expecting $10^{4}-10^6$ MeV neutrinos~\citep{2011A&A...535A.109A}. 
    The detection horizon for these events extends beyond our galaxy; current detectors can identify neutrinos from SNe in the Large and Small Magellanic Clouds, while next-generation telescopes will push this horizon to over 2\,Mpc~\citep{Nakamura:2016kkl}.
    \item \textbf{TXS 0506+056} is a landmark source in multi-messenger astronomy. It remains the only source identified in a flaring gamma-ray state coincident with a high-energy neutrino alert at $3\sigma$ coincidence.
    \begin{itemize}
        \item \textbf{2017 Flare:} It was identified following a $\sim290$ TeV neutrino alert (IC-170922A) that was coincident with a gamma-ray flaring state. The chance correlation was rejected at a $3\sigma$ level~\citep{2018Sci...361.1378I}.\\
        \item \textbf{2014-2015 Flare:} An archival search revealed an earlier, more significant neutrino flare ($13\pm5$ signal events, $3.5\sigma$ significance) from the same direction. Crucially, this neutrino flare was not accompanied by a gamma-ray flare, pointing to a gamma-ray suppressed or ``quiet'' emission state~\citep{IceCube:2018cha}.
     \end{itemize}
    \item \textbf{GRBs} have long been considered primary candidates for cosmic-ray acceleration, as they are expected to produce neutrinos during their prompt emission phase via internal shocks~\citep{Kimura:2022zyg}. However, IceCube observations of over 1,100 GRBs have placed stringent limits, suggesting that prompt GRB emission contributes less than one percent of the total diffuse neutrino flux~\citep{Abbasi:2022whi}. A follow-up on the ``brightest of all time'' GRB (GRB 221009A) provided the strongest limits to date on neutrino emission and the baryon loading factor in GRB jets~\citep{IceCube:2023rhf}.
    \item \textbf{X-ray Binaries} exhibit time-dependent outbursts; for instance, the strongest excess in time-dependent searches for binaries to date was correlated with the V404 Cyg X-ray outburst~\citep{IceCube:2022jpz}.
\end{itemize}

Analysis by the IceCube collaboration for coincident emission from these sources employ triggered searches for pre-identified locations and untriggered searches that look for clusters of events in both space and time~\citep{IceCube:2016cqr,IceCube:2020mzw}. These temporal analyses are significantly more sensitive than time-integrated searches because they reduce the background of atmospheric neutrinos by narrowing the observation window to the duration of the astrophysical flare.

\subsubsection{The Future of Neutrino Detectors}

The next decade promises transformative discoveries in neutrino astronomy, driven by planned upgrades and a new generation of observatories. IceCube-Gen2 is a planned next-generation neutrino observatory at the South Pole, designed to be nearly 10 times larger than the current IceCube detector. It will augment the existing facility with an enlarged in-ice optical array, a surface cosmic-ray detector, and a giant radio array for detecting the highest-energy neutrinos. Gen2 will be a wide-band observatory with sensitivity spanning six orders of magnitude in energy, from GeV to beyond EeV. This will allow it to probe the high-energy extension or cutoff of the astrophysical neutrino spectrum, test models of cosmogenic neutrino production, constrain the nature of ultra-high-energy cosmic ray accelerators, and resolve the neutrino sky with significantly improved source detection capabilities.

IceCube-Gen2 is part of a growing global effort to build a network of large-scale neutrino telescopes. Other key facilities under construction or in operation include KM3NeT in the Mediterranean Sea, TRIDENT in the South China Sea, Baikal-GVD in Lake Baikal, and the proposed P-ONE in the Pacific Ocean. Together, these observatories will provide full-sky coverage and usher in an era of precision neutrino astronomy.

\newpage
\section{Community Observing Plans\label{subsec:obs_plans}}

\subsection{Gamma-ray Bursts}

LGRBs remain a cornerstone of time-domain and multimessenger astrophysics, providing unique insight into relativistic jet physics, massive star death, compact object formation, and the high-redshift universe. Recent discoveries, including ultra-long GRBs, high-redshift GRBs, merger-origin long GRBs, and low-luminosity/X-ray-dominated events, demonstrate that coordinated observing strategies are essential to fully exploit their scientific return.

The prompt emission of GRBs is characterized by intense, highly variable gamma-ray radiation lasting from fractions of a second to many minutes, with temporal structure spanning milliseconds to tens of seconds \citep[see, e.g., the review by][]{piran2004}. Prompt spectra are typically non-thermal, often well described by broken power-law or cutoff power-law models \citep[e.g.,][]{band1993}, and exhibit strong spectral evolution over the burst duration. This phase encodes information about the jet launching mechanism, energy dissipation processes, magnetic field structure, and baryon loading, while prompt polarization and high-energy ($>$GeV) emission provide additional diagnostics of jet composition and radiation mechanisms \citep[see, e.g.,][]{zhang2007, kumar2015, gill2022}.

Following the prompt phase, GRBs produce long-lived afterglows across the electromagnetic spectrum, from X-ray and ultraviolet through optical, infrared, and radio wavelengths \citep{costa1997, galama1998, djorgovski1998, meszaros1997, sari1998, wijers1999, granot2002}. These afterglows arise from the interaction of the relativistic jet with the circumburst medium and typically fade as power laws over hours to months. Multi-band afterglow observations enable precise localization, redshift determination, and constraints on jet opening angles, energetics, and ambient densities  \citep[e.g.,][]{metzger1997, wijers1999, rhoads1999, sari1999, chevalier2000, frail2001, panaitescu2001, bloom2003}. Early afterglow measurements, particularly within minutes to hours of the trigger, probe jet structure and viewing geometry \citep{meszaros1998, rossi2002, granot2002b, panaitescu2003, kumar2003}, while late-time radio observations can reveal mildly relativistic ejecta and constrain the total kinetic energy budget \citep{frail2000, berger2003, frail2004}.

Many long GRBs are accompanied by luminous, broad-lined Type Ic supernovae, firmly linking a subset of GRBs to the core collapse of massive, stripped-envelope stars \citep{galama1998, Hjorth2003, HjorthBloom2012, cano2017}. These GRB-associated supernovae typically peak days to weeks after the burst and provide critical insight into the explosion geometry, nucleosynthesis, and the coupling between the relativistic jet and the stellar envelope. A growing number of merger-origin or ambiguous long-duration events have been associated with kilonovae powered by the radioactive decay of r-process nuclei synthesized in neutron-rich ejecta \citep{dellavalle2006, galyam2006, gao2015, Rastinejad2022, troja2022, levan2023, yang2024}. Kilonova emission evolves rapidly from blue to red on timescales of days \citep[e.g., ][]{li1998, Barnes2013, metzger2010} requiring prompt, multi-epoch optical and near-infrared observations to constrain ejecta mass, composition, and opacity. 


\subsubsection{High-Priority Subclasses}

Given the high discovery rate of ordinary LGRBs, community-level coordination is most effective when focused on subclasses with exceptional scientific leverage. The following subclasses are prioritized:

\begin{itemize}

    \item \textbf{Ultra-Long GRBs}: Probe prolonged central engine activity and non-standard progenitors, including extended-envelope collapsars, helium-core–black-hole mergers, and micro-tidal disruption events. Their extreme durations challenge standard GRB models.

    \item \textbf{High-Redshift Long GRBs (z $>$ 6)}: Act as beacons of the early universe, enabling studies of star formation, reionization, intergalactic medium properties, and the deaths of the first generations of massive stars.

    \item \textbf{Merger-Origin and Multimessenger Long GRBs}: Demonstrate that compact object mergers can produce long-duration $\gamma$-ray emission, breaking the traditional duration–progenitor association and linking GRB science directly to gravitational-wave and neutrino astronomy.
    

    \item \textbf{Low-Luminosity Long GRBs, XRFs, and FXTs}: Likely dominate the true volumetric rate of relativistic stellar explosions and probe weak, failed, or choked jets, cocoon emission, and shock breakout physics.
\end{itemize}

\subsubsection{Triggering Criteria \& Expected Event Rates}

Triggers for the identified subclasses are generally well defined and can be implemented through existing alert streams (e.g., GCN Notices), supplemented by rapid community vetting when necessary. LGRBs span a wide range of intrinsic luminosities, durations, and progenitor channels, leading to event rates that vary by several orders of magnitude depending on subclass. On-axis LGRBs are detected at a rate of approximately one per day by current wide-field $\gamma$-ray monitors such as Fermi and Swift \citep{wanderman2010, salvaterra2012, ghirlanda2022}, but the prioritized subclasses have substantially lower rates. The following quantifies both triggering criteria and the expected event rate for each subclass:

\begin{itemize}
    \item \textbf{Ultra-long GRBs}: Triggering Criteria: Prompt emission lasting $>$5000 s, or repeated triggering over many orbits, indicating prolonged central engine activity. Expected Event Rate: $\sim$0.5–1 per year.

\item \textbf{High-redshift GRBs}: Triggering Criteria: Secure spectroscopic redshift z $>$ 6, or strong photometric dropout requiring rapid infrared confirmation. Expected Event Rate: $\sim$0.5 per year (Swift era), increasing to $\sim$1 or more per year with SVOM and Einstein Probe.

\item \textbf{Merger-origin LGRBs}: Triggering Criteria: Spatial and temporal coincidence with gravitational-wave or high-energy neutrino alerts, or late-time kilonova-like emission inconsistent with a supernova. Expected Event Rate: $\ll 1$ per year (highly uncertain; selection-limited).


\item \textbf{Low-luminosity GRBs / X-ray flashes (FXTs)}: Triggering Criteria: Soft X-ray–dominated transient with low inferred isotropic energy, detected by EP or other soft X-ray monitors. Expected Event Rate: Volumetric rate $\sim$100–200x classical LGRBs; the observed rate of soft transients (ll-GRBs, XRFs, FXT) currently reaches several / year and expected to increase.

\end{itemize}

\begin{table}[h]
    \centering
    \caption{Summary of High-Priority GRB Subclasses and Expected Rates}
    \label{tab:GRBSubclasses}
    \begin{tabular}{|>{\raggedright\arraybackslash}p{3cm}|>{\raggedright\arraybackslash}p{4cm}|>{\raggedright\arraybackslash}p{4cm}|>{\raggedright\arraybackslash}p{5cm}|}
        \hline
        \textbf{LGRB Subclass} & \textbf{Rationale} & \textbf{Characteristics} &
        \textbf{Expected Rate}\\
        \hline
    Ultra-Long GRBs & Prolonged engine activity; exotic progenitors & 
Prompt emission $>$1000 s or repeated triggering & $\sim$0.5–1 per year\\
\hline
High-Redshift LGRBs (z $>$ 6) & Probes of early star formation, reionization, Pop III candidates & Spectroscopic or secure photometric z $>$ 6 &
$\sim$0.5 per year (Swift); $\sim$1+ per year (SVOM/EP)\\
\hline
Merger-Origin LGRBs & 
Breaks duration–progenitor paradigm; multimessenger astrophysics & 
Long GRBs associated with GW/neutrino events or kilonovae & 
$<$1 per year \\
\hline
Low-Luminosity GRBs / XRFs / FXTs & Dominant explosion channel; weak/failed jets; cocoon emission & 
Soft spectrum; low isotropic energy & Volumetric rate $\sim 100–200 \times$ classical LGRBs; observed $\sim$ 1 every few years\\

        \hline
    \end{tabular}
\end{table}


\subsubsection{Discovery \& Alert Channels}

While most LGRBs are initially discovered via wide-field $\gamma$-ray or X-ray monitors, several high-priority subclasses rely on secondary classification, multi-mission correlation, or follow-up confirmation before their significance is recognized. For this reason, discovery and alerting should be viewed as a multi-stage process, beginning with an initial trigger and followed by refinement through additional data products, cross-messenger association, or temporal evolution. The channels described below reflect current best practices and anticipated near-term capabilities.

\begin{table}[h]
    \centering
    \caption{Summary of Discovery Channels \& Alerts for GRBs}
    \label{tab:GRBdiscovery}
    \begin{tabular}{|>{\raggedright\arraybackslash}p{2cm}|>{\raggedright\arraybackslash}p{6cm}|>{\raggedright\arraybackslash}p{3cm}|>{\raggedright\arraybackslash}p{5.5cm}|}
        \hline
       \textbf{Subclass} & \textbf{Primary Discovery Channels} & \textbf{Alert Pathways} & \textbf{Notes on Identification} \\
\hline
Ultra-Long GRBs & 
Fermi-GBM, Swift-BAT, Konus-Wind, EP & 
GCN Notices +  Circulars & 
Often recognized only after prolonged or repeated triggering; IPN correlation important\\
\hline
High-Redshift LGRBs ($z > 6$) & 
Swift-BAT, SVOM-ECLAIRs, EP & 
GCN Notices → Circulars (photometric/spectroscopic $z$) & 
Requires rapid OIR/IR follow-up; classification latency of hours–days\\
\hline
Merger-Origin / Multimessenger LGRBs &
Fermi-GBM, Swift-BAT & 
GCN Notices + LVK/IceCube alerts & 
Identified via spatial/temporal coincidence; joint alert logic essential\\
\hline
Low-Luminosity LGRBs / XRFs / FXTs & 
Einstein Probe (WXT), SVOM-ECLAIRs & 
GCN Notices, Circulars& 
Often lack $\gamma$-ray trigger; classification evolves with follow-up\\
        \hline
    \end{tabular}
\end{table}

\subsubsection{Observing Strategy}

The scientific goals of LGRB observations require multiwavelength follow-up over logarithmically spaced timescales, from seconds to years. No single facility can address this alone. Instead, success depends on coordinated use of wide-field discovery instruments, rapid-response narrow-field facilities, and sensitive long-term monitoring assets.

Key strategy principles include:
\begin{itemize}
    \item 
{\bf Immediate response} to characterize prompt emission and early afterglow.
    \item 
{\bf Rapid redshift determination} to assess event priority (especially for high-z and nearby events).
    \item 
{\bf Sustained multi-epoch follow-up} to capture jet breaks, supernovae/kilonovae, and late-time environment signatures.
    \item 
{\bf Public, rapid data release} to enable broad participation and maximize scientific return.
\end{itemize}

\subsubsection{Observatory Roles and Capabilities}

Space-based observatories: Fermi, Swift, EP, and SVOM are critical for initial discovery and monitoring prompt gamma-ray and X-ray emissions. Swift provides rapid UV/optical/X-ray follow-up. Chandra and XMM-Newton are utilized for X-ray localization and deep, late-time X-ray tracking. HST and JWST provide high-resolution imaging for precise localization, late-time observations, and spectroscopy (especially JWST for IR spectra of SNe/KNe and redshift).
Ground-based optical/NIR telescopes: Keck, VLT, and Gemini perform crucial optical/NIR spectroscopy to determine redshift and characterize SNe/KNe. Optical time domain surveys like ZTF and LSST are important for discovering orphan afterglows.
Ground-based Radio/gamma-ray telescopes: VLA and ALMA provide radio afterglow observations and radio localization. LHAASO, HAWC, MAGIC, HESS, VERITAS, and CTAO are essential for detecting Very High Energy (VHE) photons above 100 GeV.


\begin{table}[h]
    \centering
        \caption{Summary of Observatory Roles and Capabilities for GRB Follow-up
}
\label{tab:GRBobsroles}
    \begin{tabular}{|>{\raggedright\arraybackslash}p{2.6cm}|>{\raggedright\arraybackslash}p{1.8cm}|>{\raggedright\arraybackslash}p{5cm}|>{\raggedright\arraybackslash}p{6.8cm}|}
        \hline
       \textbf{Observatory} & \textbf{Wavelength} & \textbf{Unique Capability} & \textbf{Ideal Use Case} \\
        \hline
\textbf{Fermi / Swift} &
$\gamma$-ray, X-ray, UV, Optical & 
Wide-field monitoring, rapid slews & 
GRB discovery, prompt emission studies, early afterglow monitoring.\\
\hline
\textbf{Chandra, XMM-Newton} & 
X-ray & 
High spatial resolution / sensitivity & 
X-ray localization, monitoring afterglow decay, detection of jet break\\
\hline
\textbf{HST} & 
UV/Optical & 
High-resolution imaging & 
Late-time optical follow-up and localization in host galaxies\\
\hline
\textbf{JWST} & 
NIR/MIR, spectroscopy & 
IR imaging and spectroscopy & 
KNe spectroscopy for merger GRBs, afterglow follow-up in the IR, redshift determination\\
\textbf{VLA, ALMA} & 
Radio, mm & 
Sensitive broadband radio, polarization & 
Monitoring of radio afterglow, radio localization, jet structure studies\\
\hline
\textbf{Keck, VLT, Gemini} & 
Optical/NIR & 
Large-aperture spectroscopy & 
Redshift, IGM, SN/KN characterization\\
\hline
\textbf{EP, SVOM}&
Visible, Soft X-ray, $\gamma$-ray &
Wide-field discovery, rapid follow-up &
FXT/GRB discovery (high-z, XRFs), early optical and X-ray monitoring \\
\hline
\textbf{UVEX} &
NUV / UV & 
UV spectroscopy and imaging& 
UV spectroscopy of llGRB/cocoon\\
\hline
\textbf{ULTRASAT} & 
NUV & 
High-cadence imaging & 
Shock breakout (from llGRB) discovery, early orphan afterglow\\
\hline
\textbf{LSST, ZTF} & 
Optical (Deep, High-cadence survey) & 
Deep surveys, Orphan afterglow discovery & 
Late-time afterglow discovery, Supernova detection, Orphan afterglow discovery\\
\hline
\textbf{LHAASO, HAWC, MAGIC, H.E.S.S., VERITAS, CTAO} & 
VHE ($>$100 GeV) & 
VHE photon detection& 
Jet composition, particle acceleration\\
\hline

    \end{tabular}
\end{table}

\subsubsection{Observing Timeline}

Table~\ref{tab:GRBobsstrategy} defines a baseline time-phased observing plan designed to capture key physical diagnostics. This timeline represents a minimum coordinated dataset; individual facilities may exceed this baseline as resources allow.

\begin{table}[h]
    \centering
        \caption{Summary of GRB Observing Strategy and Timeline}
    \label{tab:GRBobsstrategy}
    \begin{tabular}{|>{\raggedright\arraybackslash}p{3cm}|>{\raggedright\arraybackslash}p{5cm}|>{\raggedright\arraybackslash}p{8cm}|}
        \hline
       \textbf{Timescale} & \textbf{Recommended Observations} & \textbf{Primary Science Goals} \\
        \hline
        \textbf{T0 + 0 to 1 hr} &
        Continuous $\gamma$-ray/X-ray monitoring&
        Constrain the prompt emission properties\\
        \hline
        \textbf{T0 + 1 to 24 hr} & 
        Optical (including polarization) and X-ray imaging ($\geq$ 2 epochs); UVOIR spectroscopy (once) to determine redshift. & 
        Localize the afterglow, measure the early decline rate, and determine the source redshift\\
        \hline
        \textbf{T0 + 1 to 7 days} & 
        Multi-epoch coverage across X-ray/UV/optical/IR/radio; Cadence: 1/day &
        Characterize the early evolution of the afterglow and catch the expected jet break\\
        \hline
        \textbf{T0 + 1 week to 1 month} & 
        Radio/UVOIR/X-ray imaging follow-up; UVOIR spectroscopy if SN/KN detected. Cadence: once per 4 days. & 
        Catch the expected SN peak and constrain the SN type and time\\
        \hline
        \textbf{T0 + 1 months to years} & 
        Long-term X-ray/UVOIR/radio monitoring (including VLBI if applicable)
        Cadence: Uniform in log scale with interval of $\Delta \log t \approx 0.3$ dex until undetectable. & 
        Study the long-term evolution of the afterglow and surrounding environment\\
        \hline
    \end{tabular}
\end{table}

\subsubsection{Lessons Learned from Past Events}

Over the past decade, a small number of landmark events have reshaped our understanding of LGRBs and exposed both strengths and weaknesses in the current follow-up ecosystem. Notable examples include the exceptionally bright GRB 221009A, ultra-long events such as GRB 250702B, the handful of confirmed GRBs at redshifts above six, and the emergence of long-duration GRBs that appear to originate from compact object mergers rather than collapsars (GRB 19109A, GRB 211211A, GRB 230307A). The identification of orphan afterglows such as AT2019pim and low-luminosity events has further expanded the observational parameter space.
Several aspects of the existing infrastructure have proven highly effective. The modernization of the GCN has enabled rapid, automated and machine readable alert dissemination. Public and near-real-time data releases from missions such as Swift and Fermi have continued to allow the community to respond quickly and flexibly. Automated Target-of-Opportunity (ToO) triggering, particularly on Swift, has demonstrated the scientific value of minimizing human latency in the earliest phases of follow-up.
At the same time, persistent challenges remain. Rare but high-impact subclasses, such as ultra-long GRBs or nearby GRB-supernovae, are often disfavored by traditional Time Allocation Committee (TAC) and Guest Investigator (GI) processes because of their low trigger probability. Ultra-rapid response capabilities remain severely constrained at several flagship facilities; for example, HST currently permits only a single ultra-disruptive ToO per year across all time-domain science. Coordination across wavelength regimes is further hampered by the absence of a fully unified database for prompt and follow-up observations, although efforts led by the IPN aim to address this gap. Scheduling policies at some ground-based facilities can also delay time-critical observations, particularly in the radio. Finally, lessons from these events underscore a strategic vulnerability: the lack of a clearly articulated pathway toward a next-generation gamma-ray transient mission capable of sustaining discovery and localization capabilities over the coming decades.

\newpage
\subsection{Tidal Disruption Events}

TDEs are crucial astrophysical laboratories for studying quiescent supermassive black holes (SMBHs), accretion disk physics, and the co-evolution of black holes and their host galaxies. They offer a unique opportunity to observe the formation and evolution of accretion disks in real-time as a star is disrupted and its debris falls onto an SMBH. Key scientific goals include probing the black hole mass function over cosmic time, understanding accretion state transitions from super- to sub-Eddington, and investigating the mechanisms behind jet and outflow launching.

TDEs are relatively slow transients compared to other events, with timescales calibrated for a $10^6$ $M_\odot$ black hole. They have UV/optical rise times of a $\sim$week to months, with a majority being about 30 days \citep{Hammerstein23}. Their emission declines on months to years, with some exhibiting late-time plateaus that last multiple years \citep[e.g.,][]{van_velzen_late-time_2019}. Events exhibit variability on hours to days, with hours representing the dynamical timescale at inner disk and days the potential diffusion timescale for optical variability. Generally, faster-rising events have been seen to decay more quickly and have lower peak luminosities \citep[e.g.,][]{yao_tidal_2023}.

Observational campaigns have identified distinct TDE sub-classes, including nearby, jetted, repeating, and off-nuclear events, each offering unique insights. Case studies of events like AT2019qiz \citep[nearby;][]{nicholl_outflow_2020, hung_discovery_2021, patra_spectropolarimetry_2022, short_delayed_2023} and AT2022cmc \citep[jetted;][]{Andreoni2022, pasham_birth_2023, yao_-axis_2024} demonstrate the value of rapid, multiwavelength follow-up, yet also reveal significant observational gaps. The primary limitation in current strategies is the difficulty of early-time classification and color measurement, as most candidates are identified near or after their peak brightness.

A significant under-explored parameter space remains, particularly in the pre-peak phase of the event. There is a critical need for early-time (pre-peak) X-ray and UV data (both photometry and spectroscopy) to constrain disk formation \citep[e.g.,][]{Huang2024, giron_multigroup_2026}, energetics \citep[e.g.,][]{lu_missing_2018}, and the composition of the disrupted star \citep[e.g.,][]{mockler_evidence_2022, mockler_tidal_2024}. Other key areas for improvement include obtaining polarization measurements to discern emission geometry \citep[e.g.,][]{patra_spectropolarimetry_2022} and conducting early-time radio follow-up to determine when outflows are launched \citep[e.g.,][]{alexander_multi-wavelength_2025}. The consensus is to prioritize rapid-response, high-cadence, multiwavelength observations for rare, high-value targets such as TDEs occurring closer than 65 Mpc or those launching relativistic jets.

\subsubsection{High-Priority Subclasses}

While the broader population of TDEs provides valuable insight into black hole demographics and accretion physics, a small number of TDE subclasses offer disproportionately high scientific return. These events are unique probes to otherwise inaccessible regimes of black hole mass, accretion state, and environment. The subclasses identified below are therefore prioritized for coordinated, rapid, and multiwavelength follow-up within the community observing plan. 

\begin{itemize}
    \item \textbf{Nearby TDEs:} Nearby events occurring within $\sim$65 Mpc (z $<$ 0.015) provide the best opportunity to detect faint, early-time emission from initial stream shocks and accretion disk formation. It allows for high-resolution spectral analysis across the electromagnetic spectrum to constrain gas conditions, outflow velocities, and event energetics.
    \item \textbf{Jetted TDEs:} Jet launching in TDEs remains poorly understood. multiwavelength observations of jetted events can help determine when the jet is launched relative to the optical TDE and constrain the properties of the disk that launches it. Only a handful of confirmed on-axis events exist, so each new one is disproportionately valuable.
    \item \textbf{Repeated TDEs:} Over the last several years there have been multiple examples of “repeated” TDE  flares with rebrightening episodes with similar luminosities to the initial transient peak (e.g., AT2022dbl). Repeating behaviour on human timescales implies stars in close-in orbits around a stellar-mass black hole (SMBH). With a larger population of these, TDEs can help constrain EMRI rates and multi-messenger physics.
    \item \textbf{Off-nuclear TDEs:} TDEs discovered outside the centers of galaxies, such as AT2024tvd, are a unique probe of galaxy and black hole co-evolution, and could either be a sign of a ‘wandering SMBH’ that experienced a kick or has not settled into its galaxy’s center OR an inspiraling secondary SMBH.
\end{itemize}

\subsubsection{Triggering Criteria \& Expected Event Rates}

Tidal disruption events are currently primarily discovered through wide-field optical surveys and high-energy observations and are characterized by their long evolutionary timescales and diverse multiwavelength behavior. Because TDEs evolve on timescales of weeks to years, their triggering is less dependent on short alert latencies and more reliant on rapid classification, early multiwavelength follow-up, and coordinated community assessment. While optical surveys currently detect dozens of TDE candidates per year and are expected to identify hundreds to thousands annually in the Rubin Observatory era, without further spectral follow-up capabilities the number of TDEs classified in real time is unlikely to increase dramatically. The two subclasses that most strongly motivate coordinated community observing plans, such as nearby disruption and jetted systems, are much rarer. These high-priority subclasses can be efficiently identified using trigger criteria based on distance, luminosity, and early spectral and temporal properties.

\begin{itemize}
    \item \textbf{Nearby TDEs:} Trigger criterion: Candidate TDE with distance $<$ 65 Mpc (z $<$ 0.015). Expected rate: Extremely rare ($\sim0.1$/yr) given that only two such events have thus far been detected.
    \item \textbf{Jetted TDEs: }Trigger criterion: Evidence for strong X-ray/radio activity (often initially discovered in X-rays, sometimes in optical surveys). Expected rate: Only four total confirmed on-axis jetted TDEs have been detected, consistent with $\sim1\%$ of TDEs launching jets, or roughly $\sim0.2$/yr.
\end{itemize}


\begin{table}[h]
    \centering
    \caption{Summary of High-Priority TDE Subclasses and Expected Rates}
    \label{tab:TDEsubclass}
    \begin{tabular}{|>{\raggedright\arraybackslash}p{2cm}|>{\raggedright\arraybackslash}p{6.5cm}|>{\raggedright\arraybackslash}p{4.5cm}|>{\raggedright\arraybackslash}p{2.6cm}|}
        \hline
       \textbf{Subclass} & \textbf{Scientific Priority / Rationale} & \textbf{Defining Characteristics} & \textbf{Expected Rate} \\
        \hline
       Nearby TDEs & Enables early-time measurements and strong constraints on disk formation, outflows, and energetics. & Distance $< 65$ Mpc $(z < 0.015)$ & $\sim0.1$ per year \\
        \hline
        Jetted TDEs & Constraints on disk properties and jet formation. & Strong X-ray/radio emission & $\sim0.2$ per year \\
        \hline
    \end{tabular}
\end{table}

\subsubsection{Discovery \& Alert Channels}

Discovery pathways for TDEs are heterogeneous, spanning wide-field optical time-domain surveys, X-ray monitors, and soft-X-ray survey missions, with different subclasses preferentially identified in different bands. Table~\ref{tab:TDEdiscovery} summarizes the primary discovery channels, alert pathways, and key identification considerations for the high-priority TDE subclasses targeted by this observing plan.

\begin{table}[h]
    \centering
        \caption{Summary of TDE Discovery Channels \& Alerts}
    \label{tab:TDEdiscovery}
    \begin{tabular}{|>{\raggedright\arraybackslash}p{2cm}|>{\raggedright\arraybackslash}p{6cm}|>{\raggedright\arraybackslash}p{3cm}|>{\raggedright\arraybackslash}p{5.5cm}|}
        \hline
       \textbf{Subclass} & \textbf{Primary Discovery Channels} & \textbf{Alert Pathways} & \textbf{Notes on Identification} \\
        \hline
       Nearby TDEs & Optical wide-field surveys (ZTF/ATLAS/ASAS-SN/LS4; Rubin era), sometimes X-ray surveys  & TNS + AstroNotes / ATel; Slack, GCN-Circulars & Coordinates (arcsec), best distance estimate, host ID, discovery epoch \& rise constraint, preliminary classification probability \\
        \hline
        Jetted TDEs & X-ray monitors/soft-X surveys; also optical surveys (demonstrated) & TNS + AstroNotes / ATel; Slack, GCN-Circulars & Evidence for non-thermal component, X-ray hardness/variability, early radio follow-up request, redshift/host limits \\
        \hline
        Off-nuclear TDEs &Optical surveys with careful centroiding; multiwavelength confirmation & TNS + AstroNotes / ATel; Slack, GCN-Circulars & Offset measurement, host imaging needs, prompt high-res imaging request, X-ray/radio checks \\
        \hline
    \end{tabular}
\end{table}

\subsubsection{Observatory Roles and Capabilities}

TDE science requires observations across a diverse set of facilities, each contributing unique capabilities across discovery, early follow-up, and long-term monitoring.

\begin{itemize}

\item \textbf{Discovery}

\begin{itemize}
    \item Wide-field optical surveys (Rubin/LSST, ZTF, LS4) and space-based missions (Einstein Probe, UVEX/ULTRASAT, Roman) enable the discovery of TDEs across optical, UV, and X-ray bands, including early-rise events, UV-bright systems, X-ray–selected and jetted TDEs, and high-redshift populations.
\end{itemize}
\item \textbf{Early Follow-up}

\begin{itemize}
    \setlength{\itemsep}{1pt}
    \setlength{\parskip}{1pt}
    \item Swift provides rapid UV/X-ray ToOs essential for pre-peak monitoring, UV blackbody sampling, and early accretion-disk diagnostics.
    \item XMM-Newton delivers high-quality X-ray spectroscopy to probe disk formation, ultra-fast outflows, and jet launching.
    \item Ground-based optical/NIR spectroscopy (Keck, VLT, Gemini) enables rapid classification, redshift determination, and early spectral evolution studies.
    \item HST and JWST provide high-resolution UV–optical–IR spectroscopy and imaging to study outflows and stellar composition.
\end{itemize}
\item \textbf{Late-Time Monitoring}

\begin{itemize}
    \setlength{\itemsep}{1pt}
    \setlength{\parskip}{1pt}
    \item HST and JWST also provide capability to study host-galaxy structure, slower evolving dust-obscured systems, and high-redshift TDEs.
    \item Radio facilities (VLA, MeerKAT, ALMA) trace jets and outflows through late-time radio afterglows, constraining energetics, geometry, and accretion-state transitions over months to years.
\end{itemize}
\end{itemize}

\begin{table}[h]
    \centering
        \caption{Summary of Observatory Roles and Capabilities for TDE Follow-up}
    \label{tab:TDEobsroles}
    \begin{tabular}{|>{\raggedright\arraybackslash}p{2.6cm}|>{\raggedright\arraybackslash}p{1.8cm}|>{\raggedright\arraybackslash}p{5cm}|>{\raggedright\arraybackslash}p{6.8cm}|}
        \hline
       \textbf{Observatory} & \textbf{Wavelength} & \textbf{Unique Capability} & \textbf{Ideal Use Case} \\
        \hline
       \textbf{HST} & UV/Optical & High-resolution host imaging, UV spectra & UV spectral evolution (pre \& post-peak); host imaging \\
        \hline
        \textbf{JWST} & NIR/MIR & Spectroscopy, high-z capability & Dust-obscured TDEs, infrared echoes of optical/UV emission, high-redshift environments \\
        \hline
        \textbf{Swift} & UV/X-ray & Rapid follow-up ToO & Early-time monitoring, UV blackbody sampling \\
        \hline 
        \textbf{XMM-Newton} & X-ray & High-quality X-ray spectroscopy & Probing disk formation, ultra-fast outflows, jets \\
        \hline
        \textbf{Rubin} & Optical (Wide-field) & High-cadence surveys & Discovery, precursor searches \\
        \hline
        \textbf{Roman} & NIR & High-redshift discovery, NIR characterization & Probing TDEs in the early universe \\
        \hline
        \textbf{Ultrasat} & UV & UV discovery \& characterization & Finding TDEs near their peak emission band \\
        \hline
        \textbf{UVEX} & UV & UV discovery (photometry) \& rapid UV spectroscopic follow-up & Finding TDEs near their peak emission band, constraining UV spectral evolution of broad lines \\
        \hline
        \textbf{EP} & X-ray & X-ray discovery & Finding X-ray bright and jetted TDEs \\
        \hline
        \textbf{VLA, MeerKAT, ATCA} & Radio & Jet and ejecta studies & Late-time radio afterglow, early-time outflow detection \\ 
        \hline
        \textbf{ZTF, LS4} & Optical (Wide-field) & Multi-band imaging, highcadence, g-band (ZTF) & Complementary to Rubin coverage, discovery in optical. \\
        \hline
        \textbf{Ground-based Optical} & Optical & Spectroscopy and imaging & Classification, follow-up, host galaxy studies \\
        \hline
    \end{tabular}
\end{table}

\subsubsection{Observing Strategy and Timeline}
Observations should prioritize pre-peak ultraviolet and X-ray coverage to capture the earliest stages of disk formation and energy release, followed by short-cadence monitoring to probe rapid variability and structural changes in the accretion flow. When feasible, early ultraviolet spectroscopy should be obtained to diagnose temperatures, outflows, and composition during the most information-rich phase of the event. As the transient evolves, observations should transition to logarithmically spaced monitoring that tracks longer-term luminosity and spectral evolution, with late-time campaigns focused on host-galaxy characterization and X-ray timing studies (e.g., searches for quasi-periodic eruptions) once the transient emission has faded.

\begin{table}[h]
    \centering
        \caption{Summary of TDE Observing Strategy and Timeline}
    \label{tab:TDEobsstrategy}
    \begin{tabular}{|>{\raggedright\arraybackslash}p{1.7cm}|>{\raggedright\arraybackslash}p{5cm}|>{\raggedright\arraybackslash}p{5.1cm}|>{\raggedright\arraybackslash}p{4.7cm}|}
        \hline
       \textbf{Phase} & \textbf{Recommended Observations} & \textbf{Primary Science Goals} & \textbf{Facilities} \\
        \hline
        \textbf{T0 to 24 hr} & Swift-UVOT+XRT rapid ToO; optical spectroscopy + multiband photometry; rapid radio “first look”; if feasible, X-ray spectrum and UV spectroscopy & Confirm TDE nature, establish UV/X-ray baseline, catch earliest variability; initiate jet/outflow constraints & Swift; ground optical spectroscopy (Keck-class + networked 1–2m photometry); VLA/MeerKAT/ATCA; NOEMA/ALMA/SMA; XMM-Newton; HST (high-priority cases) \\
        \hline
        \textbf{T0+1 to 7 days} & Daily/near-daily UV+optical photometry; repeated optical spectra; UV spectroscopy; short-cadence X-ray sampling; early radio cadence (including upper limits); polarization if possible & Map rise physics; constrain temperature evolution; measure early outflows; search for sub-day variability & Swift; ground optical networks; X-ray observatories (as available); radio arrays; polarization-capable optical facilities; HST (high-priority cases) \\
        \hline
        \textbf{Weeks to Months} & Logarithmic multi-epoch UV/X-ray/optical/radio; one or more deep X-ray spectra; UV spectroscopy near peak when possible; continued radio follow-up & Track accretion state transitions, outflow evolution, jet evolution; build SED; connect peak timescale to BH mass/fallback & Swift; XMM-Newton; HST; VLA/MeerKAT/ATCA; NOEMA/ALMA/SMA; ground spectroscopy \\
        \hline 
        \textbf{T0+ $\sim1$ yr to years} & UV photometry for late-time evolution; X-ray monitoring for QPEs (short cadence for timing); radio monitoring for late-time flaring; infrared monitoring for dust echoes; high-res imaging; host spectroscopy/IFU after transient fades & Late-time plateaus; QPE searches; disk evolution – disk flares from state transitions or instabilities, jet turn on/precession; energy constraints from dust echoes reprocessing early UV;  host galaxy/offset studies; stellar kinematics \& morphology & Swift; Chandra/XMM (as available); HST/JWST imaging; ground IFU/long-slit host work \\
        \hline
    \end{tabular}
\end{table}

\subsubsection{Lessons Learned from Past Events}

Past TDEs provide clear demonstrations of how early, coordinated, multiwavelength observations can fundamentally shape our understanding of black hole accretion, outflows, and jet formation. Two particularly instructive examples, one nearby, one jetted, highlight both what is currently achievable and where future community observing plans can dramatically improve scientific return.

\subsubsubsection*{The Nearby TDE: AT2019qiz}

Owing to its proximity (z = 0.015, or $\sim$65 Mpc), AT2019qiz is one of the most comprehensively studied tidal disruption events to date and serves as a benchmark for nearby TDEs. Although discovered in the optical, the event was detected prior to optical peak across the electromagnetic spectrum, from X-rays (Swift-XRT) to radio wavelengths (VLA). These observations represent some of the earliest X-ray data and the earliest radio detections obtained for any TDE.

The radio dataset for AT2019qiz is particularly extensive, comprising observations from at least four independent observing programs as well as public data from VLASS Epoch 3. The X-ray coverage similarly reflects coordinated community effort, with multiple groups contributing Swift-XRT observations. As the optical luminosity declined, the X-ray emission rose and became progressively softer following optical peak—behavior not previously observed in TDEs and highly constraining for models of accretion disk emission and reprocessing.

AT2019qiz exhibits exceptionally well-sampled optical spectral evolution, with more than 30 epochs of spectroscopy. Early spectra show blueshifted H and He II emission lines that fade as Bowen fluorescence features emerge, consistent with an initially expanding photosphere that gradually becomes optically thin \citep{Nicholl24}. The event was also observed with three post-peak HST UV spectra \citep{hung21}, which revealed broad absorption line outflows whose velocities decrease over time. Notably, these spectra include the first detections of low-ionization broad absorption iron lines in a TDE, resembling features previously seen in superluminous supernovae. Independent radio detections yielded outflow velocities consistent with those inferred from the optical and UV spectra (Alexander et al. 2025).

At late times, X-ray monitoring revealed the emergence of a quasi-periodic eruption (QPE) source associated with AT2019qiz, further highlighting the long-term, multiwavelength complexity of TDE evolution \citep{Nicholl24}.
The optical light-curve rise is well constrained to approximately 30 days prior to peak, with upper limits reaching $\approx$20th magnitude in the g-band. For comparable events discovered in the Rubin/LSST era, pre-peak constraints will extend roughly four magnitudes deeper. This sensitivity will probe luminosities of $\sim$10$^{40}$–10$^{41}$ erg s$^{-1}$ in bolometric emission (corresponding to $\sim$10$^{39}$–10$^{40}$ erg s$^{-1}$ in optical bands), directly testing theoretical predictions for early-time variability driven by debris stream self-intersections and disk-forming shocks.

\subsubsubsection*{Jetted TDEs}
The four confirmed on-axis jetted TDEs, Swift J1644, Swift J2058, Swift J1112, and AT2022cmc, provide the current observational framework for understanding relativistic jet production in TDEs. Most known jetted events have been discovered in X-ray surveys, with Swift J1644 remaining the most extensively studied example. However, the recent discovery of AT2022cmc demonstrated that jetted TDEs can also be identified at optical wavelengths by wide-field time-domain surveys.
AT2022cmc was first detected by the Zwicky Transient Facility, promptly reported to the TNS, and followed by rapid multiwavelength observations spanning X-ray to radio frequencies within days. The event was detected in soft and hard X-rays (Swift-XRT, NICER, XMM-Newton, NuSTAR), UV (Swift-UVOT), radio (VLA, AMI-LA, e-MERLIN, MeerKAT), and submillimeter wavelengths (SMA, NOEMA, JCMT/SCUBA-2), making it one of the most luminous high-redshift transients observed across these bands. Optical and near-infrared spectroscopy with X-shooter confirmed its high redshift. HST observations detected the transient in optical (F606W) and near-infrared (F106W) bands, although the host galaxy was too faint to detect at that time.
The bright optical emission from AT2022cmc, which exhibited both thermal and non-thermal components, enabled the event to be placed in the broader context of the optical TDE population \citep{hammerstein26}. Future deep surveys, particularly Rubin/LSST, are expected to uncover additional jetted TDEs via their optical signatures, enabling more complete population studies.

\subsubsubsection*{Off-Nuclear TDE: AT2024tvd}

As transient-search algorithms improve and expand beyond strict nuclear selection criteria, off-nuclear TDEs are beginning to be identified. These events offer a unique probe of galaxy–black hole co-evolution and may signal either a “wandering” supermassive black hole displaced by a gravitational recoil or an inspiraling secondary SMBH in a merging system.
AT2024tvd represents the first off-nuclear TDE discovered by an optical survey. Follow-up X-ray and radio observations are consistent with an origin offset by approximately 0.8 kpc from the center of its host galaxy (Yao et al. 2025), confirming its off-nuclear nature and underscoring the importance of high-resolution imaging and multiwavelength follow-up for this emerging TDE subclass.

\subsubsection{Limitations in Current Strategies}

The primary limitation is currently early-time classification of TDE candidates. Most candidates are classified around the peak of the light curve, with only a handful being classified pre-peak with subsequently triggered pre-peak follow-up observations. This is limited by the ability to find candidates very early, and differentiate them from imposters like nuclear SNe, which can look very similar to TDEs at early times. Contemporaneous UV photometry with ULTRASAT (3-4 day cadence, roughly 23rd mag) will screen out a majority of SNe in the near future. Still, optical spectroscopy is the “gold standard” when classifying TDEs and differentiating them from SNe and AGN, but it is expensive and will be insufficient to classify the vast majority of candidates in the age of LSST.

\newpage
\subsection{X-ray Binaries}

XRBs is an umbrella term encompassing a variety of distinct astrophysical systems. These systems share some properties but probe a range of fundamental physical processes. An XRB is primarily defined by the nature of its compact object (either a black hole or a neutron star) and the mass of the companion star (high mass, $\gtrsim$10~$M_\odot$, or low mass, $\lesssim$2~$M_\odot$). Accretion is a universally important process in XRBs, which can occur via a disk, through the wind of a massive star, or sometimes be magnetically guided. Systems can exhibit jets and various types of winds in different accretion states.

The study of XRBs provides key insights into several areas, including binary stellar evolution, understanding the violent deaths of stars (via natal supernova kicks), and how material behaves in strong gravitational and magnetic fields. XRB observations offer some of the most direct measurements of fundamental compact object properties, such as their spin and mass, and for neutron stars, their equation of state. There are hundreds of known XRBs, with a population likely numbering in the thousands in our galaxy, and we also observe some in nearby galaxies like the Small Magellanic Cloud (SMC)

There are two primary configurations of XRBs: High Mass XRBs (HMXRBs) and Low Mass XRBs (LMXRBs), with several sub-classes based on the nature of the compact object. Classification is often separated by companion mass first, as this often indicates a distinct formation channel and accretion process. The two configurations include:

\begin{itemize}

\item \textbf{HMXRBs}: HMXRBs contain a companion star with a mass greater than approximately 10~$M_\odot$. They are thought to form from evolving stellar binaries where the system remains bound after a supernova.  In HMXRBs, the compact object primarily accretes material directly from the powerful stellar wind of its massive O/B, Be, or Wolf-Rayet companion. As the companion evolves and its mass loss rate increases, the system's X-ray luminosity rises. The majority of known HMXRBs contain a neutron star. Due to strong magnetic fields, the accreted material is funneled onto the magnetic poles, creating accretion columns and emitting X-rays. This can lead to observable X-ray pulsations with periods from seconds to minutes. Their radio emission is significantly lower than that of their black hole counterparts. Identifying a black hole in an HMXRB can be challenging. The absence of pulsations or Type I X-ray bursts is not conclusive. The most definitive method is a dynamical mass measurement showing the compact object's mass to be significantly greater than $\sim$2~$M_\odot$.

\item \textbf{LMXRBs}: LMXRBs feature a lower-mass companion (typically $\lesssim$2~$M_\odot$) and have a less clear formation channel, possibly involving common envelope evolution or tidal capture. Accretion is not driven by a strong stellar wind. Instead, material flows through the first Lagrange point via Roche lobe overflow, forming a multi-temperature accretion disk that emits from optical to X-ray wavelengths. Black hole LMXRBs are perhaps the most studied of the XRB population, largely due to their bright outburst phases. They are considered ideal sources for developing predefined observing plans because they undergo fairly predictable outbursts defined by accretion states, making a phenomenological framework possible. LMBHXRBs are excellent for studying accretion and its connection to jet launching, and for measuring fundamental black hole properties like mass and spin. Neutron Star LMXRBs systems differ from their black hole counterparts primarily due to the presence of a solid surface and strong magnetic fields. These systems are critical for the measurement of neutron star mass and radius, which constrains their unknown equation of state. The presence of a solid surface in NS systems causes key differences compared to black hole systems, such as the potential for Type-I X-ray bursts (thermonuclear explosions on the neutron star surface). Their radio emission is significantly weaker, and some systems harbor detectable jets even in their soft states, a marked difference from BH LMXRBs.

\end{itemize}

\subsubsection{High-Priority Subclasses}

The event rate of outbursting XRB systems is too high to make pre-defined, fleet-wide follow-up of all events feasible. Coordinated observing plans should therefore be reserved for truly unique systems that exhibit predictable outbursts, exceptional brightness, or are at close proximity to Earth. 

\begin{itemize}

\item \textbf{Predictable Outbursts}: A subset of X-ray binaries, particularly low-mass black hole X-ray binaries, exhibit repeatable, well-characterized outburst cycles driven by accretion-disk instabilities. These systems evolve through distinct accretion states with predictable spectral, timing, and jet signatures that can often be anticipated using optical and X-ray precursors. Their structured evolution makes them ideal candidates for predefined, time-phased observing plans, enabling strictly simultaneous, multiwavelength campaigns. Coordinated observations during key phases yield high-impact constraints on accretion physics, jet launching and quenching, and fundamental compact-object properties, making these systems a priority for organized community follow-up.

\item \textbf{Exceptional Brightness}: A small number of X-ray binaries undergo exceptionally luminous outbursts, occasionally reaching tens of Crab and triggering gamma-ray burst monitors such as Swift-BAT. These extreme events, exemplified by systems like V404 Cygni and MAXI J1820+070, offer a rare opportunity to apply observational techniques that are otherwise photon-limited, including fast timing, X-ray polarimetry, detailed spectral-timing studies, and spatially resolved radio imaging of jets and ejecta. Their brightness allows contributions from a wide range of facilities but also requires specialized strategies to handle saturation and dynamic-range limits. Because such outbursts are both infrequent and scientifically transformative, they merit rapid, coordinated, multi-facility responses to fully exploit their diagnostic power.

\item \textbf{Nearby Systems}: Nearby X-ray binaries provide unique opportunities for precision measurements, including parallax distances, proper motions and natal kicks, and dynamical mass constraints from optical spectroscopy. Their proximity enables resolved studies of jets and ejecta, sensitive investigations of quiescent accretion, and robust constraints on neutron-star equations of state and black-hole mass and spin distributions. Although not always the most luminous systems, nearby XRBs anchor population studies and provide critical ground-truth measurements. As such, they represent high-priority targets for coordinated observing campaigns during outburst or extended bright states.

\end{itemize}

\subsubsection{Triggering Criteria \& Expected Event Rates}

There are approximately 350 known LMXRBs and 150 known HMXRBs in the Galaxy. While the total population may be thousands, the most informative physical constraints come from outbursts. Typically, there are a handful of sources undergoing outburst at any given time, perhaps around five. These outbursts, which are accretion-driven, are relatively uncommon but can last for weeks, months, or even years.

A possible discriminator for unusual bright or nearby XRB outbursts is their ability to trigger gamma-ray instruments such as Swift-BAT. Such triggers typically indicate the onset of an extreme or unusual outburst (e.g., V404 Cygni or MAXI J1820+070). Follow-up triggered by GRB mechanisms should generally be limited to source positions near the galactic plane to help distinguish them from gamma-ray bursts.

For known sources entering an outburst, optical monitoring is often the best way to detect the rise of the outburst first, before the X-rays, providing a good precursor warning.

\subsubsection{Observing Strategy}

The sources are too diverse, and the range of science goals too broad, for there to be a single, standardized observing plan for all X-ray binaries. The optimal XRB science is achieved through strictly simultaneous and multiwavelength observations. Coordinated observations are essential to understand systems and maximize science. Because sources exhibit rapid variability and connections across wavelengths on short timescales, tight coordination is required.

Instead of attempting to predict exact timing schedules, a more promising approach is to react to well-understood system changes, such as the evolution of accretion states during a Low Mass Black Hole XRB outburst. XRB outbursts can be extremely bright, sometimes becoming among the brightest X-ray sources in the sky, and optical magnitudes are easily observable by small telescopes. This high luminosity allows for a range of techniques beyond flux monitoring. Observing extremely bright X-ray sources is challenging, though, as some X-ray binaries can be too luminous to point X-ray telescopes at. Future facilities need to prioritize the ability to observe bright sources, potentially utilizing techniques like sub-arraying in radio facilities like SKA/ngVLA when sensitivity is not the primary concern. While high brightness is a good trigger, coordination should not preclude observations of fainter systems that may still be scientifically interesting.

Organizing strictly simultaneous observations is a major challenge, as observation schedules can evolve rapidly, leading to a fairly high failure rate for campaigns put together in an ad hoc manner.

\subsubsection{Discovery \& Alert Channels}

HMXRBs were initially discovered through X-ray observations using instruments like sounding rockets and satellites like INTEGRAL and RXTE. New or known systems entering outburst can be detected by all-sky monitors such as MAXI or Swift-BAT (via GRB detection mechanisms). Optical monitoring (e.g., using LCO facilities) is essential for monitoring known sources to detect the rising optical emission that precedes the X-ray rise, signaling the start of an outburst. Upcoming observations by Rubin will be important for discovering extragalactic, particularly long-period, XRBs, while the Roman galactic bulge survey is expected to find quiescent systems.

Automated dissemination of alerts in machine-readable format, such as VOEvent, is valuable for enabling robotic follow-up (e.g., AMI-LA). Alerts for outburst starts are sometimes tracked via services like XBNEWS. While automated follow-up to GRB triggers is useful, ToO observations related to XRBs are currently disseminated through a diverse range of streams, with ATels still broadly used, which are less regularized than TNS or GCN/VOEvent reporting.

\subsubsection{Observatory Roles and Capabilities}

Successful XRB observations require continuous monitoring and the use of dedicated facilities across the electromagnetic spectrum.
 
\begin{table}[h]
    \centering
    \caption{Summary of Observatory Roles and Capabilities for XRB Follow-up}
    \label{tab:xrb:obs}
    \begin{tabular}{|>{\raggedright\arraybackslash}p{2cm}|>{\raggedright\arraybackslash}p{3cm}|>{\raggedright\arraybackslash}p{5.1cm}|>{\raggedright\arraybackslash}p{4.7cm}|}
\hline
\textbf{Observatory} & \textbf{Wavelength} & \textbf{Unique Capability} & \textbf{Ideal Use Case}\\
\hline
 
\textbf{Swift} &
Soft X-rays, Gamma-rays, UV, Optical &
Wide field of view, rapid follow-up &
Outburst detection, spectral and temporal monitoring, accretion state changes \\[6pt]

\hline
\textbf{MAXI} &
Soft X-rays &
Wide field of view &
Outburst detection, outburst monitoring, accretion state changes \\[6pt]
 
\hline
\textbf{NICER} &
Soft X-rays &
Fast timing &
Mapping QPO evolution and accretion state changes, detecting Type~I bursts \\[6pt]
 
\hline
\textbf{IXPE} &
Soft X-rays &
Polarimetry &
Accretion disc orientation measurements, accretion disk models \\[6pt]

\hline
\textbf{SVOM} &
Soft X-rays, Gamma-rays, Optical &
Wide field of view, rapid follow-up &
Outburst detection, spectral and temporal monitoring, accretion state changes \\

\hline
\textbf{VLBA, EVN, LBA} &
Radio &
High angular resolution &
Parallax for distance measurements, resolving jet ejection events, resolving the core jet \\[6pt]
 
\hline
\textbf{VLA, MeerKAT, ATCA} &
Radio &
High sensitivity, good resolution, polarization &
Monitoring ejecta and core positions/luminosities \\[6pt]
 
\hline
\textbf{AMI/ATA} &
Radio &
Automated response, flexibility &
Radio--X-ray correlation studies, accretion state monitoring \\[6pt]
 
\hline
\textbf{Small Optical Telescopes} &
Optical &
Monitoring &
Monitoring potential outbursts from known systems \\[6pt]
 
\hline
\textbf{GTC/VLT} &
Optical &
Sensitive spectroscopy &
Disk wind signatures, mass estimates via radial velocity \\[6pt]
 
\hline
\textbf{JWST} &
IR &
High sensitivity &
Timing of quiescent sources, quiescent studies \\

\hline
\end{tabular}
\end{table}

\subsubsection{Observing Timeline}
 
A specific observing framework can be applied to the well-understood outburst cycle of
Low Mass Black Hole X-ray Binaries.

 \begin{table}[h]
    \centering
    \caption{Summary of Observatory Timelines for XRB Follow-up}
    \label{tab:xrb:follow}
    \begin{tabular}{|>{\raggedright\arraybackslash}p{2cm}|>{\raggedright\arraybackslash}p{3cm}|>{\raggedright\arraybackslash}p{5.1cm}|>{\raggedright\arraybackslash}p{4.7cm}|}
\hline
\textbf{Observatory} & \textbf{Wavelength} & \textbf{Unique Capability} & \textbf{Ideal Use Case}\\
\hline
 
\textbf{T0 to T0 + 1 week} &
Rising from Quiescence / Hard State &
The outburst rises to a maximum over about one week. This phase provides data on the
onset of the DIM, the beginning of outbursts, and the
accretion--ejection connection. &
Monitoring instruments to track the rise across the EM spectrum: Swift, LCO, smaller
radio facilities, MAXI. \\[6pt]

\hline
\textbf{T0 + 1 week to T0 + weeks, months} &
Bright Hard State &
The source is typically at its brightest and can remain in this state for a long period.
This is the ideal time for coordinated, high-quality observations to understand the
accretion-jet connection, black hole properties (spin), and geometries/winds. &
High-end facilities for high signal-to-noise: VLBI, IXPE, 10\,m class telescopes. \\[6pt]

\hline
\textbf{T0 + ?} &
State Transition / Ejecta Launch &
The hard-state core jet switches off (flux drops by a factor of ${\sim}1000$), and the
source launches transient ejecta (not connected to the black hole). This phase is
important for understanding jet quenching and ejecta launch mechanisms. &
VLBI, Radio Monitoring, Timing facilities. \\[6pt]

\hline
\textbf{T0 + ?} &
Return Transition &
The core jet reignites on return to the hard state, but the reverse transition is
typically not associated with the launch of transient ejecta. &
VLBI, IXPE, 10\,m class optical telescopes, NICER. \\[6pt]

\hline
\textbf{T0 + ?} &
Final Hard State / Fading and Quiescence &
Sources fade out to quiescence. This phase is used to measure black hole masses via the
companion star's mass function and radial velocity curves. Long observations allow for
VLBI parallax measurements (distance) and proper motion studies (kicks). &
VLBI, 10\,m class optical telescopes, JWST. \\
 
\hline
\end{tabular}
\end{table}

\subsubsection{Lessons Learned from Past Events}

The most significant success in XRB observations is highlighted by the importance of strictly simultaneous observing across multiple wavelengths to maximize scientific return. Examples include:

\begin{itemize}

\item \textbf{Neutron Star Jet Constraints}: Simultaneous X-ray and radio monitoring (e.g., using ATA) captured a radio response following Type I X-ray bursts on the surface of the neutron star X-ray binary 4U1727. This demonstrated the connection between the accretion column and jet launch mechanisms in neutron stars and was used to constrain jet velocities (which were found to be lower than in black hole systems).

\item \textbf{Fundamental Jet Properties}: Simultaneous radio, sub-mm, IR, optical, and soft X-ray observations of bright LMXRBs in the hard state have allowed for modeling of jets, high-precision timing, and placing constraints on the Lorentz factor of the launched jets.

\end{itemize}

Despite this, significance challenges remain to maximizing the scientific return of XRB follow up. Areas that could could be improved include:

\begin{itemize}

\item \textbf{Coordination Challenges}: The primary limitation is the difficulty in coordinating strictly simultaneous observations across numerous facilities, especially since many observing campaigns are put together in an ad-hoc manner. The ability to pre-arrange observing schedules or have earlier warning of schedule changes is valuable. More formal channels for simultaneous observations are needed.

\item \textbf{Reporting Standardization}: Reporting for X-ray binaries is not adequately standardized, with heavy reliance currently placed on ATels, which are less regularized than VOEvent or GCN reporting. 

\item \textbf{Bright Source Observation}: There is a need for observational strategies and capabilities, such as X-ray instruments capable of handling extremely bright sources (up to 50 Crab) and the use of sub-arraying in radio facilities, to ensure bright outbursts can be fully observed across all wavelengths.

\end{itemize}

\newpage 
\subsection{Novae}
Classical and recurrent novae are among the most accessible and information-rich explosive transients in the Galaxy, offering a uniquely local laboratory for studying a broad range of astrophysical processes \citep{2008clno.book.....B, 2021ARA&A..59..391C}. A nova eruption is triggered by a thermonuclear runaway on the surface of an accreting white dwarf (WD), but modern observations demonstrate that their phenomenology extends far beyond a simple, thermally powered outburst. multiwavelength campaigns over the past decade have revealed that shocks are a fundamental and often dominant energy source, producing luminous emission from radio through GeV–TeV gamma rays and potentially powering a significant fraction of the optical light \citep{2014Sci...345..554A, 2015MNRAS.450.2739M, 2017NatAs...1..697L, 2018A&A...609A.120F, 2020NatAs...4..776A}.

As such, novae probe key questions in stellar and binary evolution (mass transfer, common envelope interaction, and envelope ejection), radiative and particle-accelerating shocks, dust formation in explosive environments, and nuclear burning on accreting WDs \citep{2015MNRAS.450.2739M, 2021ARA&A..59..391C, 2026NatAs..10..271A}. The X-ray super-soft source (SSS) phase directly constrains WD mass, envelope mass, and burning efficiency, informing the long-standing question of whether recurrent novae can grow towards the Chandrasekhar mass and contribute to the population of Type Ia supernova progenitors \citep{2013ApJ...777..136W}. The recent detection of GeV emission with Fermi-LAT, and TeV emission from the recurrent nova RS Ophiuchi, has further elevated novae as nearby laboratories for high-energy astrophysics and candidate multi-messenger sources, with potential (though as yet unobserved) neutrino emission.

Nova eruptions evolve through a sequence of physically distinct phases spanning hours to years, with dramatic changes in spectral energy distribution and dominant emission mechanisms. A short-lived fireball phase, predicted to last up to $\sim$0.5~days, may produce an X-ray/UV flash immediately following ignition and has been observed only once to date \citep[with a duration of $<$~8~hr;][]{2022Natur.605..248K}. The optical rise can be extremely rapid (hours) or very slow (weeks to months), followed by a decline characterized by the parameter t$_{2}$ (the number of days for the nova to drop by 2 magnitudes from peak), ranging from days in very fast novae to $>$100 days in slow systems.

High-energy emission traces shock formation: GeV gamma-rays often emerge near optical maximum and persist for days to weeks, sometimes with delayed onset, while hard X-rays ($>$10~keV) can appear contemporaneously, largely unaffected by absorption. Intermediate (1–-10~keV) X-rays probe shock-heated plasma and may be detectable within hours in systems with dense circumbinary material or only after weeks when shocks are embedded within dense ejecta. As the ejecta expand and thin, nuclear burning on the WD surface becomes visible as luminous super-soft X-ray emission, turning on days to months after eruption and lasting from days to years, often with initial strong variability and quasi-periodic oscillations. In parallel, UV and optical wavelengths track the declining photosphere, infrared emission may signal dust formation on timescales of days to months, and radio emission evolves from early non-thermal (shock-powered) to long-lived thermal emission from expanding ejecta.
Despite rapid progress, several critical regions of nova parameter space remain poorly sampled. The earliest moments of eruption, including the predicted fireball X-ray/UV flash, are almost entirely unexplored due to their brevity and the lack of wide-field, high-cadence coverage. The connection between shock properties (velocity contrasts, densities, geometry) and the resulting gamma-ray and X-ray luminosities is still incompletely mapped, as is the understanding of whether particle acceleration occurs through leptonic or hadronic regimes \citep[e.g.,][]{2013A&A...551A..37M, 2022NatAs...6..689A, 2022Sci...376...77H, 2023ApJ...951...62D}. Similarly, the role of shocks in dust formation, how novae eject their envelopes, the origin of the large-amplitude variability and quasi-periodic oscillations during the SSS phase, and the conditions under which novae might produce detectable neutrino emission remain open questions. Preplanned, coordinated, multi-facility observations that begin within hours of eruption and continue through late-time evolution can help address these gaps.

\subsubsection{High-Priority Subclasses}

Based on the morphology of their optical light curves, novae are grouped into several distinct subclasses. These differences likely reflect variations in the underlying physical properties of the white dwarf, companion star, and ejected material. These subclasses include classical novae, which have only been observed in eruption once (though are expected to recur eventually, over 10$^3$–10$^5$~yr), typically involving main-sequence companions and shocks internal to the ejecta; recurrent novae with decade-scale recurrence intervals often hosting massive white dwarfs and producing early, externally driven shocks; symbiotic novae where a white dwarf accretes from a red-giant companion, generating strong external shocks in dense circumbinary media (some recurrent novae are also symbiotic systems); and fast and slow novae, defined observationally by optical decline rates, correlating with ejecta velocities, shock properties, and SSS behavior.

Of these subclasses, the following classes are prioritized:

\begin{itemize}
    \item {\bf Nearby fast novae}: Fast novae evolve on timescales that demand immediate, coordinated response, making them ideal targets for pre-approved community observing plans. Slower novae evolve over months to years and can typically be accommodated through standard proposal mechanisms. Nearby fast novae are particularly valuable because their proximity maximizes brightness and signal-to-noise across the electromagnetic spectrum during the critical early phases, when shock-powered emission, gamma rays, and hard X-rays emerge.

\item {\bf Recurrent novae}: Anticipated eruptions of systems such as T Coronae Borealis are expected to be exceptionally bright and to generate powerful particle acceleration, making them prime candidates for very-high-energy gamma-ray and potential neutrino detection. Knowledge of expected recurrence times of the known Galactic recurrent novae will provide a certain level of advance warning for planning observations. The next eruption of T Coronae Borealis, for example, is predicted to be imminent and, given this system is particularly close by at $\sim$900 pc, the nova is expected to be exceptionally bright, and a prime candidate for the detection of VHE gamma-rays and, potentially, neutrinos. In general, the repeated outbursts of recurrent novae also enable direct tests of white dwarf mass growth, shock physics, and links to Type Ia supernova progenitors, justifying their designation as top-priority targets. 

\end{itemize}

\subsubsection{Triggering Criteria \& Expected Event Rates}

The estimated nova rate in the Milky Way is between 30 and 80 per year. However, due to obscuration and dust in the galaxy, only about 10--15 novae per year are actually observed, mostly out of the Galactic plane.

\begin{itemize}
    \item {\bf Nearby fast novae}: A rapid optical rise indicates evolution on timescales of days to weeks; this is the case for around \(\frac{1}{3}\) -- \(\frac{1}{2}\)  of the Galactic novae discovered. The nova should also reach a high optical peak brightness, ideally V $\lesssim$ 5--6 mag. A GeV gamma-ray detection by Fermi-LAT is highly desirable and serves as a strong indicator that powerful shocks are present early in the eruption. The event should also be a Galactic nova to ensure maximum brightness, but preferably not located in the plane of the Galaxy to avoid excessive absorption \citep[studies have shown, however, that fast novae are typically concentrated towards the plane;][]{1992A&A...266..232D}. The nova must also have a good Sun angle, to ensure that it is easily observable for an extended period, allowing time for the accumulation of a decent dataset. 

\item {\bf Recurrent novae}: There are 10 known recurrent novae in the Milky Way with recurrence timescales of $\sim$10--100 years. These, along with fast-recurring extragalactic novae like M31N~2008-12a in the Andromeda galaxy, provide (semi-)predictable targets for study (in some cases the recurrence times are quite variable: RS Oph, for example, erupted in 1898 and 1907 -- a gap of only 9 years -- though then remained in quiescence for the following 26 years). Based on previously-detected outbursts, it is likely that at least one of the Galactic recurrent novae (specifically CI~Aql, V394~CrA, T~CrB, T~Pyx) will re-erupt in the next few years, with others expected to recur in the 2030s. While extragalactic recurrent novae can be very interesting, their greater distances make detailed observations less feasible. Again, a good Sun angle will be conducive for an extended monitoring campaign.

\end{itemize}

\subsubsection{Discovery \& Alert Channels}

A network of both professional and amateur surveys discovers most novae via their optical rise, with the majority of discoveries then reported either through AAVSO (the American Association of Variable Star Observers) or TNS (the Transient Name Server). In contrast, V959~Mon was identified in gamma-rays before the optical, because the transient was too close to the Sun at eruption for ground-based observations; this event was announced through an Astronomer’s Telegram, but was an unusual situation. 

\subsubsection{Observatory Roles and Capabilities}

Since novae are detectable across the entire electromagnetic spectrum and potentially beyond (i.e., neutrinos), a wide range of observatories are relevant.


\begin{table}[h]
    \centering
    \caption{Summary of Observatory Roles and Capabilities for Nova Follow-up}
    \label{tab:novae:missions}
    \begin{tabular}{|>{\raggedright\arraybackslash}p{2cm}|>{\raggedright\arraybackslash}p{3cm}|>{\raggedright\arraybackslash}p{5.1cm}|>{\raggedright\arraybackslash}p{4.7cm}|}
\hline
\textbf{Observatory} & \textbf{Wavelength} & \textbf{Unique Capability} & \textbf{Ideal Use Case}\\
\hline
\textbf{H.E.S.S., MAGIC, VERITAS, HAWC, LHAASO} &   VHE ($>$ 100 GeV) & VHE  photon detection & Follow-up detection of early, high-energy emission; leptonic versus hadronic identification\\
\hline
\textbf{Fermi-LAT} & GeV & Wide-field monitoring & Detection of gamma-ray emission, before or 
simultaneous with X-rays. \\
\hline
\textbf{NuSTAR}, \textbf{Swift-BAT} (survey mode)& Hard X-rays ($>$10 keV) & Early X-rays from shocks & Monitoring early, hard X-ray shock emission; identification of thermal vs non-thermal 
emission. \\ 
\hline
\textbf{Swift, EP, XMM-Newton, Chandra} & X-rays (General) & Rapid slews and follow-up (Swift), wide-field discovery (EP), high-spatial resolution, spectroscopy and sensitivity (XMM/Chandra) & Swift monitoring of X-rays throughout the evolution; XMM-Newton/Chandra performing frequent  spectroscopy (RGS, LETG) to follow changing line emission and abundances, specifically during the SSS phase; search for QPOs by all X-ray missions \\

\hline
\textbf{Swift, HST, CHARA, VLT/VLTI, Gemini}  & UV/Optical & Rapid slews and follow-up (Swift), high-res imaging (HST) & Swift-UVOT monitors the UV and optical, including UV grism. Ground-based optical imaging essential for T0 + 0--1 hr.\\

\hline
\textbf{JWST, VLT/VLTI, Gemini} & IR & IR imaging and spectroscopy & High-res IR spectroscopy (0.7-5 micron) and JHK (potentially LMN) photometry necessary to study potential dust formation. \\
\hline

\textbf{VLA, VLBA, ATCA, e-MERLIN}, \textbf{LOFAR}, \textbf{MeerKAT}, \textbf{SKA} & Radio & Sensitive broadband radio & Monitoring non-thermal and later thermal radio emission to map ejecta.\\
\hline
\textbf{IceCube, KM3NeT} & Neutrinos & Multi-messenger emission & Checking for neutrinos
in the first 24 hours.\\
\hline
\end{tabular}
\end{table}

\subsubsection{Observing Strategy and Timeline}

An effective observing strategy for a nova must be dynamic and multi-phased, covering multiple timescales from the immediate aftermath of the explosion to the long-term evolution of the remnant shell. Each phase utilizes different observatories to probe the changing physical conditions of the event.

\begin{itemize}
    \item {\bf Phase 1: Early Days/Weeks}: The initial response focuses on high-cadence photometry and spectroscopy to track the eruption's brightness and velocity evolution. Rapid-response, ultra-high-resolution interferometry is critical during this phase. For V1674 Her, the CHARA Array was able to resolve the expanding fireball just 2.2 and 3.2 days after discovery, revealing a complex, asymmetric outflow that deviated significantly from a simple spherical explosion \citep{2026NatAs..10..271A}.
     \item {\bf Phase 2: Months Later}: As the ejecta expand and become transparent at longer wavelengths, radio imaging becomes paramount. Observations of V959 Mon with the VLA and EVN between 91 and 126 days post-eruption traced the evolution of colliding outflows \citep{2014Natur.514..339C}. These images clearly separated the thermal emission from the bulk ejecta and the non-thermal synchrotron emission from knots created by shocks.
     \item {\bf Phase 3: Years Later}: Long after the initial eruption has faded, high-resolution optical imaging can resolve the structure of the now-large remnant. Observations of V906 Car approximately two years after its eruption using the Zorro instrument on the Gemini telescope, along with HST observations of other novae, provide detailed maps of the cooling and expanding shell, revealing the fossil record of the initial asymmetric explosion \citep{2020NatAs...4..776A}.
\end{itemize}


\begin{table}[h]
    \caption{Summary of Nova Observing Strategy and Timeline}
    \label{tab:novae:obs}
    \begin{tabular}{|>{\raggedright\arraybackslash}p{3.5cm}|>{\raggedright\arraybackslash}p{11cm}|}
        \hline
       \textbf{Phase} & \textbf{Recommended Observations}  \\
    \hline
    
    Before T0 & Serendipitous capture of fireball phase from all-sky X-ray monitoring. \\
    \hline
    T0 + 0--1 hr & Optical imaging from the ground, whenever source is visible.\\
    \hline
    T0 + 1--24 hr & X-ray, UV, and optical observations should begin immediately (aiming for Swift exposures within a few hours, 
      and performed at least 2x within the first 24 hr), utilizing interferometers (e.g., CHARA). 
      Neutrino checks should be performed.\\
     \hline
    T0 + 1--7 d & Daily Coverage: multi-epoch coverage (at least daily observations) across the entire spectrum, from VHE to radio. \\
    \hline
    T0 + weeks-months & High Cadence for SSS: Monitor the ejecta and SSS emission. The SSS phase requires X-ray and UV photometry 
     observations every $\sim$6 hr due to its potential high variability during switch-on, then daily once the variability
     amplitude decreases. UV grism every $\sim$3 days. Longer, continuous observations (Chandra/XMM) are vital to  
     constrain QPOs once SSS has leveled out. Weekly X-ray spectroscopy (soft: LETG/RGS and hard: NuSTAR). 
     IR photometry and spectroscopy (R$>$2000) daily for at least 3 weeks in total, then weekly.\\
     \hline
    T0 + years & High-resolution imaging using HST and JWST to monitor the long-term evolution of the ejecta.\\
    \hline

\end{tabular}
\end{table}

\subsubsection{Lessons Learned from Past Events}

Recent campaigns demonstrate that rapid, coordinated, multiwavelength observations deliver transformative science for novae.

\begin{itemize}

    \item {\bf Multi-Observatory Success}: RS Ophiuchi is one of the ten known Galactic recurrent novae, erupting approximately every 15-20 years, and its most recent eruption in 2021 became one of the best-observed nova events thus far. The astronomical community rapidly targeted RS Oph with a vast array of instruments, including Swift, H.E.S.S., MAGIC, XMM-Newton, and Chandra. The resulting campaign produced an amazing dataset that included high-cadence light curves, detailed spectra, and the first-ever detection of a nova at TeV energies. Observations by Swift of the earlier (2006) eruption of RS Oph first revealed  high X-ray flux variability during the switch-on of the SSS emission, which was subsequently detected in a number of other novae, too; likewise the first SSS QPO was detected in Swift-XRT observations of RS Oph during the 2006 eruption.
    
    \item {\bf multiwavelength Coverage}: Overall, the strategy of combining simultaneous data from different observatories has been highly effective. The campaign on V906 Car, which integrated optical data from BRITE, gamma-ray data from Fermi-LAT, and hard X-ray data from NuSTAR, allowed for the construction of a comprehensive physical model of the shock emission
    
    \item {\bf Swift Response Capability}: Swift demonstrated the ability to respond extremely quickly, with observations starting as soon as T0+3.7 hours for V745 Sco without requiring the new “urgency zero” follow-ups, though this was a stand-out occurrence. In general, proactive planning will enable more of these extremely rapid responses.
    
    These same case studies also highlight clear operational gaps that could be addressed in the future:
    
    \item {\bf Solar Constraints}: Recent events highlight the difficulty posed by solar constraints. Both V462 Lup and V1935 Cen demonstrated interesting SSS X-ray activity, but they entered the solar observing constraint soon after \citep{2025ATel17436....1L, 2025ATel17459....1P}, limiting the duration of valuable data collection. Future triggers should prioritize novae with good Sun angles.
    
    \item {\bf Observation Strategy}: For variable phenomena like the SSS phase, obtaining longer continuous observations is necessary accurately to study features like QPOs, as short snapshots can lead to apparent period drift \citep{2025ApJ...995...30W}
    
    \item {\bf Observational Gaps}: There is a persistent need for more observations across the entire electromagnetic spectrum to build a larger sample of well-studied novae. For this reason, advocating for the indefinite operation of key high-energy facilities like Fermi, Swift, and NuSTAR is a high priority.

    \item {\bf Lack of Collaboration}: Greater collaboration with researchers in related fields, such as those studying supernovae and other shock-powered transients, could be useful. Sharing knowledge, observational techniques, and theoretical models across different classes of astrophysical transients would be mutually beneficial.

\end{itemize}

\newpage
\subsection{Supernovae}

Supernovae are among the most scientifically rich phenomena in time-domain and
multimessenger astrophysics, marking the violent endpoints of stellar evolution and
directly shaping the chemical, dynamical, and energetic evolution of galaxies. They
provide unique access to fundamental physical processes including core collapse, shock
breakout, neutrino production, particle acceleration, and the interaction of fast ejecta
with CSM. Because these processes unfold across an enormous
dynamic range in wavelength and time, from minutes to years and from high-energy X-rays
to long-wavelength radio, supernovae are intrinsically multi-messenger, multi-facility
targets whose full scientific return can be best realized through rapid, multiwavelength
observations.
 
Observationally, supernovae exhibit a rich sequence of characteristic phases. The earliest
moments, spanning minutes to hours after explosion, may include shock breakout and
cooling emission, luminous ultraviolet flashes, or prompt high-energy signatures. Over the
ensuing days, rapidly evolving optical and UV light curves and ``flash'' spectroscopy can
reveal the composition, density, and geometry of material lost by the progenitor in the
final stages before collapse. Weeks to months trace the transition to photospheric and
nebular phases, probing explosion energetics, nucleosynthesis, and ejecta structure, while
late-time radio and X-ray emission, often persisting for years, encodes the physics of
shocks, magnetic field amplification, and particle acceleration in the surrounding medium.
 
Despite major advances, large regions of supernova parameter space remain
under-explored. The earliest phases (the first few hours to days) are sparsely sampled at
most wavelengths, particularly in the ultraviolet, millimeter, and hard X-ray bands.
Intermediate timescales can be missed due to seasonal gaps or solar constraints, while
late-time infrared and sub-millimeter observations remain limited for all but the very
nearest events. Pre-defined community observing plans that prioritize rapid triggering,
coordinated multi-facility response, and open data sharing can address these gaps and help
ensure that the next rare, nearby, or unexpected supernova can be exploited to its fullest
scientific potential.
 
\subsubsection{High-Priority Subclasses}
 
Supernovae encompass a wide range of subclasses reflecting different progenitor systems
and explosion mechanisms, including thermonuclear Type~Ia events and the diverse family
of core-collapse supernovae (Types~II, Ib, Ic, and their variants). This plan prioritizes
two specific subclasses of events, each with potentially exceptional scientific return.
 
\begin{itemize}
    \item \textbf{FBOTs:} Typified by the event AT2018cow,
    these are extremely luminous and fast-evolving transients discovered in nearby,
    star-forming galaxies (typically $<$100~Mpc). The underlying physics appears to involve a
    unique combination of ingredients: a long-lived central engine (likely an accreting black
    hole), interaction with dense and confined circumstellar matter, and significant
    asymmetry. They are exceptionally rare, accounting for less than 0.1\% of the
    core-collapse supernova rate.
 
    \item \textbf{Very Nearby Core-Collapse Supernovae (CCSNe):} Exemplified by
    SN~2023ixf, this class includes any core-collapse supernova discovered within the local
    galactic neighborhood (e.g., $<$7~Mpc) or within our own Milky Way and its satellite
    Magellanic Clouds. These events are uniquely bright, allowing for exquisitely detailed
    study. They often exhibit signs of early interaction with CSM,
    providing a direct probe of the final stages of stellar mass loss before explosion.
\end{itemize}
 
\subsubsection{Triggering Criteria \& Expected Event Rates}
 
Based on the distinct natures of the target subclasses, the following triggers are proposed:
 
\begin{itemize}
    \item \textbf{FBOTs:} The discovery of an optical transient that exhibits a very
    fast-rising and highly luminous ultraviolet/optical light curve in a nearby galaxy. This
    initial alert must be promptly followed by confirmation of luminous emission in the X-ray
    and/or radio bands to distinguish it as a high-priority FBOT candidate.
 
    \item \textbf{Very Nearby CCSNe:} The discovery of a new optical transient by a survey
    within a very nearby galaxy (e.g., within ${\sim}7$~Mpc). This trigger also applies to any
    new supernova candidate discovered within the Milky Way or the Magellanic Clouds,
    which represent the most valuable, albeit challenging, opportunities.
\end{itemize}
 
Events meeting these criteria are considered rare opportunities, with an expected rate of
approximately once per decade.
 
\subsubsection{Discovery \& Alert Channels}
 
For the subclasses prioritized in this plan, discovery will be driven primarily by wide-field
optical surveys operating at high cadence, including Rubin-era facilities, precursor surveys
(e.g., ZTF, ATLAS, ASAS-SN), and targeted nearby-galaxy programs. In parallel, small
professional and amateur telescopes play an outsized role for the nearest events, often
providing the earliest detections, dense early-time photometry, and crucial constraints on
explosion time. For FBOTs, optical discovery alone is insufficient; rapid confirmation of
luminous X-ray and/or radio emission is required to elevate a candidate to high priority.
Future soft X-ray and UV survey missions are expected to further reduce discovery latency
and may, in some cases, provide the first alert.
 
Multimessenger channels represent an additional, high-impact discovery pathway. As
discussed under the Neutrino and Compact Binary Merger science cases, neutrino alerts
from facilities such as IceCube and Super-Kamiokande, and gravitational-wave alerts from
the LVK network, may precede or accompany electromagnetic discovery for Galactic or
extremely nearby events. These alerts can provide advance warning, ranging from minutes
to days before shock breakout, allowing optical, UV, and X-ray facilities to be
pre-positioned.
 
Once a credible candidate is identified, immediate dissemination through low-latency
community channels is essential. Initial alerts should be issued via established mechanisms
such as the TNS for formal reporting, complemented by
rapid-notice systems (e.g., GCN Circulars, Astronotes) and real-time community
coordination platforms (e.g., TDAMM Slack or equivalent).
 
\subsubsection{Observatory Roles and Capabilities}
 
The multiwavelength, multi-timescale nature of these events necessitates a carefully
orchestrated use of a wide array of ground- and space-based observatories. No single
facility can capture the full physical picture. The following tables outline the unique
capabilities and ideal roles for key facilities for the two high-priority SNe subclasses,
providing a blueprint for asset allocation once a candidate is identified.
  
\begin{table}[h]
    \centering
    \caption{Summary of Observatory Roles and Capabilities for Very Nearby Supernova Follow-up}
    \label{tab:novae:missions}
    \begin{tabular}{|>{\raggedright\arraybackslash}p{2cm}|>{\raggedright\arraybackslash}p{3cm}|>{\raggedright\arraybackslash}p{5.1cm}|>{\raggedright\arraybackslash}p{4.7cm}|}
\hline
\textbf{Observatory} & \textbf{Wavelength} & \textbf{Unique Capability} & \textbf{Ideal Use Case}\\
\hline
 
\textbf{HST}   & UV/Optical          & High-resolution imaging, UV spectroscopy                     & Late-time optical follow-up, NUV spectra \\[5pt]
\hline
\textbf{JWST}  & NIR/MIR             & High sensitivity in dusty environments, IR spectroscopy       & Probing dust-obscured SNe and environments \\[5pt]

\hline
\textbf{Swift} & UV/X-ray            & Rapid-response capability                                    & Early-time monitoring for UV/X-ray emission \\[5pt]

\hline
\textbf{Rubin} & Optical (Wide-field) & High-cadence, deep surveys                                  & Precursor searches and discovery \\[5pt]

\hline
\textbf{VLA}   & Radio (cm-band)     & Tracing ejecta and shock evolution at late times             & Jet and ejecta studies, late-time radio afterglow \\[5pt]

\hline
\textbf{ALMA}  & Radio (mm/sub-mm)   & Probing high radio frequencies, CSM and shock properties     & Early-time radio follow-up \\

\hline
\end{tabular}
\end{table}
 
\begin{table}[h]
    \centering
    \caption{Summary of Observatory Roles and Capabilities for Nearby FBOT Follow-up}
    \label{tab:novae:missions}
    \begin{tabular}{|>{\raggedright\arraybackslash}p{2cm}|>{\raggedright\arraybackslash}p{3cm}|>{\raggedright\arraybackslash}p{5.1cm}|>{\raggedright\arraybackslash}p{4.7cm}|}
\hline
\textbf{Observatory} & \textbf{Wavelength} & \textbf{Unique Capability} & \textbf{Ideal Use Case}\\
\hline
 
\textbf{XMM}        & X-rays      & High-throughput soft X-ray spectroscopy               & Multiple epochs of X-ray spectral observations \\[5pt]

\hline
\textbf{NuSTAR}     & X-rays      & Hard X-ray spectroscopy                               & Characterizing transient hard X-ray components \\[5pt]

\hline
\textbf{INTEGRAL}   & X-rays      & High-energy X-ray coverage                            & Constraining the highest-energy parts of the SED \\[5pt]

\hline
\textbf{Swift}      & UV/X-ray    & Rapid and high-cadence monitoring                     & Early discovery and high-cadence light curve monitoring \\[5pt]

\hline
\textbf{HST}        & UV/Optical  & Long-term UV monitoring, UV spectroscopy              & Tracking accretion disk evolution, composition analysis \\[5pt]

\hline
\textbf{Gemini, Keck}& Optical     & Sensitive optical spectroscopy                        & Multi-epoch spectroscopy of ejecta/CSM out to late times \\[5pt]

\hline
\textbf{JWST}       & NIR/MIR     & High-resolution imaging, IR sensitivity               & Searching for underlying clusters and dust formation \\[5pt]

\hline
\textbf{ALMA}       & Sub-mm      & High-frequency radio (up to ${\sim}1$~THz) coverage  & Snapshots of CSM/ejecta properties, shock physics \\[5pt]

\hline
\textbf{SMA}        & Sub-mm      & Detailed high-frequency radio light curves, dual-band capability & High-cadence monitoring of flux and spectral index \\[5pt]

\hline
\textbf{VLA}        & Radio (cm-band) & Late-time radio sensitivity                       & Tracing the late-time radio afterglow and evolution \\

\hline
\end{tabular}
\end{table}
 
\subsubsection{Observing Strategy and Timeline}
 
For fast-evolving transients, the timing and cadence of observations are as critical as the
choice of instrument. The first priority is to secure prompt spectroscopic observations to
confirm the nature of the transient and its redshift. This foundational data point informs all
subsequent follow-up decisions. For FBOT candidates, it is also critical to rapidly confirm
the presence of luminous X-ray and/or radio emission. This step is essential to distinguish
these rare events from other classes of fast-evolving transients.
 
The starting condition, $T_0$, is defined as the moment a source is identified as a strong
candidate by a survey, with multiwavelength confirmation of its key characteristics (fast
optical rise, luminous X-ray/radio emission) expected within 24 hours of first light.

\subsubsection*{T0 + 1--24 hours: Immediate Response}

\begin{itemize}
    \setlength{\itemsep}{1pt}
    \setlength{\parskip}{1pt}
    \item Obtain a broad-band X-ray Spectral Energy Distribution (SED) covering soft to
    hard X-rays (e.g., using \textit{Swift}, XMM, and NuSTAR).
    \item Secure an early UV spectrum with HST to capture the initial composition and
    ionization state of the ejecta.
    \item Conduct initial SMA and VLA observations to establish the radio and millimeter
    brightness and provide a baseline for future monitoring.
 
    \item Small optical facilities can provide high-cadence photometry.
\end{itemize}

\subsubsection*{T0 + 1--7 days: Intensive Monitoring}

\begin{itemize}
    \setlength{\itemsep}{1pt}
    \setlength{\parskip}{1pt}
    \item Obtain a second epoch of X-ray SED, HST UV spectroscopy, and VLA
    observations.
    \item Secure a multi-band ALMA SED to characterize the early peak of the radio
    emission.
    \item Continue intensive, daily monitoring with flexible facilities like \textit{Swift}
    (X-ray) and SMA (sub-mm) to track rapid variability.
    \item Optical facilities should continue monitoring photometric and spectroscopic
    evolution.
\end{itemize}

\subsubsection*{T0 + weeks--months: Logarithmic Follow-up}

Continue observations with a logarithmic cadence. A potential campaign could include:
 
\begin{itemize}
    \setlength{\itemsep}{1pt}
    \setlength{\parskip}{1pt}
    \item ${\sim}5$ epochs of X-ray SEDs.
    \item ${\sim}3$ epochs of HST UV observations.
    \item ${\sim}5$ epochs of ALMA multi-band observations.
    \item ${\sim}10$ epochs of VLA observations.
    \item Multiple epochs of Gemini/Keck optical spectroscopy to track the evolution of
    spectral features.
\end{itemize}

\subsubsection*{T0 + months to years: Late-Time Monitoring}
 
\begin{itemize}
    \setlength{\itemsep}{1pt}
    \setlength{\parskip}{1pt}
    \item Monitor the source on a logarithmic timescale with approximately 5--6 epochs.
    \item Obtain deep late-time Gemini/Keck spectroscopy to search for nebular features.
    \item Use HST (UV) and JWST (NIR) imaging to monitor the fading accretion disk,
    search for an underlying stellar cluster, and detect any dust formation.
    \item Continue VLA SED monitoring to track the final phases of the radio afterglow.
\end{itemize}
 
\subsubsection{Lessons Learned from Past Events}
 
\smallskip
\subsubsubsection*{\textbf{Case Study: SN~2023ixf (Nearby Core-Collapse Supernova)}}

The response to SN~2023ixf demonstrated several strengths of the existing community
ecosystem:
 
\begin{itemize}
    \item \textbf{Small Telescope Contribution:} The event benefited from extensive
    coverage by small and amateur telescopes, which provided some of the earliest detections
    and crucial constraints on the time of shock breakout.
 
    \item \textbf{Rapid Reporting:} The first professional spectrum, taken just 1.1~days
    after first light, was immediately reported to the TNS, enabling
    prompt classification.
 
    \item \textbf{\textit{Swift}'s Agility:} The \textit{Swift} Observatory was triggered
    immediately, securing its place as a critical first-response asset for UV and X-ray
    observations.
 
    \item \textbf{Broad Observatory Deployment:} The campaign successfully utilized a wide
    range of major X-ray (Chandra, XMM, NuSTAR) and radio (VLA, GMRT, NOEMA)
    observatories, providing rich multiwavelength datasets.
\end{itemize}
 
Despite these successes, the campaign also highlighted significant areas for improvement:
 
\begin{itemize}
    \item \textbf{Responsiveness:} A ${\sim}1$-day delay between the initial amateur
    discovery and the first formal report led to a corresponding delay in the first professional
    spectrum. In the LSST era, automated systems should mitigate this lag.
 
    \item \textbf{Wavelength Coverage:} There were major missed opportunities in spectral
    coverage. No Far-UV spectrum was obtained with HST, and the most critical early
    Near-UV information (at $<$3.6~days) was missed. Furthermore, the first JWST spectrum
    was not obtained until 33~days post-explosion, and spectral coverage in the crucial
    mm/sub-mm bands was lacking.
 
    \item \textbf{Coordination and Data Latency:} The campaign suffered from a need for
    better-coordinated \textit{simultaneous} observations across X-ray and radio facilities.
    The reliance on numerous, independent DDT proposals
    for a ``once in a decade event'' was inefficient and argues for a more open,
    community-driven effort with rapid, open data access from the outset.
\end{itemize}
 
\subsubsubsection*{\textbf{Case Study: AT2018cow (Fast Blue Optical Transient)}}
 
The AT2018cow FBOT campaign was a landmark in transient follow-up, showcasing the
community at its best in several respects:
 
\begin{itemize}
    \item \textbf{Rapid Discovery and Alerting:} The surprising nature of the event was
    identified and communicated very quickly, enabling broad, worldwide follow-up within
    days.
 
    \item \textbf{Community Engagement:} The event sparked extensive and prompt
    engagement from both observers and theorists, accelerating the evolution of our
    understanding of this new class of objects.
 
    \item \textbf{Prompt Data Publication:} Many groups published their initial datasets
    within months, making extensive multiwavelength data available to the public early on
    and enabling more detailed theoretical investigations.
 
    \item \textbf{Flexible Facilities:} The campaign underscored the critical role of highly
    flexible observatories like \textit{Swift} and the SMA, which can get on target in minutes
    to hours and execute high-cadence monitoring programs.
\end{itemize}
 
The unprecedented nature of AT2018cow also stressed the limits of our coordination
systems:
 
\begin{itemize}
    \item \textbf{Inter-Facility Coordination:} Achieving truly simultaneous observations
    was extremely difficult due to inconsistent ToO and DDT
    policies across different observatories. This hindered efforts to correlate variability
    across different wavelengths.
 
    \item \textbf{DDT Responsiveness:} For a fast-evolving event, some observatory
    processes are simply too slow. The 1--2~week response time for an ALMA DDT, for
    example, is insufficient to capture the earliest, most critical phases of evolution.
 
    \item \textbf{multiwavelength Infrastructure:} Community infrastructure for tracking
    and communicating transient evolution is often heavily focused on the optical domain.
    This makes it challenging to track multiwavelength behavior cohesively and identify
    cross-band correlations.
 
    \item \textbf{High-Energy Coverage:} Confirming or ruling out associated high-energy
    gamma-ray emission was not trivial. The fragmented nature of gamma-ray facility
    coverage required checking multiple instruments individually, a process that must be
    streamlined for rapid response.
\end{itemize}

\newpage
\subsection{Magnetars}

Magnetars are among the most extreme objects in the universe and serve as unparalleled laboratories for fundamental physics. These ultra-magnetized neutron stars exhibit a wide range of transient phenomena—from catastrophic giant flares to subtle state changes—that probe regimes of gravity, density, and magnetic field strength far beyond terrestrial experiments. Because many of these phenomena are rare, short-lived, and multi-channel in nature, a coordinated, multiwavelength and multi-messenger observing strategy is essential to fully exploit their scientific potential.

The study of magnetars addresses several foundational questions in astrophysics and physics. Their super-critical magnetic fields, orders of magnitude stronger than any achievable on Earth, provide a unique opportunity to test predictions of quantum electrodynamics, including vacuum birefringence and magnetic photon splitting. Observations of magnetar flares, glitches, and oscillations also probe matter at supra-nuclear densities, constraining the neutron-star equation of state and shedding light on exotic interior physics such as superfluidity and superconductivity. The enormous energy release and internal rearrangements associated with magnetar activity make these systems promising targets for gravitational-wave searches, offering a potential new probe of neutron-star interiors. In addition, powerful flares may briefly create conditions suitable for r-process nucleosynthesis, providing a possible, though unconfirmed, contribution to cosmic chemical enrichment. Finally, magnetars are now established as engines for at least some fast radio bursts, linking their activity to cosmological probes of the intergalactic medium and the origin of FRBs.

The most dramatic manifestations of magnetar activity are magnetar giant flares, which reach peak luminosities of $\sim 10^{45}\text{--}10^{47}\ \mathrm{erg\ s^{-1}}$. These events are characterized by an intense, millisecond-duration initial spike followed, for nearby events, by a pulsating X-ray tail lasting hundreds of seconds. Nuclear gamma-ray emission may persist for thousands of seconds after the flare. Theoretical models also predict a short-lived optical or infrared transient powered by radioactive decay from flare-driven nucleosynthesis—a so-called “nova brevis”—which could be detectable for both Galactic and nearby extragalactic events. Spectrally, giant flares are dominated by non-thermal, Comptonized emission produced in a hot, highly magnetized plasma.

Beyond giant flares, magnetars display a rich variety of state changes that offer insight into their magnetospheres and internal dynamics. These include episodes of intense bursting activity accompanied by enhanced persistent X-ray emission, the transient activation of pulsed radio emission lasting weeks to months, and the joint occurrence of X-ray bursts and fast radio bursts, as demonstrated by Galactic magnetars. Magnetars also exhibit extreme rotational behavior, including glitches, anti-glitches, and other timing anomalies that often coincide with changes in radiative output, providing additional diagnostics of internal and external torque mechanisms.

Despite substantial progress, key regions of magnetar parameter space remain largely unexplored. No confirmed detection of the predicted “nova brevis” emission has yet been made, and magnetar activity has not been conclusively associated with gravitational waves or neutrinos. While soft X-ray polarization measurements have begun to test QED effects in magnetar magnetospheres, extending polarization studies into the hard X-ray regime during flares remains a critical frontier. Addressing these gaps will require deliberate, rapid, and well-coordinated observing campaigns that combine high-energy coverage with multi-messenger capabilities, ensuring that future magnetar transients are captured with the breadth and depth needed to fully realize their scientific promise.

\subsubsection{High-Priority Subclasses}

Based on the key unanswered questions in the field, the two highest-priority subclasses, magnetar giant flares and magnetar undergoing active state changes, offer the greatest potential for transformative scientific breakthroughs.

\begin{itemize}
    \item \textbf{Magnetar Giant Flares}: As the most energetic explosions from magnetars, MGFs offer an excellent opportunity to probe QED physics in super-critical fields beyond 50 keV, search for gravitational waves from stellar rearrangements or outflows, and detect the signatures of r-process nucleosynthesis, and quasi-periodic oscillations to study magnetar crusts and interiors.
    \item \textbf{Significant State Changes}: This category includes events like the joint detection of FRBs with X-ray bursts and the occurrence of major timing anomalies such as glitches and anti-glitches. These events are critical for understanding the magnetar-FRB connection and the complex interplay between the magnetar's interior, crust, and magnetosphere.
\end{itemize}

\subsubsection{Triggering Criteria \& Expected Event Rates}

The triggering criteria for magnetar follow-up campaigns depend strongly on the scientific value of different classes of magnetar activity, ranging from rare giant flares to more common but highly informative state changes.

\begin{itemize}
    \item \textbf{Magnetar Giant Flares}: The MGFs volumetric rate is estimated to be $>$ 3.8~$\times$~10$^{5}$ Gpc$^{-3}$ yr$^{-1}$, with a local rate ($\sim$5 Mpc) or approximately 1 event per few years.  They are expected to occur within the Milky Way at a rate of 1 event per decade or two. The extragalactic discovery rate may be substantially higher with future MeV observatories. 
    \item \textbf{Significant State Changes}: Other State Changes (a burst storm, radio emission, large timing anomalies and/or glitches) events occur on a timescale of ``once every few years" for a given magnetar.  Joint FRB-burst detections from a particularly active source like SGR 1935 can occur as frequently as once per year. Identifying when these state changes occur require continuous monitoring from instruments like NICER, Nustar, and Swift-XRT.
\end{itemize}

\begin{table}[h]
\centering
\caption{Magnetar subclasses}
\label{tab:magnetar_subclasses}
\begin{tabular}{|>{\raggedright\arraybackslash}p{3.2cm}|>{\raggedright\arraybackslash}p{4.5cm}|>{\raggedright\arraybackslash}p{3.8cm}|>{\raggedright\arraybackslash}p{3.5cm}|}
\hline

\textbf{Subclass} & 
\textbf{Scientific Priority / Rationale} &
\textbf{Defining Characteristics} &
\textbf{Expected Rate} \\
\hline

\textbf{Magnetar Giant Flares} &
Probe QED, search for GW from stellar rearrangements, and detect the signatures of r-process nucleosynthesis. &
Distance $< 5$ Mpc \newline
Gamma-ray Flux $>$ ... &
$> 0.25$ per year extragalactic, \newline
$0.5$ per decade galactic \\
\hline

\textbf{State Change} &
Insight into magnetar magnetospheres and internal dynamics &
Outburst activity, burst storms, radio emission, glitches &
$\sim 0.25$ per year \\
\hline

\end{tabular}
\end{table}

\subsubsection{Discovery \& Alert Channels}

The primary discovery method for magnetar events, such as MGFs and state changes, is through high-energy monitoring by space missions. Instruments like Swift-BAT, Fermi-GBM, INTEGRAL, and other IPN nodes are essential for detecting hard X-ray and gamma-ray bursts and disentangling MGFs from cosmological short GRBs. Recently, Einstein Probe is providing wide-field monitoring for discovery and trigger alerts in the X-ray spectrum, but magnetar bursts are generally too hard for its intruments. For events coincident with FRBs or radio emission changes, wide-field radio telescopes, such as CHIME, DSA, CHORD, ASKAP, FAST, BURSTT and STARE, are likely to be used discover associations with magnetars.

For space based missions, the preferred and most established channel for these alerts is the GCN notices and circulars.  Challenges exist in the automated identification of MGF candidates in real-time, necessitating human in the loop identification of candidate events. Likewise, automated infrastructure to cross-correlate triggers from different instruments (e.g., a radio telescope and a gamma-ray satellite) in real-time is currently lacking and currently relies on community associations via messages disseminated through GCN circulars and the TDAMM Slack.

\subsubsection{Observatory Roles and Capabilities}

A comprehensive understanding of a magnetar event cannot be achieved with a single instrument. Success relies on a multi-observatory, multi-messenger approach that leverages the unique strengths of a coordinated network of facilities. Past discoveries have demonstrated the power of this paradigm, with crucial contributions from a wide range of instruments, including INTEGRAL, Konus-Wind, Fermi-GBM, CHIME, FAST, Insight-HXMT, NICER, and NuSTAR. Current observatories like IXPE and XRISM may provide new insights into magnetar physics. IXPE's X-ray polarimetry capabilities provide a crucial probe of quantum electrodynamics in the magnetosphere, while XRISM has the potential to resolve the spectral lines produced by scattering effects.

\subsubsection{Observing Strategy and Timeline}

The overall observing strategy for Magnetars is designed to capture the critical phases of the event’s evolution, from the initial moments of the explosion to its long-term after-effects. This strategy is informed by an analysis of what has been learned from applying similar—though often less advanced—approaches to major events in the past. The following timeline outlines the recommended sequence of observations following an initial trigger (T$_0$).

\bigskip
\textbf{Magnetar Giant Flares}:

\begin{itemize}
    \item \textbf{T$_0$ + 1 minute}: Initiate rapid hard X-ray observations. The primary goal is to capture the oscillatory tail emission, which is a key signature for confirming an event as an MGF, especially for extragalactic candidates.
    
    \item \textbf{T$_0$ + 15 minutes - hours}: Deploy UV and optical telescopes for spectroscopy and imaging. This window is optimized to search for the faint, short-lived "nova brevis" signal expected from nucleosynthesis.
    
    \item \textbf{T$_0$ + 0.5 days - weeks}: Begin a campaign of multi-epoch coverage across X-ray, radio, and gamma-ray bands. These observations will monitor the magnetar's long-term temporal and spectral evolution and search for the development of any associated nebula or afterglow emission.
\end{itemize}

\textbf{State Changes}:

\begin{itemize}
    \item \textbf{T$_0$ + 0.5 days - weeks}: Initiate multi-epoch coverage in X-ray and radio bands. The objective is to monitor temporal changes in the magnetar's emission, search for associated timing anomalies like glitches, and actively seek related phenomena such as FRBs and the emergence of pulsed radio episodes.
\end{itemize}

\subsubsection{Lessons Learned from Past Events}

\textbf{What worked well}: Several recent magnetar events demonstrate the scientific value of rapid identification, coordinated multiwavelength follow-up, and sustained monitoring campaigns. These case studies highlight how existing infrastructure and community coordination have already enabled major advances in understanding magnetar giant flares, fast radio bursts, and state-change phenomena. These include:

\begin{itemize}
    \item \textbf{2004 Galactic MGF from SGR 1806-20}: Reports of the source's unusually active state prior to the main event led to dedicated monitoring campaigns. This readiness resulted in valuable observations before, during, and after the giant flare, including contemporaneous radio observations that captured the afterglow.
    
    \item \textbf{2023 Extragalactic MGF from M82}: The event was quickly identified as a likely MGF from a nearby star-forming galaxy. This rapid classification required an understanding of how the characteristics of a giant flare differ from a short gamma-ray burst and was aided by INTEGRAL's ability to provide arcminute-scale localizations, which immediately pointed to a plausible local host.
    
    \item \textbf{SGR 1935 X-ray Bursts + FRBs}: The association between the X-ray bursts from the magnetar and the radio emission from the FRBs was identified early. Subsequent high-cadence follow-up observations were successful in connecting the source's glitches and spectral evolution directly to its FRB emission.
\end{itemize}

 \textbf{What could be improved}: At the same time, these events also exposed important gaps in the current time-domain and multimessenger ecosystem. In particular, limitations in automated event identification, low-latency alert generation, and comprehensive multiwavelength follow-up capabilities continue to constrain the scientific return from rare and rapidly evolving magnetar phenomena.

\begin{itemize}
    \item \textbf{2004 Galactic MGF from SGR 1806-20}: There was a lack of rapid follow-up observations in key UV and optical wavelengths, resulting in a missed opportunity to search for the predicted nova brevis, underscoring the critical need for the automated alert systems and rapid follow-up timelines detailed in this plan. 
    
    \item \textbf{2023 Extragalactic MGF from M82}: This event highlighted the urgent need for more automated systems capable of identifying likely MGF events in real-time from gamma-ray data, which would accelerate the follow-up process.
    
    \item \textbf{SGR 1935 X-ray Bursts + FRBs}: As with the M82 MGF, this series of events underscored the general need for automated identification to improve response times for complex, joint-detection events.

\end{itemize}

The lessons from these events reveal two overarching limitations. A lack of investment in automated infrastructure for discovering, classifying, and alerting on transient events, and a lack of systematic tracking of the current activity states of known magnetars. Addressing these systemic gaps through dedicated investment and coordination is the most critical action we can take to realize the full scientific potential of magnetar astronomy.

\newpage
\subsection{Compact Binary Mergers}

The advanced sensitivity of the LIGO–Virgo–KAGRA has established compact binary mergers as a common feature of the local universe \citep{LIGOScientific:2017ync}, but electromagnetic counterparts to these events remain exceptionally rare \citep{Niu:2025nha}. The joint gravitational-wave and electromagnetic detection of GW170817 demonstrated the extraordinary scientific return of a coordinated, rapid, multiwavelength response, fundamentally advancing our understanding of neutron stars, relativistic jets, and heavy-element production. That single event underscores the need for a prepared, community-wide observing strategy to fully exploit future detections.

Compact binary mergers provide a unique laboratory for physics under extreme conditions \citep{Baiotti:2016qnr}. Kilonova emission from GW170817 confirmed that neutron star mergers are a major site of r-process nucleosynthesis, producing heavy elements such as gold and platinum \citep{Kasen:2017sxr}. By combining gravitational-wave distances with host-galaxy redshifts, these systems act as ``standard sirens,'' offering an independent probe of the Hubble constant \citep{Hotokezaka:2018dfi}. The gravitational-wave signal, together with the nature of the electromagnetic counterpart, constrains the neutron-star equation of state by revealing whether the merger forms a prompt black hole or a longer-lived remnant \citep{Radice:2020ddv}. The associated short gamma-ray burst and its broadband afterglow established the direct link between neutron star mergers and short GRBs, while providing critical insight into jet structure and dynamics \citep{Ghirlanda:2018uyx, Mooley:2018qfh}.

The electromagnetic signatures of a merger evolve rapidly across the spectrum. A short gamma-ray burst may occur within seconds for favorably aligned systems, followed by a thermal kilonova lasting days to weeks as radioactive ejecta cool and expand \citep{Fernandez:2015use, Metzger:2019zeh}. On longer timescales, interaction between fast ejecta or a relativistic jet and the surrounding medium produces X-ray and radio afterglow emission that can peak months after the merger and persist for years \citep{Nakar:2011cw}. Capturing this full sequence requires fast response and sustained follow-up across multiple wavelengths.

The observable outcome depends strongly on the merger type. Binary neutron star mergers can produce a range of remnants, from prompt collapse to long-lived neutron stars, with significant impact on ejecta and brightness \citep{Shibata:2019wef}. Neutron star–black hole mergers may yield luminous counterparts if the neutron star is tidally disrupted, but can be electromagnetically faint if it is swallowed whole. Binary black hole mergers are unlikely to produce detectable emission except in rare, uncertain environments.

Despite recent progress, most of the merger parameter space remains unexplored. No confirmed electromagnetic counterpart to a neutron star–black hole merger has yet been observed, the predicted early ultraviolet precursor remains undetected, and GW170817 samples only a narrow range of binary properties. Addressing these gaps requires prioritizing mergers involving at least one neutron star, where the potential for transformative multi-messenger discovery is greatest.

\subsubsection{High-Priority Subclasses}

While any CBM detection is of interest, a full-scale, pre-planned community follow-up campaign should be reserved for events with the highest probability of yielding transformative scientific results. Therefore the following events are currently prioritized:

\begin{itemize}
    \item \textbf{BNS and Merger}: The prioritization of BNS mergers is driven by the confidence in the existence of detectable EM counterparts from these systems, a confidence built upon the bedrock of GW170817's unambiguous detection. In contrast, while theoretical possibilities for EM-bright binary black hole mergers exist, they remain speculative.

\item \textbf{NSBH Merger}: Although no confirmed electromagnetic counterpart has been detected to date from a NSBH merger, they are nonetheless prioritized because they probe a largely unexplored regime of multi-messenger astrophysics. Even well-constrained non-detections provide powerful limits on merger geometry and neutron-star compactness

\end{itemize}

\subsubsection{Triggering Criteria \& Expected Event Rates}

The expected rates for high-priority mergers are low and uncertain. Based on the latest from \citep{abbott2023} the predicted merger rates in the local universe for BNS and NSBH mergers is 10 to 1700 Gpc$^{-3}$ yr$^{-1}$ and 7.8 to 140 Gpc$^{-3}$ yr$^{-1}$, respectively. Despite these broad ranges, the actual rate of observable events with detectable EM counterparts is likely far less than one per year at the current LVK sensitivity, making each detection of immense importance to the community. However, additional factors must be evaluated to determine the scale of the observational follow-up campaign. These primarily include the on-sky localization and the estimated distance, although the latter is not always immediately available. A well-localized event significantly reduces the search area, making counterpart identification faster and more efficient leading such events to be prioritized. 

\begin{itemize}
    \item \textbf{BNS Merger}: GW detections with localizations within tens to a few hundred square degrees and an inferred distance within the local universe ($\lesssim$200--300 Mpc) are strongly favored. Additional weight is given to events with favorable observing geometry (good Sun angle) and any coincident gamma-ray signal, which further strengthens the case for an EM-bright counterpart.

\item \textbf{NSBH Merger}: Similar considerations apply as in the BNS case, but for NSBH mergers the mass ratio between the black hole and the neutron star becomes a critical factor. Systems with high mass ratios are more likely to result in the neutron star being swallowed whole, with little to no mass ejection and correspondingly weak or absent electromagnetic emission. Although, the detection of a coincident gamma-ray signal would constitute a compelling trigger, independent of the inferred mass ratio.
\end{itemize}

\subsubsection{Discovery \& Alert Channels}

CBMs are discovered primarily through their GW emission, detected by the global network of ground-based interferometers operated by LIGO, Virgo, and KAGRA. When a candidate event is identified, rapid parameter estimation pipelines determine whether the system likely contains at least one neutron star and produce a probabilistic sky localization. In a subset of favorable cases, high-energy EM emission detected by all-sky or wide-field monitors provides an additional or confirming discovery channel. 

Once a candidate CBM is identified, initial GW notices, including classification probabilities and sky maps, are distributed via GCN and machine-readable alert streams (e.g., VOEvent and AMON). As candidate counterparts are identified, their positions, photometry, and classifications are shared in near real time through GCN Circulars, the Transient Name Server, and coordination tools such as Treasure Map and similar platforms. In addition, new community efforts like TROVE are being developed to provide real-time, automated scoring of published candidates, helping to accelerate the vetting process and focus follow-up resources more efficiently.

While GW signals are the primary trigger, the potential for non-GW triggers also exists. For instance, a few possible kilonovae have been identified in conjunction with gamma-ray bursts, and next-generation wide-field surveys like the Vera C. Rubin Observatory's LSST may discover them independently.

\subsubsection{Observatory Roles and Capabilities}

Table~\ref{tab:cbm_obs} defines the key observatories and their strategic contributions to a follow-up campaign.

\begin{table}
    \centering
    \caption{Summary of Observatory Roles and Capabilities for Compact Binary Merger Follow-up}
    \label{tab:cbm_obs}
    \begin{tabular}{|>{\raggedright\arraybackslash}p{3cm}|>{\raggedright\arraybackslash}p{4cm}|>{\raggedright\arraybackslash}p{4cm}|>{\raggedright\arraybackslash}p{5cm}|}
        \hline
        \textbf{Observatory} & \textbf{Wavelength} & \textbf{Unique Capability} &
        \textbf{Ideal Use Case}\\
        \hline
\textbf{HST}&
UV/Optical&
High-resolution imaging&
Late-time optical follow-up\\
\hline
\textbf{JWST}&
NIR/MIR&
Unprecedented NIR/MIR sensitivity for spectroscopy of r-process features and deep late-time imaging.&
Bright, nearby events (spectroscopy) or distant events with a clear counterpart (late-time imaging)\\
\hline
\textbf{Swift}&
UV/X-ray&
Rapid follow-up&
Early-time monitoring\\
\hline
\textbf{Rubin}&
Optical (Wide-field)&
Unmatched combination of sensitivity, wide field-of-view, and rapid cadence for efficient counterpart searches in large localization volumes.&
GW-independent KN identification; deep imaging of LVK localization regions\\
\hline
Large ground based optical telescopes (\textbf{Keck}, \textbf{Gemini}, etc)&
Optical&
KN spectroscopy, deep optical imaging in multiple bands&
Events with a clear counterpart\\
\hline
\textbf{VLA, MeerKAT, ATCA, GMRT}&
Radio&
Jet and ejecta studies&
Late-time synchrotron afterglow\\
\hline
\textbf{ASKAP}&
Radio&
Wide-field radio imaging&
Radio counterpart identification; monitoring of late-time synchrotron afterglow\\
\hline
\textbf{Chandra}&
X-rays&
High spatial resolution X-ray imaging, essential for isolating faint afterglows from host galaxies&
Late-time synchrotron afterglow\\
\hline
    \end{tabular}
\end{table}

\subsubsection{Observing Strategy and Timeline}

To maximize the scientific return from a high-priority merger, this plan recommends a multi-stage observing strategy. This phased strategy is designed to directly address the key challenges identified from past events.

\begin{table}[h]
    \caption{Summary of Compact Binary Merger Observing Strategy and Timeline}
    \label{tab:cbm:obs}
    \begin{tabular}{|>{\raggedright\arraybackslash}p{3.5cm}|>{\raggedright\arraybackslash}p{11cm}|}

    \hline
   \textbf{Phase} & \textbf{Recommended Observations}  \\
    \hline

    T0 + 0--1 hr &  Immediately following the GW trigger, automated rapid-response facilities (e.g., Fermi, Swift) should search for prompt high-energy emission (gamma-rays, X-rays) from a potential relativistic jet.
    \\
    \hline
    T0 + 1--24 hr (Search Phase)& Wide-field optical and NIR imaging campaigns must commence to systematically search the GW localization region for new transient sources. Once candidates are identified, rapid-response spectroscopy is required to vet them and distinguish a true kilonova from contaminants.\\
     \hline
     T0 + 1--24 hr (Post-Identification) & Once a counterpart is unambiguously identified, an intensive campaign of UV, optical, and NIR spectroscopy and imaging must begin to characterize its initial properties before it evolves significantly.\\
     \hline
    T0 + 1--14 d (Peak Characterization) &  This is the critical period for understanding the kilonova. High-cadence, multi-epoch observations across the full electromagnetic spectrum, from UV and optical to IR, radio, and X-ray, are essential to capture the peak of the light curve and its color evolution. X-ray monitoring during this time is also essentially for observing the peak emission from an off-axis relativistic jet.
    \\
    \hline
    T0 + weeks-months (Afterglow Monitoring) & As the kilonova fades, sustained monitoring in the radio, X-ray, and optical bands is necessary to track the emergence and evolution of the synchrotron afterglow, which reveals the properties of the fastest ejecta and the surrounding environment.
    \\
    \hline

\end{tabular}
\end{table}

\subsubsection{Lessons Learned from Past Events}

The current state of our observing strategy is heavily informed by the singular success of GW170817 and the significant challenges encountered in the follow-up of subsequent, less favorable events. GW170817 represented a ``best-case scenario'': it was nearby (40 Mpc), exceptionally well-localized (28 deg$^2$), and visible in the night sky, triggering an unprecedented and highly successful global response. The discovery timeline was a model of multi-messenger coordination: a GW trigger was followed almost immediately by a Fermi gamma-ray burst detection ($\sim$1.7 seconds later). The optical counterpart was discovered by multiple teams within $\sim$11 hours, which in turn triggered intensive follow-up campaigns that led to the subsequent detection of the X-ray ($\sim$10 days) and radio ($\sim$15 days) afterglows.

The comprehensive follow-up of GW170817 was enabled by several key elements of community coordination that should be maintained and strengthened. These include:

\begin{itemize}
    
    \item \textbf{Rapid Dissemination}: The use of GCN email alerts proved to be a highly effective mechanism for the immediate circulation of triggers and counterpart coordinates, allowing the global community to react in near-real time.
    
    \item \textbf{Community Collaboration}: Observers effectively utilized public platforms like the TNS to publish candidate counterparts. This open sharing of early results enabled rapid, independent confirmation and vetting by multiple teams.

\end{itemize}

\newpage

\subsection{Neutrinos}

High-energy neutrinos provide a uniquely direct view of the most extreme particle accelerators in the universe. Because they interact only weakly and travel undeflected by magnetic fields, neutrinos carry unambiguous information about where and how hadrons are accelerated, escaping even from dense or highly obscured environments that are opaque to photons. Over the past decade, observations by IceCube have established a diffuse astrophysical neutrino flux spanning TeV–PeV energies, confirming the existence of powerful extragalactic and Galactic sources. More recently, accumulating evidence for spatial and temporal clustering has begun to reveal individual source classes as well as definitive evidence for emission from the Galactic plane. 

Neutrino observations directly probe hadronic processes, proton–proton and proton–photon interactions, that cannot be uniquely identified with electromagnetic data alone. As such, neutrinos are essential for resolving the long-standing origin of cosmic rays and for distinguishing leptonic from hadronic emission mechanisms in astrophysical sources. The discovery of a steady neutrino signal from the nearby Seyfert galaxy NGC 1068 marked a watershed moment, demonstrating that non-blazar active galactic nuclei can efficiently produce high-energy neutrinos in heavily obscured environments. Together with earlier associations between neutrinos and flaring blazars such as TXS 0506+056, these results indicate that neutrino astronomy is transitioning from discovery to population studies. 

Astrophysical neutrino sources span a broad range of temporal and spectral behaviors. Emission may be steady, as suggested for some obscured AGN, or transient and episodic, as seen in blazar flares and potentially in supernovae interacting with dense circumstellar material. Neutrino energies typically range from tens of GeV to several PeV, with spectral shapes that vary between soft, low-energy–dominated populations and harder spectra characterized by a small number of very high-energy events. Importantly, neutrino production is often accompanied by gamma-ray emission at the source, but internal absorption and extragalactic propagation effects can suppress or reprocess the gamma-ray signal, leading to “gamma-ray–quiet” neutrino sources. This decoupling underscores the need for broadband, multiwavelength follow-up rather than reliance on gamma rays alone.

Current observations and theoretical work point to several subclasses of neutrino sources. These include obscured active galactic nuclei, particularly X-ray–bright Seyfert galaxies whose dense cores can efficiently convert cosmic-ray power into neutrinos; blazars, which may produce neutrinos during rare flaring states; core-collapse supernovae with strong interaction between the shock and circumstellar material; Galactic sources such as supernova remnants, pulsar wind nebulae, and X-ray binaries; and diffuse emission from the Galactic plane. Each subclass probes different physical conditions and timescales, implying that no single search strategy is sufficient and that tailored, source-class–specific observing plans are required. 

\subsubsection{High-Priority Subclasses}

Given the rarity of neutrino detections and the substantial observational resources required for comprehensive campaigns, the following sources are prioritized: 

\begin{itemize}

    \item \textbf{Galactic Core-Collapse Supernovae (MeV Neutrinos)}: For low-energy neutrinos from a Galactic core-collapse SNe, the sole triggering criterion for a community response is a confirmed alert from the global neutrino network coordinated through the SNEWS. A coincident detection of a burst of MeV neutrinos across multiple detectors constitutes an unambiguous signature of core collapse and warrants immediate, all-hands follow-up. Individual detectors provide little or no directional information, but inter-experiment triangulation can constrain the source to regions of a few square degrees, sufficient to guide rapid electromagnetic searches. The expected event rate is extremely low, approximately one to three Galactic core-collapse supernovae per century.
    
    \item \textbf{High-Energy Neutrino Transients (TeV–PeV Neutrinos)}: For high-energy neutrinos, triggering criteria depend not only on the characteristics of the neutrino detection~\citep[see][for details of the IceCube realtime program]{IceCube:2016cqr}, but also on the identification of spatially and temporally coincident emission in other wavelengths, especially gamma-ray and X-ray bands. This plan prioritizes alerts from IceCube Neutrino Observatory that meet thresholds in:

\end{itemize}

\subsubsection{Triggering Criteria \& Expected Event Rates}

Because both the intrinsic event rates and detection characteristics vary significantly across source subclasses, from extremely rare Galactic core-collapse supernovae to more frequent but lower confidence high-energy neutrino alerts, triggering criteria must be tailored to each class to balance scientific return against the efficient use of limited follow-up resources.

\begin{itemize}

    \item \textbf{Galactic Core-Collapse Supernovae (MeV Neutrinos)}: For low-energy neutrinos from a Galactic core-collapse SNe, the sole triggering criterion for a community response is a confirmed alert from the global neutrino network coordinated through the SNEWS. A coincident detection of a burst of MeV neutrinos across multiple detectors constitutes an unambiguous signature of core collapse and warrants immediate, all-hands follow-up. Individual detectors provide little or no directional information, but inter-experiment triangulation can constrain the source to regions of a few square degrees, sufficient to guide rapid electromagnetic searches. The expected event rate is extremely low, approximately one to three Galactic core-collapse supernovae per century.
    
    \item \textbf{High-Energy Neutrino Transients (TeV–PeV Neutrinos)}: For high-energy neutrinos, triggering criteria depend not only on the characteristics of the neutrino detection, but also on the identification of spatially and temporally coincident emission in other wavelengths, especially gamma-ray and X-ray bands. This plan prioritizes alerts from IceCube Neutrino Observatory that meet thresholds in:

    \begin{itemize}
        
        \item \textbf{Astrophysical Significance}: Astrophysical probability (signalness/$p_{\rm astro}$) typically favoring events with a high likelihood of non-atmospheric origin.
        
        \item \textbf{Localization Accuracy}: Preference given to well-localized events (of order 1 square degree) that are tractable for wide-field follow-up.
        
        \item \textbf{Coincidence Probability}: Temporal or spatial coincidence with plausible electromagnetic counterparts, identified through wide-field surveys and/or monitors (e.g., ZTF, Swift-BAT, Fermi-LAT, Einstein Probe) or rapid Target-of-Opportunity observations.
        
    \end{itemize}

\end{itemize}

IceCube currently releases public high-energy muon neutrino track alerts at a rate of $\sim$30 per year with an astrophysical probability ($p_{\rm astro}>30$\%), and $\sim$10 per year for a cleaner sample with $p_{\rm astro} >50$\%~\citep{IceCube:2016cqr}. IceCube performs automatic coincidence checks against existing high-energy catalogs and monitoring data at the time an alert is issued. In particular, known gamma-ray sources from GeV surveys are cross-matched against the neutrino localization region, and this contextual information is included directly in alert notices. Archival coincidence programs also work in the opposite direction, in which historical neutrino data are analyzed in conjunction with electromagnetic transients discovered independently at other wavelengths. These efforts are complemented by community-led searches, in which wide-field surveys and follow-up facilities independently identify candidate electromagnetic counterparts within neutrino localization regions and report potential associations.

The rate of such coincident detections has historically been low, with only one flaring AGN source, TXS 0506+056 \citep{2018Sci...361.1378I}, having been identified in real-time, producing an empirical rate of 1 event per decade. This rate is expected to increase, though, with new data streams and new instruments such as KM3NeT come online.

\begin{table}[h]
    \centering
    \caption{ Summary of High-Priority Subclasses and Expected Rates}
    \label{tab:nusubclass}
    \begin{tabular}{|>{\raggedright\arraybackslash}p{3cm}|>{\raggedright\arraybackslash}p{4cm}|>{\raggedright\arraybackslash}p{4cm}|>{\raggedright\arraybackslash}p{5cm}|}
        \hline
        \textbf{Subclass} & \textbf{Scientific Rationale} & \textbf{Characteristics} & \textbf{Expected Rate}\\
        \hline
        \textbf{Galactic Core-Collapse Supernovae} & Direct probe of neutron-star formation, neutrino flavor physics, supernova explosion mechanisms, and potential exotic physics. & Brief ($\sim$10 s), intense burst of MeV neutrinos emitted at core collapse, preceding electromagnetic emission by hours to days. & Extremely rare: $\sim$1–3 Galactic events per century \\
        \hline
        \textbf{High-Energy Neutrino Transients} & Key to identifying sites of hadronic particle acceleration and resolving the origin of cosmic rays.  & Emission may be temporally coincident with electromagnetic activity (gamma-ray, X-ray, optical, or radio. &  IceCube public alerts: $\sim$10 per year with ($p_{\rm astro}>50$\%). Confirmed real-time EM associations historically rare ($\sim$1 per decade) \\
        \hline
    \end{tabular}
\end{table}

\subsubsection{Discovery and Alert Channels}

The two preferred alert channels for notifying electromagnetic observatories about neutrino events are the SNEWS and the GCN. Each serves a specific purpose:

\begin{itemize}
    
\item \textbf{SNEWS} is the primary, dedicated channel for disseminating alerts related to MeV neutrino bursts from Galactic Supernovae. SNEWS aggregates real-time data streams from participating neutrino experiments and issues alerts when statistically significant coincidences are identified. Current and planned upgrades support public alerts, sub-threshold notifications, and inter-experiment triangulation, enabling coarse localization on the sky. These alerts are disseminated rapidly to enable electromagnetic and gravitational-wave facilities to begin immediate searches for shock breakout and early emission, often hours to days before the first optical light would otherwise be detected. 

\item \textbf{GCN} is the established framework used to distribute alerts for high-energy (TeV-PeV) neutrino events, such as those from the IceCube observatory\footnote{\url{https://gcn.nasa.gov/missions/icecube}}. As part of the alert process, IceCube performs automated cross-checks against catalogs of known high-energy sources, allowing contextual information, such as the presence of a known gamma-ray emitter within the neutrino error region, to be included in the initial alert stream. Due to the range of timescales associated with the sources that could potentially produce neutrinos, there is currently no single joint statistics that can be used to assess the significance of coincident electromagnetic detection, so this final assessment inherently requires a human-in-the-loop to make that assessment. 

\end{itemize}

\subsubsection{Observatory Roles and Capabilities}

The roles and unique capabilities of individual observatories in neutrino follow-up depend strongly on the nature of the astrophysical source responsible for the neutrino emission. This dependence is particularly pronounced for TeV–PeV neutrinos, where a wide range of Galactic and extragalactic transients~\citep[see][for a recent review]{Halzen:2022pez, Arguelles:2024ncf}, including gamma-ray bursts~\citep{Kimura:2022zyg}, tidal disruption events~\citep{Hayasaki:2021jem}, interacting~\citep{,Murase:2010cu, Kheirandish:2022eox, Waxman:2024njn} or choked-jet supernovae~\citep{He:2018lwb,2016PhRvD..93h3003S}, active galactic nucleus flares~\citep{Petropoulou:2022sct}, novae~\citep{Bednarek:2022vey,Metzger:2015zka}, and X-ray binaries~\citep{Kantzas:2023oww}, may produce spatially and temporally coincident emission. Each of these source classes requires a distinct observing strategy, cadence, and wavelength emphasis, which are addressed in the corresponding source-specific community observing plans elsewhere in this document.

Accordingly, the observatory capability table in this section largely focuses on the Galactic core-collapse supernova case, where the roles of space- and ground-based facilities are relatively well defined and where coordinated, multiwavelength observations are universally relevant following a MeV neutrino trigger.

\begin{table}[h]
    \centering
    \caption{Summary of Observatory Roles and Capabilities for EM follow-up of neutrinos.}
    \label{tab:nusubclass}
    \begin{tabular}{|>{\raggedright\arraybackslash}p{3cm}|>{\raggedright\arraybackslash}p{4cm}|>{\raggedright\arraybackslash}p{4cm}|>{\raggedright\arraybackslash}p{5cm}|}
        \hline
        \textbf{Observatory} & \textbf{Wavelength} & \textbf{Unique Capability} &
        \textbf{Ideal Use Case}\\
        \hline
        \textbf{HST} & UV/Optical & High-resolution imaging & Late-time optical follow-up of a Galactic Supernova.\\
        \hline
        \textbf{JWST} & NIR/MIR  & Spectroscopy and observation of dust-obscured supernovae. &  Late-time follow-up of a Galactic Supernova. \\
        \hline
        \textbf{Swift} & UV/X-ray & Rapid ToO response. &Identify and monitor counterparts. For MeV events, can search for the shock breakout if the neutrino localization is sufficiently precise. \\
        \hline
        \textbf{Rubin} & Optical (Wide-field) & ToO follow-up for high-energy alerts and deep survey capability. & Identify and monitor counterpart candidates and provide photometric classification. For MeV events, can conduct deeper searches for dust-obscured supernovae that are missed by smaller, initial surveys.\\
        \hline
        \textbf{VLA} & Radio & Studies of jets and ejecta. & Follow-up of identified sources to study late-time non-thermal emission from supernovae and TDEs, or a potential brightening of the radio core.\\
        \hline
        \textbf{ULTRASAT} & UV & ToO response for both high-energy and MeV neutrino alerts. & Supernova shock breakout detection; observation of Tidal Disruption Events.\\
        \hline
        \textbf{VLT} & Optical/NIR & Spectroscopic follow-up. & Classification of sources (e.g., determining supernova type).\\
        \hline
        \textbf{Fermi-LAT} and ground-based Gamma-ray Telescopes & Gamma ray &  Non-thermal EM transient & Identification of sources (e.g., confirming association of neutrino with astrophysical object).\\
        \hline
    \end{tabular}
\end{table}

\subsubsection{Observing Strategy and Timeline}

As with the previous section, the specific observing strategies depend heavily on the nature of the astrophysical source responsible for the neutrino emission. 

\bigskip
\textbf{Strategy for Cosmic Accelerators (MeV)}

\begin{itemize}
    \item \textbf{T0 + 0–24 hours}: The immediate priority is to trigger rapid-response observations with wide-field instruments across multiple wavelengths (infrared, optical, UV, and X-ray) to detect the fleeting shock breakout and precisely localize the source. Infrared instruments, such as WINTER and PRIME, are especially critical for this search, as the supernova may be heavily obscured by Galactic dust.
    
    \item \textbf{T0 + 24 hours to months}: Once the source is identified and localized, the campaign transitions to a long-term monitoring phase. This involves obtaining regular observations across all available wavelengths to track the spectral evolution of the supernova, as well as the evolution of its ejecta and its interaction with the surrounding environment.  
\end{itemize}

\textbf{Strategy for Cosmic Accelerators (TeV-PeV)}

\begin{itemize}
    \item \textbf{T0 + 0–24 hours}: The initial response requires rapid triggers of wide-field-of-view instruments (ZTF, Swift-BAT, Swift-XRT, MAGIC) to survey the multi-square-degree uncertainty region. The goal of this phase is to identify potential counterparts and perform photometric classification of candidates.
    
    \item \textbf{T0 + 24 hours to 7 days}: Following the wide-field search, the focus shifts to the most promising candidates identified. This phase is dedicated to obtaining spectroscopic observations to classify these candidates and determine their nature.
    
    \item \textbf{T0 + 1–14 days}: For high-interest candidates, a multi-epoch monitoring program should be initiated. This multi-epoch, multiwavelength coverage is critical for constructing the SED, a key diagnostic tool for testing leptohadronic emission models as discussed in the case of TXS 0506+056.
    
    \item \textbf{T0 + weeks to months}: For confirmed and scientifically compelling counterparts, a long-term strategy is employed to monitor the evolution of the source's ejecta and its surrounding environment, which can provide crucial insights into the underlying physics.
\end{itemize}

\subsubsection{Lessons Learned from Past Events}

Current observing strategies are the direct result of decades of experience, refined by the successes and shortcomings of past multi-messenger campaigns.

\subsubsubsection*{Case Study: SN198A (MeV Event)}

Supernova 1987A stands as the only supernova ever observed in neutrinos, making it the foundational case study for MeV neutrino astronomy \citep{1987PhRvL..58.1490H,1987PhRvL..58.1494B}. The neutrino burst was detected by several instruments and arrived approximately two hours before the first light from the explosion was observed.

\begin{itemize}
    \item \textbf{What Worked Well}: Despite the technological limitations of the era, the observation of neutrinos from SN1987A was a landmark scientific achievement. The data led to the crucial conclusion that neutrinos are the main driver of the actual supernova explosion and also allowed for the first estimation of the total energy release of the explosion.
    \item \textbf{What Could Be Improved}: The response to SN1987A also highlighted major operational and technological gaps that have since been addressed. At the time, there was no real-time neutrino detection, no network of observatories, and no automated trigger system. The neutrinos were only discovered in archival data on tapes after the optical discovery of the supernova was announced. Because there was no early warning from the neutrino signal, the critical shock breakout phase of the supernova was not observed. These limitations have been directly addressed by the development of the SNEWS and the establishment of formal data-sharing agreements among the world's neutrino observatories, ensuring that the next Galactic Supernova will trigger an immediate, coordinated global response.
\end{itemize}

\subsubsubsection*{Case Study: TXS 0506+056 and IceCube-170922A (TeV-PeV Event)}

The 2017 detection of the high-energy neutrino IceCube-170922A in coincidence with a flare from the blazar TXS 0506+056 was a pivotal moment in multi-messenger astronomy \citep{2018Sci...361.1378I}.

\begin{itemize}

\item \textbf{What Worked Well}: The key to this event was the quick dissemination of the neutrino coordinates~\citep{2017GCN.21916....1K} and the prompt detection of a flaring gamma-ray source by the Fermi-LAT satellite within the neutrino's localization region~\citep{2017ATel10791....1T}. This triggered a large, multiwavelength observing campaign~\citep[e.g.,][]{2017ATel10845....1F}. The inclusion of X-ray data from this campaign proved to be crucial for successfully modeling the source's spectral energy distribution \citep[see e.g.][]{2018ApJ...864...84K,Gao:2018mnu}.

\item \textbf{What Could Be Improved}: The primary shortcoming was a multi-day delay in the official announcement (an Astronomer's Telegram, or ATel) from the Fermi team about the flare. This process has since been improved; checks against known Fermi sources are now automatically included in the initial IceCube GCN alerts to accelerate identification.

\end{itemize}

\subsubsubsection*{Case Study: SN 2023uqf and IceCube-231004A (TeV-PeV Event)}

SN 2023uqf demonstrates a successful strategy to search for counterparts to high-energy neutrino events through follow-up observations.

\begin{itemize}

\item \textbf{What Worked Well}: Following the alert for neutrino IceCube-231004A, the ZTF was used to survey the 4.3 square-degree neutrino footprint~\citep{Stein:2022rvc}. This systematic search successfully identified a supernova candidate, SN 2023uqf~\citep{Stein:2025uxi}. The discovery was followed by rapid spectroscopic observations with the Nordic Optical Telescope (NOT) and the Keck-I telescope, which confirmed the nature of the counterpart. This event established a new, highly effective template for high-energy neutrino follow-up: rapid, wide-field optical surveying to pinpoint candidates, followed by immediate spectroscopic deployment for definitive classification.

\end{itemize}

\newpage
\section{Workshop Agenda\label{subsec:workshop_details}}
\url{https://s3df.slac.stanford.edu/data/fermi/tdamm/agenda.html}
\centering
\includegraphics[width=1\textwidth, page=1]{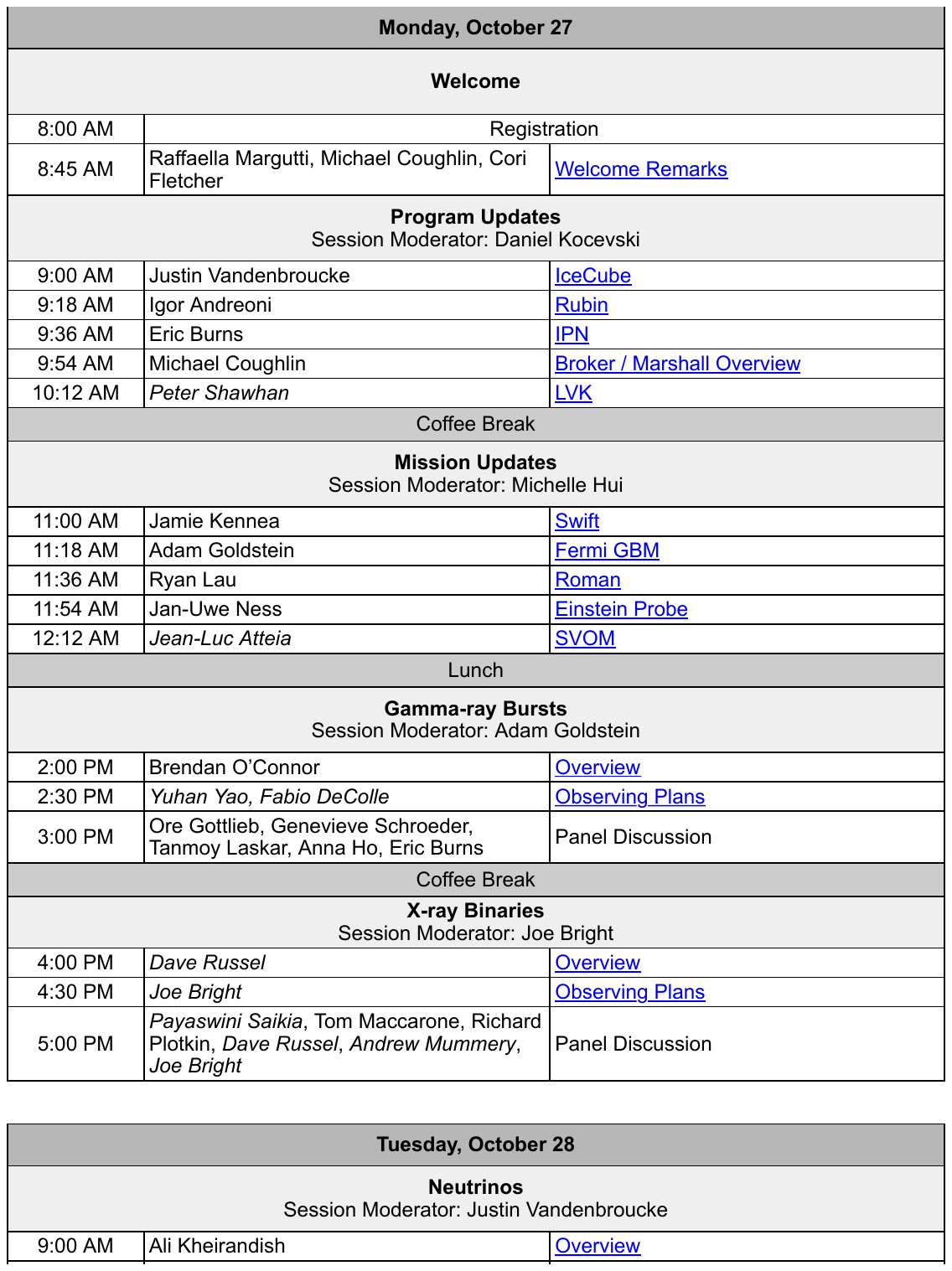}
\newpage
\includegraphics[width=1\textwidth, page=2]{4th_TDAMM_Workshop_Schedule.pdf}
\newpage
\includegraphics[width=1\textwidth, page=3]{4th_TDAMM_Workshop_Schedule.pdf}
\newpage
\includegraphics[width=1\textwidth, page=4]{4th_TDAMM_Workshop_Schedule.pdf}

\newpage
\section{Acronym List}

\begin{longtable}{ll}
\hline
\textbf{Acronym} & \textbf{Definition} \\
\hline
\endfirsthead
\hline
\textbf{Acronym} & \textbf{Definition} \\
\hline
\endhead
AAS & American Astronomical Society \\
ACROSS & Astrophysics Cross-Observatory Science Support \\
AEON & Astronomical Event Observatory Network \\
AGN & Active Galactic Nuclei \\
ALMA & Atacama Large Millimeter/submillimeter Array \\
AMXP & Accreting Millisecond X-ray Pulsar \\
API & Application Programming Interface \\
ATCA & Australia Telescope Compact Array \\
ATel & Astronomer's Telegram \\
ATLAS & Asteroid Terrestrial-impact Last Alert System \\
AXIS & Advanced X-ray Imaging Satellite \\
BAL & Broad Absorption Line \\
BAT & Burst Alert Telescope \\
BBH & Binary Black Hole \\
BH & Black Hole \\
BH--NS & Black Hole--Neutron Star \\
BNS & Binary Neutron Star \\
BOOM & Burst and Outburst Observations Monitor \\
CBM & Compact Binary Merger \\
CCSN & Core-Collapse Supernova \\
CHIME & Canadian Hydrogen Intensity Mapping Experiment \\
CHORD & Canadian Hydrogen Observatory and Radio-transient Detector\\
CMB & Cosmic Microwave Background \\
COSI & Compton Spectrometer and Imager \\
COTS & Commercial Off-the-Shelf \\
CSM & Circumstellar Material \\
CTA/CTAO & Cherenkov Telescope Array / Cherenkov Telescope Array Observatory \\
DDT & Director's Discretionary Time \\
DIM & Disk Instability Model \\
DOI & Digital Object Identifier \\
DSA & Deep Synoptic Array \\
EM & Electromagnetic \\
EMRI & Extreme Mass-Ratio Inspiral \\
EP & Einstein Probe \\
FBOT & Fast Blue Optical Transient \\
FRA & Fast Response Analysis \\
FRB & Fast Radio Burst \\
FWHM & Full Width at Half Maximum \\
FXT & Fast X-ray Transient; also Follow-up X-ray Telescope in the Einstein Probe context \\
GBM & Gamma-ray Burst Monitor \\
GCN & General Coordinates Network; historically Gamma-ray Coordinates Network \\
GeV & Giga-electronvolt \\
GI & Guest Investigator \\
GR & General Relativity \\
GRB & Gamma-ray Burst \\
GRM & Gamma-Ray Monitor \\
GW & Gravitational Wave \\
HAWC & High-Altitude Water Cherenkov Observatory \\
HEROIC & Hopskotch Enabled Real-Time Observatory Information and Coordination \\
HID & Hardness-Intensity Diagram \\
HLTDS & High Latitude Time-Domain Survey \\
HMXB & High-Mass X-ray Binary \\
HMXRB & High-Mass X-ray Binary \\
HST & Hubble Space Telescope \\
IGWN & International Gravitational-Wave Network \\
IMBH & Intermediate-Mass Black Hole \\
IMBH--WD & Intermediate-Mass Black Hole--White Dwarf \\
IPN & InterPlanetary Network \\
IR & Infrared \\
ISCO & Innermost Stable Circular Orbit \\
ISM & Interstellar Medium \\
JWST & James Webb Space Telescope \\
KAGRA & Kamioka Gravitational Wave Detector \\
LAST & Large Array Survey Telescope \\
LAT & Large Area Telescope \\
LCO & Las Cumbres Observatory \\
LEO & Low-Earth Orbit \\
LGRB & Long Gamma-ray Burst \\
LHAASO & Large High Altitude Air Shower Observatory \\
LIGO & Laser Interferometer Gravitational-Wave Observatory \\
LMXB & Low-Mass X-ray Binary \\
LMXRB & Low-Mass X-ray Binary \\
LSST & Legacy Survey of Space and Time \\
LVK & LIGO--Virgo--KAGRA \\
MAGIC & Major Atmospheric Gamma Imaging Cherenkov Telescopes \\
MAXI & Monitor of All-sky X-ray Image \\
MBH & Massive Black Hole \\
MeV & Mega-electronvolt \\
MGF & Magnetar Giant Flare \\
MIR & Mid-Infrared \\
MMX-MEGANE & Mars-moon Exploration with GAmma rays and NEutrons \\
MWA & Murchison Widefield Array \\
MXT & Micro-channel X-ray Telescope \\
NASA & National Aeronautics and Space Administration \\
NED & NASA/IPAC Extragalactic Database \\
NED-GWF & NASA/IPAC Extragalactic Database Gravitational-Wave Follow-up \\
NED-LVS & NASA/IPAC Extragalactic Database Local Volume Sample \\
NGC & New General Catalogue \\
NICER & Neutron star Interior Composition Explorer \\
NIR & Near-Infrared \\
NOIRLab & NSF's National Optical-Infrared Astronomy Research Laboratory \\
NOT & Nordic Optical Telescope \\
NS & Neutron Star \\
NS--NS & Neutron Star--Neutron Star \\
NSBH & Neutron Star--Black Hole \\
NSF & National Science Foundation \\
NUV & Near-Ultraviolet \\
PeV & Peta-electronvolt \\
QPE & Quasi-Periodic Eruption \\
QPO & Quasi-Periodic Oscillation \\
SCiMMA & Scalable Cyberinfrastructure for Multi-Messenger Astrophysics \\
SED & Spectral Energy Distribution \\
SFXT & Superfast X-ray Transient \\
SGR & Soft Gamma Repeater \\
SKA & Square Kilometre Array \\
SMA & Submillimeter Array \\
SMBH & Supermassive Black Hole \\
SMC & Small Magellanic Cloud \\
SN & Supernova \\
SNe & Supernovae \\
SNEWS & SuperNova Early Warning System \\
SNR & Supernova Remnant \\
SOC & Science Organizing Committee \\
SSS & Super-Soft Source \\
SVOM & Space Variable Objects Monitor \\
TAC & Time Allocation Committee \\
TDAMM & Time-Domain and Multi-Messenger Astrophysics \\
TDE & Tidal Disruption Event \\
TDRSS & Tracking and Data Relay Satellite System \\
TeV & Tera-electronvolt \\
tMSP & Transitional Millisecond Pulsar \\
TNR & Thermonuclear Runaway \\
TNS & Transient Name Server \\
TOM & Target and Observation Manager \\
ToO & Target of Opportunity \\
TROVE & Tool for Rapid Object Vetting and Examination \\
UCXB & Ultracompact X-ray Binary \\
ULGRB & Ultra-long Gamma-ray Burst \\
ULTRASAT & Ultraviolet Transient Astronomy Satellite \\
ULX & Ultraluminous X-ray Source \\
USRA & Universities Space Research Association \\
UV & Ultraviolet \\
UVEX & Ultraviolet Explorer \\
UVOT & UV/Optical Telescope\\
VFXT & Very Faint X-ray Transient \\
VHE & Very High Energy \\
VLA & Very Large Array \\
VLBI & Very Long Baseline Interferometry \\
VLT & Very Large Telescope \\
VT & Visible Telescope \\
WD & White Dwarf \\
WFD & Wide-Fast-Deep \\
WFI & Wide Field Instrument \\
WXT & Wide-field X-ray Telescope \\
XMM & X-ray Multi-Mirror Mission \\
XRB & X-ray Binary \\
XRF & X-ray Flash \\
XRT & X-Ray Telescope \\
ZTF & Zwicky Transient Facility \\
\hline
\end{longtable}

\end{appendices}



\newpage
\bibliography{main}{}
\bibliographystyle{aasjournalv7}

\end{document}